\documentclass[a4paper,10pt,twoside]{book}

\usepackage{amsmath}
\usepackage{graphicx}
\usepackage{amssymb}
\usepackage[numbers,square]{natbib}

\newlength\encuadernacion \newlength\longB 
\def\be{\begin{equation}}
\def\ee{\end{equation}}
\def\bea{\begin{eqnarray}}
\def\eea{\end{eqnarray}}

\def\nn{\nonumber \\}
\def\e{{\rm e}}
\def\tR{\tilde{R}}
\def\bx{\textbf{x}}
\def\mG{\mathcal{G}}
\def\mV{\mathcal{V}}
\def\hN{\bar{N}}
\def\hK{\bar{K}}
\newcommand{\mA}{\mathcal{A}}
\def\hnabla{\hat{\nabla}}

\title{\begin{huge}{\bf On Friedmann-Lema\^{\i}tre-Robertson-Walker cosmologies in non-standard gravity}\end{huge}}
\author{
\\
{\begin{large}\bf Diego S\'aez G\'omez\end{large}}\\
\\
PhD thesis
}
\date{under the supervision of
\\
\vskip0.65cm
Dr.\ Emilio Elizalde Rius and Dr.\ Sergei D. Odintsov\\
\vskip3cm
\begin{figure}[h]
\begin{center}
\begin{tabular}{cc}
\includegraphics[scale=0.7]{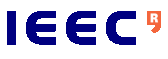} & \hspace{5 cm}
\includegraphics[scale=0.7]{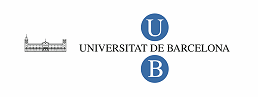} 
\end{tabular}
\end{center}
\end{figure}
\begin{figure}[h]
 \centering
\includegraphics[scale=0.4]{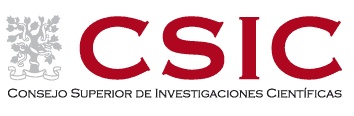} 
\end{figure}
\vskip 0.6 cm Barcelona\\
2011
}
\begin{document}

\frontmatter
\maketitle
\newpage
\
\newpage

\newpage
\begin{flushright}
\ \\ \par
\vspace{7cm}
{\it A mis padres, Dionisio y Teresa}\\
\vskip 0.2cm
\end{flushright}
\
\newpage
\
\newpage
{\huge 
Resumen} 
\vskip1cm
La presente memoria de tesis tiene como objeto el estudio de soluciones cosmol\'ogicas descritas por una m\'etrica FLRW (en general tomada espacialmente plana) en el contexto, fundamentalmente, de extensiones de la Relatividad General, pero tambi\'en en presencia de algunos de los candidatos m\'as populares de la energ\'ia oscura. \\
Pr\'acticamente desde los tiempos de Cop\'ernico, la noci\'on fundamental sobre nuestro Universo es la ausencia de una posici\'on privilegiada de cualquier observador, lo que traducido en t\'erminos modernos da lugar al llamado {\it Principio Cosmol\'ogico}, que asume el hecho de que nos encontremos en un Universo homog\'eneo e is\'otropo, lo que traducido en palabras significa que la m\'etrica es la misma en todo punto del espacio y que \'este parece el mismo sea cual sea la direcci\'on desde la que se observe. Esta aseveraci\'on, que puede parecer pretenciosa  si tenemos en cuenta que el centro del Sol es muy diferente al medio intergal\'actico, podr\'ia tener validez a escalas cosmol\'ogicas, es decir, mucho m\'as all\'a de ni siquiera los c\'umulos de galaxias. A pesar de que hoy en d\'ia hay sectores de la comunidad cient\'ifica que cuestionan  dicha validez, las observaciones de las galaxias lejanas y especialmente la gran isotrop\'ia del CMB, siguen mostr\'andose favorables a mantener el principio. El modelo de descripci\'on del Universo, el llamado modelo del Big Bang, que tiene en dicho principio su base fundamental, ha sabido predecir con bastante acierto algunos de los aspectos m\'as fundamentales de la evoluci\'on cosmol\'ogica, como puede ser la abundancia de los elementos. Dicho modelo predice un Universo en expansi\'on que comenz\'o tras una gran explosi\'on. Sin embargo, los detractores de dicho modelo cuestionan ciertos aspectos de la teor\'ia, pues el modelo a su vez no da respuesta a problemas fundamentales como la {\it planitud} o el problema del horizonte (bien contrastados en las observaciones), para los cuales es necesario ``parchear'' la teor\'ia con nuevos mecanismos (inflaci\'on en este caso), que a su vez introducen  nuevos problemas. A pesar de todo,  el modelo del Big Bang sigue siendo hoy la mejor descripci\'on  del Universo que poseemos, y los problemas que la teor\'ia pueda contener, pueden ser vistos como la incompletitud de nuestras teor\'ias, y podr\'ian servir para descubrir  nueva f\'isica que pudiera dar lugar a futuras predicciones.\\ 
En el a\~{n}o 1998, a trav\'es de  observaciones de Supernovas IA, se lleg\'o a la conclusi\'on de que el Universo no solo  estaba en expansi\'on como hab\'ia demostrado Hubble setenta a\~{n}os atr\'as, sino que adem\'as dicha expansi\'on se estaba acelerando. Mucho antes de que \'este descubrimiento tuviera lugar otro problema diferente sacud\'ia el mundo de la f\'isica te\'orica, el llamado problema de la {\it constante cosmol\'ogica}. La cuesti\'on aqu\'i se deb\'ia a si la contribuci\'on de la densidad de energ\'ia de vac\'io que predec\'ia la f\'isica cu\'antica deb\'ia tomarse en consideraci\'on en las ecuaciones de Einstein a la hora de estudiar un determinado sistema como el cosmol\'ogico. Dicha contribuci\'on aparec\'ia en las ecuaciones en la forma precisamente de una constante, que a nivel cosmol\'ogico pod\'ia producir una expansi\'on acelerada. Esto podr\'ia parecer el final feliz de un problema abierto, la aceleraci\'on del Universo no ser\'ia sino el resultado de la presencia de una constante cosmol\'ogica, cuyo origen estar\'ia en el vac\'io cu\'antico. Sin embargo, lejos de resolver el problema, quiz\'as lo agrav\'o, pues la diferencia entre la densidad de vac\'io predicha por las teor\'ias cu\'anticas, y aqu\'ella medida en las observaciones es de 122 \'ordenes de magnitud!\\ Esto di\'o lugar a que en la \'ultima d\'ecada se hayan propuesto multitud de posibles candidatos para explicar dicha aceleraci\'on del Universo, todos ellos englobados bajo el nombre de eneg\'ia oscura, siendo el m\'as popular de estos modelos el llamado modelo $\Lambda$CDM, que sigue considerando la constante cosmol\'ogica como \'unica responsable. Sin embargo, la constante cosmol\'ogica, adem\'as de los problemas ya expuestos, posee una ecuaci\'on de estado constante $p_{\Lambda}=-\rho_{\Lambda}$, mientras que las observaciones apuntan cada vez m\'as a un car\'acter din\'amico de dicha ecuaci\'on de la energ\'ia oscura. Esta \'epoca de expansi\'on acelerada comenzar\'ia cuando la densidad de materia bari\'onica/oscura descendiese, y la de la energ\'ia ocura pasar\'ia entonces a dominar.\\
Sin embargo, adem\'as de esta \'epoca de aceleraci\'on, es necesario considerar la inflaci\'on tambi\'en, la cual supone otra fase de aceleraci\'on al comienzo del Universo, y cuyo principal candidato es el campo escalar conocido como inflat\'on. Todo ello hace pensar de forma natural que ambas fases pudieran estar relacionadas de alg\'un modo, y que ambas hayan sido el producto de un mismo mecanismo. \\

En este contexto de cosmolog\'ia actual, la presente tesis tiene como objetivo fundamental la reconstrucci\'on de modelos cosmol\'ogicos que a trav\'es de un mismo  mecanismo puedan explicar tanto la inflaci\'on como la aceleraci\'on actual del Universo, y dar una respuesta fehaciente la evoluci\'on del Universo. Para ello, a partir de soluciones FLRW y haciendo uso de las ecuaciones de campo, se reconstruir\'a la acci\'on precisa que reproduce una soluci\'on determinada. En este sentido se investigar\'a en primer lugar campos escalares del tipo ``quintessence/phantom'', los cuales ya han sido considerados como candidatos para energ\'ia oscura, pero que aqu\'i tambi\'en se considerar\'a la posibilidad de que puedan servir para explicar inflaci\'on. Varios ejemplos de esta reconstrucci\'on se muestran en el cap\'itulo dos. Adem\'as, aqu\'i tambi\'en se considerar\'a la posibilidad de utilizar varios escalares de este tipo para poder producir una transici\'on suave a una fase de ``phantom'' (donde la ecuaci\'on de estado ser\'ia menor que -1), pues dicha transici\'on se ha demostrado inestable cuando hay solo un campo escalar cuyo t\'ermino cin\'etico se hace cero en alg\'un punto. Por otro lado, se explorar\'an teor\'ias del tipo Brans-Dicke, donde la gravedad viene definida por el campo m\'etrico y por un campo escalar fuertemente acoplado, estas teor\'ias tienen un gran inter\'es, pues aparte del hecho de que puedan explicar la energ\'ia oscura, son equivalentes a las llamadas teor\'ias  de gravedad modificada $f(R)$, lo cual supone una buena herramienta para estudiar \'estas \'ultimas tal y como se explorar\'a en el cap\'itulo cuatro. Adem\'as, se estudiar\'a la relaci\'on, siempre discutida, entre estas teor\'ias escalar-tensor definidas en el marco de Jordan y el de Einstein, donde interesantes resultados ser\'an obtenidos. Se explora tambi\'en la posibilidad de un Universo con una evoluci\'on oscilatoria, donde el llamado problema de la coincidencia (el hecho de que las densidades de materia y energ\'ia oscura sean pr\'acticamente iguales en el presente, y no antes ni despu\'es) puede ser f\'acilmente rebatido, puesto que el Universo se mover\'ia por fases c\'iclicas.\\ \\
Al margen de esto, el principal objetivo de la memoria es el estudio de teor\'ias de gravedad modificada, especialmente las llamadas teor\'ias $f(R)$ y gravedad de Gauss-Bonnet, como respuestas a la energ\'ia oscura conjuntamente con inflaci\'on. Dado que los principios que sustentan la Relatividad General, el de equivalencia, covariancia y relatividad,  no hacen una gran restricci\'on a la forma que han de tener las ecuaciones de campo, la elecci\'on de \'estas puede quedar supeditada a las observaciones y experimentos. Ya en su momento Einstein eligi\'o sus ecuaciones por el hecho de que, aparte de la conservaci\'on de la energ\'ia, en el l\'imite no relativista se recuperaba la expresi\'on de Newton. Del mismo modo, uno puede pensar que la energ\'ia oscura y la inflaci\'on no son m\'as que malas definiciones de nuestras ecuaciones que, bajo ciertas correcciones, pueden ser explicadas. En este sentido se mostrar\'an diversas reconstrucciones de teor\'ias que puedan explicar las m\'etricas de FLRW que se propongan. Especialmente se har\'a \'enfasis en la reconstrucci\'on del modelo de $\Lambda$CDM  sin constante cosmol\'ogica pero con t\'erminos geom\'etricos adicionales, as\'i como de modelos donde exista una transici\'on ``phantom''. Adem\'as, este tipo de teor\'ias podr\'ian dar una explicaci\'on certera al problema de la constante cosmol\'ogica, pues la energ\'ia de vac\'io es contrarrestada por los t\'erminos geom\'etricos, dando lugar al valor observado. Tambi\'en se estudiar\'an modelos donde fluidos adicionales son considerados conjuntamente con las correcciones de las ecuaciones para explicar los comportamientos de la evoluci\'on del Universo. En este contexto se explorar\'a tambi\'en la relaci\'on entre los marcos de Jordan y el de Einstein. \\
Sin embargo, hay que ser cuidadoso a la hora de modificar dichas ecuaciones, pues tales correcciones pueden dar resultados exitosos a escalas cosmol\'ogicas pero catastr\'oficas violaciones a escalas locales como la Tierra o el Sistema Solar. En este sentido se estudiar\'an los llamados modelos ``viables o realistas'', donde las modificaciones de la Relatividad General solo son importantes a grandes escalas, y la ley de Einstein se recupera a escalas menores, un poco en la misma l\'inea que el mecanismo de ``camale\'on'' en las teor\'ias escalar-tensor.  Por tanto, se estudiar\'a la reconstrucci\'on de soluciones cosmol\'ogicas (espaciostiempo FLRW planos esencialemnte) en el marco de teor\'ias $f(R)$ y Gauss-Bonnet, donde se mostrar\'a como, de forma natural, las \'epocas de inflaci\'on y aceleraci\'on actual, pueden ser explicadas en el marco de dichas modificaciones sin necesidad de invocar componentes ex\'oticos. \\
Adem\'as, y dada la reciente propuesta por parte de Ho\v{r}ava de una teor\'ia gravitatoria que parece ser renormalizable, distintos aspectos de \'esta ser\'an estudiados. Dicha teor\'ia, conocida ya como gravedad de Ho\v{r}ava-Lifshitz, asume una anisotrop\'ia en la coordenada temporal respecto a las espaciales, de manera similar a las anisotrop\'ias utilizadas  en las teor\'ias de Lifshitz en materia condensada, lo que supone la renormalizaci\'on de la teor\'ia pero a la vez la ruptura de la covariancia, dando lugar a violaciones de los principios que sustentan (y que est\'an bien comprobados por los experimentos a escalas muy por debajo de la escala de Planck) la relatividad de Einstein. Sin embargo, se supone que en el l\'imite infrarojo (IR) la teor\'ia cl\'asica se recupera, siendo el mecanismo para ello no del todo claro, aunque se han realizado intentos para comprenderlo. El \'ultimo de ellos es introducir una nueva simetr\'ia $U(1)$ en el modelo, lo que forzar\'ia en el l\'imite IR a recuperar la Relatividad General, en este sentido en la presente memoria se exploran extensiones de dicho modelo. Otro modo, estudiado en la presente tesis, y que evita las inestabilidades del modo de esp\'in-0 que la teor\'ia de  Ho\v{r}ava produce en el l\'imite IR alrededor de la soluci\'on de vac\'io de la teor\'ia, es considerar una acci\'on, donde ``de Sitter'' ser\'ia estable y considerada como la soluci\'on natural en vac\'io en lugar del espaciotiempo de Mikowski. \\
Al margen de los problemas intr\'insecos de la teor\'ia, la gravedad de Ho\v{r}ava-Lifshitz sufre del mismo problema que la Relatividad General a nivel cosmol\'ogico, esto es que no puede explicar por s\'i sola la aceleraci\'on del Universo, tanto actual como de la \'epoca inflacionaria. De modo que en analog\'ia con las extensiones estudiadas en gravedad $f(R)$, la acci\'on de Ho\v{r}ava-Lifshitz puede extenderse a acciones m\'as generales $f(\tilde{R})$ para reconstruir soluciones cosmol\'ogicas que reproduzcan los efectos de la energ\'ia ocura, y que incluso lo unifiquen con inflaci\'on. En este contexto, varios modelos y ejemplos son estudiados. Adem\'as se explorar\'a las correcciones Newtonianas debidas al modo escalar introducido por la funci\'on $f(\tilde{R})$ (que no por la anisotrop\'ia temporal), y la reconstrucci\'on de una acci\'on libre de singularidades cosmol\'ogicas.\\

Por \'ultimo, respecto a las singularidades, los dos \'ultimos cap\'itulos de esta memoria estar\'an dedicados al estudio de posibles efectos semicl\'asicos alrededor de ellas. Las singularidades son bastante comunes en soluciones de la Relatividad General, as\'i como en sus extensiones, y suponen puntos del espaciotiempo donde nuestras teor\'ias se suponen no v\'alidas y es necesario una teor\'ia cu\'antica consistente de la gravedad. Algunos efectos de origen cu\'antico como el efecto Casimir o la anomal\'ia conforme, ser\'an explorados alrededor de singularidades cosmol\'ogicas del tipo ``Big Rip''. Se demostrar\'a que dichos efectos  semicl\'asicos apenas tienen relevancia y que la singularidad no es evitada en la mayor parte de los casos. Adem\'as, se estudiar\'a la validez de los l\'imites  sobre la entrop\'ia del Universo, como aqu\'ellos propuestos por Verlinde, y la validez de la f\'ormula de Cardy-Verlinde, que da una derivaci\'on directa de los l\'imites din\'amicos de la entrop\'ia, y la correspondencia que se establece entre las ecuaciones de FLRW y las de una Teor\'ia de Campos Conforme (CFT) 2d. Todo ello  ser\'a estudiado en contextos m\'as generales.
\newpage
{\huge 
Resum} 
\vskip1cm
La present mem\`oria de tesis t\'e com a objecte l'estudi de solucions cosmol\`ogiques descrites per una m\`etrica FLRW (en general espacialment plana) en el context, fonamentalment, d'extensions de la Relativitat General, per\`o tamb\'e en pres\`encia d'alguns dels candidats m\'es populars de l'energia fosca. \\
Practicament des dels temps de Cop\`ernic, la noci\'o fonamental sobre al nostre Univers \'es  l'abs\`encia d'una posici\'o privilegiada de qualsevol observador, el que en termes moderns d\'ona lloc a l'anomenat {\it Principi Cosmol\`ogic},  que assumeix el fet que ens trobem en un Univers homogeni i is\`otrop, el que tradu\"{\i}t en paraules planeres vol dir que la m\`etrica \'es la mateixa en tot punt de l'espai i que aquest sembla el mateix sigui la que  sigui la direcci\'o des de la qual s'observi. Aquesta asseveraci\'o, que pot semblar pretenciosa si tenim en compte que el centre del Sol \'es molt diferent al medi intergal\`actic, podria tenir validesa a escales cosmol\`ogiques, \'es a dir, molt m\'es enll\`a dels c\'umuls de gal\`axies. Tot i que avui en dia hi ha sectors de la comunitat cient\'ifica que q\"{u}estionen la seva validesa, les observacions de les gal\`axies llunyanes i especialment la gran isotropia del CMB, segueixen mostrant-se favorables a mantenir el principi. El model de descripci\'o de l'Univers, l'anomenat model del Big Bang, que t\'e en aquest principi la seva base fonamental, ha sabut predir amb for\c{c}a encert alguns dels aspectes m\'es fonamentals de l'evoluci\'o cosmol\`ogica, com pot ser l'abund\`ancia dels elements. Aquest model prediu un Univers en expansi\'o que va comen\c{c}ar despr\'es d'una gran explosi\'o. No obstant aix\`o, els detractors d'aquest model q\"{u}estionen certs aspectes de la teoria, ja que el model  no d\'ona resposta a problemes fonamentals com la {\it planitud} o el problema de l'horitz\'o (ben contrastats a les observacions), per als quals cal ``apeda\c{c}ar'' la teoria amb nous mecanismes (inflaci\'o en aquest cas), a la vegada introdueixen nous problemes. Malgrat tot, el model del Big Bang segueix sent avui la millor descripci\'o  que posse\"{i}m de l'Univers, i els problemes que la teoria pugui contenir, poden ser entesos com deguts a la imcompletesa de les nostres teories, i podrien servir per descobrir nova f\'isica que pogu\'es donar lloc a futures prediccions.\\
L'any 1998, a trav\'es d'observacions de Supernoves IA, es va arribar a la conclusi\'o que l'Univers no nom\'es estava en expansi\'o, com havia demostrat Hubble setanta anys abans, sin\'o que a m\'es aquesta expansi\'o s'estava accelerant. Molt abans que aquest descobriment tingu\'es lloc, un altre problema diferent sacsej\`a el m\'on de la f\'isica te\`orica, l'anomenat problema de la constant cosmol\`ogica. La q\"{u}esti\'o era si la contribuci\'o de la densitat d'energia del buit que predeia la f\'isica qu\`antica s'havia de tenir en compte en les equacions d'Einstein a l'hora d'estudiar un determinat sistema com el cosmol\`ogic. Aquesta contribuci\'o apareixia en les equacions en la forma precisament d'una constant, que a nivell cosmol\`ogic podia produir una expansi\'o accelerada. Aix\`o podria semblar el final feli\c{c} d'un problema obert, l'acceleraci\'o de l'Univers no seria sin\'o el resultat de la pres\`encia d'una constant cosmol\`ogica, amb origen  en el buit qu\`antic. No obstant aix\`o, lluny de resoldre el problema, m\'es aviat el va agreujar, ja que la difer\`encia entre la densitat de buit predita per les teories qu\`antiques, i la mesurada per les observacions, \'es de 122 ordres de magnitud! \\ 
Aix\`o va donar lloc a que en l'\'ultima d\`ecada s'hagin proposat multitud de possibles candidats per explicar aquesta acceleraci\'o de l'Univers, tots ells englobats sota el nom d'energia fosca, sent el m\'es popular d'aquests models l'anomenat model $\Lambda$CDM, que segueix considerant la constant cosmol\`ogica com a \'unica responsable de l'acceleraci\'o. No obstant aix\`o, la constant cosmol\`ogica, a m\'es dels problemes ja exposats, t\'e una equaci\'o d'estat constant $p_{\Lambda}=-\rho_{\Lambda}$, mentre que les observacions apunten cada vegada m\'es a un car\`acter din\`amic d'aquesta equaci\'o de l'energia fosca. L'\`epoca d'expansi\'o accelerada comen\c{c}ar\'ia quan la densitat de mat\`eria bari\`onica/fosca baix\'es, i la de l'energia fosca passaria llavors a dominar. \\
Per\`o, a m\'es d'aquesta \`epoca d'acceleraci\'o, cal considerar la inflaci\'o tamb\'e, que suposa una altra fase d'acceleraci\'o al comen\c{c}ament de l'Univers, on el principal candidat \'es el camp escalar conegut com inflat\'o. Tot aix\`o fa pensar de forma natural que les dues fases puguin estar relacionades d'alguna manera, i que ambdues hagin estat el producte d'un mateix mecanisme.\\

En el context de la cosmologia actual, aquesta tesi t\'e com a objectiu fonamental la reconstrucci\'o de models cosmol\`ogics que mitjan\c{c}ant un \'unic mecanisme puguin explicar tant la inflaci\'o com l'acceleraci\'o actual de l'Univers, i donar una resposta fefaent a l'evoluci\'o de l'Univers. Per aix\`o, a partir de solucions FLRW i fent \'us de les equacions de camp, es reconstruir\`a l'acci\'o precisa que reprodueix una soluci\'o determinada. En aquest sentit s'investigar\`a en primer lloc camps escalars del tipus ``quintessence/phantom'', els quals ja han estat considerats com a candidats per a l'energia fosca, per\`o que tamb\'e es considerar\`a la possibilitat que puguin servir per explicar inflaci\'o. A m\'es, aqu\'i tamb\'e es considerar\`a la possibilitat d'utilitzar diversos escalars d'aquest tipus per poder produir una transici\'o suau a una fase de ``phantom'' (on l'equaci\'o d'estat seria menor que -1), ja que aquesta transici\'o s'ha demostrat inestable quan hi ha nom\'es un camp escalar, el terme cin\`etic es fa zero en algun punt. D'altra banda, s'exploren teories del tipus Brans-Dicke, on la gravetat ve definida pel camp m\`etric i per un camp escalar fortament acoblat, aquestes teories tenen un gran inter\`es, perqu\`e a banda del fet que puguin explicar l'energia fosca, s\'on equivalents a les anomenades teories de gravetat modificada $f(R)$, la qual cosa suposa una bona eina per estudiar aquestes \'ultimes tal com s'explorar\`a en el cap\'itol quatre. A m\'es, s'estudiar\`a la relaci\'o, sempre discutida, entre aquestes teories escalar-tensor definides en el marc de Jordan i el d'Einstein, on obtindrem interessants resultats. S'explorar\`a tamb\'e la possibilitat d'un Univers amb una evoluci\'o oscilatoria, on l'anomenat problema de la coincid\`encia (el fet que les densitats de mat\`eria i energia fosca siguin pr\`acticament iguals en el present, i no abans ni despr\'es) pot ser f\`acilment rebatut, ja que l'Univers es mouria per fases c\'icliques.\\ \\

Al marge d'aix\`o el principal objectiu de la mem\`oria \'es l'estudi de les teories de gravetat modificada, especialment les anomenadas teories $f(R)$ i la gravetat de Gauss-Bonnet, com a resposta a la energ\'ia fosca conjuntament amb inflaci\'o. Donat que els principis que sustenten la relativitat general, els d'equival\`encia, covari\`ancia i relativitat, no imposen una gran restricci\`o sobre la forma que han de tenir les equacions de camp, l'elecci\'o d'aquestes pot quedar supeditada a les observacions i experiments. Cal recordar que Einstein va escollir les ecuacions pel fet que, a part de la conservaci\'o de la energia, en el limit no relativista hom recuperava la expressi\'o de Newton. De la mateixa manera, un pot pensar que l'energ\'ia fosca i la inflaci\'o no son m\'es que males definicions  de la teoria que, sota certes correccions, poden ser explicades. En aquest sentit, es mostraran diverses reconstruccions de teories que puguin explicar per si soles les metriques que es proposin. Especialment, es posara \'enfasis en la reconstrucci\'o del model $\Lambda$CDM sense constant cosmol\`ogica pero amb termes geometrics adicionals, aix\'i com de models on existeixi una transici\'o ``phantom''. A m\'es, aquest tipus de teories podrien donar una explicaci\'o precisa al problema de la constant cosmol\`ogica, doncs en aquest cas l'energia de buit es contrarestada pels termes geom\`etrics donant lloc al valor observat. Tamb\'e s'estudiaran models on fluids adicionals son considerats conjuntament amb les correccions de les ecuacions, per explicar els comportaments de l'evoluci\'o de  l'Univers. En aquest context, s'explorar\`a tamb\'e la relaci\'o entre els marcs de Jordan i el d'Einstein.
No obstant, cal ser cur\'os a l'hora de modificar aquestes ecuacions, doncs les dites correccions poden donar resultats exitosos a escales cosmol\`ogiques pero catastr\`ofiques violacions a escales locals com la Terra o el Sistema Solar. En aquest sentit, s'estudiaran els anomenats ``models viables'', on les modificacions de la Relativitat General nom\'es son importants a gran escala i la llei d'Einstein es recupera a escales menors, en la mateixa l\'inia que l'anomenat ``mecanisme de camale\'o''. S'estudiar\`a la reconstruci\'o de solucions cosmol\`ogiques en el marc de teories $f(R)$, on es mostrar\`a com de forma natural les \'epoques  d'inflaci\'o i d'acceleraci\'o actual poden ser explicades sense necessitat d'invocar continguts ex\`otics.\\
A m\'es, i donada l'\'ultima proposta per part de Ho\v{r}ava d'una teoria gravitat\`oria que sembla ser renormalizable, n'estudiarem diferents aspectes. Aquesta teoria anomenada com gravetat de Ho\v{r}ava-Lifshitz, assumeix una anisotrop\'ia en la coordenada temporal respecte a les espacials, de manera similar a les anisotropies utilitzades en les teories de Lifshitz en mat\`eria condensada, lo que suposa la renormalitzaci\'o de la teoria, per\`o a l'hora la ruptura de la covari\`ancia donant lloc a violacions dels principis que sustenten la Relativitat. No obstant, es suposa que en el limit IR, la teoria cl\'asica emergeix. El mecanisme no \'es, per aix\'o,  del tot clar, tot i que s'han realitzat diferents intents d'explicar-lo. L'\`ultim d'aquests \'es introduir una nova simetria en el model, que for\c{c}aria en el limit IR a recuperar la Relativitat General. En aquest sentit, extensions d'aquest model son explorades en la present mem\`oria. Una altra posible soluci\'o que podr\'ia evitar les inestabilitats del mode d'esp\'i-0 que es produeix en el limit IR al voltant de la soluci\'o del buit de la teoria, seria considerar una acci\'o, on "de Sitter" fos estable i que seria considerada com la soluci\'o natural al buit en lloc de Mikowski. \\
Al marge dels problemes intr\'insecs de la teoria,  la gravetat de Ho\v{r}ava pateix del mateix problema que la Relativitat General a nivell cosmol\`ogic, aix\`o \'es que no pot explicar l'acceleraci\'o del Univers, tant l'actual com la de l'epoca inlaccion\`aria. Aix\'i doncs, en analog\'ia amb les extensions estudiades de la Relativitat General, l'acci\'o es pot estendre a $f(\tilde{R})$ per reconstruir solucions cosmol\`ogiques que reprodueixin els efectes de l'energ\'ia fosca i l'unifiquin amb inflaci\'o. Diferents models i exemples son estudiats, a m\'es s'exploraran les correccions newtonianes (nom\'es per al mode escalar de $f(\tilde{R})$, i no per a la correcci\'o de l'anisotropia temporal) i la reconstrucci\'o d'una acci\'o lliure de singularitats cosmol\`ogiques. \\
Per \`ultim, respecte a les singularitats, els dos darrers cap\'itols d'aquesta mem\`oria estan dedicats a  l'estudi de possibles efectes semicl\`asics al voltant d'elles. Les singularitats s\'on bastant comuns en solucions de la Relativitat General, aix\'i com en les seves extensions, i suposen punts de l'espaitemps, on les nostres teories es suposen no v\`alides. Alguns efectes d'origen qu\`antic com  l'efecte Casimir, o l'anomolia conforme seran explorats al voltant de singularitats cosmol\`ogiques del tipus ``Big Rip''. Es demostrar\`a que els efectes esmentats, gaireb\'e no tenen rellev\`ancia i la singularitat no \'es evitada en la major part dels casos. A m\'es, s'estudiar\`a la validesa dels l\'imits sobre l'entropia de l'Univers al voltant d'aquestes singularitats. Tamb\'e s'estudiar\`a la f\'ormula de Cardy-Verlinde, que d\'ona una derivaci\'o directa dels limits din\`amics de l'entropia, i la validesa de la correspond\`encia que s'estableix entre les equacions de FLRW i les d'una Teoria de Camps Conforme (CFT) en 2d, en contextes m\'es generals.
\newpage
{\bf Acknowledgements}\\

I would like to thank  my supervisors Emilio Elizalde and Sergei D. Odintsov, who have introduced me to the research world and have stimulated my research by means of proposals and questions. The major part of this work has been performed at the Institut de Ci\`encies de l'Espai (CSIC), and I am very grateful to all the staff of the institute. I acknowledge Dr.~Iver Brevik and Dr.~K\aa{}re Olaussen for hosting me  at NTNU (Trondheim, Norway) and working together during a pleasant short stage, as well as I also acknowledge  Dr.~Salvatore Capozziello for our collaboration and for hosting me at the Universit\'a di Napoli (Italy). I would like also to thank for fruitful collaborations  Dr.~Shin'ichi Nojiri, Dr.~Peter K.~S. Dunsby, Dr.~Valerio Faraoni, Dr.~Olesya Gorbunova, Dr.~Rituparno Goswami, Dr.~Josef Kluso\v{n}, Dr.~Ratbay Myrzakulov, Dr.~Valery V. Obukhov and Dr.~Anca Tureanu. This work has been financed by an FPI fellowship from the Spanish Ministry of Science.\\

Of course, I would like to thank all my colleagues from the Institute, but especially I am very grateful to  Carlos L\'opez Arenillas, without his support probably I had never performed this task, and our discussions have improved my understanding of the World. I also would like to mention  Felipe Alv\'es, Jacobo Asorey, Priscilla Ca\~{n}izares, Sante Carloni, Jorge Carretero, Elsa de Cea, Andreu Font, Ane Garc\'es, Daniela Hadasch, Antonio L\'opez Revelles, Isabel Molto, Delfi Nieto and Santiago Serrano, I have spent greats years in our Institute with them.  \\
Also I would like to thank my collegues and friends, Mariam Bouhmadi L\'opez and \'Alvaro de la Cruz Dombriz for good moments  during several Physics Meetings. \\ 
I  also  remember my friends from my years at the UAM, Daniel Garc\'ia Figueroa, Julia Garc\'ia Trilla, Gonzalo Santoro, Bel\'en Serrano, Alvaro Subias and Juan Valent\'in-Gamazo.\\
 
I am very grateful to my teachers from high school, Rafael Calder\'on Fern\'andez and Antonia Mena Ramos, who were the first to stimulate my interest in physics and explained me the basic concepts on  nature.\\ 

I want to thank all my family, specially my brothers, sisters and grandmother, who have supported me even though they did not understand very well what I was doing. I would like also to thank my friends who have made my life easier, Alfonso Alonso P\'erez, Javier Bartolom\'e G\'omez, Hector Gonz\'alez, Jes\'us Gonz\'alez Mas, Mar\'ia Mendoza Bermejo, Eduardo P\'erez Lozano, ...among other people from Segovia, Madrid and Barcelona.  \\

My best thanks go to Lucie Hudcov\'a, who has followed me in this world, especially during the hardest moments, and she has made me look at life always with a smile.
\begin{flushright}
\vspace{1cm}
{\it Some people say, ``How can you live without knowing?'' I do not know what they mean. I always live without knowing. That is easy. How you get to know is what I want to know.}\\
\vskip 0.2cm
Richard Feynman\\
\vskip 0.2cm
{\it La paci\`encia comen\c{c}a amb ll\`agrimes i, al final, somriu.}\\
\vskip 0.2cm
Ramon Llull 
\end{flushright}
\newpage
\vspace{3cm}
The present thesis, aimed to obtain the title of Philosophy Doctor in Physics, is based on the following papers published in referred journals, pre-prints and conference proceedings:      \cite{2010GReGr..42.1513B,2010EPJC...69..563B,2010PhRvD..82b3519D,2010CQGra..27i5007E,2010EPJC...70..351E,2008PhRvD..77j6005E,2009PhRvD..80d4030E,2010OAJ.....3...73G,2010arXiv1012.0473K,2011GReGr.tmp...14M,2009PhLB..681...74N,2008arXiv0812.1980S,2009GReGr..41.1527S,2009GrCo...15..134S,2010arXiv1012.4605S,2010JPhCS.229a2066S,2010arXiv1011.2090S}.

\tableofcontents

\mainmatter

\chapter{Introduction}

\section{Modern cosmology: from Hubble law to dark energy}

Let us  start with a brief historical review of the main facts and discoveries occurred during the last century in the field of cosmology and  gravitational physics. The theories constructed and observations performed along the last one hundred years, are the basis of the research nowadays. The history of Cosmology as a science can be looked as the construction of a pyramid that you do not know how high will be at the end, and whose basis has to be so robust to support even if increased. Thus, step by step cosmology is being built, where new problems arise on each step, so that the theory has to be  amplified (sometimes just papered over it) with new elements in order to achieve the explanations of the problems arisen, what usually drive us to new physical aspects. The main question, probably with no right answer at  present, is how strong is the basis in order to resist the weight to add new aspects and therefore, new problems. Among other questions, some of the main task of theoretical physics nowadays has something to be with cosmology, whose unresolved problems have motivated, among a lot of published papers, a lot of PhD thesis as the current one. This justifies the need to understand well the origins of modern cosmology, whose theory, observations and experiments are the basis of the research nowadays\footnote{On modern cosmology, see for example the textbooks Ref.~\cite{Mukahnov1999,Peacock1999,Peebles1993}.}. \\

One could say that modern cosmology started with the appearance of Einstein theory of gravity in 1915, Ref.~\cite{1915..Einstein,1916..Einstein}, which abandoned the notion of gravity as  a force, and instead of it, the so-called General Relativity (GR) proposed  a geometrical point of view for gravity. As at present, one believed in the validity of the so-called {\it Cosmological Principle}\footnote{Actually at that moment it was believed on the validity of Copernican Principle, and it was not until 1933 when modern Cosmological Principle was proposed by Edward Milne, which  supposed a generalization of the preceding one.}, which basically establishes that the Universe is the same in every point (at large scales), what from a mathematical point of view, it means that the Universe is homogeneous and isotropic. Nevertheless, the first attempt to construct a realistic cosmological model in the frame of GR was based on a very different idea from what we think now, it was believed on an static Universe with no beginning and end. In order to achieve such kind of solution, Einstein introduced for the first time, the cosmological constant (cc) in his field equations (see Ref.~\cite{1917..Einstein}), what yields, under certain conditions, an unstable static Universe  (what for Einstein would be {\it my biggest blunder of my life}). At the same time, Willem de Sitter found an expanding cosmology with the presence of a cosmological constant, but this kind of cosmology was ignored along the years. Later in 1922, Alexander Friedmann found a solution of the Einstein's field equations that suggests an expanding Universe. Almost simultaneously, Georges Lemaitre proposed for the first time, a creation event as the beginning of the Universe expansion, being the first model  that later would be known as the Big Bang model\footnote{The term Big Bang was later proposed by  Fred Hoyle in 1950.}, and proposed the distance-redshift relation that would explain the expansion of the space. These proposals, together with the metric given by Howard Percy Robertson and Arthur Geoffrey Walker, give the name of what we know nowadays as the Friedmann-Lema\^{\i}tre-Robertson-Walker solution of the gravity equations, and which (we believe) is the best fit to describe our Universe evolution. However, the crucial fact occurred in 1929, when Edwin Hubble showed the linear relation between the distance and  redshift\footnote{Hubble's law established a linear redshift-distance linear relation, which is valid to distance of some hundreds of Mpc, but has to be corrected at larger scales}. By using the observational data accounted by Vesto Slipher on galaxies spectra some years before, he showed the fact of the expansion of the Universe. \\

From here on, the discover of the expansion of the Universe gave a strong support to the development of the Big Bang model proposed by Lemaitre. The measures on the abundances of the elements as the Hydrogen, Helium or Lithium were well predicted by the Big Bang model, by means of the initial hot state of the Universe, when primordial nucleosynthesis occurred. As the Universe was growing, it became colder and colder, such that the radiation decoupled, and at that moment, it was emitted what we call the Cosmic Microwave Background (CMB), discovered by the astronomers  Arno Penzias and Robert Wilson in 1965. Then, these two facts among others, made the Big Bang model as the most popular model to explain the Universe evolution\footnote{Note that an alternative to the Big Bang model, the so-called Steady State Theory, was important until the discovery of CMB in 1965 and definitely refused when the anisotropies of CMB were measured by COBE  in 1992.}. However, this does not mean that the model was absent of problems. Actually, during the sixties and seventies, main unresolved questions of the model  were pointed out:
\begin{itemize}
\item The horizon problem, which present the problem of the causal connection between far away sides of the Universe.
\item The flatness problem. Big Bang model does not predict a flat Universe while the observations on the CMB suggested that the Universe was almost flat.
\item Other important problems: Baryogenesis, the observed asymmetry between matter-antimatter is absent of any  explanation. The Monopole problem, it consists on the production of magnetic monopoles at the initial states of the Universe, which is not observed. 
\end{itemize}
These were probably the main troubles of the Big Bang model at that time. However, the horizon and flatness problems could be explained at the beginning of the eighties by  postulating an initial super-accelerated phase called inflation, proposed initially by Alan Guth in 1981, Ref.~\cite{1981..Guth} (what nowadays is known as old inflation), and  later by Andrei Linde in 1982, Ref.~\cite{1982..Linde}. While the baryogenesis problem is still unresolved (for a review on the topic see Ref.~\cite{1997hep.ph....7419D}), although I will not enter on this trouble as this is not the target of the present thesis. Almost simultaneously, it was pointed out that the rotational curves of galaxies could not be explained in the frame of GR if no new and unobserved matter is included in our galaxies and then, in our Universe. This new matter, which is believed to be ten times the amount of observed barionic matter, was called Dark Matter, and among the astrophysical problems, it is also necessary in order to explain the structure formation of the Universe, the CMB anisotropies or the observed gravitational lensing. Then, we have another question to resolve, to find out what actually  Dark Matter is. 

Then, we had with these ingredients, an inflationary-Big Bang model, with relatively successful, but not absent of serious problems. However, our pyramid hided another surprise inside, in 1998  by means of measuremens of the luminosity of Supernovae IA, it was discovered that the Universe is not just expanding but accelerating its expansion, what supposed the beginning of the so-called dark energy models in order to explain such atypical behavior of the Universe at large scales. 

It is difficult to explain in a brief introduction all the implications of these (not essential) modifications of the standard Big Bang model. However, one has to note that inflation, late-time acceleration or CMB observations are products of the basis of our model, what means that under a different model probably we would have never reached, along the last century, such conclusions. Then, now the question arises, is our pyramid on the Big Bang model so robust to support all these troubles? while we try to answer this difficult question, let us try to study and find  a natural explanation to the problems that the model still has.

\subsection*{Inflationary cosmology}

As it has been commented above, inflation was needed in order to explain two intrinsic problems of the Standard model of cosmology: the horizon and flatness problems. It consists in an initial accelerated phase after the origin of the Universe, that makes the Universe expansion to proceed beyond its proper horizon. This epoch ends in a hot state, assumed to be produced by the so-called {\it reheating}, and leads into the known radiation and matter dominated epochs. The first attempt, pointed out  by Alan Guth Ref.~\cite{1981..Guth}, was based on a first order transition phase of our Universe from a false vacuum to a true one. The responsible for such transition is the so-called inflaton, which is immersed initially in a false vacuum, acting as an effective cosmological constant and producing a de Sitter accelerating expansion. Then, inflation ends when the inflaton tunnels to the true vacuum state  nucleating bubbles. The problem (already pointed by Guth in his original paper) is that in order to get an initial hot state after inflation ends, the nucleated bubbles  should be enough close to produce radiation by means of collisions between the walls of the bubbles. However,  as inflation has to last long enough to resolve the initial problems, the bubbles stay far from each other, and the  collisions between bubbles become very rare, what gives no possibility to reheat the Universe and get the initial Big Bang state, where primordial nucleosyntesis has to be realized.

\begin{figure}[h]
 \centering
 \includegraphics[width=3in,height=2in]{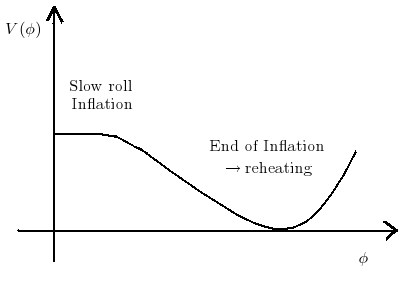}
 \caption{A typical potential for slow roll inflation. Other kind of potentials can also produce inflation}
 \label{fig0}
\end{figure}

Nevertheless, after this initial model, it was Linde who proposed a more successful one, where inflation is driven by a scalar field $\phi$, which produces a second order phase transition. In this model, known as slow-rolling model, inflation occurs  because the potential of the scalar field $V(\phi)$ has a flat section, working as an effective cc, where the field slow rolls. Then, the potential is assumed to have an steep well where the inflaton decays starting to roll faster to a minimum situated in $V(\phi)=0$, breaking the slow-roll and ending the inflationary period (an example of such kind of potentials is given in Fig.\ref{fig0}). When the field reaches the minimum, it starts to oscillate around it, producing particles and reheating the expanded Universe. This can resolve the problems of the old inflation, and also gives an explanation of the origin of density perturbations by means of quantum fluctuations of the scalar field, and which later produce the formation of large scale structures as well as the anisotropies of the CMB. However, some questions are still unresolved, as the mechanism to start inflation (initial conditions problem) or how reheating works, what is not yet completely clear.

As it will be shown along the thesis, inflation can be also reproduced purely in terms  of gravity, with no need to invoke scalar or other  kind of fields. On this way, Starobinsky proposed in 1980 Ref.~\cite{1980..Starobinsky} a model for cosmology free of initial singularity, where higher order terms as $R^2$ are involved in the action. This model can reproduce inflation, what can be seen easily by the fact of the mathematical equivalence between some kind of scalar-tensor theories and modified gravities of the type $f(R)$. Nevertheless, the same problems as in scalar field inflation remain in modified gravity. 

\begin{figure}[h]
 \centering
 \includegraphics[width=3in,height=2in]{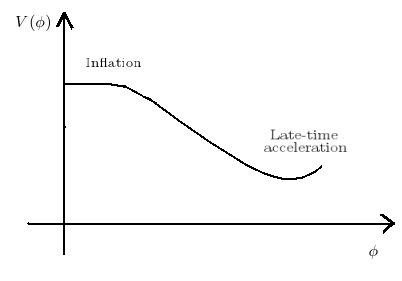}
 \caption{Example of an scalar potential for the unification of inflation and late-time acceleration.}
 \label{fig1}
\end{figure}

Another important aspect that will be studied in the present thesis is the possibility to unify inflation and late-time acceleration under the same mechanism (scalar fields, modified gravity or perhaps a combination of both). The majority of the scientific community, as a consequence of recent observations, has accepted that our Universe expansion is accelerated nowadays, so one could ask if both accelerated phases could be related, and produced by the same mechanism. In this way, in Chapter 2 and 3,  the possibility to construct an scalar field potential that could reproduce both epochs is shown, as in Fig.~\ref{fig1}, where the scalar field has decayed after inflation but the minimum is now non-zero such that the residual of inflation acts as a cc that becomes dominant at late-times  when the radiation/matter densities decrease.  

In the same way,  modified gravities as $f(R)$ or Gauss-Bonnet models, can reproduce both epochs in a natural way with no need of any extra field, and remaining invariant the rest of cosmological aspects, mainly the radiation/matter epochs, as it will be shown along the thesis. 

\subsection*{Late-time acceleration}

Since 1998, when  observations of Supernova IA, realized by two independent groups (see Refs.~\cite{1998AJ....116.1009R,1999ApJ...517..565P}), pointed to the idea of an accelerating expansion,  such hypothesis has been accepted by the majority of scientific community.  Since then, a lot of models and candidates, which are included under the name of Dark Energy (DE), has been proposed in order to explain the phenomena (for reviews on DE candidates see Refs.~\cite{Copeland_et_al_06,Odintsov_Nojiri_2006hep.th....1213N,2007PhLB..651..224N,Padmanachan_2006AIPC..861..179P,Starobinsky_2006IJMPD..15.2105S}), being the $\Lambda$CDM model, where a cc acts as a perfect fluid with negative pressure,  the most popular one.

Supernova data pointed to  a transition from decelerated to accelerated expansion (assuming an spatially flat metric), such that a Universe with a current content of dust-matter $\Omega^{0}_m=1$ does not fit the observations, while it works when $\Omega^{0}_m\sim0.3$ and the rest of content is filled by a perfect fluid with negative pressure, actually with an Equation of State (EoS) $p=w\rho$, where $w<-1/3$ is required. This requires the violation of at least one of the energy conditions, which for a perfect fluid can be written as, 
\begin{itemize}
 \item Strong energy condition: $\rho+3p\geq0 \longrightarrow w\geq-1/3$.
\item Weak energy condition: $\rho+p\geq0 \longrightarrow w\geq-1$.
\end{itemize}
Then, in order to satisfy the requirement to have accelerating expansion, one has to violate at least the Strong energy condition, but may be also the weak one (Phantom case). As it has been mentioned, the simplest model that satisfied such requirement is a cc $\Lambda$, which behaves with an EoS $p_{\Lambda}=-\rho_{\Lambda}$. However, other candidates as scalar-tensor theories or modified gravities, analyzed along the present thesis, could be good candidates from a theoretical point of view, as it is shown below, or at least they could just contribute to the acceleration in combination with other candidates. \\
 \begin{figure}[h]
 \centering
 \includegraphics[width=3in,height=4in]{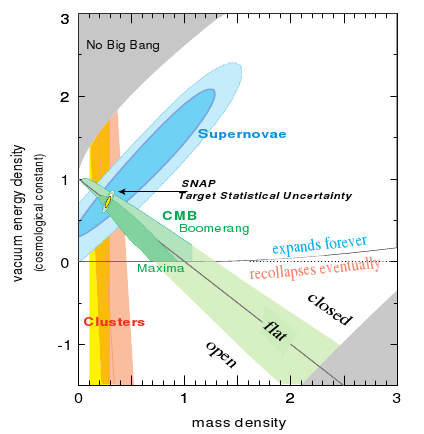}
 \caption{Confidence regions for dust-matter and dark energy densities constrained from the observations of Supernovae IA, CMB and galaxy clustering. From Ref.~\cite{2002SPIE.4835..146A}.}
 \label{fig2}
\end{figure}
Before analyze the reasons to explore different models beyond $\Lambda$CDM, let us continue with the arguments which support the idea of an accelerating Universe apart from the luminosity distance from Supernova. Then, we have that the age of the Universe is very sensible to the content of the Universe, where the oldest stellar populations observed are about $13  Gyrs$ old, so that the age of the Universe has to be $t_u>13 Gyrs$. Then, a Universe absent of dark energy violates such constraint, while  for example for $\Lambda$CDM model, we have   $t_u=13.73 Gyrs$. Also the observations related to the CMB and large scale structure support the idea of a Universe filled with dark energy. From the data collected by WMAP3, we have that the primordial fluctuations are nearly scale invariant, which agree very well with inflationary predictions, and the position of the first peak constraints the spatial curvature of the Universe to be $\Omega_K^{0}=0.015^{+0.02}_{-0.016}$ (see Ref.~\cite{2003ApJS..148..175S}), almost spatially flat, where it is assumed an EoS parameter for dark energy $w_{DE}=-1$. Then, the density for dark energy would be $\Omega_{\Lambda}=0.72^\pm0.04$. In \ref{fig2}, the confidence areas are plotted using Supernova IA, CMB and large scale galaxy clustering. We can see that a non-flat Universe is almost refused and the most probable solution is a Universe filled with $30\%$ of dust matter (barionic and cold dark matter) and $70 \%$ of the so-called Dark Energy.

Finally, it is important to note that all these measurements are sensible to the choice of model, such that in a FLRW Universe, the theory needs dark energy while in other kind of metrics, it may not. 

\subsubsection{Why not just $\Lambda$CDM?}

As it was mentioned above, $\Lambda$CDM model was the first model proposed to explain dark energy and probably  it is going to be the simplest (at least aesthetically) model shown at the present thesis. It is described by the action,
\be
S=\frac{1}{16\pi G}\int d^4x \sqrt{-g}(R-2\Lambda) + \int  d^4x \sqrt{-g} L_m\ ,
\label{1}
\ee
where $g=\det[g_{\mu\nu}]$, being $g_{\mu\nu}$ the metric, $R$ the Ricci scalar, $L_m$ the lagrangian for the matter content (radiation, baryons, cold dark matter..), and $\Lambda$ the  cc. This action can well reproduced cosmic acceleration and fit the observational data commented above by fixing the value of $\Lambda$ to be approximately the current Hubble parameter,
\be
\Lambda\sim H_0^2= (2.13h 10^{-42})^2 GeV^2 \longrightarrow \rho_{\Lambda}=\frac{\Lambda}{8\pi G}\sim10^{-47} GeV^4\ .
\label{2}
\ee
However, this presents several problems. The main one is the so-called {\it cosmological constant problem}\footnote{For a review on the ``old`` cc problem, see Ref~\cite{1989..Weinberg}.}, which was known much before late-time acceleration was discovered. In particle physics, the cc appears as a natural vacuum energy density, which has to be included in Einstein's field equation to contribute to gravity. The problem arises because of the large difference between the observed value of the cc (\ref{2}), and the vacuum energy density predicted by particle physics, suffering of a severe fine-tuning problem. The vacuum energy density can be calculated as the sum of zero-point energies of quantum fields with mass $m$,
\be
\rho_{vac}=S=\frac{1}{2}\int_0^{\infty} d^3k \sqrt{k^2+m^2}=\frac{1}{4\pi^2}\int_0^{\infty} dk k^2\sqrt{k^2+m^2}\propto k^4\ ,
\label{3}
\ee
which exhibits an ultraviolet divergence. Nevertheless, one expects that quantum field theory is valid up to a cut-off $k_{max}$, what makes the integral (\ref{3}) converge. It is natural to take such cut-off for GR as the Planck scale $m_{Pl}$, what yields an vacuum energy density,
\be
\rho_{vac}\sim10^{74} GeV^4\ . 
\label{4}
\ee
That means 122 orders of magnitude larger than the observed value (\ref{3}). Even if one takes the cut-off of scales for QCD, it is still very large compared with (\ref{3}). There have been attempts to resolve this problem by many and different ways (supersymmetry, strings, anthropic principle...), but there is not yet a successful answer, probably because of the absence of a consistent quantum theory of gravity. However, a possibility explored along this thesis may be a classical adjustment of the value of vacuum energy density, produced by scalar fields or extra gravity terms in the action, what may provide a natural relaxing mechanism of the large value (\ref{4}) (see Refs.\cite{2005PhLB..624..147S,2009PhLB..678..427B}).  
The second question is not a theoretical problem as the one above, but it could be important in the  future when more observational data are obtained. It is the question on the dynamics of the EoS for dark energy. For the case of a cc, we have an static EoS $p_{\Lambda}=-\rho_{\Lambda}$, while the EoS for dark energy may be dynamic and even inhomogeneous $p_{\Lambda}=w(t)\rho_{\Lambda}+f(t)$, what can be well modeled in scalar-tensor theories or modified gravities.

Another question that remained ``unresolved'' in the $\Lambda$CDM model, is the so-called coincidence problem, which establishes the need to explain why precisely nowadays the matter and dark energy densities are more or less equal (of the same order). Note that the quotation marks on the word ''unresolved`` is just to emphasize the fact that this problem could not be a real physical problem, but just the product of a real coincidence! For example, along the Universe history there was a period when  radiation and dust matter densities were completely equal (during the transition between both eras). Then, if we were living during that moment, would we think of this as of a coincidence problem? probably we would not, in spite of the fact that, at the present, we look to such period  as a natural evolution of our Universe. However, if we insist on the validity of the open question, we have to point out that a cc does not give a natural explanation of such coincidence, where another kind of candidates could contribute with a more natural answer, as for example an oscillating Universe (this possibility is explored in Chapter 3).

Finally, it is natural to relate, as it was mentioned above, the two accelerated epochs of the Universe evolution, on a unification of cosmic and inflationary eras under the same mechanism, what needs, apart from a vacuum energy density,  another contribution from an scalar field, or extra terms in the gravitational field equations. This possibility is well explored and analyzed along the present report, as a real possibility that could resolve the problems commented above.

Probably the question of the dark energy, what adds up to the cc problem, is one of the main unresolved task of nowadays physics. As it was commented above, this question has produced a lot of publications where a plenty of different models have been proposed. Along this work, it will be analyzed, among other models, scalar-tensor theories and mainly modified gravities, exploring the difference between them. However, other attempts to resolve the mysteries of dark energy, which will not be analyzed here, have been considered in the literature, as for example with vectors  (see \cite{2009PhRvD..80f3512B}) or Yang-Mills fields (see Ref.~\cite{2010PhRvD..82f3504E}).

\section{Scalar-tensor theories}

Scalar-tensor theories have become the most popular ones, after $\Lambda$CDM, to model late-time acceleration. One of the main reasons is the simplicity to construct models using an scalar field, as  it is well known that scalar-tensor theories can well reproduce any cosmology (see Ref.~\cite{1991ClassQuantGrav..8..667}). Also the success to reproduce inflation with an scalar field (the majority of models for inflation proposed during the eighties and nineties, contain an scalar field), makes from scalar fields a  good candidates for dark energy. Another important reason, mentioned above, is that nowadays the observations do not exclude the possibility to have dark energy with a dynamical EoS, and not static as it occurs in $\Lambda$CDM. In the era of precision cosmology, the observations constraint the value of $w$ to be close to $-1$, but tell us little about the time evolution of $w$. In scalar-tensor theories usually the EoS parameter is going to be time-dependent, and it can be easily fixed to be close to $-1$ at the current epoch. 
Apart from the cosmological considerations, scalar fields naturally arise in particle physics as well as in string theory, where they play important roles, as the Higgs boson, in spite of what a spin-0 particle  has never been detected.

A wide range of models have been proposed using scalar fields. In this  thesis, we will analyze models with an action of the type,
\begin{equation}
S=\int dx^{4}\sqrt{-g}\left[ \frac{f(\phi)}{2\kappa^{2}}R
 - \frac{1}{2} \omega (\phi)
\partial_{\mu} \phi \partial^{\mu }\phi -V(\phi )+L_{m}\right]\ ,
\label{5}
\end{equation}
where $f(\phi)$ determines the coupling between the scalar and metric field, $\omega(\phi)$ is the kinetic term and $V(\phi)$ the scalar potential. Depending on the values of $(f,\omega,V)$, the nature of the scalar field will be very different. However, scalar fields are not just restricted to the action (\ref{5}), but there are other important theories that include other kind of scalar fields. Some of them are k-essence, which generalizes quintessence field and present a non canonical scalar field, the tachyon  (a  superluminal particle), or the Chaplygin gas, which also has been considered as a candidate to dark energy. Nevertheless, along the first part of the thesis, where scalar-tensor theories are analyzed, we will focus on scalar fields of the type quintessence and phantom, as well as we explore non-minimally coupling theories of the type of Brans-Dicke theory.

Note that quintessence or phantom behaviors can well be reproduced not just by an scalar field  but also by other possible fields or modifications of gravity, as it will be also explored in the Part II of the current thesis.

\subsection*{Quintessence} 

Of this models, a special mention is required for quintessence, which is probably the most known scalar field candidate for dark energy. First proposed by Caldwell {\it et al} in Ref.~\cite{1998PhRvL..80.1582C}, it has been well studied during the past ten years. Quintessence can be described by the action (\ref{5}) with $f(\phi)=1$, where the kinetic term $\omega(\phi)>0$ is canonical, and with a specific potential, whose choice will restrict the behavior of the scalar field, and then, the cosmological evolution. For a suitable choice of the scalar potential, late-time acceleration can be well reproduced. This kind of model will be studied in Part 1, where  the possibility to reproduce not just dark energy, but also inflation through the same scalar field is explored. However, let us now study some qualitative properties of this kind of models. By writing the scalar field as a perfect fluid, its effective EoS yields,
\be
w_{\phi}=\frac{p_{\phi}}{\rho_{\phi}}=\frac{\dot{\phi}^2-2V(\phi)}{\dot{\phi}^2+2V(\phi)}=-1+\frac{2\dot{\phi}^2}{\dot{\phi}^2+2V(\phi)}\ .
\ee
here for simplicity we have taken $\omega(\phi)=1$, what just corresponds to a redefinition of the scalar field. On the other hand, the second derivative of the scale factor can be written as,
\be
\frac{\ddot{a}}{a}\propto -(\rho+3p)\ ,
\label{6}   
\ee
In order to have an accelerating expansion for a perfect fluid with $p=w \rho$, it is necessary to have an EoS parameter $w<-1/3$. Then, for the case of a quintessence field, looking at the EoS (\ref{6}), we see that $V(\phi)>\dot{\phi}^2$ has to be satisfied in order to have cosmic acceleration. Otherwise, if $V(\phi)<<\dot{\phi}^2$, the EoS parameter gives $w_{\phi}\sim1$, which corresponds to a contracting Universe. In the case of slow-roll inflation, we have that the slow-roll limit imposes  $V(\phi)>\dot{\phi}^2$, which yields $w_{\phi}\sim-1$, that is basically an effective cosmological constant.

In part 1, it will be studied how this kind of fields could be used to reconstruct, not only dark energy or inflation separately, but both events unified under the same scalar field. This unification can well be achieved, as it will be shown, by choosing an appropriate kinetic and potential terms for the scalar field. Then,  canonical scalar fields can be used to reproduce, at least in an effective way, the entire history of the Universe.  

\subsection*{Phantom} 

Phantom fluids, first named by Caldwell in Ref.~\cite{2002PhLB..545...23C} can be easily described by scalar fields. For the action (\ref{5}), we can take a minimally coupled field, $f(\phi)=1$, where its  kinetic term $\omega(\phi)$ becomes negative. This kind of fields yields an EoS parameter,
\be
w_{\phi}<-1\ .
\label{7}
\ee
The kind of fluids with such EoS parameter has several consequences both at the cosmological level and in microphysics. Among the strong energy condition, it violates also the weak one, while a cosmological constant or a quitessence field only violate the strong condition. The phantom field described by the action (\ref{5}) owns a negative energy, what can  lead to imaginary effective masses. From a microscopical point of view, it is well known that phantom fluids own serious problems, where its presence could introduce catastrophic quantum instabilities of the vacuum energy (see Refs.~\cite{2002PhLB..545...23C,2004PhRvD..70d3528C}). However, one could study such kind of fluids dealing with them in an effective description, just valid to some energy scale, or where they are just a consequence of extra geometrical terms in the action (as it will be explored). From a cosmological point of view, we have that a phantom fluid produces a super-acceleration of the expansion that ends in a singularity, called Big Rip\footnote{Note that phantom models free of singularities can be reconstructed. This aspect will be explored along the thesis.}. Let us consider a simple phantom model to show its consequences, by considering a constant EoS paremeter less than $-1$. Then, a general solution for a flat FLRW metric yields,
\be
H(t)\propto\frac{1}{t_s-t}\ , \quad \rho_{\phi}\propto\frac{1}{(t_s-t)^2}
\label{8}
\ee
where $H=\frac{\dot{a}}{a}$ is the Hubble parameter, and $t_s$ is the so-called Rip time, defined as the time when the future singularity occurs. Then, for $t\rightarrow t_s$, the Hubble parameter diverges producing a curvature singularity. Note that as the Universe evolves, the acceleration is increased and the phantom energy density becomes larger, such that sufficiently close to the Big Rip singularity, the strength of the phantom would be so large that breaks local systems as galaxies or solar systems, and even molecules and atoms. However,  close to the singularity  quantum effects should become important, and could affect the occurrence of the singularity, an aspect  studied in the last chapters of the thesis, where semiclassical effects are taken into account.

Nevertheless, one could ask why not to restrict our cosmological models to EoS parameters greater than $-1$, such that phantom fluids would be avoided, but  observations have pointed to the real possibility that  dark energy behaves as a phantom fluid, what makes them of enough importance to be studied.

\subsection*{Non-minimally scalar-tensor theories}

In this kind of theories, the scalar field is strongly coupled to the metric field through the Ricci scalar in the action. Then, for the action (\ref{5}) we would have $f(\phi)$ not a constant, but a function of the field $\phi$. The first proposed model of non-minimally coupled scalar-tensor theory was constructed by Brans and Dicke in 1961 \cite{1961..brans.Dicke}, which was the first real attempt of an alternative to General Relativity with the main objective to incorporate Mach's principle into the theory of gravity, which is not satisfied in GR as it was well known already at that time. The model proposed by Brans and Dicke was described by an action of the same type as (\ref{5}), where $f(\phi)=\phi$, $\omega(\phi)=\frac{\omega}{\phi}$ and with a null potential $V(\phi)=0$.  In spite of that is a theory that presents various problems, as corrections to the Newtonian potential or violation of the Birkhoff's theorem, it gave an alternative theory that provided a way to test GR in the experiments. Nowadays, these theories have recovered some interest due to their cosmological implications, where dark energy can be well reproduced. Also it can be a very useful instrument to analyze modified gravity theories as $f(R)$ gravity due to the mathematical equivalence between them, as it will be studied in the thesis. 

However, theories with a non-minimally scalar field could violate equivalence principle, which is very well tested at least at small scales (Earth, solar system, galaxies...). It is well known that in the Jordan frame, on which the action (\ref{5}) is defined, a test particle will follow a geodesic, while in the Einstein frame (related to the Jordan's one by a conformal transformation), the path of a test particle moving in a given spacetime will suffer of corrections on its geodesic due to the appearance of a ''fifth force``, produced by the scalar field\footnote{Note that the physical equivalence between the Jordan and the Einstein frame is not completely clear, and big discussions have been going on this topic with no successful answer yet. For a review on it see Ref.~\cite{1999FCPh...20..121F} and references therein. For recent results, see Ref.~\cite{2010PhLB..689..117C,2011..Capoziello..DSG}.}. In Brans-Dicke theory, this violation in the Einstein frame can be written by the geodesic equation of a test particle, what yields a fifth force correction, 
\be
\frac{d^2x^{\mu}}{d\tau^2}+\Gamma^{\mu}_{\nu\sigma} \frac{dx^{\nu}}{d\tau}\frac{dx^{\sigma}}{d\tau}=-\beta\partial^{\mu}\phi\ ,
\label{9}
\ee
where $\beta$ is a free parameter of the theory related to $\omega$ in (\ref{5}). Then, assuming that Einstein's frame is physical, this fifth force can be checked by the experiments. In 2003, the Casini spacecraft restricted the value of the parameter to be $\beta<1.6\times10^{-3}$ (see Ref.\cite{Bertotti:Nature_2003}). However, another solution to the problem of fifth forces and corrections to the Newtonian potential, is the so-called ''chameleon mechanism``, proposed in Ref.~\cite{Khoury:2003rn}, where the scalar field depends on the scale, such that it becomes important at large scales, where dark energy is needed, and it is neglible at local scales, satisfying the gravitational experiments. Here, the models will be analyzed where the coupling is given by $f(\phi)=1+g(\phi)$, being $g(\phi)$ a function that can be restricted at small scales by the chameleon mechanism, where GR is recovered in the action (\ref{5}). It will be shown how this kind of theories can well reproduce late-time acceleration, as well as its relation with the results obtained in the Jordan and Einstein frame, and its possible physical implications. It will be also used as an auxiliary field to reconstruct modified theories of gravity able to reproduce  cosmic acceleration.

\section{Modifying General Relativity: towards a complete theory of gravity}

Along the present thesis, the analysis on Extended Theories of Gravity (ETG) is  one of main points to be studied. In this brief introduction, it will be argued the motivations for an extension of General Relativity to more complex and also general, laws of gravity. GR has been along the last one hundred years a very successful theory in order to explain the macroscopic world, on this sense GR gives a good description of nature at scales of the Earth or the Solar system, as the experiments have well proved. However, as it was commented above, GR presents some shortcomings when it is tested at large scales, these problems can be interpreted as unknown forms of matter/energy or as a failure of the theory. It has been well known since long time ago, that  rotational curves of galaxies can not be fitted in GR unless dark matter or some modification is taken into account, as well as at cosmological scales it is necessary to incorporate a perfect fluid with negative pressure to explain late-time acceleration. Then, we have to decide if dark matter and/or dark energy could be a sufficiently satisfactory explanation or if our theory of gravity actually presents an scale of validity, and has to be modified beyond some limit. There is not a right answer to this serious dilemma yet, or at least a consensus over the scientific community to decide for one or another option. Nevertheless, there is an historical analogy to this current problem that could serve as an example of how to proceed: at the end of the nineteenth century, it was believed (with almost no doubt) that a fluid called aether existed, and everything was immersed on it, light or the planets were moving through it, and it served as the inertial frame to refer the movements of any test body. The end of this history is well known by everyone. In the same way, dark energy and specially dark matter, have been searched over the last years with no result, what has opened the real possibility to investigate theories beyond GR in order to explain such phenomena\footnote{Some modified gravities try to explain the origin of dark matter as a pure gravitational effect. Several  attempts have been performed on this sense, as MONDs or even $f(R)$ gravity. At the present work, we are not focusing on this aspect, but on dark energy and inflation.}. It will be shown that generalizing Einstein's field equations, any cosmological solution can be reproduced, and then, late-time acceleration and even its unification with inflation can well be explained in the context of ETGs. 

However, as it was pointed in the above section and it will be shown in the first part at the present thesis,  scalar fields can easily reproduce the cosmological evolution, and do not introduce so big changes as modified gravity does (except non-minimally scalar-tensor theories, which from a mathematical point of view are equivalents to some kind of modified gravities). However, as it has been pointed, quintessence or phantom fields are exotic forms of matter/energy never observed directly, and with an unknown origin. On the other hand, modified gravities offer an alternative explanation, which seems to be more natural and could avoid the coincidence problem unifying inflation and late-time acceleration under the same mechanism. Also, it has been recently shown that some ''viable'' modified gravities could keep  GR at small scales almost unchanged, avoiding the terrible problems of violation of local tests of gravity that usually affect to ETGs. However, one has to say that modified gravities have to be seen  not as the ultimate theory of gravity that resolves all the theoretical and observational problems, but as an effective description of a probably more complex theory.

Let us deeper analyze  the statements and motivations to consider GR as a good description of nature at some scale, but which fails in some limits, where corrections are needed. The principles on which General Relativity is based, and  which are well tested at least at macroscopic local scales, have to be satisfied by any extension of GR on those scales. General Relativity  is based on three assumptions\footnote{For a complete analysis on General Relativity see for example the textbooks Refs.~\cite{Wald1984,Weinberg1972}}:
\begin{itemize}
\item The Principle of Relativity. The laws of nature are the same for every observer, what means that there is not  preferred system of reference.
\item The Principle of Equivalence, which establishes the equivalence between accelerated observers and those immersed in a gravitational field. In a sample form, it imposed the equivalence between the inertial and  gravitational masses.
\item The Principle of General Covariance. It  implies  that every law's equation of nature has to be covariant, what means invariant under general transformation of coordinates. 
\end{itemize}
However, these assumptions tell us little about the form of the field equations. In order to restrict our theory of gravity, we can  impose two additional conditions, that will constraint the field equations, 
\begin{itemize}
 \item The energy has to be conserved, what implies that the action of the theory has to be invariant under general transformations of coordinates.
\item At the non-relativistic limit, Newtonian law has to be recovered, or basically, at some limit, Special Relativity has to be recovered.
\end{itemize}
It is interesting to point that the first proposal for gravitational field equations by Einstein was in 1913, and it did not satisfy the above conditions. He proposed probably the most simple construction using the Ricci tensor with the following equations for gravity,
\be
R_{\mu\nu}=\kappa^2T_{\mu\nu}\ ,  
\label{10}
\ee
where $R_{\mu\nu}$ is the Ricci tensor, $T_{\mu\nu}$ is the energy-momentum tensor and  $\kappa^2$ the coupling constant. It is easy to see that the field equations (\ref{10}) do not satisfy the energy conservation as the covariant derivative of the equation is not null, $D_{\mu}R^{\mu\nu}\neq 0$, for a general spacetime. After this first attempt, Einstein arrived to the well-known field equations for GR,
\be
G_{\mu\nu}=R_{\mu\nu}-\frac{1}{2}g_{\mu\nu}R=\kappa^2T_{\mu\nu}\ ,  
\label{11}
\ee
which now satisfies $D_{\mu}G^{\mu\nu}=0$. Also on the Newtonian limit of the theory, the field equations (\ref{11}) reduce to the Newtonian law in a non-relativistic limit. It is straightforward to show that for a weak and static limit, the field equation (\ref{11}) yields to,
\be
G_{\mu\nu}\sim\nabla^2g_{00}=-\kappa^2T_{00}\ ,
\label{12}
\ee
which coincides with the Newtonian law. Then, one can impose the above two conditions in order to construct a realistic theory of gravity. Instead of the field equations (\ref{11}), GR can be expressed in a more elegant way through  the action principle,
\be
S=\frac{1}{16\pi G}\int d^4x \sqrt{-g}R\ .
\label{13}
\ee
This action, called the Hilbert-Einstein action, is invariant under general transformation of coordinates, what yields to conserved field equations. Then, one could propose an invariant action, more general than (\ref{13}), but which could evolve more complex invariants as the terms $R^2, R^{\mu\nu}R_{\mu\nu}, R^{\sigma\lambda\mu\nu}R_{\sigma\lambda\mu\nu}, f(G)...$. However, these additional terms could imply corrections in the Newtonian law (see Ref.~\cite{1978GReGr..9.353}), although can be neglected.  Then, the kind of theories studied here will be of the form,
\be
S=\frac{1}{16\pi G}\int d^4x \sqrt{-g}f(R,G)\ .
\label{14}
\ee
Here $f(R,G)$ is an arbitrary function of the Ricci scalar and the Gauss-Bonnet term, the arbitrariness conveys the ignorance of our knowledge on the right theory of gravity, but which can be constrained to satisfy the requirements of the observational data among the requirements written above, and could provide an explanation for the cosmic acceleration with no additional fields (see Ref.~\cite{2002Capozziello..IntJModPhys11}). The action (\ref{14}) is invariant under general transformation of coordinates but yields corrections on the Newtonian law, which can be controlled by specific functions and parameters of the theory (see for example Ref.~\cite{2007PhRvD..76j4019C}). However, in this case the field equations are fourth order instead of second order in GR, what implies more complexity on the equations, a not well formulated Cauchy problem unless some conditions are imposed (see Ref.~\cite{2011Capozziello_Faraoni}) or massive gravitons. However, these problems can be resolved by restricting the action(\ref{14}) and the free parameters of the theory, what constraints the gravity law. 

From a quantum  point of view, there is not  a successful and consistent theory of quantum gravity yet, i.e., a  renormalizable and unitary theory. However, much before higher order theories of gravity became of interest for low energy scales (late cosmology), they were of great interest in order to construct a quantum theory of gravity. It is well known that GR is a non-renormalizable theory, but when extra geometrical terms are added in the action, the theory turns to be renormalizable at one-loop, but it fails at two-loops. It has been shown that when quantum corrections are taken, the low effective action admits higher order terms (see Ref.~\cite{Buchbinder.Odintsov.Shapiro..1992}). Then, it is natural to think that gravity admits extra terms in the action coming from quantum corrections. However, recently a new proposal on quantum gravity, performed by Ho\v{r}ava \cite{2009PhRvD..79h4008H}, claims to be power counting renormalizable, and where instead to new terms on the action, it imposes non-covariant equations. This new theory has became very popular as it presents a new path on the way to find a consistent theory of quantum gravity. This theory will be discussed in the next section, but among serious problems that it has, it suffers of the same cosmological problems as GR does.

Mathematically, ETGs provide very interesting properties to be analyzed. Among the theories of higher order which are intrinsically interesting to study, modified gravities introduces new ways to study gravitational physics. On this sense, in $F(R)$ theories, one can proceed in two ways, called the metric and Palatini formalism respectively. The metric approach assumed that the connection $\Gamma_{\mu\nu}^{\lambda}$ depends and is defined in terms of the metric field $g_{\mu\nu}$, which is the only field of the theory. While in the Palatini's approach, one does not assume that the connection $\Gamma_{\mu\nu}^{\lambda}$ is already defined but it is dealt with a dynamical field as the metric tensor. In GR, both approaches are equivalent, as in Palatini the connection naturally appears defined in terms of $g_{\mu\nu}$, but in higher order theories they are not equivalent, and one has to distinguish between both approaches. Along the present thesis, it will be analyzed only the metric approach. For a review on Palatini's $F(R)$ gravities see Ref.~\cite{2010RvMP...82..451S} and references therein.
 
As we will see, the action (\ref{14}) can be chosen such that, among the problems commented, it can reproduce late-time acceleration and  even inflation together, with no need of an extra exotic fluid. As the extra geometrical terms in the action can reproduce cosmic acceleration, what basically means that behaves as an effective perfect fluid with an EoS parameter $w\sim-1$ at the current epoch, it could produce a relaxation on the value of the vacuum energy density, resolving the cc problem by a simple mechanical way (see Ref.~\cite{2009PhLB..678..427B}).
Some gravity theories will be studied, where the action is given  by the one of Hilbert-Einstein action (\ref{13}) plus a function of the Ricci and/or Gauss-Bonnet scalars that would become important at large scales but reduces to  GR for local scales. Note that for higher order gravities as (\ref{14}), the presence of various de Sitter solutions in the equations can provide a way to unify inflation and late-time acceleration under purely gravitational terms, and even it could predict future phases of accelerated expansion, what avoids the coincidence problem. Also it is important to point that, in general, the extra terms in the gravitational action can be modeled as a perfect fluid, whose effective EoS parameter will not be a  constant in general, and it could vary along the Universe evolution, even crossing the phantom barrier, providing a way to interpret EoS parameters less than -1. As has been pointed out, these theories are mathematically equivalent to non-minimally scalar-tensor theories, what can be very useful for the analysis and reconstruction of the action for modified gravity. Also the relation between Einstein and Jordan frames provide a way to reconstruct modified gravities from scalar field models, in spite of the fact that  cosmological solutions will not coincide in both frames, what could give an explanation of the (non-)physical equivalence between them. Moreover, it is well known that phantom scalar fields have not a representation in the Jordan frame as the scalar kinetic term is negative, what implies a complex conformal transformation that yields an action that becomes complex in the Jordan frame \cite{2007PhLB..646..105B}. This  problem will be analyzed, and this kind of theories will be studied with the presence of fluids with different EoS.

Hence, it will be here shown that theories of gravity can explain the cosmological shortcomings, and at the same time, remain unchanged the principles of GR at local scales, where the gravitational test on GR are well known. Also,  the reconstruction of any kind of cosmological solution will be analyzed, including $\Lambda$CDM model, which gives interesting properties in modified gravity. this kind of theories with the presence of fluids with different EoS, and its cosmological evolution, will be studied.

\subsection*{Ho\v{r}ava-Lifshitz gravity}

In this introduction a special mention  is due to the so-called Ho\v{r}ava-Lifshitz (HL) gravity for it is a new theory that has risen to a lot of interest in the scientific community. This new theory, proposed just two years ago by Ho\v{r}ava Ref.~\cite{2009PhRvD..79h4008H} claims to be power counting renormalizable by losing the covariance of the theory. It is well known that GR is not renormalizable, the main problem in perturbative renormalization comes from the fact that the gravitational coupling $G$ is dimensionful, with a negative dimension $[G]=-2$ in mass units, what gives uncontrolled UV divergences. This can be cured by introducing higher terms of the scalar curvature in the action, but at the same time it provides an additional problem on the uniqueness. The main success of HL theory is that avoids these problems, and gives a power counting renormalizable theory by introducing an anisotropic scaling property between space and time characterized by a dynamical critical exponent z. This kind of anisotropy is  common in condensed matter physics, introduced firstly by Lifshitz \cite{1941Lifshitz} to study some kind of phase transitions,  it is imposed a different degree between the spatial and time components. In HL gravity, this scaling property can be written as,
\be
x^i=b x^i\, , \quad t=b^zt\, .
\label{15}
\ee
The theory becomes power counting renormalizable in 3+1 spacetime dimension when the critical exponent is $z=3$, while GR is recovered when $z=1$. This scaling property makes the theory to not be invariant under diffeomorphism but under some restricted transformations called foliation-preserving diffeomorphisms,
\be
\delta x^i=\zeta(x^i,t)\, , \quad \delta t=f(t)\, .
\label{16}
\ee
Then, the theory breaks  Lorentz invariance, what would have very serious consequences as different observers would measure different values of the speed of light $c$. However, it is assumed that the full diffeomorphism is recovered in some limit, where the critical exponent flows to $z=1$ and the symmetries of GR are recovered, although the mechanism for this transition is not clear yet. 

Respect the action of the theory, let us write the scalar Ricci in the ADM decomposition \cite{2008GReGr..40.1997A},
\be
R=K_{ij}K^{ij}-K^2+R^{(3)}+2\nabla_{\mu}(n^{\mu}\nabla_{\nu}n^{\nu}-n^{\nu}
\nabla_{\nu}n^{\mu})\, ,
\label{17}
\ee
here $K=g^{ij}K_{ij}$, $K_{ij}$ is the extrinsic curvature, $R^{(3)}$ is the spatial scalar curvature, and $n^{\mu}$ a unit vector perpendicular to a hypersurface of constant time. The action proposed in \cite{2009PhRvD..79h4008H} accounts the anisotropic scaling properties (\ref{15}) by introducing  additional parameters $(\mu,\lambda)$ in the theory,
\be
S=\frac{1}{2\kappa^2}\int dtd^3x\sqrt{g^{(3)}}N \tilde{R}\, , \quad
\tilde{R}= K_{ij}K^{ij}-\lambda K^2 + R^{(3)}+
2\mu\nabla_{\mu}(n^{\mu}\nabla_{\nu}n^{\nu}-n^{\nu}\nabla_{\nu}n^{\mu})-
L^{(3)}(g_{ij}^{(3)})\, .
\label{18}
\ee
Note that the term in front of $\mu$ can be dropped out as it turns out to be a total derivative, although it becomes important for extensions of the action (\ref{18}). Then, the term $L^{(3)}(g_{ij}^{(3)})$ is chosen to be the variation of an action in order to satisfy an additional symmetry, the detailed balanced, introduced by Ho\v{r}ava in Ref.~\cite{2009PhRvD..79h4008H}. However, from a cosmological point of view, HL theory suffers from the same problems as GR, additional components are needed to explain late-time acceleration or inflation. Then, following the same procedure as in GR, the HL action can be extended to more general actions \cite{2010CQGra..27r5021C},
\be
S=\frac{1}{2\kappa^2}\int dtd^3x\sqrt{g^{(3)}}N F(\tilde{R})\ .
\label{19}
\ee
It will be shown that this kind of action can well reproduce the cosmological history and avoid corrections to the Newtonian law.\footnote{Here we do not discuss the gravity sector corresponding to 
the Ho\v{r}ava gravity but we show
that the scalar mode, which also appears in the usual $F(R)$ gravity,
can decouple from gravity and matter, and then the scalar mode does not give
a measurable correction to Newton's law.} Even how the matter/radiation epochs of the cosmic evolution yields to an accelerated epoch  modeled by the kind of theories described by (\ref{18}).

Nevertheless, the theory proposed by Ho\v{r}ava, or its extension, described by the action (\ref{19}) contain several problems that makes the theory to not be a good consideration. The lose of symmetries due to the anisotropic scaling (\ref{15}) introduces an additional spin-0 mode that can rise some problems. In order to satisfy the observations, such scalar mode has to decouple in the IR limit, where GR has to be recovered. However, this is not the case here, where the scalar mode introduces strong instabilities around flat spacetime. One possible solution for this trouble is to consider an stable de Sitter spacetime as the natural vacuum solution of the theory, what can be achieved by some specific actions $F(\tilde{R})$. Another solution, proposed in Ref.~\cite{2010PhRvD..82f4027H}, is to include an additional local $U(1)$ symmetry, what could force to the additional parameters $(\lambda,\mu)$ to be equal to 1, neglecting the spin-0 mode in the IR limit. The extension of this kind of HL gravity will be studied here. 

Hence, the new proposal as candidate for  quantum field theory of gravity seems to be a good promise, in spite of the serious problems that owns itself. Here,  their applications to cosmology will be explored, and the possible solutions to the intrinsic problems of the theory as well as the possibility to reproduce inflation and late-time acceleration under the theory with no additional components. 

\section{Future singularities in FLRW universes}

A wide range of solutions in GR or ETGs contain singularities. Some examples can be found in spherical symmetric solutions in vacuum as  Schwarzschild or Kerr metrics, which contain a curvature singularity at the origin of the radial coordinate, or the FLRW metric, where in general contains an initial singularity, the so-called Big Bang. The singularities can be viewed as the limits of our law of gravity, where a consistent theory of quantum gravity or a theory of ``everything`` as String theory is needed. However, the singularities can be treated in a semiclassical approach introducing some corrections coming from quantum field theory in the field equations, to see if some corrections could cure the singularity. The last part of the present thesis is devoted to this subject, (future) singularities  of FLRW metrics  will be analyzed  when some effects are taken into account as the effective cosmological Casimir effect or the conformal anomaly. Also,  the validity of holographic principle will be studied, where a bound on the Universe entropy is imposed, whose validity  around future singularities is studied and interesting results are obtained. In parallel, it is explored the possibility of a generalization of Cardy-Verlinde (CV) formula Ref~\cite{2000hep.th....8140V}. CV formula suggests a more fundamental origin for the Friedman equation, which coincides with a two-dimensional conformal field theory for a closed FLRW metric with some particular requirement when a bound on the  entropy proposed by Verlinde Ref~\cite{2000hep.th....8140V} is reached.    

Let us define the list of possible singularities contained in FLRW universes. It has been pointed out that, apart Big Bang initial singularity, some kind of perfect fluids, specially phantom fluids, could drive the Universe to future singularities of different type. In Ref.~\cite{2005PhRvD..71f3004N},  following classification for future singularities was obtained,\footnote{Note that we have omitted here other kind of future singularities as the so-called ''Big Crunch'' as we are interested to study the future state of an accelerating Universe and not a contracting one.}:
\begin{itemize}
\item Type I (``Big Rip''): For $t\rightarrow t_s$, $a\rightarrow
\infty$ and $\rho\rightarrow \infty$, $|p|\rightarrow \infty$.
\item Type II (``Sudden''): For $t\rightarrow t_s$, $a\rightarrow
a_s$ and $\rho\rightarrow \rho_s$, $|p|\rightarrow \infty$.
(see Refs.\cite{2004CQGra..21L..79B,2008PhRvD..78d6006N})
\item Type III: For $t\rightarrow t_s$, $a\rightarrow a_s$
and $\rho\rightarrow \infty$, $|p|\rightarrow \infty$.
\item Type IV: For $t\rightarrow t_s$, $a\rightarrow a_s$ and
$\rho\rightarrow \rho_s$, $p \rightarrow p_s$
but higher derivatives of Hubble parameter diverge.
\end{itemize}
Here $\rho$ and $p$ are referred to the energy and pressure densities of the  fluid responsible for the occurrence of the future singularity. The most usual future singularity in FLRW metrics is the first type of the above list, the so-called Big Rip singularity, which is usually produced by a phantom fluid whose  EoS parameter $w<-1$, and whose origin may be an scalar field or pure geometrical additional terms in the action. As the singularities are regions or points of a spacetime, where the classical theory fails, it is natural to introduce some quantum corrections in the field equations. Here, the consequences of considering an effective Casimir effect will be analyzed (dealing the Universe as an spherical shell) near the singularity, on one side, and we will account for the effects of the effective action produced by conformal anomaly, on the other side. It will be shown that for a general case, those effects incorporated in the theory do not cure the singularity, such that additional semiclassical terms should be taken into account. Also, it will be shown that ``universal'' bounds of the entropy are violated near and even far away from the singularity, losing its universal nature.  

On the other hand,  the construction of ETGs free of future singularities  is also studied. It is well known that the majority of higher order gravities hide solutions for a FLRW Universe that contain some class of the above future singularities. It was found that an action with $F(R)=R^2$ is free of singularities and even any standard modified  gravity where a term proportional to $R^2$ is added, can cure the singularity (see Refs.~\cite{2005CQGra..22L..35A,2010AIPC.1241.1094N}). This analysis is extended to the Ho\v{r}ava-Lifshitz gravity, where  the terms that can avoid the occurrence of the future singularity are found, too.

\section{Organization of the thesis}

The thesis is composed of three main blocks, each of them dedicated to cosmological models and aspects of FLRW metrics. As it has been pointed out in this brief introduction, the main purpose of the present thesis is the reconstruction of cosmic evolution by several different ways, but where modified gravities have an special interest. In parallel to this main objective, other related topics are studied, which have something to do with FLRW Universes. On those instances where explicitly specified, we will deal with spatially flat FLRW metrics. 

In the first block,  scalar-tensor theories are analyzed; there quintessence/phantom scalar fields as well as Brans-Dicke-like theories are studied. Unification of inflation and late-time acceleration can be easily achieved on this context, even the problem on the transition to a phantom epoch (big instability) can be resolved by not allowing a transition of the scalar field, in this way the observational current data can be better fitted using more than one scalar field. It is also explored in the context of non-minimally scalar-tensor theories the possibility to reproduce the Universe evolution and the related solutions in Jordan and Einstein frames, where it is shown that both solutions are in general very different, and for example the occurrence of a singularity in one frame can be avoided in the other, what could be a signal of the non-physical equivalence between both frames. In this first block, it is also explored the possibility of an oscillating Universe, where the accelerated phases of the Universe would be repeated until a perturbation or a future singularity ends with this periodic behavior of the Universe. It is shown that such oscillations can be reproduced easily by an scalar field. This could resolve the so-called coincidence problem, where the period of the oscillations can be fixed with the age of the Universe. Also, the possibility to have an interacting dark energy is studied, which exchanges energy with dust matter. And how the interacting term could yield an accelerated expansion.

The second block is dedicated to the analysis of Extended Theories of Gravity, specially extensions of General Relativity where more general functions of the Ricci and/or Gauss-Bonnet scalars are involved in the action. It is shown through out the equivalence of $f(R)$ gravities and non-minimally scalar-tensor theories that this kind of modified gravity admits almost any class of solutions for a flat FLRW Universe. It is also shown that $f(R)$ gravity can be dealt with as an effective perfect fluid with a dynamical EoS. Some phantom cosmological  models free of future singularities are constructed on this frame. It is shown that unification of inflation and late-time acceleration can easily be achieved in these theories, what provides a natural explanation for accelerating expansion phases of the Universe. A reconstruction  method for cosmological solutions is also implemented, where no auxiliary field are used. In this case, $\Lambda$CDM model is constructed in modified gravities with no cosmological constant presence in the action, and the solution is a series of powers of the Ricci scalar. In the context of viable modified gravities, the ones that have a good behavior at local scales,  the cosmological evolutions in comparison with those of the $\Lambda$CDM model are analyzed, as well as in the case when an additional phantom fluid is taken into account, the scalar-tensor representation for this class of theories is obtained and analyzed. Finally, as in $f(R)$ theories,  models are studied where the modification of GR comes from functions of the Gauss-Bonnet invariant $G$, where it is shown that the use of a combination of Ricci/Gauss-Bonnet terms provide a theory of gravity.

The recent proposal of quantum gravity, Ho\v{r}ava-Lifshitz gravity, is explored in the third block. This theory is power-counting renormalizable at the price of breaking Lorentz invariance. Apart from some intrinsic problems of the theory, as its transition to recover the full diffeomorphisms, HL gravity suffers problems of the same kind as GR when FLRW metrics are considered. On this sense, an exotic fluid or a cc is needed to explain cosmic acceleration. Extensions of the action proposed by Ho\v{r}ava are studied, where the Newtonian limit is studied. It is also found the condition to have a theory free of future singularities. The stability of the cosmic solutions is explored, where some general conditions on the action for HL gravity are obtained. Several examples are analyzed to illustrate it. By following these stability conditions, one can reconstruct the entire Universe history. Finally, one of the intrinsic problems of the theory is analyzed, the scalar mode associated to the lose of full diffeomorphism that affects the vacuum solution of the theory. The solution to this problem can come from adding an additional U(1) symmetry or from fixing the vacuum solution as de Sitter. Both possibilities are explored.

Finally, in the last part of the thesis, other aspects of FLRW metrics are studied. Generalizations of the Cardy-Verlinde formula, where different types of fluids are involved, are explored. It is shown that only for special cases, the original CV formula is recovered. Also dynamical entropy bounds are analyzed close to future singularities, where some semiclassical effects are taken into account. The inclusion of an effective Casimir effect is studied close to the singularity and even along the Universe evolution, its effects are studied.

The thesis ends with the general conclusions corresponding to the results obtained in the present work. 

\part{Dark energy and inflation from scalar-tensor theories and inhomogeneous fluids}
\chapter[Inflation and acceleration in scalar-tensor theories]{Reconstructing inflation and cosmic
acceleration with phantom and canonical scalar fields}

\footnote{This Chapter is based on the publications: \cite{2008PhRvD..77j6005E,2008arXiv0812.1980S}}The increasing amount and precision of observational data
demand that theoretical cosmological models be as
realistic as possible in their description of the evolution of
our universe. The discovery of late-time cosmic acceleration
brought into this playground
a good number of dark energy models which
aim at describing the observed accelerated expansion, which seems
to have started quite recently on the redshift scale. Keeping in
mind the possibility, which is well compatible with the
observational data, that the effective equation of state
parameter $w$ be less than $-1$, phantom
cosmological models \cite{2004PhRvD..70b3509A,2008JCAP...02..015A,Andrianov:2005tm,2005PhRvD..72f4017A,2005CQGra..22..143B,2006GReGr..38.1609B,2006PhRvD..73j3520B,2007PhLB..646..105B,2008JCAP...03..002B,2005hep.th....5186C,2003PhRvL..91g1301C,2008PhLB..661....1C,2003PhRvL..91u1301C,2004MPLA...19.2479C,2006AnPhy.321..771D,2003PhRvD..68j3519D,2004PhRvD..70d3539E,2005PhRvD..71j3504E,2007IJTP...46.2366F,2006IJMPD..15..199F,2005CQGra..22.3235F,2006astro.ph..5682G,2004PhLB..586....1G,2005PhRvD..72l5011G,2006PhRvD..74l7304G,2005PLB_Hao_Li,2010MNRAS.405.2639J,2004PhRvD..70d1303J,2005PhRvD..71h4011L,2002JHEP...08..029M,Nesseris:2004uj,2003PhLB..562..147N,2004PhLB..595....1N,2005PhRvD..72b3003N,2005PhRvD..71f3004N,Perivolaropoulos:2005yv,Samart:2007xz,2004MPLA...19.1509S,2009AdHEP2009....1S,2008PhLB..668..177S,2005PhLB..624..147S,2007arXiv0707.1376S,2007JCAP...06..010S,Tsujikawa:2004dp,2004PhLB..586....5S,2005PhRvD..71b3515V,Wei:2005fq,2007JCAP...10..014W,2006NuPhB.759..320Y,2008JCAP...03..007Z,2006MPLA...21..231Z} share a
place in the list of theories capable to explain dark
energy. Ideal fluid and scalar field
quintessence/phantom models still remain among the easiest and most
popular constructions. Nevertheless, when working with these
models, one should
bear in mind that such theories are at best effective descriptions of
the early/late universe, owing to a number of well-known problems.

Even in such a situation, scalar field models still remain quite
popular candidates for dark energy. An additional problem
with these theories--which traditionally has not been
discussed sufficiently--is that a good
mathematical theory must not be limited to the description of
a single side of the cosmic evolution:  it
should  rather provide a unified description of the whole
expansion history of the universe,
from the inflationary epoch to the onset of cosmic acceleration,
and beyond. Note that a similar drawback is also typical of
inflationary models, most
of which have problems with ending inflation and also fail
to describe realistically the late-time universe.

The purpose of this chapter is to show that, given a certain scale
factor (or Hubble parameter) for the universe expansion
history, one can in fact reconstruct it from a specific scalar
field theory. Using multiple scalars, the
reconstruction becomes easier due to the extra
freedom brought by the arbitrariness in the
scalar field potentials and kinetic factors. However, there are
subtleties in these cases that can be used advantageously, and
this makes the study of those models even more interesting.

Specifically, in this chapter we overview the reconstruction
technique for scalar theories with one, two, and an arbitrary
number $n$ of fields, as well as we consider a generalization of a type of Brans-Dicke theory, where the scalar field appears non-minimally coupled to the gravitational field, what could produce violations of gravitational tests at local scales. In order to satisfy the observational constraints  the so-called chameleon mechanism is studied (see Ref.~\cite{Khoury:2003rn}). Many explicit examples are
presented in which a unified, continuous  description of the
inflationary era and of the late-time cosmic acceleration epoch
is obtained in a rather simple and natural way.

\section{Unified inflation and late time acceleration in scalar
theory}

Let us consider a universe filled with matter with equation
of state $p_{m}=w_{m}\rho_{m}$ (here $w_{m}$ is a
constant) and a scalar field
which only depends on time. We will show that it is possible to
obtain both inflation and accelerated expansion at late times
by using a single scalar field $\phi$ (see also \cite{2006PhLB..632..597C} and
\cite{2006GReGr..38.1285N}).
In this case, the action is
\begin{equation}
S=\int dx^{4}\sqrt{-g}\left[ \frac{1}{2\kappa^{2}}R
 - \frac{1}{2} \omega (\phi)
\partial_{\mu} \phi \partial^{\mu }\phi -V(\phi )+L_{m}\right]\ ,
\label{1.eq:1}
\end{equation}
where $ \kappa^2 =8\pi G$, $V(\phi )$ being the scalar potential
and $ \omega (\phi)$ the kinetic function, respectively,
while $L_{m}$ is the matter
Lagrangian density.
Note that for convenience the kinetic factor is introduced.
At the final step of calculations, scalar field maybe always redefined so
that kinetic factor is absorbed in its definition.
 As we work in a spatially
flat Friedmann-Robertson-Walker (FLRW) spacetime, the metric is
given by
\begin{equation}
ds^{2}=-dt^{2}+a^2(t) \sum_{i=1}^{3} dx_{i}^{2}\ .
\label{1.eq:1.2}
\end{equation}
The corresponding FLRW equations are written as
\be
H^{2} = \frac{\kappa ^{2}}{3}\left( \rho _{m}
+\rho_{\phi}\right)\ , \quad \quad \dot H = -\frac{\kappa ^{2}}{2}\left(
\rho _{m}+p_{m}+\rho _{\phi }+p_{\phi }\right)\ ,
\label{1.eq:1.3}
\ee
with $\rho_{\phi}$ and $p_{\phi}$ given by
\be
\rho _{\phi } = \frac{1}{2} \omega (\phi )\, {\dot \phi}^{2}
+V(\phi)\ ,\quad \quad
p_{\phi } = \frac{1}{2} \omega (\phi ) \, {\dot \phi}^{2}-V(\phi)\ .
\label{1.eq:1.4}
\ee
Combining Eqs.~(\ref{1.eq:1.3}) and (\ref{1.eq:1.4}), one obtains
\be
\omega (\phi ) \, \dot{\phi ^{2}}
= -\frac{2}{\kappa^{2}}\dot{H}-(\rho _{m}+p_{m})\ ,\quad \quad
V(\phi ) = \frac{1}{\kappa ^{2}}
\left( {3H}^{2}+\dot{H} \right) -\frac{\rho_{m}-p_{m}}{2}\ .
\label{1.eq:1.5}
\ee
As the matter is not coupled to the scalar field, by using
energy conservation one has
\be
\dot{\rho _{m}}+3H(\rho _{m}+p_{m})=0\ ,\quad
\dot{\rho _{\phi }} +3H(\rho _{\phi }+p_{\phi })=0\ .
\label{1.eq:1.6}
\ee
 From the first equation, we get $\rho_{m}=\rho _{m0}a^{-3(1+ w_{m})}$.
We now consider the theory in which $V(\phi)$ and $\omega(\phi)$ are
\be
\omega (\phi ) = -\frac{2}{\kappa ^{2}}f^{\prime }(\phi )
 -(w_{m}+1)F_{0} \e^{-3(1+w_{m})F(\phi )}\ ,\quad
V(\phi ) = \frac{1}{\kappa ^{2}}
\left[ {3f(\phi)}^{2}+f^{\prime }(\phi ) \right]
+\frac{w_{m}-1}{2}F_{0}\, \e^{-3(1+w_{m})F(\phi )}\ ,
\label{1.eq:1.6a}
\ee
where $f(\phi) \equiv F'(\phi )$, $F$ is an arbitrary (but
twice differentiable) function of $\phi$, and $F_{0}$ is an
integration constant. Then, the following solution is found
(see
\cite{2006PhRvD..73d3512C, 2006PhLB..632..597C,2006GReGr..38.1285N,2006PhRvD..74h6009N}):
\be
\label{1.ST1}
\phi =t\ , \quad H(t)=f(t)\ ,
\ee
which leads to
\begin{equation}
a(t)=a_{0}\e^{F(t)}, \qquad a_{0}=\left(
\frac{\rho _{m0}}{F_{0}}\right) ^{\frac{1}{3(1+w_{m})}}.
\label{1.eq:1.7}
\end{equation}
We can study this system by analyzing the effective
EoS parameter which, using the FLRW
equations, is defined as
\begin{equation}
w_{\rm eff} \equiv \frac{p}{\rho }=-1-\frac{2\dot{H}}{{3H}^{2}}
\; ,
\label{1.eq:1.8}
\end{equation}
where
\be
\rho = \rho _{m}+\rho _{\phi }\ ,\quad
p = p_{m}+p_{\phi }\ . \label{1.1.8b}
\ee
Using the formulation above, one can present  explicit
examples of reconstruction as follows.

\subsection*{Example 1}

As a first example, we consider the following model:
\begin{equation}
f(\phi )=h_{0}^{2}\left( \frac{1}{t_{0}^{2}-\phi ^{2}}
+ \frac{1}{\phi^{2}+t_{1}^{2}}\right)\ . \label{1.eq:1.9}
\end{equation}
Using the solution~(\ref{1.eq:1.7}), the Hubble parameter and the
scale factor are given by
\be
H=h_{0}^{2}\left(
\frac{1}{t_{0}^{2}-t^{2}}+\frac{1}{t^{2}+t_{1}^{2}}\right)\ ,\quad
a(t)=a_{0}\left( \frac{t+t_{0}}{t_{0}-t}\right)
^{\frac{h_{0}^{2}}{2t_{0}}
}e^{\frac{h_{0}^{2}}{t_{1}}\arctan\frac{t}{t_{1}}}\ .
\label{1.eq:1.9a}
\ee
As one can see, the scale factor vanishes at $t=-t_{0}$, so we
can fix that
point as corresponding to the creation of the universe. On the
other hand, the kinetic function and the scalar
potential are given by Eqs.~(\ref{1.eq:1.6a}),
hence
\bea
\omega (\phi ) &=& -\frac{8}{\kappa^{2}}
\frac{h_{0}^{2}(t_{1}^{2}+t_{0}^{2})
\left( \phi ^{2}-\frac{t_{1}^{2}+t_{0}^{2}}{2}\right)
\phi}{(t_{1}^{2}
+\phi^{2})^{2}(t_{0}^{2}-\phi ^{2})^{2}}-(w_{m}+1)F_{0}
\e^{-3(w_{m}+1)F(\phi )}\ ,\nn
&&\nonumber \\
V(\phi ) &=& \frac{h_{0}^{2}(t_{1}^{2}+t_{0}^{2})}
{\kappa^{2}(t_{1}^{2}+\phi^{2})^{2}
(t_{0}^{2}-\phi ^{2})^{2}}\left[ 3h_{0}^{2}(t_{1}^{2}+t_{0}^{2})+4
\phi \left( \phi ^{2}-\frac{t_{1}^{2}+t_{0}^{2}}{2}\right) \right]
+\frac{w_{m}-1}{2}F_{0}\ e^{-3(w_{m}+1)F(\phi )}\ ,
\eea
where $F_{0}$ is an integration constant and
\be
\label{1.ST2}
F(\phi )=\frac{h_{0}^{2}}{2t_{0}}\ln\left( \frac{\phi
+t_{0}}{t_{0}-\phi }\right) +\frac{h_{0}^{2}}{t_{1}}\arctan \frac{\phi }{t_{1}}\ .
\ee
Then, using Eq.~(\ref{1.eq:1.8}), the effective EoS
parameter is written as
\begin{equation}
w_{\rm eff}=-1-\frac{8}{3h_{0}^{2}}\frac{t(t-t_{+})(t+t_{-})}{
(t_{1}^{2}+t_{0}^{2})^{2}} \;, \label{1.eq:1.10}
\end{equation}
where $t_{\pm }=\pm \sqrt{\frac{t_{0}^{2}-t_{1}^{2}}{2}}$. There
are two phantom phases that occur when $ t_{-}<t<0$ and
$t>t_{+}$, and
another two non-phantom phases for $-t_{0}<t<t_{-}$ and
$0<t<t_{+}$, during which $
w_{\rm eff}>-1$ (matter/radiation-dominated epochs). The first
phantom phase can be interpreted as an inflationary epoch, and
the second one as corresponding to the current accelerated
expansion, which will end in a Big Rip singularity when $
t=t_{0}$. Note that superacceleration ({\em i.e.}, $\dot{H}>0$)
is due to the negative sign of the kinetic
function $\omega (\phi)$, as for ``ordinary'' phantom fields
(to which one could reduce by redefining the scalar $\phi$).

\subsection*{Example 2}

As a second example, we consider the choice
\begin{equation}
f(\phi)=\frac{H_0}{t_s-\phi}+\frac{H_1}{\phi^2}\ .
\label{1.1.17}
\end{equation}
We take $H_0$ and $H_1$ to be  constants and $t_s$ as the Rip
time, as specified
below. Using (\ref{1.eq:1.6a}), we find that the kinetic
function and the scalar potential are
\bea
\omega(\phi) &=& -\frac{2}{\kappa^2}\left[
\frac{H_0}{(t_s-\phi)^2}-\frac{2H_1}{\phi^2}
\right] -(w_m+1)F_0\left(
t_s-\phi\right)^{3(1+w_m)H_0}\exp\left[
\frac{3(1+ w_m)H_1}{\phi}\right]\ , \nn
&&\nonumber \\
V(\phi) &=& \frac{1}{\kappa^2}\left[
\frac{H_0(3H_0+1)}{(t_s-\phi)^2}
+\frac{H_1}{\phi^3}\left(\frac{H_1}{\phi}-2 \right)\right]
+\frac{w_m-1}{2}
F_0\left(t_s-\phi \right)^{3(1+ w_m)H_0}
\e^{\frac{3(1+ w_m)H_1}{\phi}}\ ,
\label{1.1.18}
\eea
respectively. Then, through the solution~(\ref{1.eq:1.7}), we
obtain the Hubble parameter and the scale factor
\be
H(t)=\frac{H_0}{t_s-t}+\frac{H_1}{t^2}\ ,\quad
a(t)=a_0\left( t_s-t\right)^{-H_0} \e^{-\frac{H_1}{t}}\ .
\label{1.1.19}
\ee
Since $a(t)\to 0^{+} $ for $t\to 0$, we can fix $t=0$ as the
beginning of the universe. On the other hand, at $t=t_s$ the
universe reaches a
Big Rip singularity, thus we keep $t<t_s$. In order to study the different stages that our
model will pass through, we calculate the acceleration parameter and the first
derivative of the Hubble parameter. They are
\be
\dot H=\frac{H_0}{(t_s-t)^2}-\frac{2H_1}{t^3}\ ,\quad
\frac{\ddot a}{a}=H^2+\overset{.}{H}=\frac{H_0}{(t_s-t)^2}(H_0+1)
+ \frac{H_1}{t^2}\left(\frac{H_1}{t^2}-\frac{2H_1}{t}
+\frac{2H_0}{t_s-t}
\right)\ .
\label{1.1.20}
\ee
As we can observe, for $t$ close to zero, $\ddot a/a>0$, so that
the universe is accelerated during some time. Although this is
not a phantom
epoch, since $\dot H<0$, such stage can be interpreted as
corresponding to the
beginning of inflation. For $t>1/2$ but $t\ll t_s$, the universe
is in a
decelerated epoch ($\ddot a/a<0$). Finally, for $t$ close to $t_s$,
it turns out that $\dot H>0$, and then the universe is
superaccelerated, such acceleration being of phantom nature and
ending in a Big Rip singularity at $t=t_s$.

\subsection*{Example 3}

Our third example also exhibits unified inflation and
late time
acceleration, but in this case we avoid phantom phases and, therefore, Big Rip
singularities. We consider the following model:
\begin{equation}
f(\phi )=H_{0}+\frac{H_{1}}{\phi ^{n}}\ , \label{1.1.11}
\end{equation}
where $H_{0}$ and $H_{1}>0$ are constants and $n$ is a positive
integer (also constant). The case $n=1$ yields an initially
decelerated universe and a late time acceleration phase. We
concentrate on cases corresponding to $n>1$ which gives, in
general, three epochs: one of early acceleration (interpreted as
inflation), a second decelerated phase and,
finally, accelerated expansion at late times. In this model, the
scalar potential and the kinetic parameter are given, upon
use of Eqs.~(\ref{1.eq:1.6a}) and (\ref{1.1.11}), by
\bea
\omega (\phi ) &=&\frac{2}{\kappa ^{2}}\frac{nH_{1}}
{\phi^{n+1}}
 -(w_{m}+1)F_{0} \, \e^{-3(w_{m}+1)\left( H_{0}\phi
 -\frac{H_{1}}{(n-1)
\phi^{n-1}}\right) }, \label{1.1.12} \\
&&\nonumber\\
V(\phi ) &=&\frac{1}{\kappa ^{2}}\frac{3}{\phi ^{n+1}}\left[
\frac{\left(
H_{0}\phi ^{n/2}+H_{1}\right)^{2}}{\phi ^{n-1}}-\frac{nH_{1}}{3}\right]
+\frac{w_{m}-1}{2}F_{0} \, \e^{-3(w_{m}+1)\left( H_{0}\phi -
\frac{ H_{1}}{(n-1)\phi ^{n-1}}\right) }.
\eea
Then, the Hubble parameter given by the solution (\ref{1.eq:1.7})
can be written as
\be
H(t) = H_{0}+\frac{H_{1}}{t^{n}}\ , \label{1.1.13} \quad
a(t) = a_{0}\exp \left[
H_{0}t -\frac{H_{1}}{(n-1)t^{n-1}}\right]\ .
\ee
We can fix $t=0$ as the beginning of the universe because at
this point $a\to 0$, so $t>0$. The effective EoS
parameter~(\ref{1.eq:1.8}) is
\begin{equation}
w_{\rm eff}=-1+\frac{2nH_{1}t^{n-1}}{\left(
H_{0}t^{n}+H_{1}\right) ^{2}}\ .
\label{1.1.14}
\end{equation}
Thus, when $t\to 0$ then $ w_{\rm eff}\to -1$ and we
have an acceleration epoch, while for $t\to \infty $,
$w_{\rm eff} \to -1$ which
can be interpreted as late time
acceleration. To find the phases of acceleration and deceleration for $t>0$,
we study $\ddot a/a$, given by:
\begin{equation}
\frac{\ddot a}{a}=\dot H +H^{2}=-\frac{nH_{1}}{t^{n+1}}
+\left( H_{0}+\frac{H_{1}}{t^{n}}\right) ^{2}\ . \label{1.1.15}
\end{equation}
For sufficiently large values of $n$
we can find two positive zeros of this function, which means
two corresponding
phase transitions. They happen, approximately, at
\begin{equation}
t_{\pm }\approx \left[ \sqrt{nH_{1}} \,\, \frac{\left( 1\pm
\sqrt{1-\frac{4H_{0} }{n}} \, \right)}{2H_{0}} \, \right]^{2/n}\ ,
\label{1.1.16}
\end{equation}
so that, for $0<t<t_{-}$, the universe is in an accelerated
phase interpreted as an inflationary epoch; for $t_{-}<t<t_{+}$
it is in a decelerated phase (matter/radiation dominated); and,
finally, for $t>t_{+}$ one obtains late time acceleration, which
is in agreement with the current cosmic expansion.

\subsection*{Example 4}

As our last example, we consider another model unifying
early universe inflation and the accelerating expansion
of the present universe. We may choose $f(\phi)$ as
\be
\label{1.uf0}
f(\phi)=\frac{H_i + H_l c  \e^{2\alpha\phi}}{1 + c
\e^{2\alpha\phi}}\ ,
\ee
which gives the Hubble parameter
\be
\label{1.uf1}
H(t)=\frac{H_i + H_l c \e^{2\alpha t}}{1 + c \e^{2\alpha t}}\ .
\ee
Here $H_i$, $H_l$, $c$, and $\alpha$ are positive constants.
In the early universe ($t\to -\infty$), we find that $H$
becomes a constant
$H\to H_i$ and at late times ($t\to +\infty$), $H$ becomes a
constant again $H\to H_l$.
Then $H_i$ could be regarded as the effective cosmological
constant driving inflation, while $H_l$ could be a small
effective constant generating the late acceleration.
Then, we should assume $H_i \gg H_l$.
Hence, if we consider the model with action
\bea
\label{1.uf2}
S &=& \int d^4 x \sqrt{-g}\left\{\frac{R}{2\kappa^2} -
\frac{\omega(\phi)}{2}\partial_\mu \phi
\partial^\mu \phi - V(\phi)\right\}\ ,\nn
&&\nonumber \\
\omega(\phi) &\equiv& - \frac{f'(\phi)}{\kappa^2}
= \frac{2\alpha\left(H_i - H_l\right)c
\e^{2\alpha\phi}}{\kappa^2
\left(1 + c\e^{2\alpha\phi}\right)^2}\ ,\nn
&&\nonumber \\
V(\phi) &\equiv& \frac{3f(\phi)^2 + f'(\phi)}{\kappa^2}
= \frac{3 H_i^2 + \left\{6 H_i H_l - 2\alpha\left( H_i - H_l
\right)\right\}c\e^{2\alpha\phi}
+ c^2 H_l^2  \e^{4\alpha\phi}}{\kappa^2 \left(1 +
c\e^{2\alpha\phi}\right)^2}\ ,
\eea
we can realize the Hubble rate given by (\ref{1.uf1}) with $\phi=t$.
If we redefine the scalar field as
\be
\label{1.uf2b}
\varphi \equiv \int d\phi \sqrt{\omega(\phi)}
= \frac{\e^{\alpha\phi}}{\kappa}\sqrt{\frac{2(a-b)c}{\alpha}}\ ,
\ee
the action $S$ in (\ref{1.uf2}) can be rewritten in the canonical
form
\be
\label{1.uf2c}
S =  \int d^4 x \sqrt{-g}\left\{\frac{R}{2\kappa^2} -
\frac{1}{2}\partial_\mu \varphi
\partial^\mu \varphi - \tilde V(\varphi)\right\}\ ,
\ee{equation}
where
\be
\tilde V(\varphi) = V(\phi)
= \frac{3 H_i^2 + \frac{\kappa^2 \alpha \left\{6 H_i H_l
 - 2 \alpha\left( H_i - H_l \right)\right\}}{2\left( H_i - H_l
\right)}\varphi^2
+ \frac{3  \kappa^4 H_l^2 \alpha^2}{4\left(H_i -
H_l\right)^2}\varphi^4}
{\kappa^2  \left(1 + \frac{\kappa^2 \alpha}{2\left( H_i - H_l
\right)}\varphi^2\right)^2}\; .
\ee
One should note that $\phi\to - \infty$ corresponds to
$ \varphi\to 0$ and $V\sim 3H_i^2/\kappa^2$, while  $\phi\to +
\infty$  corresponds to $\varphi\to \infty$ and $V\sim
3H_l^2/\kappa^2$, as expected.
At early times ($\varphi\to 0$), $\tilde V(\varphi)$ behaves  as
\be
\label{1.uf3c}
\tilde V(\varphi) \sim \frac{3H_i^2}{\kappa^2} \left\{
1 - \frac{\kappa^2 \alpha \left(3 H_i + \alpha
\right)}{3H_i^2}\varphi^2
+ {\cal O}\left(\varphi^2\right)\right\}\ .
\ee
At early times, $\phi<0$ and therefore, from Eq.~(\ref{1.uf2b}),
we find
$\kappa\varphi \sqrt{\alpha/2\left(H_i - H_l\right)c}
\ll 1$, from which it follows that
\bea
\label{1.uf3d}
&& \frac{1}{3\kappa^2}\frac{{\tilde V}'(\varphi)}{\tilde
V(\varphi)^2}
\sim \frac{4\alpha^2 \kappa^2\left(3 H_i +
\alpha\right)^2}{27H_i^4}\varphi^2
< \frac{8\alpha \left(3 H_i + \alpha\right)^2 \left(H_i -
H_l\right)c}{27H_i^4}\ ,\nn
&&\nonumber \\
&& \frac{1}{3\kappa^2}\frac{\left|{\tilde V}''(\varphi)
\right|}{\tilde V(\varphi)}
\sim \frac{2\left(3H_i + \alpha\right)\alpha}{9H_i^2}\ .
\eea
Then, if $\alpha\ll H_i$, the slow-roll conditions can be
satisfied.

We may include matter with constant EoS parameter $w_m$. Then
$\omega(\phi)$ and $V(\phi)$
are modified as
\bea
\label{1.uf3}
\omega(\phi) &\equiv& - \frac{f'(\phi)}{\kappa^2}
 - \frac{w_m + 1}{2}g_0 \e^{-3(1+w_m)g(\phi)}\ ,\nn
&&\nonumber \\
V(\phi) &\equiv& \frac{3f(\phi)^2 + f'(\phi)}{\kappa^2}
+ \frac{w_m -1}{2}g_0 \e^{-3(1+w_m)g(\phi)}\ ,\nn
&&\nonumber \\
g(\phi) &\equiv& \int d\phi f(\phi) = H_l \phi + \frac{H_i -
H_l}{2}
\ln \left(c + \e^{-2\alpha \phi}\right)\ .
\eea
The matter energy density is then given by
\be
\label{1.uf4}
\rho_m = \rho_0 a^{-3(1+w_m)} = g_0 \e^{-3(1+w_m)g(t)}
= g_0 \left( c + \e^{-2\alpha t}\right)^{-3(1+w_m)(H_i-H_l)/2} \e^{-3(1+w_m)H_l t}\ .
\ee
In the early universe $t\to -\infty$, $\rho_m$ behaves as
\be
\label{1.uf5}
\rho_m \sim g_0 \e^{3(1+w_m)\left(2\alpha - H_l\right) t}\ .
\ee
On the other hand, the energy density of the scalar field behaves as
\be
\label{1.uf6}
\rho_\varphi = \frac{1}{2}{\dot\varphi}^2 + \tilde V(\varphi)
= \frac{\omega(\phi)}{2}{\dot\phi}^2 + V(\phi) \to \frac{3H_i^2}{\kappa^2}\ .
\ee
Then, if $2\alpha < H_l$ (and $w_m>-1$), the matter contribution
could be neglected in comparison with the
scalar field contribution.

Now let the present time be $t=t_0$. Then, we find that
\bea
\label{1.uf7}
\Omega_m &\equiv& \frac{\kappa^2 \rho_m}{3 H^2} \nn
&&\nonumber \\
&=& \frac{\kappa^2 g_0 \e^{4\alpha t_0}
\left( c + \e^{-2\alpha t_0}\right)^{-3(1+w_m)(H_i-H_l)/2\alpha + 2} \e^{-3(1+w_m)H_l t_0}}
{3\left( H_i + H_l c \e^{2\alpha t_0}\right)^2}\;,
\eea
and $\Omega_\phi = 1 - \Omega_m$. If we assume $\alpha t_0 \gg
1$, we find
\be
\label{1.uf8}
\Omega_m \sim \frac{\kappa^2 g_0}{3H_l^2}
c^{-3(1+w_m)(H_i-H_l)/2\alpha} \e^{-3(1+w_m)H_l t_0} \ .
\ee
Hence, we may choose the parameters so that $\Omega_m \sim
0.27$, which could be consistent with the observed data.
This model provides a quite realistic picture of the unification
of the inflation with the present cosmic speed-up.

\section[Non-minimally coupled scalar theory]{Accelerated expansion in the non-minimally
curvature-coupled scalar theory}

In the preceding section we have considered an action,
(\ref{1.eq:1}), in which the
scalar field is minimally coupled to gravity. In the present
section, the scalar
field couples to gravity through the Ricci scalar
(see \cite{2000hep.th....9053F} for a review on cosmological
applications). We begin from the action
\begin{equation}
S=\int d^{4}x\sqrt{-g} \left[(1+f(\phi))\frac{R}
{\kappa^{2}}-\frac{1}{2} \omega (\phi ) \partial_{\mu }\phi
\partial^{\mu }\phi -V(\phi )\right]\ ,
\label{1.2.1}
\end{equation}
where $f(\phi )$ is an arbitrary function of the scalar field
$\phi $. Then,
the effective gravitational coupling depends on $\phi $, as
$\kappa_{eff}=\kappa [1+f(\phi)]^{-1/2}$. One can work in the
Einstein frame,
by performing the scale transformation
\begin{equation}
g_{\mu \nu }=[1+f(\phi )]^{-1} \widetilde{g}_{\mu \nu }\ .
\label{1.2.2}
\end{equation}
The tilde over $g$ denotes an Einstein
frame quantity. Thus, the action~(\ref{1.2.1}) in such a frame
assumes the form \cite{2007PhRvD..75b3501F}
\begin{equation}
S=\int d^{4}x\sqrt{-\widetilde{g}}\left\{
\frac{\widetilde{R}}{2 \kappa ^{2}}
 -\left[ \frac{ \omega (\phi )}{2(1+f(\phi))}+\frac{6}
{\kappa^{2}(1+f(\phi ))}
\left( \frac{ d(1+f(\phi)^{1/2})}{d\phi }\right) ^{2}\right]
\partial_{\mu }\phi
\partial^{\mu } \phi -\frac{V(\phi)}{[ 1+f(\phi )]^{2}}\right\}\
.
\label{1.2.3}
\end{equation}
The kinetic function can be written as $W(\phi )=\frac{ \omega
(\phi )}{1+f(\phi)}
+\frac{3}{\kappa ^{2}(1+f(\phi ))^{2}}
\left( \frac{df(\phi )}{d\phi }\right) ^{2}$, and the extra term in the
scalar potential can be absorbed by defining the new potential
$U(\phi )=\frac{ V(\phi )}{ \left[ 1+f(\phi ) \right]^2}$, so
that we recover
the action~(\ref{1.eq:1}) in the
Einstein frame, namely
\begin{equation}
S=\int dx^{4} \sqrt{-\widetilde{g}}\left(
\frac{\widetilde{R}}{ \kappa ^{2}}-
\frac{1}{2} \, W( \phi ) \, \partial _{\mu }\phi \partial ^{\mu}
\phi -U(\phi)\right)\ .
\label{1.2.4}
\end{equation}
We assume that the metric is FLRW and spatially flat in this
frame
\begin{equation}
d\widetilde{s}^{2}=-d\widetilde{t}^{2}
+\widetilde{a}^{2}(\widetilde{t})\sum_{i}dx_{i}^{2}\ ,
\label{1.2.5a}
\end{equation}
then, the equations of motion in this frame are given by
\begin{eqnarray}
&& \widetilde{H}^{2}=\frac{\kappa ^{2}}{6}\rho _{\phi } \;,\\
&&\nonumber \\
&& \dot{\widetilde{H}}=-\frac{\kappa ^{2}}{{4}}\left( \rho
_{\phi } + p_{\phi}\right)\ , \label{1.2.6}\\
&&\nonumber \\
&& \frac{d^2 \phi}{d\tilde{t}^2} +3\tilde{H} \,
\frac{d\phi}{d\tilde{t}} +\frac{1}{2W(\phi)}\left[ W'(\phi)
\left( \frac{d\phi}{d\tilde{t}}\right)^2 +2U'(\phi) \right] =0
\;,
\end{eqnarray}
where $\rho _{\phi }=\frac{1}{2}W(\phi ){\dot \phi}^{2} +
U(\phi )$, $p_{\phi }=\frac{1}{2}W(\phi ){\dot \phi}^{2} -
U(\phi )$, and
the Hubble parameter is $\widetilde{H}
\equiv \frac{1}{\widetilde{a}}
\frac{d\widetilde{a}}{d\widetilde{t}}$. Then,
\be
W(\phi )\dot{\phi}^{2} =-4\dot{\widetilde{H}}\ ,\quad \quad
U(\phi )=6\widetilde{H}^{2}+2\dot{\widetilde{H}}. \label{1.2.7}
\ee
Note that $ \dot{\widetilde{H}} >0$ is equivalent to $W<0$;
superacceleration is due to the ``wrong'' (negative) sign of the
kinetic energy, which is the distinctive feature of a phantom
field. The scalar field could be redefined to eliminate
the factor $W(\phi)$, but this would not correct the sign of the
kinetic energy.

If we choose $W(\phi)$ and $U(\phi)$ as $ \omega(\phi)$ and
$V(\phi)$ in~(\ref{1.eq:1.6a}),
\be
W (\phi ) = -\frac{2}{\kappa ^{2}}g^{\prime }(\phi )\ ,\quad
U(\phi ) = \frac{1}{\kappa ^{2}} \left[ {3 g(\phi )}^{2}+ g'
(\phi ) \right]\ ,
\label{1.ST4}
\ee
by using a function $g(\phi)$ instead of $f(\phi)$
in~(\ref{1.eq:1.6a}), we find a solution as in
(\ref{1.ST1}),
\begin{equation}
\phi =\widetilde{t}\ , \quad \widetilde{H}(\widetilde{t})=g(\widetilde{t})\ .
\label{1.2.8}
\end{equation}
In (\ref{1.ST4}) and hereafter in this section,  we have dropped
the matter contribution for simplicity.

We consider the de Sitter solution in this frame,
\begin{equation}
\widetilde{H}=\widetilde{H}_{0}=\mbox{const.\ } \to \
\widetilde{a}( \widetilde{t})
=\widetilde{a}_{0}e^{\widetilde{H}_{0}\widetilde{t}_{{}}} \;.
\label{1.2.9}
\end{equation}
We will see below that accelerated expansion can be obtained
in the original
frame corresponding to the Einstein frame~(\ref{1.2.5a}) with
the solution~(\ref{1.2.9}),
by choosing an appropriate function $f(\phi )$. From (\ref{1.2.9})
and
the definition
of $W(\phi )$ and $U(\phi )$, we have
\be
W(\phi )=0 \ \to \ \omega (\phi )=-\frac{3}{\left[ 1+f(\phi)
\right] \kappa^{2}}
\left[ \frac{df(\phi )}{d\phi }\right]^{2}\ ,\quad
U(\phi )=\frac{6}{\kappa ^{2}}\widetilde{H}_{0}^{2}\ \to \
V(\phi )=\frac{6}{\kappa ^{2}}\widetilde{H}_{0}^{2}[1+f(\phi )]^{2}\ . \label{1.2.10}
\ee
Thus, the scalar field has a non-canonical kinetic term in
the original frame, while in the Einstein frame the latter can
be
positive, depending on $W(\phi)$. The correspondence between
conformal frames can be made explicit through the conformal
transformation~(\ref{1.2.2}). Assuming a spatially flat FLRW metric
in the
original frame,
\begin{equation}
ds^{2}=-dt^{2}+a^{2}(t)\sum_{i=1}^3 dx_{i}^{2}\ , \label{1.2.10a}
\end{equation}
then, the relation between the time coordinate and the scale
parameter in these frames is given by
\be
t=\int \frac{d\widetilde{t}}{([1+f(\widetilde{t})]^{1/2}}\ ,
\quad a(t)=[1+f(\widetilde{t})]^{-1/2} \,
\widetilde{a}(\widetilde{t})\ .
\label{1.2.11}
\ee
Now let us discuss the late-time acceleration in the model under
discussion.

\subsection*{Example 1}

As a first example, we consider the coupling function
between the scalar field and the Ricci scalar
\begin{equation}
f(\phi )=\frac{1-\alpha \phi }{\alpha \phi }\ , \label{1.2.12}
\end{equation}
where $\alpha $ is a constant. Then, from~(\ref{1.2.10}), the
kinetic function $ \omega(\phi)$ and the
potential $V(\phi)$ are
\be
\omega(\phi)=-\frac{3}{\kappa^{2}\alpha^{2}}
\, \frac{1}{\phi^{3}}\ ,
\quad \quad
V(\phi)=\frac{6\widetilde{H}_{0}}{\kappa^{2}
\alpha^{2}}\frac{1}{\phi^{2}}\ ,
\label{1.2.13}
\ee
respectively. The solution for the current example is found to
be
\be
\phi (t)=\widetilde{t} =\frac{1}{\alpha }\left(
\frac{3\alpha}{2} \, t\right)^{2/3}\ , \quad \quad
a(t)=\widetilde{a}_{0}\left( \frac{3\alpha }{2} \,
t\right)^{1/3}
\exp \left[ \frac{\widetilde{H}_{0}^{{}}}{\alpha } \left(
\frac{3\alpha}{2} \, t\right)^{2/3}\right]\ .
\label{1.2.14}
\ee
We now calculate the acceleration parameter to study the
behavior of the scalar parameter in the original frame,
\begin{equation}
\frac{\ddot a}{a}=-\frac{2}{9}\frac{1}{t^{2}}+\widetilde{H}_{0}
\left( \frac{2}{3\alpha }\right)^{1/3}\left[ \frac{1}{t^{4/3}}
+ \widetilde{H}_{0}\left( \frac{2}{3\alpha }\right)^{1/3}\frac{1}{t^{2/3}}\right]\ .
\label{1.2.15}
\end{equation}
We observe that for small values of $t$ the acceleration is
negative; after that we get accelerated expansion for large
$t$; finally, the universe ends with zero acceleration as $t\to
\infty $. Thus, late time accelerated expansion is reproduced by
the action~(\ref{1.2.1}) with the function $f(\phi )$ given by
Eq.~(\ref{1.2.12}).

\subsection*{Example 2}

As a second example, consider the function
\begin{equation}
f(\phi )=\phi -t_{0}\ .
\label{1.2.16}
\end{equation}
 From (\ref{1.2.11}), the kinetic term and the scalar potential are, in this case,
\be
\omega(\phi)=-\frac{3}{\kappa ^{2}}\frac{1}{(1+\phi-t_{0})}\ ,
\quad \quad
V(\phi)=\frac{6\widetilde{H}_{0}^{2}}{\kappa ^{2}}(1+\phi-t_{0})\ .
\label{1.2.17}
\ee
The solution in the original (Jordan) frame reads
\be
\phi(t)=\frac{t^{2}}{4}+t_{0}-1\ , \quad
a(t)= \frac{2\widetilde{a}_{0}}{t} \exp\left[\widetilde{H}_{0}
\left(\frac{t^{2}}{4} +t_{0}-1\right) \right]\ ,
\ee
and the corresponding acceleration is
\begin{equation}
\frac{\ddot a}{a}=\frac{1}{t^{2}}\left[\frac{\widetilde{H}_{0}t^{2}}{2}
+\left(\frac{\widetilde{H}_{0}t^{2}}{2}-1 \right)^{2}+1 \right]\ .
\end{equation}
Notice that this solution describes acceleration at every time $t$ and, for
$t\to \infty$, the acceleration tends to a constant value, as in de Sitter spacetime,
hence similar to what happens in the Einstein frame.
Thus, we have proved here that it is possible to reproduce accelerated expansion in both frames,
by choosing a convenient function for the coupling $f(\phi)$.

\subsection{Chameleon mechanism}

The kind of theories described by the above action (\ref{1.2.1}) have problems when matter is included and one perform a conformal transformation and the Einstein frame is recovered, then the local gravity tests may be violated by a faith force that appeared on a test particle and the violation of the Equivalence Principle is presented. This kind of problems are well constrained by the experiments to a certain value of the coupling parameter as it is pointed below. Recently, a very interesting idea originally proposed in Ref. \cite{Khoury:2003rn} avoids the constrains from local gravity tests in such a way that the effects of the scalar field are neglible at small scales but it acquires an important role for large scales, whose effects  may produce the current acceleration of the Universe \\
Let us start by rewriting the action in the Einstein frame (\ref{1.2.3}) in a similar form as the original Brans-Dicke action by redefining the scalar field $\phi$ and rewriting the kinetic term $\omega(\phi)$ in terms of the coupling $(1+f(\phi))$, then the action is given by:
\be
S_E=\int d^4x\sqrt{-\widetilde{g}}\left[\frac{\widetilde{R}}{2\kappa^2}-\frac{1}{2×}\partial_\mu\sigma\partial^\mu\sigma-U(\sigma)+\e^{4\beta\sigma}\widetilde{L_m} \right]\ , 
\label{1.C3.1}
\ee  
where:
\be
\e^{-2\beta\sigma}=1+f(\phi)\ ,
\label{1.C3.2}
\ee
here $\beta$ is a constant. As it is seen in the action (\ref{1.C3.1}), the matter density lagrangian couples to the scalar field $\sigma$, such that a massive test particle will be under a fifth force, and the equation of motion will be:
\be
\ddot{x}^\mu +\widetilde{\Gamma}^\mu_{\lambda\nu}\dot{x}^\lambda \dot{x}^\nu=-\beta\partial^\mu\sigma\ ,
\label{1.C3.3}
\ee
where $x^\mu$ represents the four-vector describing the path of a test particle moving in the metric $\widetilde{g}^(\mu\nu)$, and $\widetilde{\Gamma}^\mu_{\nu\lambda}$ are the Christoffel symbols for the metric $\widetilde{g}^{\mu\nu}$. From the equation (\ref{1.C3.3}), the scalar field $\sigma$ can be seen as a potential from a force given by:
\be
F_{\sigma}=-M\beta\partial^\mu\sigma\ ,
\label{1.C3.4}
\ee
where $M$ is the mass of a test particle. Then, this kind of theories reproduces a fifth force which it has been tested by the experiments to a limit $\beta<1.6 \times 10^{-3}$ (Ref. \cite{Bertotti:Nature_2003}). The aim of the chameleon mechanism is that it makes the fifth force neglible for small scales passing the local tests. Such mechanism works in the following way, by varying the action (\ref{1.C3.1}) with respect the scalar field $\sigma$, the equation of motion for the scalar field  is obtained:
\be
\bigtriangledown^2\sigma=U_{,\sigma}-\beta \e^{4\beta\sigma}\widetilde{g}^{\mu\nu}\widetilde{T}_{\mu\nu}\ , 
\label{1.C3.5}
\ee
where the energy-momentum tensor is given by $\widetilde{T}^{\mu\nu}=\frac{2}{\sqrt{-\widetilde{g}}×}\frac{\partial L_m}{\partial\widetilde{g}^{\mu\nu}×}$. For simplicity we restrict to dust matter $\widetilde{g}^{\mu\nu}\widetilde{T}_{\mu\nu}=-\widetilde{\rho}_m$, where the energy density may be written in terms of the conformal transformation (\ref{1.C3.2}) as $\rho_m=\widetilde{\rho}_m\e^{3\beta\sigma}$. Then, The equation for the scalar field (\ref{1.C3.5}) is written in the following way:
\be
\bigtriangledown^2\sigma=U_{,\sigma}+\beta\rho_m \e^{\beta\sigma}\ , 
\label{1.C3.6}
\ee
here it is shown that the dynamics of the scalar field depends on the matter energy density. We may write the right side of the equation (\ref{1.C3.6}) as an effective potential $U_{eff}=U(\sigma)+\rho_m\e^{\beta\sigma}$. Then, the behavior of the scalar field will depend on the effective potential, and the solutions for the equation (\ref{1.C3.6}) are given by studying $U_{eff}$. By imposing to the scalar potential $U(\sigma)$ to be a monotonic decreasing function, the effective potential will have a minimum that will govern the solution for the scalar field (for more details see \cite{Khoury:2003rn}), this minimum is given by:
\be
 U_{,\sigma}(\sigma_{min})+\beta\rho_m \e^{\beta\sigma_{min}}=0\ ,
\label{1.C3.7}
\ee
which depends on the local matter density. At this minimum, the scalar field mass will be given by:
\be
m^2_{\sigma}=U_{,\sigma\sigma}(\sigma_{min})+\beta^2\rho_m \e^{\beta\sigma_{min}}\ .
\label{1.C3.8}
\ee
Then, because of the characteristics of the scalar potential $U(\sigma)$, larger values of the local density $\rho_m$ corresponds to small values of $\sigma_{min}$ and large values of $m_{\sigma}$, so it is possible  for sufficiently large values of the scalar field mass to avoid Equivalence Principle violations and fifth forces on the Earth. As the energy density $\rho_m$ becomes smaller, the scalar field mass $m_{\sigma}$ decreases and $\sigma_{min}$ increases, such that at large scales (when $\rho\sim H^2_0$), the effects of the scalar field become detected, where the accelerated expansion of the Universe may be a possible effect. Then, one may restrict the original scalar potential $V(\phi)$ and the coupling $(1+f(\phi))$ trough the mechanism shown above. The effective potential $U_{eff}$ is written in terms of $\phi$ by the equation (\ref{1.C3.2}) as:
\be
U_{eff}(\phi)= U(\phi)+\frac{\rho_m}{(1+f(\phi))^{1/2}×}\ ,
\label{1.C3.9}
\ee
where $U(\phi)=V(\phi)/(1+f(\phi))$. Then, by giving a function $f(\phi)$ and a scalar potential $V(\phi)$, one may construct using the conditions described above on the mass $m_\sigma$, a cosmological model that reproduces the current accelerated expansion and at the same time, avoid the local test of gravity.

\section{Reconstruction of non-minimally coupled scalar field
theory}

We now consider the reconstruction problem in the non-minimally
coupled scalar field theory, or the Brans-Dicke theory. We begin
with the same scalar-tensor theory with constant
parameters $\phi_0$ and $V_0$:
\be
\label{1.BD1}
S=\int d^4 x \sqrt{-g} \left[ \frac{R}{2\kappa^2}
 - \frac{1}{2}\partial_\mu \phi \partial^\mu \phi
 - V(\phi) \right] \ ,\quad \quad V(\phi)=V_0 \,
\e^{-2\phi/\phi_0}\ ,
\ee
which admits the exact solution
\be
\label{1.BD2}
\phi =\phi_0\ln \left|\frac{t}{t_1}\right|\ ,\quad
H=\frac{\kappa^2\phi_0^2}{2 t}\ , \quad
t_1^2 \equiv \frac{\gamma \phi_0^2 \left(\frac{3\gamma
\kappa^2\phi_0^2}{2} - 1\right)}{2V_0}\ .
\ee
We choose $\phi_0^2 \kappa^2>2/3$ and $V_0>0$ so that
$t_1^2>0$. For this solution, the metric is given by
\be
\label{1.BD3}
ds^2 = - dt^2 + a_0^2
\left(\frac{t}{t_0}\right)^{\kappa^2\phi_0^2}\sum_{i=1}^3
\left(dx^i\right)^2\ ,
\ee
which can be transformed into the conformal form:
\be
\label{1.BD4}
ds^2 = a_0^2 \left(\frac{\tau}{ \tau_0}
\right)^{-\frac{\kappa^2 \phi_0^2}{\frac{\kappa^2\phi_0^2}{2} - 1}}
\left( - d\tau^2 + \sum_{i=1}^3 \left(dx^i\right)^2\right)\ .
\ee
Here
\be
\label{1.BD5}
\frac{\tau}{\tau_0} =
 - \left(\frac{t}{t_0}\right)^{-\frac{\kappa^2\phi_0^2}{2} +1 }\ ,\quad
\tau_0\equiv \frac{t_0}{a_0 \left(\frac{\kappa^2\phi_0^2}{2} - 1\right)}\ .
\ee
and therefore, by using (\ref{1.BD2}), one finds
\be
\label{1.BD6}
\frac{\tau}{\tau_0}=
 - \left(\frac{t_1}{t_0}\right)^{-\frac{\kappa^2}{2} +1}
\e^{- \frac{1}{-\frac{\kappa^2}{2} +1} \frac{\phi}{\phi_0}}\ .
\ee

We now consider an arbitrary cosmology given by the metric
\be
\label{1.BD7}
d{\tilde s}^2 = f(\tau)\left( - d\tau^2
+ \sum_{i=1}^3 \left(dx^i\right)^2\right)\ ,
\ee
where $\tau$ is the conformal time. Since
\be
\label{1.BD8}
ds^2=\frac{1}{f(\tau)}a_0^2 \left(\frac{\tau}{\tau_0}
\right)^{-\frac{\kappa^2\phi_0^2}{\frac{\kappa^2\phi_0^2}{2} - 1}}
d\tilde s^2 = \e^{\varphi} d\tilde s^2\ , \quad
\e^{\varphi} \equiv \frac{a_0^2
\left(\frac{t_1}{t_0}\right)^{\kappa^2\phi_0^2} \e^{\kappa^2 \phi_0 \phi}}
{f\left( - \tau_0
\left(\frac{t_1}{t_0}\right)^{-\frac{\kappa^2}{2} +1}
\e^{- \frac{1}{-\frac{\kappa^2}{2} +1}
\frac{\phi}{\phi_0}}\right)}\ ,
\ee
if we begin with the action in which $ g_{\mu\nu}$ in
(\ref{1.BD1}) is replaced by $\e^{\varphi}{\tilde g}_{\mu\nu}$,
\be
\label{1.BD9}
S= \int d^4 x \sqrt{-\tilde g} \, \e^{\varphi(\phi) }\left\{
\frac{R}{2\kappa^2} - \frac{1}{2}\left[
1 - 3\left(\frac{d\varphi}{d\phi}\right)^2\right]
\partial_\mu \phi \partial^\mu \phi
 - \e^{\varphi(\phi)} V(\phi) \right\}\ ,
\ee
we obtain the solution~(\ref{1.BD7}).

\subsection*{Example}

By using the conformal time $\tau$, the metric of de Sitter
space
\be
\label{1.dS}
ds^2 = - dt^2 + \e^{2H_0 t}\sum_{i=1}^3 \left(dx^i\right)^2\ ,
\ee
can be rewritten as
\be
\label{1.dS2}
ds^2 = \frac{1}{H_0^2\tau^2}\left( - d\tau^2 +
\sum_{i=1}^3 \left(dx^i\right)^2\right)\ .
\ee
Here $\tau$ is related to $t$ by $\e^{- H_0 t}=- H_0\tau$.
Then $t\to -\infty$ corresponds to $\tau \to +\infty$ and $t\to
+\infty$ corresponds to $\tau \to 0$.

As an example of $f(\tau)$ in (\ref{1.BD7}), we may consider
\be
\label{1.dS3}
f(\tau)=\frac{\left(1 + H_L^2\tau^2\right)}{H_L^2\tau^2\left(1
+ H_I^2 \tau^2\right)}\ ,
\ee
where $H_L$ and $H_I$ are constants. At early times in the
history of the
universe $\tau\to \infty$ (corresponding to $t\to -\infty$),
$f(\tau)$ behaves as
\be
\label{1.dS4}
f(\tau)\to \frac{1}{H_I^2\tau^2}\ .
\ee
Then the Hubble rate is given by a constant $H_I$, and therefore
the universe is asymptotically de Sitter space, corresponding to
inflation. On the other hand, at late times $\tau\to 0$
(corresponding to $t\to + \infty$), $f(\tau)$ behaves as
\be
\label{1.dS5}
f(\tau)\to \frac{1}{H_L^2\tau^2}\ .
\ee
Then the Hubble rate is again a constant $H_L$, which may
correspond to the late time acceleration of the universe.
This proves that our reconstruction program can be applied directly to the
non-minimally coupled scalar theory.

\section{Late time acceleration and inflation with several
scalar fields}

In this section we begin by considering a model with two scalar
fields minimally coupled to gravity (see \cite{2004PhRvD..70d3539E,2006GReGr..38.1285N,2007JPhCS..66a2005N}).
Such models are used, for example, in reheating scenarios after
inflation.

An additional degree of freedom appears in this case, so that
for a given solution we may
choose different conditions
on the scalar fields, as shown below. It is possible to restrict
these conditions by studying the perturbative regime for each solution. The action
we consider is
\begin{equation}
S=\int\sqrt{-g}\left[\frac{R}{2\kappa^{2}}- \omega(\phi)
\partial_{\mu}\phi\partial^{\mu}\phi-\sigma(\chi)\partial_{\mu}\chi
\partial^{\mu}\chi-V(\phi,\chi)\right]\ , \label{1.3.1}
\end{equation}
where $ \omega(\phi)$ and $\sigma(\chi)$ are the kinetic terms,
which depend
on the fields $\phi$ and $\chi$, respectively. We again assume a flat
FLRW metric.
The Friedmann equations are written as
\be
H^{2}=\frac{\kappa^{2}}{3}\left[\frac{1}{2}
\omega(\phi)\dot{\phi}^{2}
+ \frac{1}{2}\sigma(\chi)\dot{\chi}^{2}+V(\phi,\chi)\right]\ ,\quad
\dot{H}=-\frac{\kappa^{2}}{2}\left[ \omega(\phi)\dot{\phi}^{2}
+\sigma(\chi)\dot{\chi}^{2}\right]\ . \label{1.3.3}
\ee
By means of a convenient transformation, we can always redefine
the scalar fields so that we can write $\phi=\chi=t$. The
scalar field equations are given by
\be
\omega(\phi)\ddot{\phi}+\frac{1}{2} \omega'(\phi)\dot{\phi}^{2}
+3H \omega(\phi)\dot{\phi}+\frac{\partial V(\phi,\chi)}{\partial
\phi}=0\ , \quad
\sigma(\chi)\ddot{\chi}+\frac{1}{2}\sigma'(\chi)\dot{\chi}^{2}
+3H\sigma(\chi)\dot{\chi}+\frac{\partial V(\phi,\chi)}{\partial \chi}=0\ . \label{1.3.4}
\ee
Then, for a given solution $H(t)=f(t)$, and combining
the first Friedmann equation with each scalar field equation, respectively, we find
\be
\omega(\phi)=-\frac{2}{\kappa^{2}}\frac{\partial
f(\phi,\chi)}{\partial \phi}\ , \quad
\sigma(\chi)=-\frac{2}{\kappa^{2}}\frac{\partial f(\phi,\chi)}{\partial \chi}\ , \label{1.3.5}
\ee
where the function $f(\phi,\chi)$ carries down to $f(t,t)\equiv f(t)$, and is defined
as
\begin{equation}
f(\phi,\chi)=-\frac{\kappa^{2}}{2}\left[\int \omega(\phi)d\phi
+\int\sigma(\chi)d\chi\right]\ .\label{1.3.6}
\end{equation}
The scalar potential can be expressed as
\begin{equation}
V(\phi,\chi)=\frac{1}{\kappa^{2}}\left[3f(\phi,\chi)^{2}
+\frac{\partial f(\phi,\chi)}{\partial \phi}
+\frac{\partial f(\phi,\chi)}{\partial \chi}\right]\ ,\label{1.3.7}
\end{equation}
and the second Friedmann equation reads
\begin{equation}
 -\frac{2}{\kappa^{2}}f'(t)= \omega(t)+\sigma(t). \label{1.3.8}
\end{equation}
Then, the kinetic functions may be chosen to be
\be
\omega(\phi)=-\frac{2}{\kappa^{2}}\left[f'(\phi)+g(\phi)\right]\ ,\quad
\sigma(\phi)=\frac{2}{\kappa^{2}}g(\chi)\ ,\label{1.3.9}
\end{equation}
where $g$ is an arbitrary function. Hence, the scalar
field potential is
finally obtained as
\begin{equation}
V(\phi,\chi)=\frac{1}{\kappa^{2}}\left[3f(\phi,\chi)^{2}
+f'(\phi)+g(\phi)-g(\chi)\right]\ . \label{1.3.9a}
\end{equation}

\subsection*{Example 1}

We can consider again the solution (\ref{1.1.17})
\begin{equation}
f(t)=\frac{H_{0}}{t_{s}-t}+\frac{H_{1}}{t^{2}}\ . \label{1.3.10}
\end{equation}
This solution, as already seen in Sec.~II, reproduces unified
inflation and late time acceleration in a scalar field model
with matter given by action (\ref{1.eq:1}). We may
now understand this solution as
derived from the two-scalar field model (\ref{1.3.1}), where a
degree of freedom is added so that we can choose various types
of scalar kinetic and potential
terms, as shown below for the solution~(\ref{1.3.10}). Then,
from Eqs.~(\ref{1.3.9}) and~(\ref{1.3.10}), the kinetic terms
follow:
\be
\omega(\phi)=-\frac{2}{\kappa^{2}}\left[\frac{H_{0}}{
(t_{s}-\phi)^{2}} -\frac{2H_{1}}{\phi^{3}}+g(\phi)\right]\ ,
\quad \sigma(\phi)=\frac{2}{\kappa^{2}}g(\chi)\ . \label{1.3.11}
\ee
The function $f(\phi,\chi)$ is
\begin{equation}
f(\phi,\chi)=\frac{H_{0}}{t_{s}-\phi}+\frac{H_{1}}{\phi^{2}}
+\int d\phi \, g(\phi)-\int d\chi \, g(\chi),\label{1.3.12}
\end{equation}
while the scalar potential is
\begin{equation}
V(\phi,\chi)=\frac{1}{\kappa^{2}}\left[{3f(\phi,\chi)}^{2}
+\frac{H_{0}}{(t_{s}-\phi)^{2}}-\frac{2H_{1}}{\phi^{3}}+g(\phi)
 -g(\chi)\right]\ . \label{1.3.14}
\end{equation}
This scalar potential reproduces the solution~(\ref{1.3.10}) with
an arbitrary function $g(t)$.

It is possible to further restrict $g(t)$ by studying
the stability of the system considered. To this end, we define
the functions
\begin{equation}
X_{\phi}=\dot{\phi}\ ,\quad
X_{\chi}=\dot{\chi}\ ,\quad
Y=\frac{f(\phi,\chi)}{H}\ . \label{1.3.15}
\end{equation}
Then, the Friedmann and scalar field equations can be written as
\bea
&& \frac{dX_{\phi}}{dN}=-\frac{1}{2H}\frac{
\omega'(\phi)}{ \omega(\phi)}
\left(X_{\phi}^{2}-1\right)-3\left(X_{\phi}-Y\right)\ , \\
&&\nonumber \\
&& \frac{dX_{\sigma}}{dN}=-\frac{1}{2H}
\frac{\sigma'(\chi)}{\sigma(\chi)}\left(X_{\chi}^{2}
 -1\right)-3\left(X_{\chi}-Y\right)\ ,\nn
&&\nonumber \\
&& \frac{dY}{dN}=\frac{\kappa^{2}}{2H^{2}}\left[
\omega(\phi)X_{\phi}(YX_{\phi}-1)
+\sigma(\chi)X_{\chi}(YX_{\chi}-1)\right] \ , \label{1.3.16}
\eea
where $\frac{d}{dN}=\frac{1}{H}\frac{1}{dt}$. At
$X_\phi=X_\chi=Y=1$,
we consider the perturbations
\begin{equation}
X_{\phi}=1+\delta X_{\phi}\ ,\quad X_{\chi}=1+\delta X_{\chi}\ ,\quad
Y=1+\delta Y, \label{1.3.17}
\end{equation}
then
\begin{equation}
\frac{d}{dN}\left(\begin{array}{c}
\delta X_{\phi}\\
\delta X_{\chi}\\
\delta Y\end{array}\right)=M\left(\begin{array}{c}
\delta X_{\phi}\\
\delta X_{\chi}\\
\delta Y\end{array}\right)\ ,\qquad\qquad
M=\left(\begin{array}{ccc}
 -\frac{ \omega'(\phi)}{H \omega(\phi)}-3 & 0 & 3\\
0 & -\frac{\sigma'(\chi)}{H\sigma(\chi)}-3 & 3\\
\kappa^{2}\frac{ \omega(\phi)}{2H^{2}} &
\kappa^{2}\frac{\sigma(\chi)}{2H^{2}} & \kappa^{2}\frac{
\omega(\phi)+\sigma(\chi)}{2H^{2}}\end{array}\right). \label{1.3.18}
\end{equation}
The eigenvalue equation is given by
\bea
&& \left(\frac{ \omega'(\phi)}{H \omega(\phi)}+3+\lambda\right)
\left(\frac{\sigma'(\chi)}{H\sigma(\chi)}+3+\lambda\right)
\left(\frac{\kappa^{2}}{2H^{2}}(
\omega(\phi)+\sigma(\phi))-\lambda\right) \nn
&&\nonumber \\
&& +\frac{3\kappa^{2}
\omega(\phi)}{2H^{2}}\left(\frac{\sigma'(\chi)}{H\sigma(\chi)}+3+
\lambda\right)+\frac{3\kappa^{2}\sigma(\chi)}{2H^{2}}
\left(\frac{ \omega'(\phi)}{H \omega(\phi)}+3+\lambda\right)=0\ .
\label{1.3.18a}
\eea
To avoid divergences in the eigenvalues, we choose the
kinetic functions to satisfy
\begin{equation}
\omega(\phi)\neq 0\ ,\quad
\sigma(\chi)\neq 0\ , \label{1.3.19}
\end{equation}
hence, the eigenvalues in Eq.~(\ref{1.3.18a}) are finite.
Summing up, under these conditions, the solution (\ref{1.3.10}) has no infinite
instability when the transition from the non-phantom to the phantom phase occurs.

As an example, we may choose $g(t)=\alpha/t^{3}$, where
$\alpha$
is a constant that satisfies $\alpha>2H_{1}$. Then, the $f(\phi,\chi)$
function (\ref{1.3.12}) is given by
\begin{equation}
f(\phi,\chi)=\frac{H_{0}}{t_{s}-\phi}-\frac{(\alpha-2H_{1})}{2\phi^{2}}
+\frac{\alpha}{2\chi^{2}}\ . \label{1.3.20a}
\end{equation}
As a result, the kinetic terms (\ref{1.3.11}) are expressed as
\be
\omega(\phi)=-\frac{2}{\kappa^{2}}\left[ \frac{H_{0}}{(t_{s}
 -\phi)^{2}}
 -\frac{2H_{1}}{\phi^{3}}+\frac{\alpha}{\phi^{3}}\right] \ ,\quad
\sigma(\chi)=\frac{2}{\kappa^{2}}\frac{\alpha}{\chi^{3}}\ ,
\label{1.3.21}
\ee
and the potential reads
\begin{equation}
V(\phi,\chi)=\frac{1}{\kappa^{2}}\left[ 3f(\phi,\chi)^{2}
+\frac{H_{0}}{(t_{s}-\phi)^{2}}
+\frac{\alpha-2H_{1}}{\phi^{3}}-\frac{\alpha}{\chi^{3}}\right] \ .
\label{1.3.22}
\end{equation}
This potential reproduces the solution (\ref{1.3.10}) that unifies inflation
and late time acceleration in the context of scalar-tensor
theories,
involving two scalar fields. Notice that the extra degree of
freedom gives
the possibility to select a different kinetic and scalar
potential in such a manner that we get the same solution.

In the case in which the condition (\ref{1.3.19}) is not imposed,
the kinetic terms~(\ref{1.3.9}) may have zeros for $0<t<t_{s}$, so
that the perturbation analysis performed above ceases to be
valid
because some of the eigenvalues
could diverge.

\subsection{General case: $n$ scalar fields}

As a generalization of the action~(\ref{1.3.1}), we now
consider the corresponding one for $n$
scalar fields,
\begin{equation}
S=\int d^{4}x\sqrt{-g}
\left[\frac{1}{2\kappa^{2}}R-\frac{1}{2}\sum_{i=1}^{n}
\omega_{i}(\phi_{i})\partial_{\mu}\phi_{i}\partial^{\mu}\phi_{i}
 - V(\phi_{1},\phi_{2},\cdots,\phi_{n})\right]\ .
\label{1.3.23}
\end{equation}
The associated Friedmann equations are
\be
H^{2}=\frac{\kappa^{2}}{3}\left[\sum_{i=1}^n
\frac{1}{2}\omega_{i}
(\phi_{i})\dot{\phi_{i}}^{2}+V(\phi_{1},\cdots ,\phi_{n})\right]\ ,\quad
\dot{H}=-\frac{\kappa^{2}}{2}\left[\sum_{i=1}^n
\omega_{i}(\phi_{i})\dot{\phi_{i}}^{2}\right]\ .
\label{1.3.24}
\ee
We can proceed analogously to the case of two scalar fields
so that the kinetic terms are written as
\begin{equation}
\sum_{i=1}^n \omega_{i}(t)=-\frac{2}{\kappa^{2}}f'(t)\ ,
\label{1.3.25}
\end{equation}
hence,
\be
\omega_{i}(\phi_{i})=
 -\frac{2}{\kappa^{2}}\frac{df(\phi_{1},\cdots ,
\phi_{n})}{d\phi_{i}}\ ,\quad
V(\phi_{1},\cdots,\phi_{n})=\frac{1}{\kappa^{2}}\left[3f(\phi_{1},\cdots , \phi_{n})^{2}
+\sum_{i=1}^n \frac{df(\phi_{1},\cdots ,
\phi_{n})}{d\phi_{i}}\right]\ , \label{1.3.26}
\ee
where $f(t,t,\cdots,t)\equiv f(t)$. Then the following solution is found
\be
\phi_{i}=t\ , \quad H(t)=f(t)\ . \label{1.3.27}
\ee
 From (\ref{1.3.25}) we can choose, as done above, the kinetic
terms to be
\be
\omega_{1}(\phi_{1})=
 -\frac{2}{\kappa^{2}}\left[f'(\phi_{1})+g_{2}(\phi_{1})
+\cdots +g_{n}(\phi_{1})\right]\ ,\quad
\omega_{2}(\phi_{2})=\frac{2}{\kappa^{2}}g_{2}(\phi_{2})\ ,\cdots ,\
\omega_{n}(\phi_{n})=\frac{2}{\kappa^2}g_{n}(\phi_{n})\ .
\label{1.3.28}
\ee
Then, there are $n-1$ arbitrary functions that reproduce the
solution~(\ref{1.3.27}) so reconstruction may be successfully done. They
could be chosen so that dark
matter is also represented by some of the scalar fields appearing in
the action~(\ref{1.3.23}).

\section[Inflation and cosmic acceleration from two scalar fields]{Reconstruction of inflation and cosmic acceleration from
two-scalar theory}

In the present section, inflation and cosmic acceleration are
reconstructed
separately, by means of a two scalar field model that reproduces
some of the cosmological constraints at each epoch. We explore
an inflationary model in which  the scalar
potential, given for a pair of scalar fields, exhibits an extra
degree of freedom and can be chosen in such way that slow-roll
conditions are satisfied. Also, the cosmic acceleration is
reproduced with a pair of scalar fields plus an ordinary matter
term, in which the values of the observed
cosmological density parameter ($\Omega _{\rm DE}\simeq 0.7$)
and of the EoS parameter ($w_{\rm DE}\simeq -1$) are reproduced
in a quite natural way. For the case of a single scalar, the
reconstruction for similarly distant epochs was given
in Ref.~\cite{2008JCAP...05..009N}.

\subsection*{Inflation}

In the previous sections, models describing inflation and
late-time accelerated expansion have been constructed by using
certain convenient scalar-tensor theories. In this
section, we present an inflationary model with two scalar fields,
which can be constructed in such a way that the inflationary
conditions are carefully accounted for. For this purpose, we use
some of the techniques given in the previous
section. The action during the inflationary epoch is written as
\begin{equation}
S=\int \sqrt{-g}\left[ \frac{R}{2\kappa
^{2}}-\frac{1}{2}\partial_{\mu}
\phi \partial ^{\mu }\phi -\frac{1}{2}\partial_{\mu }\chi
\partial^{\mu}\chi -V(\phi ,\chi )\right] \ .
\label{1.5.1}
\end{equation}
We will show that a general solution can be constructed, where
the scalar field  potential is not completely specified because
of the extra degree of freedom  represented by the second scalar
field added, in a  way similar to the situation occurring in
the previous section. Considering a spatially flat FLRW metric,
the Friedmann and scalar field equations are obtained by using
the Einstein equations and varying
the action with respect to both scalar fields:
\begin{equation}
H^{2}=\frac{\kappa ^{2}}{3}\left[
\frac{1}{2}\dot{\phi}^{2}+\frac{1}{2}
\dot{\chi}^{2}+V(\phi ,\chi )\right]\ , \quad \ddot{\phi}+3H\dot{\phi}
+\frac{\partial V(\phi,\chi )}{\partial \phi }=0\ , \quad
\ddot{\chi}+3H\dot{\chi}
+\frac{ \partial V(\phi ,\chi )}{ \partial \chi }=0\ .
\label{1.5.2}
\end{equation}
We assume that the slow-roll conditions are satisfied, that is,
$\ddot\phi \ll 3H\dot\phi$ $\left(\ddot\chi \ll 3H
\dot\chi\right)$, and ${\dot\phi }^{2} \ll V(\phi ,\chi )$
$\left(\dot\chi^{2} \ll V(\phi ,\chi )\right)$,
in order for inflation to occur. Then, Eqs.~(\ref{1.5.2}) take the
form
\begin{equation}
H^{2}\approx \frac{\kappa ^{2}}{3}V(\phi ,\chi )\ , \quad
3H\dot{\phi}
+\frac{\partial V(\phi,\chi )}{\partial \phi }\approx 0\ , \quad
3H\dot{\chi}+\frac{\partial V(\phi ,\chi )}{\partial \chi
}\approx 0
\label{1.5.3a}
\end{equation}
and the slow-roll conditions read
\be
\frac{1}{3\kappa ^{2}}\frac{V_{,i}V^{,i}}{V^{2}} \ll 1\ ,\quad
\frac{1}{3\kappa ^{2}}\frac{\sqrt{V_{,ij}V^{,ij}}}{V} \ll 1 \ .
\label{1.5.4}
\ee
Here,  $V_{,i}$ denotes the partial derivative of $V$ with
respect to
one of the scalar fields ($i=\phi ,\chi $). As done
previously, a scalar potential $ V(\phi ,\chi )$ can be
constructed, although in this case the conditions for inflation
need to be taken into account. From (\ref{1.5.3a}), the potential is
given by
\begin{equation}
V(\phi ,\chi )=\frac{3}{\kappa ^{2}}\, H^{2}(\phi ,\chi ) \ .
\label{1.5.5}
\end{equation}
We can choose this potential such that
\begin{equation}
V(\phi ,\chi )=\frac{3}{\kappa ^{2}}\left[ f^{2}(\phi ,\chi )
+g_{1}(\phi)-g_{2}(\chi )\right] \ .
\label{1.5.6}
\end{equation}
The three components $f$, $g_{1}$,  and $g_{2}$ are arbitrary
functions, and $g(N)=g_{1}(N)=g_{2}(N)$ where, for convenience,
we use the number of e-folds $N \equiv \ln \frac{a(t)}{a_{i}}$
instead of the cosmic time, and $a_{i}$ denotes the
initial value of the scale factor before inflation. Then, the following solution
is found:
\begin{equation}
H(N)=f(N) \ .
\label{1.5.7}
\end{equation}
Hence, Eqs.~(\ref{1.5.3a}) may be expressed as a set
of differential equations with the number of e-folds as
independent variable,
\begin{equation}
3 f^2(N) \frac{d\phi }{d N}+\frac{\partial V(\phi ,\chi
)}{\partial \phi
}\approx 0\ ,\quad 3 f^2(N)
\frac{d\chi }{dN}+\frac{\partial V(\phi ,\chi )}{\partial \chi
}\approx 0 \ .
\label{1.5.8}
\end{equation}

To illustrate this construction, let us use a simple example.
The  following scalar potential, as a function of the number of
e-folds $N$, is
considered:
\begin{equation}
V(\phi ,\chi ) =\frac{3}{\kappa ^{2}}\left[ H_{0}^{2}N^{2\alpha
}\right] \ ,
\label{1.5.9}
\end{equation}%
where $\alpha $ and $H_{0}$ are free parameters. By specifying
the arbitrary function $g(N)$, one can find a solution for the
scalar fields.  Let us choose, for the sake of simplicity,
$g(N)=g_{0}N^{2\alpha }$, where $g_{0}$ is a constant,
and $f(\phi ,\chi )=f(\phi )$ ({\em i.e.}, as a function of the
scalar field
$\phi$ only). Then, using Eqs.~(\ref{1.5.8}), the solutions for
the scalar fields are found to be
\be
\phi (N)=\phi _{0}-\frac{1}{\kappa H_{0}}\sqrt{2\alpha (H_{0}^{2}+g_{0})}\
\ln N \ ,\quad
\chi (N)=\chi _{0}-\frac{1}{\kappa H_{0}}\sqrt{2g_{0}\alpha N} \ ,
\ee
and the scalar potential can be written as
\begin{equation}
V(\phi ,\chi )=\frac{3}{\kappa ^{2}}\left[ \left( H_{0}+g_{0}\right) \exp
\left( \kappa \frac{\sqrt{2\alpha }H_{0}}{\sqrt{H_{0}^{2}+g_{0}}}\left(
\phi_{0}-\phi \right) \right) -g_{0}\left( \frac{\kappa ^{2}H_{0}^{2}
(\chi_{0}-\chi )^{2}}{2g_{0}\alpha }\right) ^{2\alpha }\right]\ .
\end{equation}
We are now able to impose the slow-roll conditions  by
evaluating the slow-roll parameters
\bea
\frac{1}{3\kappa ^{2}}\frac{V_{,i}V^{,i}}{V^{2}} &=&
\frac{2\alpha}{3H_{0}^{2}}\left(
\frac{ (H_{0}+g_{0})^{2}}{H_{0}^{2}+g_{0}}
+\frac{4g_{0}}{N}\right) \ll 1 \ ,\nn
&&\nonumber \\
\frac{1}{3\kappa ^{2}}\frac{\sqrt{V_{,ij}V^{,ij}}}{V} &=&
\frac{2}{3}\alpha ^{2}
\sqrt{\frac{(H_{0}+g_{0})^{2}}{(H_{0}^{2}
+g_{0})^{2}}+\frac{16g_{0}(4\alpha
 -1)^{2}}{4g_{0}^{2}\alpha ^{2}}\frac{1}{N^{2}}} \ll 1\ .
\eea
Hence, we may choose conveniently the free parameters so that
the slow-roll conditions are satisfied, and therefore, inflation
takes place.
 From these expressions we see that the desired conditions will be obtained,
in particular, when $\alpha$ is sufficiently small and/or $H_0$ and $N$ are
large enough. All these regimes help to fulfill the slow-roll
conditions,
in a quite natural way.

\subsection*{Cosmic acceleration with a pair of scalar fields}

It is quite reasonable, and rather aesthetic,  to think that the
cosmic acceleration could be driven by the same mechanism as
inflation. To this purpose, we apply the same model with two
scalar fields, with the aim of reproducing late-time
acceleration in a universe filled with a fluid with EoS
$p_{m}=w_{m}\rho _{m}$. The free parameters given by the model
could be adjusted to fit the observational data, as shown below.
We begin with the action representing this model,
\begin{equation}
S=\int d^{4}x\sqrt{-g}\left[ \frac{R}{2\kappa
^{2}}-\frac{1}{2}(\partial \phi )^{2}-\frac{1}{2}(\partial \chi
)^{2}-V(\phi ,\chi )+L_{m}\right] \ .
\label{1.52.1}
\end{equation}
By assuming a spatially flat FLRW metric, one obtains the
Friedmann equations
\begin{equation}
H^{2}=\frac{\kappa ^{2}}{3}\left[
\frac{1}{2}(\dot{\phi})^{2}+\frac{1}{2}(
\dot{\chi})^{2}+V(\phi ,\chi )+\rho _{m}\right] \ ,\quad \dot{H}
=-\frac{\kappa }{2}\left[ \frac{1}{2}\dot{\phi}^{2}+\frac{1}{2}\dot{\chi}^{2}
+V(\phi,\chi )+\rho _{m}\right] \ . \label{1.52.2}
\end{equation}
Variation of the action~(\ref{1.52.1}) yields the scalar field
equations
\begin{equation}
\ddot{\phi} +3H\dot{\phi}+V_{,\phi }=0\ , \quad
\ddot{\chi}+3H\dot{\chi}
+V_{,\chi}=0\ .
\label{1.52.3}
\end{equation}
This set of independent equations may be supplemented by a
fifth one, the conservation of matter-energy density,
$\rho _{m}$,
\begin{equation}
\dot{\rho _{m}}+3H\rho _{m}(1+ w_{m})=0\ .
\label{1.52.4}
\end{equation}
As done in~\cite{2008JCAP...05..009N}, we perform the substitutions
\begin{equation}
\Omega _{m}=\frac{\rho _{m}}{3H^{2}/\kappa ^{2}}\ ,\quad
\Omega _{sf}=
\frac{\frac{1}{2}(\dot{\phi})^{2}+\frac{1}{2}(\dot{\chi})^{2}
+V(\phi ,\chi )}{3H^{2}/\kappa ^{2}}\ ,\quad
\epsilon =\frac{\dot{H}}{H^{2}}\ ,\quad
w_{sf} \equiv \frac{p_{sf}}{\rho
_{sf}}=\frac{\frac{1}{2}(\dot{\phi})^{2}+\frac{1}{2}
(\dot{\chi})^{2}-V(\phi ,\chi )}{\frac{1}{2}(\dot{\phi})^{2}
+\frac{1}{2}(\dot{\chi})^{2}+V(\phi ,\chi )}\ .
\label{1.52.5}
\end{equation}
For convenience, we consider both scalar fields together under
the subscript $sf$, so that the corresponding density parameter
is $ \Omega _{sf}=\Omega _{\phi}+\Omega _{\chi }$.
Then, using the transformations~(\ref{1.52.5}), the
Friedmann equations and the scalar equation read
\be
\Omega _{m}+\Omega _{sf} = 1 \ ,\quad
2\epsilon +3(1+ w_{sf})\Omega _{sf}+3(1+ w_{m})\Omega _{m} = 0\ ,\quad
\Omega _{sf}^{\prime }+2\epsilon \Omega _{sf}+3\Omega _{sf}(1+ w_{sf})
=0\ .
\label{1.52.6}
\ee
Here, the prime denotes differentiation with respect to  the
number of e-folds $N \equiv \ln \frac{a(t)}{const}$. We can now
combine Eqs.~(\ref{1.52.5}) to write the EoS parameter for
the scalar fields as a function of $\Omega_{sf}$ and of the time
derivative of the scalar fields, namely
\begin{equation}
w_{sf}=\frac{\kappa ^{2}(\phi ^{\prime 2}+\chi ^{\prime 2})
 -3\Omega_{sf}}{3\Omega _{sf}}\ .
\label{1.52.7}
\end{equation}
Hence, it is possible to find an analytic solution for the equations
(\ref{1.52.6}), for a given evolution of the scalar fields, as will be seen
below. Before doing that, it is useful to write the effective EoS parameter,
which is given by
\begin{equation}
w_{\rm eff}=-1-\frac{2\epsilon }{3}\ ,\quad \rho_{\rm eff}=\rho _{m}
+\rho_{sf}\ ,\quad p_{\rm eff}=p_{m}+p_{sf} \;,
\label{1.52.8}
\end{equation}
and the deceleration parameter
\begin{equation}
q=-\frac{\ddot{a}}{aH^{2}}=-1-\epsilon \ . \label{1.52.9}
\end{equation}%
As usual, for $q>0$ the universe is in a decelerated phase, while $q<0$
denotes an accelerated epoch, such that for $w_{\rm eff}<-1/3$ the
expansion is accelerated. To solve the equations, we consider a
universe
which, at present, is filled with a pressureless component
($w_{m}=0$) representing ordinary matter, and  two scalar
fields which  represent a dynamical dark energy and a dark
matter
species. To show this, we make the following
assumption on the evolution of the scalar fields, which are given as
functions of $N$:
\begin{equation}
\phi (N)=\phi _{0} +\frac{\alpha }{\kappa ^{2}}N\ ,\quad \chi
(N)=\chi _{0}
+ \frac{\beta }{\kappa ^{2}}N\ .
\label{1.52.10}
\end{equation}
Then, Eqs.~(\ref{1.52.6}) can be solved,  and the scalar
field density parameter takes the form
\be
\Omega _{sf} =\Omega _{\phi }+\Omega _{\chi }
=1-\frac{\lambda }{k\mathrm{e}^{\lambda N}+3} \ ,
\label{1.52.11}
\ee
where $k$ is an integration constant and $\lambda=  3-(\alpha^2
+ \beta^2)$. It is possible to introduce an arbitrary function
$g(N)$, to express the energy
density parameter for each scalar field in the following way:
\be
 \Omega_{\phi } = 1-\frac{\lambda }{k\mathrm{e}^{\lambda
N}+3}-g(N) \ ,\quad
\Omega _{\chi } = g(N) \ .
\label{1.52.12}
\ee
The function $g(N)$ may be chosen in such a way that the
scalar field $\chi$ represents
a cold dark matter contribution at present ($w_{\chi}\simeq 0$),
and the scalar field $\phi$ represents the dark energy
responsible for
the accelerated expansion of our universe. On the other hand,
using Eqs.~(\ref{1.52.6}), $\epsilon=\frac{H'}{H}$ is obtained as
\be
\epsilon=-\frac{3}{2}\left\{ 1-\frac{k\lambda \left( k\e^{\lambda N}
+\alpha^2\beta^2\right) }{\left[ (\alpha^2+\beta^2)\e^{-\lambda N}
+3k\right]\left(k\e^{\lambda N} +3 \right) }\right\} \ .
\label{1.52.13}
\ee
 Then, it is possible to calculate the effective parameter of
EoS given
by Eq.~(\ref{1.5.8})
\be
w_{\rm eff}=-1-\frac{2}{3}\epsilon
= -\frac{k\lambda \left( k\e^{\lambda N} +\alpha^2\beta^2\right) }
{\left[ (\alpha^2+\beta^2)\e^{-\lambda N}+3k\right]\left(k\e^{\lambda N} +3 \right) }\ .
\label{1.52.14}
\ee
We have four free parameters ($N$, $k$, $\alpha$, $\beta$) that may
 be adjusted to fit  the constraints derived from
observations. With this purpose, we use the observational input
$\Omega_{m}\simeq 0.03$, referred to baryonic matter, and normalize the number
of e-folds $N$, taking $N=0$ at present, then the integration constant $k$ may be written
as a function of $\alpha$ and $\beta$ as
\be
\Omega_m (N=0)=0.03\ \to\ k=\frac{2.01-(\alpha^2 + \beta^2)}{0.03}\ .
\label{1.52.15}
\ee
The free parameter $\beta$ may be fixed in such a way that the scalar field $\chi$
represents cold dark matter at present, {\em i.e.},
$w_{\chi}\simeq 0$
and $\Omega_{\chi}\simeq0.27$, and its EoS parameter is written
as
\be
w_{\chi}=\frac{\kappa^{2}{\chi'}^{2}
 -3\Omega_{\chi}}{3\Omega_{\chi}}=\frac{\beta^2 -3g(N)}{3g(N)} \ .
\label{1.52.16}
\ee
For convenience, we choose $g(N)=\frac{\beta^2}{3}\e^{-N}$, then the energy density
and EoS parameter are given by
\be
w_{\chi}=\frac{1-\e^{-N}}{\e^{-N}}\ ,\quad
\Omega_{\chi}=\frac{\beta^2}{3}\e^{-N}\ .
\label{1.52.17}
\ee
Hence, at present ($N=0$), the expressions~(\ref{1.52.17}) can be
compared with
the observational values and the $\beta$ parameter is given by
\be
w_{\chi}(N=0)=0\ ,\quad
\Omega_{\chi}(N=0)\simeq 0.27\ \to\ \beta^2=0.81\ .
\label{1.5.18}
\ee
Finally, the energy density of $\phi$ expressed by
Eq.~(\ref{1.52.12}) takes the form
\be
\Omega _{\phi } =1-\frac{\lambda }{k\mathrm{e}^{\lambda N}+3}-\frac{\beta^2}{3}\e^{-N}\ .
\ee
The value for $\alpha$ could be taken so that
$\Omega_{\phi}\simeq 0.7$
and $w_{\phi}\simeq -1$ at present. Hence, it has been shown that
cosmic acceleration can be reproduced with a pair of scalar
fields,
where due to the presence of the extra scalar that can
be identified with the
dark matter component.

It is interesting  to point out that one can unify these
realistic descriptions of the inflationary and late-time
acceleration eras within a single theory. However, the
corresponding potential looks quite complicated. The easiest
way would be to use step ($\theta$-function) potentials in order
to  unify the whole description in the easiest way (as was
pioneered in \cite{2008PhRvD..77d6009C}).

We may also construct a model unifying early  universe inflation
and the present accelerated expansion era. To
this end we can choose $f(\phi)$ in (\ref{1.uf0}), which gives the Hubble
parameter~(\ref{1.uf1}). Then, using (\ref{1.3.9}), one can define
$\omega(\phi)$ and $\sigma(\chi)$ with the help of an arbitrary
function $g$. After defining then $f(\phi,\chi)$
with Eq.~(\ref{1.3.6}), we can construct the potential
$V(\phi,\chi)$ using Eq.~(\ref{1.3.7}). Finally, we obtain the
two scalar-tensor theory (\ref{1.3.1})
reproducing the Hubble rate (\ref{1.uf1}), which describes both
inflation and the accelerated
expansion.

\section{Discussions}

Modelling both early inflation and late-time
acceleration within the context of a single theory has,
undoubtedly, much aesthetic appeal and seems a worthy goal,
which we attempted here.  To summarize, we have
developed the reconstruction program for
the expansion history of the universe  by using a single or
multiple
(canonic and/or phantom) scalar fields. Already in the case of
single scalar, we  have  presented many examples
which prove that it is possible to unify early-time
inflation with late-time acceleration. 

 The reconstruction
technique has then been  generalized to the case of
a scalar non-minimally coupled to the Ricci curvature, and
to non-minimal (Brans-Dicke-type) scalars.
Again, various explicit examples of unification of early-time
inflation and late-time acceleration have been presented in such
formulations. The chameleon mechanism is studied for our specific model, such that local constraints are satisfied 

Finally, the case of several minimally coupled  scalar fields
has  been considered in the description of the realistic
evolution of  the Hubble parameter, and we have  shown that it
is  qualitatively easier to achieve the realistic
unification of late and early epochs in such a model in such
a way as to satisfy the cosmological
bounds coming from the observational data. We have shown that the perturbations could diverge in the case of a single scalar field that crosses the phantom barrier. This trouble is easily resolved by including a second scalar field, and imposing on each one a canonical (phantom) behavior.
Using the
freedom of choosing these scalar functions, one can constrain
the theory in an observationally acceptable way. For instance,
slow-roll conditions and stability conditions may be satisfied
in different ways for different scalar functions,  while the
scale factor remains the same. This may be  used also to obtain
the correct perturbations structure,
{\em etc}.

\chapter[Oscillating Universe from inhomogeneous EoS and coupled DE]{Oscillating Universe from inhomogeneous EoS and coupled dark energy}

\footnote{This Chapter is based on the publication: \cite{2009GrCo...15..134S}.}In the present chapter, we describe an oscillating Universe produced by an  ideal dark fluid, what allows the possibility to unify early and late time acceleration under the same mechanism, in such a way that the Universe history may be reconstructed completely. On the other hand, it is important to keep in mind that these models represent just an effective description that owns a number of well-known problems, as the end of inflation. Nevertheless, they may represent a simple and natural way to resolve the coincidence problem, one of the possibilities may be an oscillating Universe (Refs. \cite{2008arXiv0802.1294A,2007EPJC...51..179B,2006PhLB..634..101F,2006PhLB..637..139N,2005GReGr..37.2201Y}), where the different phases of the Universe are reproduced due to its periodic behavior. The purpose of this chapter is to show that, from inhomogeneous EoS for a dark energy fluid, an oscillating Universe is obtained, and several examples are given to illustrate it. The possibility of an interaction between dark energy fluid, with homogeneous EoS, and matter is studied, which also reproduces that kind of periodic Hubble parameter, such case has been studied and is allowed by the observations (see \cite{2007PhRvD..76b3508G,Gumjudpai:2005ry}). The possible phantom epochs are explored, and the possibility that the Universe may reach a Big Rip singularity (for a classification of future singularities, see Ref. \cite{2005PhRvD..71f3004N}).  \\

\section{Inhomogeneous equation of state for dark energy}

\begin{figure}
 \centering
 \includegraphics[width=3in,height=2in]{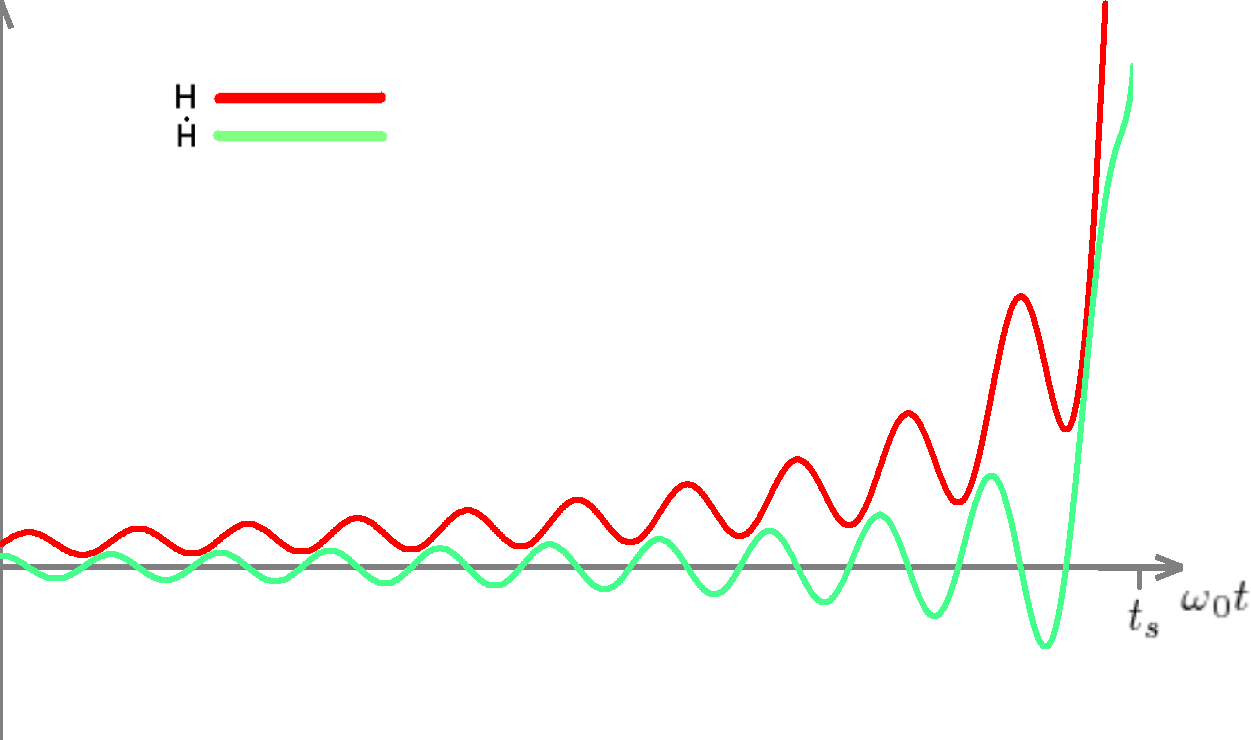}
 
 \caption{The Hubble parameter H and $\dot{H}$ for a value $w=-1.1$. Phantom phases occurs periodically, and a Big Rip singularity takes place at Rip time $t_s$.}
 \label{2.fig1}
\end{figure}

Let us consider firstly a Universe filled with a dark energy fluid, neglecting the rest of possible components (dust matter, radiation..), where its EoS depends on the Hubble parameter and its derivatives, such kind of EoS has been treated in several articles\cite{2007EPJC...52..223B,2007EPJC...51..179B,2006PhRvD..73d3512C,2005PhRvD..72b3003N,2006PhLB..639..144N}. We show that for some choices of the EoS, an oscillating Universe resulted, which may include phantom phases. Then, the whole Universe history, from inflation to cosmic acceleration, is reproduced in such a way that observational constraints may be satisfied\cite{2005PhRvD..72j3503J,2007LNP...720..257P}. We work in a spatially flat FLRW Universe. 
The Friedmann equations, considering now a perfect fluid, are obtained:
\be
H^2=\frac{\kappa^2}{3}\rho, \quad \quad \dot{H}=-\frac{\kappa^2}{2}\left( \rho+p\right)\ . 
\label{2.1.2}
\ee
At this section, the EoS considered is given to have the general form:
\be
p= w\rho+g(H, \dot{H}, \ddot{H},..;t)\ ,
\label{2.1.4}
\ee
where $w$ is a constant and $g(H, \dot{H}, \ddot{H},..;t)$ is an arbitrary function of the Hubble parameter $H$, its derivatives and the time $t$, (such kind of EoS has been treated in Ref. \cite{2005PhRvD..72b3003N}). Using the FLRW equations (\ref{2.1.2}) and (\ref{2.1.4}), the following  differential equation is obtained:
\be
\dot{H}+\frac{3}{2}(1+w)H^2+\frac{\kappa^2}{2}g(H, \dot{H}, \ddot{H},..;t)=0\ .
\label{2.1.5}
\ee
Hence, for a given function $g$, the Hubble parameter is calculated by solving the equation (\ref{2.1.5}). It is possible to reproduce an oscillating Universe by an specific EoS (\ref{2.1.4}). To illustrate this construction, let us consider the following $g$ function as an example:
\be
g(H, \dot{H}, \ddot{H})=-\frac{2}{\kappa^2}\left( \ddot{H}+\dot{H}+\omega_0^2H +\frac{3}{2}(1+w))H^2-H_0\right)\ , 
\label{2.1.6}
\ee
 where $H_0$ and $\omega_0^2$ are constants. By substituting (\ref{2.1.6}) in (\ref{2.1.5}) the Hubble parameter equation acquires the form:
\be
\ddot{H}+\omega_0H=H_0\ ,
\label{2.1.7}
\ee
which is the classical equation for an harmonic oscillator. The solution is found:
\be
H(t)=\frac{H_0}{\omega_0^2}+H_1 \sin(\omega_0t+\delta_0)\ ,
\label{2.1.8}
\ee
where $H_1$ and $\delta_0$ are  integration constants. To study the system, we calculate the first derivative of the Hubble parameter, which is given by $\dot{H}=H_1\cos(\omega_0t+\delta_0)$, so the Universe governed by the dark energy fluid (\ref{2.1.6}) oscillates between phantom and non-phantom phases with a frequency given by the constant $\omega_0$, constructing inflation epoch and  late-time acceleration under the same mechanism, and Big Rip singularity avoided \\
As another example, we consider the following EoS (\ref{2.1.4}) for the dark energy fluid:
\be
p= w\rho+\frac{2}{\kappa^2}Hf'(t)\ .
\label{2.1.9}
\ee
In this case  $g(H;t)=\frac{2}{\kappa^2}Hf'(t)$, where $f(t)$ is an arbitrary function of the time $t$, and the prime denotes a derivative on $t$. The equation (\ref{2.1.5}) takes the form:
\be
\dot{H}+Hf'(t)=-\frac{3}{2}(1+w)H^2\ .
\label{2.1.10}
\ee
This is the well-known Bernoulli differential equation. For a function $f(t)=-ln\left(H_1 + H_0 \sin\omega_0t \right)$, where $H_1>H_0$ are arbitrary constants, then the following solution for (\ref{2.1.10}) is found:
\be
H(t)=\frac{H_1+H_0 \sin \omega_0t}{\frac{3}{2}(1+w)t+k}\ ,
\label{2.1.11}
\ee
here, the $k$ is an integration constant. As it is seen, for some values of the free constant parameters, the Hubble parameter tends to infinity for a given finite value of $t$.  The first derivative of the Hubble parameter is given by: 
\be
\dot{H}=\frac{\frac{H_0}{\omega_0}(\frac{3}{2}(1+w)t+k)\cos\omega_0t-(H_1+H_0 \sin \omega_0t)\frac{3}{2}(1+w)}{(\frac{3}{2}(1+w)t+k)^2}\ .
\label{2.1.12}
\ee
As it is shown in fig.\ref{2.fig1}, the Universe has a periodic behavior, it passes through phantom and non-phantom epochs, with its respectives transitions. A Big Rip singularity may take place depending on the value of $w$, such that it is avoided for $w\geq-1$, while if $w<-1$ the Universe reaches the singularity in the  Rip time given by $t_s=\frac{2k}{3|1+w|}$.

\section{Dark energy ideal fluid and dust matter}

\subsection{No coupling between matter and dark energy}
\begin{figure}
 \centering
 \includegraphics[width=3in,height=2in]{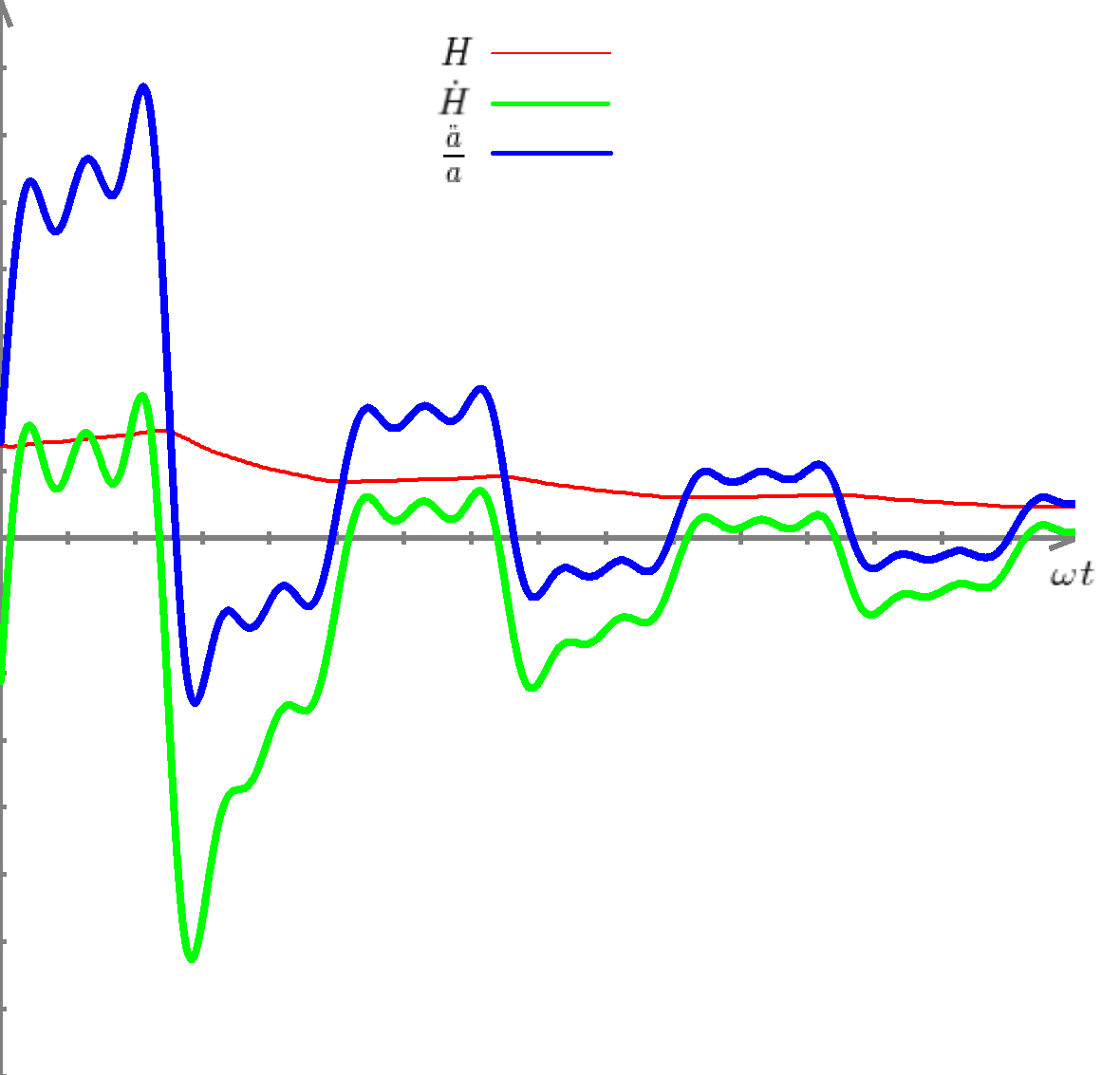}
 
 \caption{The Hubble parameter, its derivative and the acceleration parameter are represented for the ``step''  model for values $w_0=-0.2$ and $w_1=1$.}
 \label{2.fig:2}
\end{figure}

Let us now explore a more realistic model by introducing a matter component with EoS  given by $p_m=w_m\rho_m$, we consider an inhomogeneous EoS for the dark energy component\cite{2006PhRvD..73d3512C,2005PhRvD..72b3003N}. It is shown below that an oscillating Universe  may be obtained by constructing an specific EoS. In this case, the FLRW equations (\ref{2.1.2}) take the form:
\be
H^2=-\frac{\kappa^2}{3}(\rho+\rho_m), \quad \quad \dot{H}=-\frac{\kappa^2}{2}\left( \rho+p+\rho_m+p_m\right)\ . 
\label{2.1.13}
\ee
At this section, we consider a matter fluid that doesn't interact with the dark energy fluid, then the energy conservation equations are satisfied for each fluid separately:  
\be
\dot{\rho_m}+3H(\rho_m+ p_m)=0, \quad \quad \dot{\rho}+3H(\rho+ p)=0\ .
\label{2.1.14}
\ee
It is useful to construct an specific solution for the Hubble parameter by defining the effective EoS with an effective parameter $w_{eff}$:
\be
w_{eff}=\frac{p_{eff}}{\rho_{eff}}, \quad \rho_{eff}=\rho+\rho_m, \quad p_{eff}=p+p_m\ ,
\label{2.1.15}
\ee
and the energy conservation equation $\dot{\rho}_{eff}+3H(\rho_{eff}+ p_{eff})=0 $ is satisfied. We consider a dark energy fluid which is described by the EoS describes by the following expression:
\be
p=-\rho+\frac{2}{\kappa^2}\frac{2(1+w(t))}{3\int(1+w(t))dt}-(1+w_m)\rho_{m0}\e^{-3(1+w_m)\int dt \frac{2}{3\int (1+w(t))}}\ ,
\label{2.1.16}
\ee
here $\rho_{m0}$ is a constant, and $w(t)$ is an arbitrary function of time $t$. Then the following solution is found:
\be
H(t)=\frac{2}{3\int dt(1+w(t))}\ . 
\label{2.1.17}
\ee
And the effective parameter (\ref{2.1.15}) takes the form $w_{eff}=w(t)$. Then, it is shown that a solution for the Hubble parameter may be constructed from EoS (\ref{2.1.16}) by specifying a function $w(t)$. \\
\\Let us consider an example\cite{2006PhLB..637..139N} with the following function for $w(t)$:
\be
w=-1+w_0\cos \omega t\ .
\label{2.1.18}
\ee
In this case, the EoS for the dark energy fluid, given by (\ref{2.1.16}), takes the form:
\be
p=-\rho +\frac{4}{3\kappa^2}\frac{\omega w_0\cos\omega t}{w_1+w_0\sin\omega t}-(1+w_m)\rho_{m0}\e^{-3(1+w_m)\frac{2w}{3(w_1+w_0 \sin \omega t)}}\ ,
\label{2.1.19}
\ee
where $w_1$ is an integration constant. Then, by (\ref{2.1.17}), the Hubble parameter yields:
\be
H(t)=\frac{2\omega}{3(w_1+w_0 \sin \omega t)}\ .
\label{2.1.20}
\ee 
The Universe passes through phantom and non phantom phases since the first derivative of the Hubble parameter has the form:
\be
\dot{H}=-\frac{2\omega^2 w_0\cos\omega t}{3(w_1+w_0\sin\omega t)^2}\ .
\label{2.1.21}
\ee
In this way, a Big Rip singularity will take place in order that $\vert w_1\vert<w_0$, and it is avoided when $\vert w_1\vert>w_0$. As it is shown, this model reproduces unified inflation and cosmic acceleration in a natural way, where the Universe presents a periodic behavior. In order to reproduce accelerated and decelerated phases, the acceleration parameter is studied, which is given by:
\be
\frac{\ddot{a}}{a}=\frac{2\omega^2}{3(w_1+w_0 \sin \omega t)^2}\left( \frac{2}{3}-w_0\cos\omega t\right) \ .
\label{2.1.21bis}
\ee
Hence, if $w_0>2/3$ the different phases that Universe passes are reproduced by the EoS (\ref{2.1.19}), presenting a periodic evolution that may unify all the epochs by the same description. 
\\
\\As a second example, we may consider a classical periodic function, the step function:
\begin{equation}
w(t)=-1+
\left\{ 
\begin{array}{lr}
w_{0}&0<t<T/2\\
w_{1}&T/2<t<T 
\end{array}
\right.\ ,
\label{2.1.22}
\end{equation}
and $w(t+T)=w(t)$. It is useful to use a Fourier expansion such that the function (\ref{2.1.22}) become continuous. Approximating to third order, $w(t)$ is given by:
\be
w(t)=-1+\frac{(w_0+w_1)}{2}+\frac{2(w_0-w_1)}{\pi}\left( \sin\omega t +\frac{\sin3\omega t}{3}+\frac{\sin5\omega t}{5}\right) \ .
\label{2.1.23}
\ee
Hence, the EoS for the Dark energy ideal fluid is given by (\ref{2.1.16}), and the solution (\ref{2.1.17}) takes the following form:
\[
H(t)=\frac{2}{3} \left[w_2+\frac{(w_0+w_1)}{2}t \right.   
\]
\be
\left. -\frac{2(w_0-w_1)}{\pi\omega}\left( \cos\omega t+ \frac{\cos3\omega t}{9}+\frac{\cos5\omega t}{25}\right) \right]^{-1}\ .
\label{2.1.24}
\ee
The model is studied by the first derivative of the Hubble parameter in order to see the possible phantom epochs, since:
\[
\dot{H}=-\frac{3}{2}H^2\left[ \frac{(w_0+w_1)}{2} \right.
\]
\be
\left. -\frac{2}{\pi}(w_0-w_1)\left( \sin\omega t+\frac{\sin3\omega t}{3}+\frac{\sin5\omega t}{5}\right)  \right] \ .
\label{2.1.25}
\ee 
Then, depending on the values from $w_0$ and $w_1$ the Universe passes through phantom phases. To explore the different epochs of acceleration and deceleration that the Universe passes on,  the acceleration parameter is calculated:
\[
\frac{\ddot{a}}{a}=H^{2}+\dot{H}=
\]
\be
H^2\left[1-\frac{3}{2}\left( \frac{(w_0+w_1)}{2}-\frac{2}{\pi}(w_0-w_1)\left( \sin\omega t+\frac{\sin3\omega t}{3}+\frac{\sin5\omega t}{5}\right)\right)  \right]\ .   
\label{2.1.26}
\ee
Then, in order to get  acceleration and deceleration epochs, the constants parameters $w_0$ and $w_1$ may be chosen such that $w_0<2/3$ and $w_1>2/3$, as it is seen by (\ref{2.1.22}). For this selection, phantom epochs take place in the case that $w_0<0$. In any case, the oscillated behavior is damped by the inverse term on the time $t$, as it is shown in fig.\ref{2.fig:2}, where the acceleration parameter is plotted for some given values of the free parameters. This inverse time term makes reduce the acceleration and the Hubble parameter such that the model tends to a static Universe. \\

We  consider now a third example where a classical damped oscillator is shown, the function $w(t)$  is given by:
\be
w(t)=-1+\e^{-\alpha t}w_0\cos\omega t\ ,
\label{2.1.27}
\ee
here $\alpha$ and $w_0$ are two positive constants. Then, the EoS for the dark energy ideal fluid is constructed from (\ref{2.1.16}). The solution for the Hubble parameter (\ref{2.1.17}) is integrated, and takes the form:
\be
H(t)=\frac{2}{3}\frac{\omega^2+\alpha^2}{w_1+w_0\e^{-\alpha t}(\omega\sin\omega t-\alpha\cos\omega t)}\ ,
\label{2.1.28}
\ee
where $w_1$ is an integration constant. The Hubble parameter oscillates damped by an exponential term, and for big times, it tends to a constant $H(t\longrightarrow\infty)=\frac{2}{3}\frac{\omega^2+\alpha^2}{w_1}$, recovering the cosmological constant model. The Universe passes through different phases as it may be shown by the accelerated parameter:
\be
\frac{\ddot a}{a}=H^2\left(1-\frac{3}{2}\e^{-\alpha t}w_0\cos\omega t\right)\ .
\label{2.1.29}
\ee
It is possible to restrict $w_0>\frac{2}{3}$ in order to get deceleration epochs when the matter component dominates. On the other hand, the Universe also passes through phantom epochs, since the Hubble parameter derivative gives:
\be
\dot{H}=-\frac{3}{2}H^2\e^{-\alpha t}w_0\cos\omega t\ .
\label{2.1.30}
\ee
Hence, the example (\ref{2.1.27}) exposes an oscillating Universe with a frequency given by $\omega$ and damped by a negative exponential term, which depends on the free parameter $\alpha$, these may be adjusted such that the phases agree with the phases times constraints by the observational data. 

\subsection{Dark energy and coupled matter}
In general, one may consider a Universe filled with a dark energy ideal fluid whose Eos is given $p=w\rho$, where $w$ is a constant, and matter described by $p_m=\omega_m\rho_m$,  both interacting with each other. In order to preserve the energy conservation, the equations for the energy density are written as following:
\be
\dot{\rho_m}+3H(\rho_m+p_m)= Q, \quad \quad
\dot{\rho}+3H(\rho+p_)= -Q\ ,
\label{2.1.31}
\ee
here $Q$ is an arbitrary function. In this way, the total energy conservation is satisfied $\dot{\rho}_{eff}+3H(\rho_{eff}+p_{eff})=0$, where $\rho_{eff}=\rho+\rho_m$ and $p_{eff}=p+p_m$, and the FLRW equations (\ref{2.1.13}) doesn't change. To resolve this set of equations for a determined function $Q$, the second FLRW equation (\ref{2.1.13}) is combined with the conservation equations (\ref{2.1.31}), this yields:
\[
\dot{H}=-\frac{\kappa^2}{2}\left[(1+w_m)\frac{\int Q\exp(\int dt3H(1+w_m))}{\exp(\int dt3H(1+w_m))} \right.
\]
\be
\left.+(1+w)\frac{-\int Q\exp(\int dt3H(1+w))}{\exp(\int dt3H(1+w))} \right]\ .
\label{2.1.32}
\ee 

In general, this is difficult to resolve for a function Q. As a particular simple case is the cosmological constant where the dark energy EoS parameter $w=-1$ is considered,  the equations become very clear, and  (\ref{2.1.31}) yields $\dot{\rho}=-Q$, which is resolved and the  dark energy density is given by:
\be
\rho(t)=\rho_0-\int dt Q(t)\ ,  
\label{2.1.33} 
\ee
where $\rho_0$ is an integration constant. Then, the Hubble parameter is obtained by introducing (\ref{2.1.33}) in the FLRW equations, which yields:
\be
\dot{H}+\frac{3}{2}(1+w_m)H^2=\frac{\kappa^2}{2}(1+w_m)\left(\rho_0-\int dt Q \right)\ . 
\label{2.1.34}
\ee
Hence, Hubble parameter depends essentially on the form of the coupling function $Q$. This means that a Universe model may be constructed from the coupling between matter and dark energy fluid, which is given by $Q$, an arbitrary function. It is shown below that some of the models given in the previous section by an inhomogeneous EoS dark energy fluid, are reproduced by a dark energy fluid with constant EoS ($w=-1$), but coupled to dust matter. By differentiating equation (\ref{2.1.34}), the function $Q$ may be written in terms of the Hubble parameter and its derivatives: 
\be
Q=-\frac{2}{\kappa^2}\frac{1}{1+w_m}\left(\ddot{H}+3(1+w_m)H\dot{H} \right)\ .
\label{2.1.35}
\ee 
\\
As an example, we use the solution(\ref{2.1.8}):
\be
H(t)=H_0+H_1 \sin(\omega_0t+\delta_0)\ .
\label{2.1.36}
\ee
Then, by the equation (\ref{2.1.34}), the function $Q$ is given by:
\be
Q(t)=\frac{2}{\kappa^2(1+w_m)}\left[ H_0\omega^2\sin\omega t +3(1+w_m)h_0\omega\cos\omega t(H_1+H_0\sin\omega t)) \right] \ . 
 \label{2.1.37}
\ee
Then, the oscillated model (\ref{2.1.36}) is reproduced by a coupling between matter and dark energy, which also oscillates. Some more complicated models may be constructed for complex functions $Q$. As an example let us consider the solution (\ref{2.1.20}):
\be
H(t)=\frac{2\omega}{3(w_1+w_0 \sin \omega t)}\ .
\label{2.1.38}
\ee 
The coupling function (\ref{2.1.35}) takes the form:
\[
Q(t)=-\frac{4}{3\kappa^2}\frac{\omega^3 w_0}{(1+w_m)(w_1+w_0 \sin \omega t)^3}
\]
\be
\left[\sin\omega t(w_1+w_0 \sin \omega t)^2+2w_0\cos^2\omega t-2(1+w_m)\cos\omega t \right]\ .
\label{2.1.39}
\ee
This coupling function reproduces a oscillated behavior that unifies the different epochs in the Universe. Hence, it have been shown that for a constant EoS for the dark energy with $w=-1$, inflation and late-time acceleration are given in a simple and natural way. 

\section{Scalar-tensor description}
Let us now consider the solutions shown in the last sections through  scalar-tensor description, such equivalence has been constructed in Ref. \cite{2006PhLB..634...93C}. We assume, as before, a flat FLRW metric, a Universe filled with a ideal matter fluid with EoS given by $p_m=w_m\rho_m$, and no coupling between matter and the scalar field. Then, the following action is considered:
\be
S=\int dx^{4}\sqrt{-g}\left[ \frac{1}{2\kappa^{2}}R
 - \frac{1}{2} \omega (\phi)
\partial_{\mu} \phi \partial^{\mu }\phi -V(\phi )+L_{m}\right]\ ,
\label{2.1.40}
\ee
here $\omega(\phi)$ is the kinetic term and $V(\phi)$ represents the scalar potential. Then, the corresponding FLRW equations are written as:
\be
H^{2} = \frac{\kappa ^{2}}{3}\left( \rho _{m}
+\rho_{\phi}\right)\ , \quad \quad \dot H = -\frac{\kappa ^{2}}{2}\left(
\rho _{m}+p_{m}+\rho _{\phi }+p_{\phi }\right)\ ,
\label{2.eq:1.41}
\ee 
where $\rho_{\phi}$ and $p_{\phi}$ given by:
\be
\rho _{\phi } = \frac{1}{2} \omega (\phi )\, {\dot \phi}^{2}
+V(\phi)\ ,\quad \quad
p_{\phi } = \frac{1}{2} \omega (\phi ) \, {\dot \phi}^{2}-V(\phi)\ .
\label{2.eq:1.42}
\ee
By assuming:
\bea
\omega (\phi ) = -\frac{2}{\kappa ^{2}}f^{\prime }(\phi )
 -(w_{m}+1)F_{0} \e^{-3(1+w_{m})F(\phi )}\ , \nn 
&&\nonumber \\
V(\phi ) = \frac{1}{\kappa ^{2}}
\left[ {3f(\phi)}^{2}+f^{\prime }(\phi ) \right]
+\frac{w_{m}-1}{2}F_{0}\, \e^{-3(1+w_{m})F(\phi )}\ .
\label{2.eq:1.43}
\eea
The following solution is found:
\be
\phi =t\ , \quad H(t)=f(t)\ ,
\label{2.1.44}
\ee
which yields:
\be
a(t)=a_{0}\e^{F(t)}, \qquad a_{0}=\left(
\frac{\rho _{m0}}{F_{0}}\right) ^{\frac{1}{3(1+w_{m})}}.
\label{2.eq:1.45}
\ee
Then, we may assume solution (\ref{2.1.28}), in such case the $f(\phi)$ function takes the form:
\be
f(\phi)=\frac{2}{3}\frac{\omega^2+\alpha^2}{w_1+w_0\e^{-\alpha \phi}(\omega\sin\omega \phi-\alpha\cos\omega \phi)}\ ,
\label{2.1.46}
\ee
And by (\ref{2.eq:1.43}) the kinetic term and the scalar potential are given by:
\bea
\omega(\phi)=\frac{3}{\kappa^2}f^2(\phi)w_0\e^{-\alpha\phi}\cos\omega\phi-(1+w_m)F_0\e^{-3(1+w_m)F(\phi)}, \nn
V(\phi)=\frac{3f^2(\phi)}{\kappa^2}\left(1-\frac{1}{2}w_0\e^{-\alpha\phi}\cos\omega\phi \right) +\frac{w_m-1}{2}F_0\e^{-3(1+w_m)F(\phi)},
\label{2.1.47}
\eea
where $F(\phi)=\int d\phi f(\phi)$ and $F_0$ is an integration constant. Then, the periodic solution (\ref{2.1.28}) is reproduced in the mathematical equivalent formulation in scalar-tensor theories by the action (\ref{2.1.40}) and  explicit kinetic term and scalar potential, in this case, is given by (\ref{2.1.47}).

\section{Discussions}

In this chapter, a Universe model has been presented that reproduces in a natural way the early and late-time acceleration by a periodic behavior of the Hubble parameter.  The  late-time transitions are described by this model: the transition from deceleration to acceleration, and the possible transition from non-phantom to phantom epoch. The observational data does not restrict yet the nature and details of the EoS for dark energy, then the possibility that Universe behaves periodically is allowed. For that purpose, several examples have been studied in the present chapter, some of them driven by an inhomogeneous EoS for dark energy, which just represents an effective description of nature of dark energy, and others by a coupling between dark energy and matter which also may provide another possible constraint to look for.

\part{On modified theories of gravity and its implications in Cosmology}
\chapter[F(R) gravity from scalar-tensor \& inhom. EoS dark energy]{Modified f(R) gravity from scalar-tensor theories and inhomogeneous EoS dark energy}

\footnote{This Chapter is based on the publication: \cite{2009GReGr..41.1527S}.}In this chapter modified f(R) theories of gravity are introduced in the context of cosmology which avoid the need to introduce dark energy, and may give an explanation about the origin of the current accelerated expansion and even on the expansion history of the Universe (for recent reviews and books on the topic, see \cite{2011Capozziello_Faraoni,2010arXiv1011.0544N,2010RvMP...82..451S}). \\
In this sense, the cosmic acceleration and the cosmological properties of metric formulation f(R) theories have been studied in Refs.\cite{2005CQGra..22L..35A,2008PhLB..660..125A,2007PhRvD..75j4016B,2006IJMPD..15..767B,2007PhLB..646..105B,2007PhRvD..76h4005B,2006PhRvD..74f4028B,2002Capozziello..IntJModPhys11,2005PhRvD..71d3503C,2008GReGr..40..357C,2006PhLB..639..135C,2008PhRvD..77b4024C,2009GReGr..41.1757C,2004PhRvD..70d3528C,2006PhRvD..73f4029C,2008IJTP...47..898C,2005JCAP...02..010C,2006PhRvD..73h4007C,2006PhRvD..74h7501D,Faraoni:2007yn,2007CQGra..24.3637D,2007PhRvD..76f3504F,2007PhRvD..75h4010L,2006PhRvL..96d1103M,2007PhRvD..76d4008M,2007PhRvD..76f4021M,Odintsov_Nojiri_2006hep.th....1213N,2003PhRvD..68l3512N,2003PhLB..576....5N,2007PhLB..651..224N,2007PhRvD..75b3511O,2007JCAP...01..010R,2007LNP...720..403W}. Recently the main focus has been improviding a f(R)-theory that reproduces the whole history of the Universe, including the early accelerated epoch (inflation) and  the late-time acceleration at the current epoch, Refs. \cite{2008PhRvD..77d6009C,2006PhRvD..74h6005N,2007PhLB..657..238N,2008PhRvD..78d6006N,2008PhRvD..77b6007N}, where the possible future singularities have been studied in the context of f(R)-gravity(see Ref.~\cite{2008PhRvD..78d6006N}). It is important to remark that the main problem that this kind of theories found at the begining of its development was the local gravitational test; nowadays several viable models have been proposed, which pass the solar system tests and reproduce the cosmological history (see \cite{2007PhLB..654....7A,2008PhRvD..77j7501C,2007PhRvD..76f4004H,2006PhRvD..74h6005N,2007PhLB..657..238N,2008arXiv0801.4843N,2008PhRvD..78d6006N,2008PhRvD..77b6007N,2008PhRvD..77b3503P,2007JETPL..86..157S}). \\
The reconstruction of f(R)-gravity is shown, to be possible in the cosmological context by using an auxiliary scalar field and then  various examples are given where the current accelerated expansion is reproduced and also  the whole history of the Universe. Also  the analogy between the so-called dark fluids, whose EoS is inhomogeneous and which have been investigated as effective descriptions of dark energy in Chapter 3, and the f(R)-theories  is investigated. \\

\section{Reconstruction of f(R)-gravity}

In this section, it will be shown how f(R) theory may be reconstructed in such a way that cosmological solutions can be  obtained. Let us start with the action for f(R)-gravity:
\be
S=\frac{1}{2\kappa^2}\int d^4x\sqrt{-g}\left(f(R)+L_m\right)\ . \label{3.1.1}
\ee
Here $L_m$ denotes the lagrangian of some kind of matter. The field equations are obtained by varying the action on $g_{\mu\nu}$, then they are  given by:
\be
R_{\mu\nu}f'(R)-\frac{1}{2}g_{\mu\nu}f(R)+g_{\mu\nu}\Box f'(R)-\nabla_{\mu}\nabla_{\nu}f'(R)=\kappa^2T^{(m)}_{\mu\nu}\ ,
\label{3.1.2}
\ee
where $T^{(m)}_{\mu\nu}$ is the energy-momentum tensor for the matter that filled the Universe. We assume a flat FLRW metric:
\be
ds^2=-dt^2+a^2(t)\sum^{3}_{i=1}dx_{i}^2 \ .
\label{3.1.2a}
\ee
Then, if $T^{(m)}_{\mu\nu}$ is a perfect fluid, the modified Friedmann equations  for the Hubble parameter $H(t)=\frac{\dot{a}}{a}$, take the form:
\bea
\frac{1}{2}f(R)-3(H^2+\dot{H})f'(R)+18f''(R)(H^2\dot{H}+H\ddot{H})=\kappa^2\rho_{m}\ , \nn
\frac{1}{2}f(R)-(3H^2+\dot{H})f'(R)-\Box f'(R)=-\kappa^2p_m\ ,   
\label{3.1.3}
\eea
where the Ricci scalar is given by $R=6(2H^2+\dot{H})$. Hence,  by the equations (\ref{3.1.3}), any cosmology may be reproduced for a given function $f(R)$. Nevertheless, in general it is very difficult to get an exact cosmological solution directly from (\ref{3.1.3}). It is a very useful  technique developed in \cite{2003PhRvD..68l3512N,2006PhRvD..74h6005N}, where an auxiliary scalar field without kinetic term is introduced, then the action (\ref{3.1.1}) is rewritten as follows:
\be
S=\frac{1}{2\kappa^2}\int d^4x\sqrt{-g}\left(P(\phi)R+Q(\phi)\right)\ ,
\label{3.1.4}
\ee
where the scalar field $\phi$ has no kinetic term. By  variation on the metric tensor $g_{\mu\nu}$ the field equation is obtained:
\be
\frac{1}{2}g_{\mu\nu}\left( P(\phi)R+Q(\phi)\right)+P(\phi)R_{\mu\nu}+g_{\mu\nu}\Box P(\phi)-\nabla_{\mu}\nabla_{\nu}P(\phi)= \kappa^2T^{(m)}_{\mu\nu}\ .
\label{3.1.5}
\ee
The action (\ref{3.1.4}) gives an additional equation for the scalar field $\phi$, obtained directly from the action by varying it with respect to $\phi$:
\be
P'(\phi)R+Q'(\phi)=0\ ,
\label{3.1.6}
\ee   
here the primes denote derivatives respect $\phi$. This equation may be resolved  with the scalar field as a function of $R$, $\phi=\phi(R)$, and then, replacing this result in the action (\ref{3.1.4}), the action (\ref{3.1.1}) is recovered,
\be
f(R)=P\left(\phi(R)\right)R+Q\left(\phi(R)\right)\ .
\label{3.1.7}
\ee
Hence, any cosmological model could be solved by the field equation (\ref{3.1.5}), and then by (\ref{3.1.6}) and (\ref{3.1.7}) the function $f(R)$ is obtained. For the metric (\ref{3.1.2a}), the Friedmann equations read:
\bea
3H\frac{dP(\phi)}{dt}+3H^2P(\phi)+\frac{1}{2}Q(\phi)-\kappa^2\rho_m=0\ , \nn
\frac{d^2P(\phi)}{dt^2}+2H\frac{dP(\phi)}{dt}+(2\dot{H}+3H^2)P(\phi)+\kappa^2p=0\ .
\label{3.1.8}
\eea
We redefine the scalar field such that it is chosen to be the time coordinate $\phi=t$. The perfect fluid define by the energy-momentum tensor $T^{(m)}_{\mu\nu}$ may be seen as a sum of the different components (radiation, cold dark matter,..) which filled our Universe and whose  equation of state (EoS) is given by $p_m=w_m\rho_m$, then by the energy momentum conservation $\dot{\rho}_m +3H(1+w_m)\rho_m=0$, it gives:
\be
\rho_m=\rho_{m0}\exp\left(-3(1+w_m)\int dtH(t)\right)\ .
\label{3.1.9}
\ee
Hence, taking into account the equations (\ref{3.1.8}) and (\ref{3.1.9}) the Hubble parameter may be calculated as a function of the scalar field $\phi$, $H=g(\phi)$. By combining the equations (\ref{3.1.8}), the function $Q(\phi)$ is deleted, and it yields:
\be
2\frac{d^2P(\phi)}{d\phi^2}-2g(\phi)\frac{dP(\phi)}{d\phi}+4g'(\phi)P(\phi)+(1+w_m)\exp\left[-3(1+w_m)\int d\phi g(\phi) \right] =0\ .
\label{3.1.10}
\ee
By resolving this equation for a given function $P(\phi)$, a cosmological solution $H(t)$ is found, and the function $Q(\phi)$ is obtained by means of the equation given by (\ref{3.1.8}):
\be
Q(\phi)=-6(g(\phi))^2P(\phi)-6g(\phi)\frac{dP(\phi)}{dt}\ .
\label{3.1.11}
\ee
If we neglect the contribution of matter, then the equation (\ref{3.1.10}) is a first order differential equation on $g(\phi)$, and it can be easily resolved. The solution found is the following:
\be
g(\phi)=-\sqrt{P(\phi)}\int d\phi \frac{P''(\phi)}{2P^{2/3}(\phi)}+ kP(\phi)\ ,
\label{3.1.12}
\ee
where $k$ is an integration constant. As an example to show this construction, let us choose the following function that, as it is shown below, reproduce  late-time acceleration:
\be
P(\phi)=\phi^\alpha\ , \quad \mbox{where} \quad \alpha>1\ .
\label{3.1.13}
\ee
Then, by the result (\ref{3.1.12}), the following solution is found:
\be
g(\phi)=k\phi^{\alpha/2}+\frac{\alpha(\alpha-1)}{\alpha+2}\frac{1}{\phi}\ ,
\label{3.1.14}
\ee
where $k$ is an integration constant. By the expression (\ref{3.1.11}), the function $Q(\phi)$ is given by:
\be
Q(\phi)=-6\left[(k\phi)^2+\left(k+\frac{\alpha(2\alpha+1)}{\alpha+2} \right)\phi^{\frac{3\alpha-2}{2}}+\frac{\alpha^2(\alpha-1)(2\alpha+1)}{(\alpha+2)^2}\phi^{\alpha-2}  \right]\ .
\label{3.1.15}
\ee
The function (\ref{3.1.14}) gives the following expression for the Hubble parameter:
\be
H(t)=kt^{\alpha/2}+\frac{\alpha(\alpha-1)}{\alpha+2}\frac{1}{t}\ .
\label{3.1.16}
\ee
This solution may reproduce a Universe that passes through two phases for a conveniently choice of $\alpha$. For small times the Hubble and the  acceleration parameter take the expressions:
\bea
H(t)\sim\frac{\alpha(\alpha-1)}{\alpha+2}\frac{1}{t}\ , \nn
\frac{\ddot{a}}{a}\sim -\frac{\alpha(\alpha-1)}{\alpha+2}\left(1-\frac{\alpha(\alpha-1)}{\alpha+2} \right) \frac{1}{t^2}\ ,
\label{3.1.17}
\eea
where if $1+\sqrt{3}<\alpha\leq2$, the Universe is in a decelerated epoch for small times, which may be interpreted as the radiation/matter dominated epochs. When $t$ is large, the Hubble parameter takes the form:
\be
 H(t)=kt^{\alpha/2}\ .
\label{3.1.18}
\ee
This clearly gives an accelerated expansion that coincides with the current expansion that Universe experiences nowadays. Finnally, the expression for $f(R)$ (\ref{3.1.7}) is calculated by means of (\ref{3.1.13}) and (\ref{3.1.15}), and by the expression of the Ricci scalar $R=6(2g^2(\phi)+g(\phi))$, which is used to get $\phi(R)$. For simplicity, we study the case where $\alpha=2$, which gives:
\be
\phi=\sqrt{\frac{R-2k\pm\sqrt{R(1-2k)}}{2}}\ .
\label{3.1.19}
\ee  
By inserting this expression into (\ref{3.1.13}) and (\ref{3.1.15}), the function $f(R)$ is obtained:
\be
f(R)=\left[R-6(k(k+1)+5/2) \right]\frac{R-2k\pm\sqrt{R(1-2k)}}{2}+const\ .
\label{3.1.20}
\ee
Thus, with this expression for the function $f(R)$, the current cosmic acceleration is reproduced with the solution (\ref{3.1.16}). In general, as it is seen in the following example, it is very difficult to reconstruct the function $f(R)$ for the whole expansion history, and even more difficult for the kind of models that unify inflation and cosmic acceleration, in this cases it is convenient to study the asymptotic behaviour of the model, and then by resolving the equations, the expression for $f(R)$ is obtained .\\ \\

As a second example, one could try to reconstruct the whole Universe  history, from inflation to  cosmic acceleration by the f(R)-gravity. In this case, we proceed in the inverse way than above, by suggesting a function $g(\phi)$, and  trying to reconstruct the expression $f(R)$ by calculating $P(\phi)$ and $Q(\phi)$ by means of equations (\ref{3.1.8}). We study an example suggested in the above chapter, where:
\be
g(\phi)=\frac{H_1}{\phi^2}+\frac{H_0}{t_s-\phi}\ .
\label{3.1.21}
\ee
For this function the Hubble parameter takes the form $H(t)=\frac{H_1}{t^2}+\frac{H_0}{t_s-t}$. To reconstruct the form of $f(R)$ we have to resolve the equation (\ref{3.1.10}). For simplicity we study the assimptotically behaviour of $H(t)$ in such a way that allow us to resolve easily the equation (\ref{3.1.10}) for $P(\phi)$. Then, for small $t$ ($t<<t_s$), the Hubble and acceleration parameters read:
\be
H(t)\sim\frac{H_1}{t^2}\, \quad \frac{\ddot{a}}{a}\sim \frac{H_1}{t^2}\left(\frac{H_1}{t^2}-\frac{2H_1}{t}+\frac{2H_0}{t_s-t} \right)\ .
\label{3.1.22}
\ee
As it is observed, for t close to zero, $\frac{\ddot{a}}{a}>0$, so the Universe is in an accelerated epoch during some time, which may be interpreted as the inflation epoch, and for $t>1/2$ ($t<<t_s$), the Universe enters in a decelerated phase, interpreted as the radiation/matter dominated epoch. The equation (\ref{3.1.10}) for $P(\phi)$, neglecting the matter component, is given by:
\be
\frac{d^2P(\phi)}{d\phi^2}-\frac{H_1}{�\phi^2}\frac{dP(\phi)}{d\phi}-\frac{4H_1}{\phi^3}P(\phi)=0\ .  
\label{3.1.23}
\ee
The solution of (\ref{3.1.23}) is:
\be
P(\phi)=k\phi^{-4}+\frac{20k}{�H_1}\phi^{-3}+\frac{120k}{�H^2_1}\phi^{-2}+\frac{240k}{�H^3_1}\phi^{-1}+\frac{120k}{�H_1^4}\ , 
\label{3.1.24}
\ee
where $k$ is an integration constant. The function $Q(\phi)$ given by (\ref{3.1.11}), takes the form:
\be
Q(\phi)=-6\frac{H_1}{�\phi^2}\left[kH_1\phi^{-6}+6k\phi^{-4}-\frac{120k}{�H^{3}_1}\phi^{-2} \right]\ .
\label{3.1.25}
\ee
By using  the expression  for the Ricci scalar, the relation $\phi(R)\sim(12H^2_1/R)^{1/4}$ is found,  then, the function $f(R)$ for small values of $\phi$ is approximately:
\be
f(R)\sim\frac{k}{24H_1}R^2\ .
\label{3.1.26}
\ee
Hence, by the expression (\ref{3.1.26}) the early cosmological behaviour of the Universe (\ref{3.1.22}), where a first accelerated epoch (inflation) occurs and after it,   a decelerated phase comes (radiation/matter dominated epochs),  is reproduced. Let us now investigate the large values for $t$ ($t$ close to the Rip time $t_s$). In this case the Hubble and the acceleration parameters for the solution (\ref{3.1.21}) take the form:
\be
g(\phi)\sim\frac{H_0}{�t_s-\phi} \quad \rightarrow H(t)\sim\frac{H_0}{�t_s-t}\ \quad \mbox{and} \quad \frac{\ddot{a}}{�a}\sim \frac{H_0(H_0+1)}{�(t_s-t)^2}\ .
\label{3.1.27}
\ee
As it is observed, for large $t$ the solution (\ref{3.1.21}) gives an accelerated epoch which enters in a phantom phase ($\dot{H}>0$) and ends in a Big Rip singularity at $t=t_s$. In this case the equation for $P(\phi)$ reads:
\be
\frac{d^2P(\phi)}{d\phi^2}-\frac{H_0}{�t_s-\phi}\frac{dP(\phi)}{d\phi}+\frac{2H_0}{(t_s-\phi)^2}P(\phi)=0\ .
\label{3.1.28}
\ee
The possible solutions of the equation (\ref{3.1.28}) depend on the value of the constant $H_0$, as it follows:
\begin{enumerate}

 \item If $H_0>5+2\sqrt{6}$ or $H_0<5-2\sqrt{6}$, then the following solution for $P(\phi)$ is found: 
\bea
P(\phi)=A(t_s-\phi)^{\alpha_+}+B(t_s-\phi)^{\alpha_-}, \nn 
\mbox{where} \quad \alpha_{\pm}=\frac{H_0+1\pm\sqrt{H_0(H_0+10)+1}}{�2}\ ,
\label{3.1.29}
\eea
Then, through the expression (\ref{3.1.11}), the function $Q(\phi)$ is calculated. In this case, for $t$ close to $t_s$, the Ricci scalar takes the form $R=\frac{6H_0(2H_0+1)}{�(t_s-t)^2}$, and hence it takes large values, diverging at the Rip time $t=t_s$. The function $f(R)$, for a large $R$, takes the form:
\be
f(R)\sim R^{1-\alpha_-/2}\ .
\label{3.1.30}
\ee

\item If  $5-2\sqrt{6}<H_0<5+2\sqrt{6}$, the solution of (\ref{3.1.28}) is given by:
\[
 P(\phi)=(t_s-\phi)^{-(h_0+1)/2}\left[A\cos\left((t_s-\phi)\ln\frac{-H^2_0+10H_0-1}{2�} \right) \right.
\]

\be
\left.
+B\sin\left((t_s-\phi)\ln\frac{-H^2_0+10H_0-1}{2�} \right)  \right]\ .
\label{3.1.31}
\ee
Then, for this choice of the constant $H_0$, and by means of the equation (\ref{3.1.7}) the form of the function $f(R)$ is found:
\[
 f(R)\sim R^{(H_0+1)/4}\left[A\cos\left(R^{-1/2}\ln\frac{-H^2_0+10H_0-1}{2�} \right) \right.
\]
\be
 \left. +B\sin\left(R^{-1/2}\ln\frac{-H^2_0+10H_0-1}{2�} \right)  \right]
\ .
\label{3.1.32}
\ee
\end{enumerate}
Hence, the expressions (\ref{3.1.30}) and (\ref{3.1.32}) for $f(R)$ reproduce the behaviour of the Hubble parameter for large $t$ given in (\ref{3.1.27}), where a phantom accelerated epoch ocurrs, and the Universe ends in a Big Rip singularity for $t=t_s$. As  is shown, this model is reproduced by (\ref{3.1.26}) for small $t$ when the curvature $R$ is large, and by (\ref{3.1.30}) or (\ref{3.1.32}) when $t$ is large. For a proper choice of the power $\alpha$, the solution for large $t$ is given by (\ref{3.1.30}), which in combination with the solution (\ref{3.1.26}) for small $t$, it looks like standard gravity $f(R)\sim R$ for intermediate $t$. On the other hand, for negavite powers in (\ref{3.1.30}) and in combination with (\ref{3.1.26}), this takes a similar form than the model suggested in \cite{2003PhRvD..68l3512N}, $f(R)\sim R+R^2+1/R$, which is known that passes qualitatively most of the solar system tests. As this is an approximated form, it is reasonable to think that this model follows from some non-linear gravity of the sort Ref.\cite{2008PhRvD..77b6007N}, which  may behave as $R^2$ for large $R$. The stability of this kind of models (for a detailed discussion see Ref.~\cite{2007PhRvD..75h3504A}), whose solutions are given by (\ref{3.1.26}) and (\ref{3.1.32}), is  studied in Ref. \cite{2006PhRvD..74h6005N}, where the transition between epochs is well done, and then, the viable cosmological evolution may be reproduced by these models.  The quantitative study of the transition between different cosmological epochs will be analyzed later.       \\ \\
Suming up it has been shown that any cosmology may be reproduced by $f(R)$-gravity by using an auxiliary scalar field and resolving the equations (\ref{3.1.10}) and (\ref{3.1.11}) to reconstruct such function of the Ricci scalar. To fix the free parameters in the theory, it would be convenient to contrast the model with the observational data as the supernova data by means of the evolution of the scale parameter which is obtained in the models shown above.

\section{F(R)-gravity and dark fluids}

In this section  the mathematical equivalence between f(R) theories, that could reproduce a given cosmology as it was seen above, and the standard cosmology with a dark fluid included whose EoS has inhomogeneous terms that depend on the Hubble parameter and its derivatives, is investigated. Let us start with the modified Friedmann equations (\ref{3.1.3}) written in the following form:
\bea
3H^2=\frac{1}{f'(R)}\left(\frac{1}{2}f(R)-\partial_{tt}f'(R)- \Box f'(R)\right) -3\dot{H}\ , \nn
-3H^2-2\dot{H}=\frac{1}{f'(R)�}\left(\Box f'(R)-\frac{1}{2}f(R)+H\partial_tf'(R)\right)-\dot{H}\ ,
\label{3.2.1}
\eea
where we have neglected the contributions of any other kind of matter. If we compare Eqs. (\ref{3.2.1}) with the standard Friedmann equations ($3H^2=\kappa^2\rho$ and $-3H^2-2\dot{H}=\kappa^2p$), we may identify both right sides of Eqs. (\ref{3.2.1}) with the energy and pressure densities of a perfect fluid, in such a way that they are given by:
\bea
\rho=\frac{1}{\kappa^2�}\left[ \frac{1}{f'(R)�}\left(\frac{1}{2�}f(R)-\partial_{tt}f'(R)- \Box f'(R)\right) -3\dot{H}\right]  \nn
p=\frac{1}{\kappa^2�}\left[\frac{1}{f'(R)�}\left(\Box f'(R)-\frac{1}{2�}f(R)+H\partial_tf'(R)\right)-\dot{H}\right]\ . 
\label{3.2.2}
\eea
Then, Eqs. (\ref{3.2.1}) take the form of the usual Friedmann equations, where the parameter of the EoS for this dark fluid is defined by:
\be
w=\frac{p}{\rho�}=\frac{\frac{1}{f'(R)�}\left(\Box f'(R)-\frac{1}{2�}f(R)+H\partial_tf'(R)\right)-\dot{H}}{\frac{1}{f'(R)�}\left(\frac{1}{2�}f(R)-\partial_{tt}f'(R)- \Box f'(R)\right) -3\dot{H}}\ .
\label{3.2.3}
\ee
And the corresponding EoS  may be written as follows:
\be
p=-\rho-\frac{1}{\kappa^2}\left(4\dot{H}+\frac{1}{f'(R)}\partial_{tt}f'(R)-\frac{H}{f'(R)}\partial_{t}f'(R)\right)\ .
\label{3.2.4}
\ee
The Ricci scalar is a function given by $R=6(2H^2+\dot{H})$, then $f(R)$ is a function on the Hubble parameter $H$ and its derivative $\dot{H}$. The inhomogeneus EoS for this dark fluid (\ref{3.2.4}) takes the form of the kind of dark fluids studying in several works (see \cite{2007EPJC...52..223B,2007EPJC...51..179B,2006PhRvD..73d3512C,2005PhRvD..72b3003N,2006PhLB..639..144N}), and particularly the form of the EoS for dark fluids investigated in Chapter 3, which is written as follows:
\[
p=-\rho+g(H,\dot{H}, \ddot{H}...), \quad \mbox{where}     
\]
\be
g(H,\dot{H}, \ddot{H}...)=-\frac{1}{\kappa^2}\left(4\dot{H}+\partial_{tt}(\ln f'(R))+(\partial_t\ln f'(R))^2-H\partial_t\ln f'(R)\right)\ .
\label{3.2.5}
\ee
Then, as  constructed in Chapter 3, by combining the Friedmann equations, it yields the following differential equation:
\be
\dot{H}+\frac{\kappa^2}{2}g(H,\dot{H}, \ddot{H}...)=0\ .
\label{3.2.6}
\ee
Hence, for a given cosmological model, the function $g$ given in (\ref{3.2.5}) may be seen as a function of cosmic time $t$, and then by the time-dependence of the Ricci scalar, the function $g$ is rewritten in terms of $R$. Finnally, the function $f(R)$ is recovered by the expression (\ref{3.2.5}). In this sense, Eq. (\ref{3.2.6}) combining with the expression (\ref{3.2.5}) results in:
\be
\frac{dx(t)}{dt}+x(t)^2-H(t)x(t)=\dot{H}(t)\ ,
\label{3.2.7}
\ee
where $x(t)=\frac{d(lnf'(R(t))}{dt}$. Eq. (\ref{3.2.7}) is a type of Riccati equation, that may be solved by a given Hubble parameter. Hence,  $f(R)$-gravity is constructed from standard cosmology where a perfect fluid with an inhomogeneus EoS is included. To show this, let us consider a simple example:
\be
H(t)=\frac{H_1}{t}\ , \quad H_1,H_0>0\ . 
\label{3.2.8}
\ee
This model describes a power-law solution, very comoon in cosmology as it can reproduces radiation/matter epochs as well as accelerating expansion. Then, by inserting (\ref{3.2.8}) inb the differential equation (\ref{3.2.7}), and after some calculations, the solution yields,
\be
f(R)=\kappa_1^2R^{1-\frac{H_1+1+\sqrt{1+H_1(10+H_1)}}{4}}+\lambda\ ,
\label{3.2.9}    
\ee
where $\kappa_1^2$ is a constant that depends on $H_1$ while $\lambda$ is an integration constant. Then, by the function (\ref{3.2.9}) the model (\ref{3.2.8}) is reproduced. The analog dark fluid that reproduces this behaviour, may be found by inserting the function (\ref{3.2.9}) in the EoS for the fluid given by (\ref{3.2.5}). Then, it could be studied as an effective fluid with its evolution and compared with other models. However, note that for a general example, the non-linear equation (\ref{3.2.7}) has to be resolved numerically, and just simple examples can be analytycally solved.
 Thus, as it is shown,  f(R)-gravity may be written as dark fluids with particular dependence on the Hubble parameter and its derivatives through its EoS (\ref{3.2.5}), so the same model may be interpretated in several ways.
 
\section{Discussions}
f(R)-gravity theories may provide an alternative description of the current accelerated epoch of our Universe and even on the whole expansion history. As it is pointed in several works, one may construct this kind of theories in accordance with the local test of gravity and with the observational data, which provide that, at the current epoch, the effective parameter of the EoS is close to -1. The next step should be to compare the different cosmological tests,  as the supernovae luminosity distance or the positions of the CMB peaks, with the F(R) models. On the other hand, we have shown two different ways of reconstruct the f(R)-gravity in the context of cosmology, in the first one, an auxiliary scalar field is used, and in the second one, the mathematical equivalence between f(R)-gravity and dark fluids with inhomogeneus EoS  shows  that while the expansion history of the Universe may be interpreted as a perfect fluid whose EoS has dependence on the cosmological evolution, this effect may be caused by the modification of the classical theory of gravity. However, there is not any complementary probe to distinguish between both descriptions of the evolution of the Universe, and thus, such kind of modified gravity is completly allowed. Hence, f(R)-gravity is an acceptable solution to the cosmological problem, that may provide new interesting constraints to look for. In the next chapters, the reproduction of $\Lambda$CDM model will be studied in the context of modified theories of gravity, and even transition to phantom epochs.

\chapter[Cosmological reconstruction of realistic F(R) gravities]{Cosmological reconstruction of realistic F(R) gravities}

\footnote{This Chapter is based on the publications: \cite{2010PhRvD..82b3519D,2009PhLB..681...74N}}In the preceeding chapter, it was explored the reconstruction of cosmological solutions in $F(R)$ gravity by means of auxiliary fields. It
turns out that in most cases this reconstruction is done in the
presence of the auxiliary scalar which may be excluded at the final step so
that any FLRW cosmology may be realized within specific reconstructed $F(R)$
gravity. However, the weak point of a so developed reconstruction scheme is
that the final function $F(R)$ represents usually some polynomial in 
positive/negative powers of scalar curvature. On the same time, the viable
models have strongly non-linear structure.

In the present chapter we develop the new scheme for cosmological
reconstruction of $F(R)$ gravity in terms of e-folding (or, redshift $z$) so
that there is no
need to use more complicated formulation with auxiliary
scalar \cite{2009PhLB..679..282B,2009PhRvD..79h3014B,2008JCAP...10..045B,2006PhLB..639..135C,2010arXiv1005.1840C,2006PhRvD..74h6005N,2007JPhA...40.6725N}. Using such technique
the number of examples are presented where $F(R)$ gravity is reconstructed
so that it gives the well-known cosmological evolution: $\Lambda$CDM
epoch, deceleration/acceleration epoch which is equivalent to presence of
 phantom and non-phantom matter, late-time acceleration with the
crossing of phantom-divide line, transient phantom epoch and oscillating
universe. It is shown that some generalization of such technique for
viable $F(R)$ gravity is possible, so that local tests are usually
satisfied. In this way, modified gravity unifying inflation,
radiation/matter dominance and dark energy eras may be further
reconstructed in the early or in the late universe so that the future
evolution may be different.
This opens the way to non-linear reconstruction of realistic $F(R)$ gravity.
Moreover, it is demonstrated that cosmological reconstruction
of viable modified gravity may help in the formulation of non-singular
models in finite-time future. The reconstruction suggests the way to change
some cosmological predictions of the theory in the past or in the future
so that it becomes easier to pass the available observational data.
Finally, we show that our method works also for modified gravity with
scalar theory and any requested cosmology may be realized within such
theory too.

\section{Cosmological reconstruction of modified $F(R)$ gravity}

From the starting action of $F(R)$ gravity, the field equation corresponding to the first FLRW equation is:
\be
\label{4.Hm1}
0 = - \frac{F(R)}{2} + 3 \left( H^2 + \dot H\right) F'(R)
 - 18 (\left(4 H^2 \dot H + H\ddot H\right) F''(R) + \kappa^2 \rho\ .
\ee
with $R=6\dot H + 12 H^2$. We now rewrite Eq.(\ref{4.Hm1}) by using a new variable
(which is often called e-folding) instead of the cosmological time $t$,
$N=\ln \frac{a}{a_0}$.
The variable $N$ is related with the redshift $z$ by
$\e^{-N}=\frac{a_0}{a} = 1 + z$.
Since $\frac{d}{dt} = H \frac{d}{dN}$ and therefore
$\frac{d^2}{dt^2} = H^2 \frac{d^2}{dN^2} + H \frac{dH}{dN} \frac{d}{dN}$,
one can rewrite (\ref{4.Hm1}) by
\be
\label{4.RZ4}
0 = - \frac{F(R)}{2} + 3 \left( H^2 + H H'\right) F'(R)
 - 18 (\left(4 H^3 H' + H^2 \left(H'\right)^2 + H^3 H''\right) F''(R) + \kappa^2 \rho\ .
\ee
Here $H'\equiv dH/dN$ and $H''\equiv d^2 H/dN^2$.
If the matter energy density $\rho$ is given by a sum of the fluid
densities with constant EoS
parameter $w_i$, we find
\be
\label{4.RZ6}
\rho=\sum_i \rho_{i0} a^{-3(1+w_i)} = \sum_i \rho_{i0} a_0^{-3(1+w_i)} \e^{-3(1+w_i)N}\ .
\ee
Let the Hubble rate is given in terms of $N$ via the function $g(N)$ as
\be
\label{4.RZ7}
H=g(N) = g \left(- \ln\left(1+z\right)\right)\ .
\ee
Then scalar curvature takes the form: $R = 6 g'(N) g(N) + 12 g(N)^2$,
which could be solved with respect to $N$ as
$N=N(R)$. Then by using (\ref{4.RZ6}) and (\ref{4.RZ7}), one can rewrite
(\ref{4.RZ4}) as
\bea
\label{4.RZ9}
0 &=& -18 \left(4g\left(N\left(R\right)\right)^3 g'\left(N\left(R\right)\right)
+ g\left(N\left(R\right)\right)^2 g'\left(N\left(R\right)\right)^2
+ g\left(N\left(R\right)\right)^3g''\left(N\left(R\right)\right)\right) \frac{d^2 F(R)}{dR^2} \nn
&& + 3 \left( g\left(N\left(R\right)\right)^2
+ g'\left(N\left(R\right)\right) g\left(N\left(R\right)\right)\right) \frac{dF(R)}{dR}
 - \frac{F(R)}{2}
+ \sum_i \rho_{i0} a_0^{-3(1+w_i)} \e^{-3(1+w_i)N(R)}\ ,
\eea
which constitutes a differential equation for $F(R)$, where the variable
is scalar curvature $R$.
Instead of $g$, if we use $G(N) \equiv g\left(N\right)^2 = H^2$,
the expression (\ref{4.RZ9}) could be a little bit simplified:
\bea
\label{4.RZ11}
0 &=& -9 G\left(N\left(R\right)\right)\left(4 G'\left(N\left(R\right)\right)
+ G''\left(N\left(R\right)\right)\right) \frac{d^2 F(R)}{dR^2}
+ \left( 3 G\left(N\left(R\right)\right)
+ \frac{3}{2} G'\left(N\left(R\right)\right) \right) \frac{dF(R)}{dR} \nn
&& - \frac{F(R)}{2}
+ \sum_i \rho_{i0} a_0^{-3(1+w_i)} \e^{-3(1+w_i)N(R)}\ .
\eea
Note that the scalar curvature is given by $R= 3 G'(N) + 12 G(N)$.
Hence, when we find $F(R)$ satisfying the differential equation
(\ref{4.RZ9}) or (\ref{4.RZ11}),
such $F(R)$ theory
admits the solution (\ref{4.RZ7}). Hence, such $F(R)$ gravity realizes
above cosmological solution.

\section{$\Lambda$CDM model in F(R) gravity}

Let us reconstruct the $F(R)$ gravity which reproduces the $\Lambda$CDM-era. In the Einstein gravity, the FLRW equation for the $\Lambda$CDM cosmology
is given by
\be
\label{4.RZ13}
\frac{3}{\kappa^2} H^2 = \frac{3}{\kappa^2} H_0^2 + \rho_0 a^{-3}
= \frac{3}{\kappa^2} H_0^2 + \rho_0 a_0^{-3} \e^{-3N} \ .
\ee
Here $H_0$ and $\rho_0$ are constants. The first term in the r.h.s. corresponds to the
cosmological constant and the second term to the cold dark matter (CDM).
The (effective) cosmological constant $\Lambda$
in the present universe is given by $\Lambda = 3 H_0^2$.
Then one gets
\be
\label{4.RZ14}
G(N) = H_0^2 + \frac{\kappa^2}{3} \rho_0 a_0^{-3} \e^{-3N} \ ,
\ee
and $R = 3 G'(N) + 12 G(N) = 12 H_0^2 + \kappa^2\rho_0 a_0^{-3} \e^{-3N}$,
which can be solved with respect to $N$ as follows,
\be
\label{4.RZ16}
N = - \frac{1}{3}\ln \left(\frac{ \left(R - 12 H_0^2\right)}{\kappa^2 \rho_0 a_0^{-3}}\right)\ .
\ee
By considering the homogeneous part of Eq.(\ref{4.RZ11}), it takes the following form:
\be
\label{4.RZ17}
0=3\left(R - 9H_0^2\right)\left(R - 12H_0^2\right) \frac{d^2 F(R)}{d^2 R}
 - \left( \frac{1}{2} R - 9 H_0^2 \right) \frac{d F(R)}{dR} - \frac{1}{2} F(R)\ .
\ee
By changing the variable from $R$ to $x$ by $x=\frac{R}{3H_0^2} - 3$,
Eq.(\ref{4.RZ17}) reduces to the hypergeometric differential equation:
\be
\label{4.RZ19}
0=x(1-x)\frac{d^2 F}{dx^2} + \left(\gamma - \left(\alpha + \beta + 1\right)x\right)\frac{dF}{dx}
 - \alpha \beta F\ .
\ee
Here
\be
\label{4.RZ20}
\gamma = - \frac{1}{2}\ ,\alpha + \beta = - \frac{1}{6}\ ,\quad \alpha\beta = - \frac{1}{6}\ ,
\ee
Solution of (\ref{4.RZ19}) is given by Gauss' hypergeometric function $F(\alpha,\beta,\gamma;x)$:
\be
\label{4.RZ22}
F(x) = A F(\alpha,\beta,\gamma;x) + B x^{1-\gamma} F(\alpha - \gamma + 1, \beta - \gamma + 1,
2-\gamma;x)\ .
\ee
Here $A$ and $B$ are constant.
Thus, we demonstrated that modified $F(R)$ gravity may describe the
$\Lambda$CDM epoch without the need to introduce the effective
cosmological constant. However, a deeper study on the solution (\ref{4.RZ22}) is necessary. Let us write the scale factor $a$ in terms of the Ricci scalar through out (\ref{4.RZ16}), it yields
\be
a(R)=\left(\frac{\kappa^2\rho_0}{R-4\Lambda}\right)^{(1/3)}\ ,
\label{4.aR}
\ee
where $\Lambda=3H_0^2$. In order to ensure the positivity of the scale factor, the Ricci scalar has to be restricted to $R\geq4\Lambda$. We can see that the equation (\ref{4.RZ17}) has two singular points at $R=3\Lambda$ and $R=4\Lambda$. From the allowed range for the Ricci scalar, we have that one of the poles is out of the range while the other one is located at the boundary. Nevertheless, the argument of the solution (\ref{4.RZ22}), given by $x=\frac{R}{\Lambda} - 3$, ensure the convergency of the solution when $|x|\leq1$, what restricts the value of the Ricci scalar to $R\leq4\Lambda$, otherwise the function is either divergent or complex valued. Them, in order to ensure a real and finite gravitational action, we have to choose the integration constants $A=B=0$. Hence, we have to consider just the particular solution coming from the inhomogeneous part of the equation (\ref{4.RZ13}), which will depend on the kind of fluid presenced in the Universe. By considering dustlike matter ($w=0$), and substituting in the Friedmann equation (\ref{4.RZ13}), we get the particular solution,
\be
F(R)=R-2\Lambda\ ,
\label{4.PartSol}
\ee
which is the well known action for General Relativity. Then, the only physical solution for a model trying to reproduce a Hubble parameter behaving as $\Lambda$CDM solution turns to be the Einstein-Hilbert action with positive cosmological constant. Nevertheless, other approximate $\Lambda$CDM solutions can be well reproduced in $F(R)$ gravity as it is shown along the pesent work. 

\section{Reconstruction of approximate $\Lambda$CDM solutions}

As an another example, we reconstruct $F(R)$ gravity reproducing the
system with non-phantom matter
and phantom matter in the Einstein gravity, whose FLRW equation is given by
\be
\label{4.RZ23}
\frac{3}{\kappa^2} H^2 = \rho_q a^{-c} + \rho_p a^c\ .
\ee
Here $\rho_q$, $\rho_p$, and $c$ are positive constants.
When $a$ is small as in the early universe, the first term in the r.h.s.
dominates and it behaves
as the universe described by the Einstein gravity with a matter whose EoS
parameter is $w=-1 + c/3>-1$,
that is, non-phantom like. On the other hand, when $a$ is large as in the
late universe, the second
term dominates and behaves as a phantom-like matter with $w= - 1 - c/3 < -1$.
Then since $G(N) \equiv g\left(N\right)^2 = H^2$, we find
\be
\label{4.RZ24}
G = G_q \e^{-cN} + G_p \e^{cN}\ ,\quad G_q \equiv \frac{\kappa^2}{3}\rho_q a_0^{-c}\ ,\quad
\quad G_p \equiv \frac{\kappa^2}{3}\rho_p a_0^c\ .
\ee
Then since $R = 3 G'(N) + 12 G(N)$,
\be
\label{4.RX25}
\e^{cN}=\frac{R\pm \sqrt{R^2 - 4\left(144 - 9c^2\right)}}{2\left(12 + 3 c\right)}\ ,
\ee
when $c\neq 4$ and
\be
\label{4.RX26}
\e^{cN}=\frac{R}{24G_p}\ ,
\ee
when $c=4$.
In the following, just for simplicity, we consider $c=4$ case. In the case, the
non-phantom matter corresponding to the first term in the r.h.s. of
(\ref{4.RZ23}) could be radiation with $w=1/3$. Then Eq.(\ref{4.RZ11}) in this case
is given by
\be
\label{4.RX27}
0 = -6 \left(\frac{24 G_p G_q}{R} + \frac{R}{24}\right) R \frac{d^2 F(R)}{d R^2}
+ \frac{9}{2}\left( - \frac{24 G_p G_q}{R} + \frac{R}{24}\right) \frac{d F(R)}{d R}
 - \frac{F(R)}{2}\ .
\ee
By changing variable $R$ to $x$ by $R^2 = - 576 G_p G_q x$,
we can rewrite Eq.(\ref{4.RX27}) as
\be
\label{4.RX29}
0 = (1-x)x\frac{d^2 F}{dx^2} + \left( \frac{3}{4} + \frac{x}{4}\right) \frac{dF}{dx}
 - \frac{F}{2}\ ,
\ee
whose solutions are again given by Gauss' hypergeometric function (\ref{4.RZ22}) with
\be
\label{4.RX30}
\gamma = \frac{3}{4}\ ,\quad \alpha + \beta + 1 = - \frac{1}{4}\ ,\quad
\alpha \beta = \frac{1}{2}\ .
\ee
Let us now study a model where the dominant component is phantom-like one.
Such kind of system can
be easily expressed in the standard General Relativity when a phantom
fluid is considered, where
the FLRW equation reads $H^2(t)=\frac{\kappa^2}{3}\rho_\mathrm{ph}$,
Here the subscript $ph$ denotes the phantom nature of the fluid.
As the EoS for the fluid is given by $p_\mathrm{ph}=w_\mathrm{ph}\rho_\mathrm{ph}$
with $w_\mathrm{ph}<-1$,
by using the conservation equation $\dot{\rho}_\mathrm{ph}+3H(1+w_\mathrm{ph})\rho_\mathrm{ph}=0$,
the solution
for the FLRW equation $H^2(t)=\frac{\kappa^2}{3}\rho_\mathrm{ph}$ is well known,
and it yields $a(t)=a_0(t_s-t)^{-H_0}$,
where $a_0$ is a constant, $H_0=-\frac{1}{3(1+w_\mathrm{ph})}$ and $t_s$ is the so-called Rip time.
Then, the solution describes the Universe that ends at the
Big Rip singularity in
the time $t_s$. The same behavior can be achieved in $F(R)$ theory with no need to introduce
a phantom fluid. The equation (\ref{4.RZ11}) can be solved and the expression for the $F(R)$ that
reproduces the solution is reconstructed. The expression for the Hubble parameter
as a function of the number of e-folds is given by
$H^2(N)=H^2_0 \e^{2N/H_0}$.
Then, the equation (\ref{4.RZ11}), with no matter contribution, takes the form:
\be
R^2\frac{d^2F(R)}{dR^2}+AR\frac{dF(R)}{dR}+BF(R)=0\ ,
\label{4.RX34}
\ee
where $A=-H_0(1+H_0)$ and $B=\frac{(1+2H_0)}{2}$. This equation is the well known Euler
equation whose solution yields
\be
F(R)=C_1R^{m_+}+C_2R^{m_-}\ , \quad \text{where} \quad m_{\pm}=\frac{1-A\pm\sqrt{(A-1)^2-4B}}{2}\ .
\label{4.RX35}
\ee
Thus, the phantom dark energy cosmology $a(t)=a_0(t_s-t)^{-H_0}$ can be
also obtained in the frame of $F(R)$ theory and no phantom fluid is needed.

We can consider now the model where the transition to the phantom epoch
occurs. It has been
pointed out that $F(R)$ could behave as an effective cosmological
constant, such that its current
observed value is well reproduced. One can reconstruct the model where
late-time acceleration is reproduced
by an effective cosmological constant and then the phantom barrier is
crossed. Such transition,
which may take place at current time, could be achieved in $F(R)$
gravity. The solution considered can be expressed as:
\be
H^2=H_1\left( \frac{a}{a_0}\right)^{m}+H_0=H_1 \e^{m N}+H_0\ ,
\label{4.RX36}
\ee
where $H_1$,$H_0$ and $\alpha$ are positive constants. This solution can be constructed
in GR when a cosmological constant and a phantom fluid are included. In
the present case,
the solution (\ref{4.RX36}) can be achieved just by an $F(R)$ function, such that the transition
from non-phantom to phantom epoch is reproduced.
Scalar curvature can be written in terms of the number of e-folds again.
Then, the equation (\ref{4.RZ11}) takes the form:
\be
x(1-x)F''(x)+\left[x\left( -\frac{6+m}{6m}\right) -\frac{1}{3m} \right] F'(x)-\frac{m +4}{m}F(x)=0\ ,
\label{4.RX38}
\ee
where $x=\frac{1}{3H_0(m+4)}(12H_0-R)$.
The equation (\ref{4.RX38}) reduces to the
hypergeometric differential equation (\ref{4.RZ22}), so the solution is given, as in some
of the examples studied above, by the Gauss' hypergeometric function (\ref{4.RZ23}), whose parameters
for this case are given by
\be
\gamma= -\frac{1}{3m}\ , \quad \alpha+\beta=-\frac{3m+2}{2m}\ , \quad \alpha\beta=\frac{m+4}{2m}\ ,
\label{4.RX39}
\ee
and the obtained $F(R)$ gravity produces the FLRW cosmology with the
late-time crossing of the phantom
barrier in the universe evolution.

Another example with transient phantom behavior in $F(R)$ gravity can be
achieved by following the same
reconstruction described above. In this case, we consider the following Hubble parameter:
\be
H^2(N)=H_0\ln\left(\frac{a}{a_0} \right)+H_1=H_0N+H_1\ ,
\label{4.RX40}
\ee
where $H_0$ and $H_1$ are positive constants. For this model, we have a contribution of
an effective cosmological constant, and a term that will produce
a superaccelerating phase
although no future singularity will take place(compare with
earlier model \cite{2005CQGra..22L..35A} with transient phantom era). The solution for
the model (\ref{4.RX40})
can be expressed as a function of time
\bea
H(t)=\frac{a_0H_0}{2}(t-t_0) \
\label{4.RX41}
\eea
Then, the Universe is superaccelerating, but as it can be seen from
(\ref{4.RX41}), in spite of its phantom nature,
no future singularity occurs.
The differential reconstruction equation can be obtained as
\be
a_2x\frac{d^2F(x)}{dx}+(a_1x+b_1)\frac{dF(x)}{dx}+b_0F(x)=0\ ,
\label{4.RX44}
\ee
where we have performed a variable change $x=H_0N+H_1$, and the constant parameters are
$a_2=H^2_0$, $a_1=-H_0$ $b_1=-\frac{H^2_0}{2}$ and $b_0=2H_0$. The equation (\ref{4.RX44}) is
a kind of the degenerate hypergeometric equation, whose solutions are given by the Kummers
series $K(a,b;x)$:
\be
F(R)=K\left(-2, -\frac{1}{2H_0};\frac{R-3H_0}{12} \right)\ .
\label{4.RX45}
\ee
Hence, such $F(R)$ gravity has cosmological solution with the transient
phantom behavior which does not evolve to future singularity.

Let us now consider the case where a future contracting Universe is reconstructed in this kind
of models. We study a model where the universe is currently accelerating,
then the future contraction
of the Universe occurs. The following solution for the Hubble parameter is considered,
\be
H(t)=2H_1(t_0-t)\ ,
\label{4.RS1}
\ee
where $H_1$ and $t_0$ are positive constants. For this example, the Hubble parameter (\ref{4.RS1})
turns negative for $t>t_0$, when the Universe starts to contract itself, while for $t\ll t_0$,
the cosmology is typically $\Lambda$CDM one.
Using notations $\tilde{H_0}=4H_1 t_0^2 $ and $\tilde{H_1}=4H_1$ and
repeating the above calculation, one gets:
\be
F(R)=K\left(-8\tilde{H_1}, -\frac{\tilde{H_1}}{8}; \frac{12\tilde{H_0}-3\tilde{H_1}-R}{12\tilde{H_1}} \right)\ .
\label{4.RS7}
\ee
Hence, the oscillating cosmology (\ref{4.RS1}) that describes the
asymptotically contracting Universe with a current
accelerated epoch can be found in specific $F(R)$ gravity.

Thus, we explicitly demonstrated that $F(R)$ gravity reconstruction is
possible for any cosmology under consideration without the need to
introduce the auxiliary scalar. However, the obtained
modified gravity has typically polynomial structure with terms which
contain positive and
negative powers of curvature as in the first such model unifying
the early-time inflation and late-time acceleration \cite{2003PhRvD..68l3512N,2004GReGr..36.1765N}.
As a rule such models do not pass all the local gravitational tests.
Some generalization of above cosmological reconstruction is necessary.

\subsection{Cosmological solutions in $f(R)$ gravity with the
presence of an inhomogeneous EoS fluid}

We consider now a Universe governed by some specific $f(R)$ theory in
the presence of a perfect fluid, whose equation of state is given by,
\be
p=w(a)\rho+\zeta(a)\;,
\label{4.S14}
\ee
where $w(a)$ and $\zeta(a)$ are functions of the scale factor $a$,
which could correspond to the dynamical behavior of the fluid and to
its viscosity. Let us write the FLRW equations for f(R) as following,
\bea
H^2=\frac{1}{3}(\rho'+\rho_{f(R)})\ ;, \\
2\dot{H}+3H^2=-(p'+p_{f(R)})\;,
\label{4.S15}
\eea
where $\rho'=\frac{\rho}{f'(R)}$ and $p'=\frac{p}{f'(R)}$. The
pressure and energy density with the subscript $f(R)$ contains the
terms corresponded to $f(R)$ and are defined as
\bea
\rho_{f(R)}=\frac{1}{f'}\left(\frac{Rf'-f}{2}-3H\dot{R}f''\right)\;,
\\
p_{f(R)}=\frac{1}{f'}\left(\dot{R}^2f'''+2H\dot{R}f''+\ddot{R}f''+\frac{1}{2
}(f-Rf')\right)\;.
\label{4.S16}
\eea
Then, by combining both FLRW equations, and using the equation of state defined in
(\ref{4.S14}), we can write
\be
\zeta(a)=\left(w(a)\rho_{f(R)}-p_{f(R)}-2\dot{H}-3(1+w(a))H^2\right)f'(R(a))\;.
\label{4.S17}
\ee
As $\zeta(a)$ just depends on the Hubble parameter and its
derivatives, for some specific solutions, any kind of cosmology can
be reproduced. Let us consider the example,
\be
\frac{3}{\kappa^2}H^2=G_{\rho}a^{-c}+G_{q}a^{d}\;,
\label{4.S18}
\ee
where $G_{\rho}$ and $G_q$ are constants. We can check that in this
solution, the first term in the r.h.s. corresponds to a fluid with
EoS $w_p=-1+c/3>-1$, while the second term, it has an equation of state
$w_q=-1-d/3<-1$, which corresponds to a phantom fluid. We could
consider a viable $f(R)$ function proposed in
\cite{2008PhRvD..77b6007N}, which is given by
\be
F(R)=\frac{\alpha R^{m+l}-\beta R^n}{1+\gamma R^l}.
\label{4.S19}
\ee
This function is known to pass the local gravity tests and could
contribute to drive the Universe to an accelerated phase. Then, by
introducing (\ref{4.S18}) and (\ref{4.S19}) in the expression for
$\zeta(a)$ in (\ref{4.S17}), we obtain the equation of state for the inhomogeneous
fluid that, together with $f(R)$, reproduces the solution
(\ref{4.S18}), which for $c=3$ and $d\geq0$ reproduces the $\Lambda$CDM
model, and probably drives the Universe evolution into a phantom
phase in the near future.

\section{Cosmological evolution of viable $F(R)$ gravity}

In this section, we show how the cosmological reconstruction may be applied
to viable modified gravity which passes the local gravitational tests.
In this way, the non-linear structure of modified $F(R)$ gravity may be
accounted for, unlike the previous section where only polynomial $F(R)$
structures may be reconstructed.
Let us write $F(R)$  in the following form:
$F(R) = F_0(R) + F_1(R)$.
Here we choose $F_0(R)$ as a known function like that of GR or one of
viable $F(R)$ models introduced in \cite{2007PhLB..654....7A,2008PhRvD..77j7501C,2007PhRvD..76f4004H,2007PhRvD..75h4010L},
or viable $F(R)$ theories unifying inflation with dark 
energy \cite{2008PhRvD..77d6009C,2007PhLB..657..238N}, for example
\be
\label{4.RR4}
F_0(R) = \frac{1}{2\kappa^2}\left( R - \frac{\left(R-R_0\right)^{2n+1} + R_0^{2n+1}}{f_0
+ f_1 \left\{\left(R-R_0\right)^{2n+1} + R_0^{2n+1}\right\}}\right)\ .
\ee

Using the procedure similar to the one of second section, one gets the
reconstruction equation corresponding to (\ref{4.RZ11})
\bea
\label{4.RR5}
0 &=& -9 G\left(N\left(R\right)\right)\left(4 G'\left(N\left(R\right)\right)
+ G''\left(N\left(R\right)\right)\right) \frac{d^2 F_0(R)}{dR^2}
+ \left( 3 G\left(N\left(R\right)\right)
+ \frac{3}{2} G'\left(N\left(R\right)\right) \right) \frac{dF_0(R)}{dR} \nn
&& - \frac{F_0(R)}{2} \nn
&& -9 G\left(N\left(R\right)\right)\left(4 G'\left(N\left(R\right)\right)
+ G''\left(N\left(R\right)\right)\right) \frac{d^2 F_1(R)}{dR^2}
+ \left( 3 G\left(N\left(R\right)\right)
+ \frac{3}{2} G'\left(N\left(R\right)\right) \right) \frac{dF_1(R)}{dR} \nn
&& - \frac{F_1(R)}{2}
+ \sum_i \rho_{i0} a_0^{-3(1+w_i)} \e^{-3(1+w_i)N(R)}\ .
\eea
The above equation can be regarded as a differential equation for $F_1(R)$.
For a given $G(N)$ or $g(N)$ (\ref{4.RZ7}), if one can solve
(\ref{4.RZ11})) as $F(R)=\hat F(R)$, we also find the solution of
(\ref{4.RR5}) as $F_1(R) = \hat F (R) - F_0(R)$.
For example, for $G(N)$ (\ref{4.RZ14}), by using (\ref{4.RZ22}), we find
\be
\label{4.RR7}
F_1(R) = A F(\alpha,\beta,\gamma;x) + B x^{1-\gamma} F(\alpha - \gamma + 1, \beta - \gamma + 1,
2-\gamma;x) - F_0(R)\ .
\ee
Here $\alpha$, $\beta$, $\gamma$, and $x$ are given by $x=\frac{R}{3H_0^2} - 3$ and (\ref{4.RZ20}).
Using $F_0(R)$ (\ref{4.RR4}) one has
\be
\label{4.RR8}
F_1(R) = A F(\alpha,\beta,\gamma;x) + B x^{1-\gamma} F(\alpha - \gamma + 1, \beta - \gamma + 1,
2-\gamma;x) - \frac{1}{2\kappa^2}\left( R - \frac{\left(R-R_0\right)^{2n+1} + R_0^{2n+1}}{f_0
+ f_1 \left\{\left(R-R_0\right)^{2n+1} + R_0^{2n+1}\right\}}\right)\ ,
\ee
which describes the asymptotically de Sitter universe.
Instead of $x=\frac{R}{3H_0^2} - 3$ and (\ref{4.RZ20}), if we choose
$\alpha$, $\beta$, $\gamma$, and $x$ as $R^2 = - 576 G_p G_q x$ and in
(\ref{4.RX30}), $F_1(R)$ (\ref{4.RR8}) shows the
asymptotically phantom universe behavior, where
$H$ diverges in future.

One may start from $F_0(R)$ given by hypergeometric function (\ref{4.RZ22})
with $x=\frac{1}{3H_0(m+4)}(12H_0-R)$ and (\ref{4.RX39}).
In such a model, there occurs Big Rip singularity.
Let $\tilde F(R)$ be $F(R)$ again given by hypergeometric function (\ref{4.RZ22}) with
$x=\frac{R}{3H_0^2} - 3$ and (\ref{4.RZ20}):
\bea
\label{4.RRR2}
&& \tilde F(R) = \tilde A F(\tilde\alpha, \tilde\beta, \tilde\gamma; \tilde x)
+ \tilde B \tilde x^{1 - \tilde\gamma} F(\tilde\alpha - \tilde\gamma + 1, \tilde\beta
 - \tilde\gamma + 1, 2 - \tilde\gamma ; \tilde x)\ , \nn
&& \tilde x=\frac{R}{3H_0^2} - 3\ ,\quad
\tilde \gamma = - \frac{1}{2}\ , \tilde\alpha + \tilde \beta = - \frac{1}{6}\ ,
\quad \tilde\alpha \tilde\beta = - \frac{1}{6}\ .
\eea
If we choose $F(R)=\tilde F(R)$, the $\Lambda$CDM model emerges.
Then choosing $F_1(R) = \tilde F(R) - F_0(R)$,
the Big Rip singularity, which occurs in $F_0(R)$ model, does not appear
and
the universe becomes asymptotically de Sitter space. Hence, the
reconstruction method suggests the way to create the non-singular modified
gravity models \cite{2005CQGra..22L..35A,2009PhLB..679..282B,2009PhRvD..79h3014B,2008JCAP...10..045B,2009PhRvD..79l4007C,2008PhRvD..78h3515D,2009PhRvD..79b4009K,2008PhRvD..78d6006N,2009arXiv0904.3445S}. Of course, it should be
checked that reconstruction term is not large (or it affects only the
very early-time/late-time universe) so that the theory passes the local
tests as it was before the adding of correction term.

Gauss' hypergeometric function $F(\alpha,\beta,\gamma;x)$ is defined by
\be
\label{4.RZ25}
F(\alpha,\beta,\gamma;x) = \frac{\Gamma(\gamma)}{\Gamma(\alpha) \Gamma(\beta)}\sum_{n=0}^\infty
\frac{\Gamma(\alpha+n) \Gamma(\beta+n)}{\Gamma(\gamma+n) n!}z^n\ .
\ee
Since
\be
\label{4.RZ27}
\alpha_0,\beta_0 = \frac{- 3m - 2 \pm \sqrt{m^2 - 20m +4}}{4m}<0\ ,\quad
\tilde\alpha, \tilde\beta = \frac{ -1 \pm 5 }{12}\ ,
\ee
when $R$ is large, $F_1(R)$ behaves as
$F_1(R) \sim R^{\left(3m+2 + \sqrt{m^2 - 20m +4}\right)/4m}$.
In spite of the above expression, since the total $F(R)=F_0(R) + F_1(R)$ is
given by $\tilde F(R)$ (\ref{4.RRR2}), the Big Rip type singularity does
not occur.
The asymptotic behavior of $F_1(R)$ cancels the large $R$ behavior in
$F_0(R)$ suggesting the way to present the non-singular cosmological
evolution.

We now consider the case that $H$ and therefore $G$ oscillate as
\be
\label{4.RR9b}
G(N) = G_0 + G_1 \sin \left( \frac{N}{N_0} \right)\ ,
\ee
with positive constants $G_0$, $G_1$, and $N_0$.
Let the amplitude of the oscillation is small but the frequency is large:
\be
\label{4.RR10}
G_0 \gg \frac{G_1}{N_0}\ ,\quad N_0 \gg 1\ .
\ee
When $G_1=0$, we obtain de Sitter space, where the scalar curvature is a constant $R=12 G_0$.
Writing $G(N)$ as
\be
\label{4.RR11}
G = \frac{R}{6} - G_0\ ,
\ee
by using (\ref{4.RZ11}), one arrives at general relativity:
\be
\label{4.RR12}
F(R) = c_0 \left( R - 6G_0 \right)\ .
\ee
Instead of (\ref{4.RR11}), using an arbitrary function $\tilde F$, if we write
\be
\label{4.RR12B}
G = G_0 + \tilde F (R) - \tilde F(12 G_0)\ ,
\ee
we obtain a general $F(R)$ gravity, which admits de Sitter space solution.
When $G_1 \neq 0$, under the assumption (\ref{4.RR10}), one may identify
$F(R)$ in (\ref{4.RR12}) with $F_0(R)$.
We now write $G(N)$ and the scalar curvature $R$ as
\be
\label{4.RR13}
G(N) = \frac{R}{6} - G_0 + \frac{G_1}{N_0}g(N)\ ,\quad R= 12 G_0 + \frac{3G_1}{N_0} r(N) \ ,
\ee
with adequate functions $g(N)$ and $r(N)$.
Then since $R = 6 g'(N) g(N) + 12 g(N)^2$ and from (\ref{4.RR10}), we find
\be
\label{4.RR14}
g(N) = - \left(N_0 \sin\frac{N}{N_0}+\frac{1}{2}\cos\frac{N}{N_0} \right)\ ,
\quad r(N)= 4N_0\sin\frac{N}{N_0}+\cos\frac{N}{N_0}
\ee
By assuming
\be
\label{4.RR15}
F(R) = c_0 \left( R - 6G_0 + \frac{G_1^2}{N_0^3} f(R) \right)\ ,
\ee
and identifying
\be
\label{4.RR16}
F_1(R) = \frac{c_0 G_1^2}{N_0^3} f(R)\ ,
\ee
from (\ref{4.RR5}), one obtains
\be
\label{4.RR17}
0 = G_0 \frac{df}{dr} - \sin \left(\frac{N}{N_0}\right)
+ o\left(\frac{G_1}{N_0},N_0\right)\ ,
\ee
which can be solved as
\be
\label{4.RR18}
f(R) = - \frac{1}{2G_0} \left(\cos^{-1} r \mp r \sqrt{1 - r^2 } \right)\ .
\ee
Then at least perturbatively, one can construct a model which exhibits the
oscillation of $H$.

Before going further, let us find $F(R)$ equivalent to the Einstein
gravity with a perfect fluid with
a constant EoS parameter $w$, where $H$ behaves as
\be
\label{4.RR19}
\frac{3}{\kappa^2} H^2 = \rho_0 \e^{-3(w+1)}\ .
\ee
Then
\be
\label{4.RR20}
G(N) = \frac{\kappa^2 \rho_0}{3}\e^{-3(w+1)}\ ,\quad
R(N) = \left(1 - 3w\right) \kappa^2 \rho_0 \e^{-3(w+1)}\ ,
\ee
which could be solved as
\be
\label{4.RR21}
N = - \frac{1}{3(w+1)} \ln \frac{R}{\left(1 - 3w\right) \kappa^2 \rho_0 }\ .
\ee
Therefore Eq.(\ref{4.RZ11}) has the following form:
\be
\label{4.RR22}
0= \frac{3(1+w)}{1-3w}R^2 \frac{d^2 F(R) }{dR^2} - \frac{1+3w}{2(1-3w)} R \frac{dF(R)}{dR}
 - \frac{F(R)}{2}\ ,
\ee
whose solutions are given by a sum of powers of $R$
\be
\label{4.RR23}
F(R)=F_+ R^{n_+} + F_- R^{n_-} \ .
\ee
Here $F_\pm$ are constants of integration and $n_\pm$ are given by
\be
\label{4.RR24}
n_\pm = \frac{1}{2} \left\{
\frac{7 + 9w}{6(1+w)} \pm \sqrt{
\left( \frac{7 + 9w}{6(1+w)} \right)^2 + \frac{2(1-3w)}{3(1+w)}
}\right\}\ .
\ee
If $w>-1/3$, the universe is decelerating but if $-1<w<-1/3$, the universe is accelerating
as in the quintessence scenario.

By using the solution (\ref{4.RZ22}), which mimics $\Lambda$CDM model,
 and the solution
(\ref{4.RR23}), one may consider the following model:
\be
\label{4.RR25}
F(x) = \left\{ A F(\alpha,\beta,\gamma;x) + B x^{1-\gamma} F(\alpha - \gamma + 1, \beta - \gamma + 1,
2-\gamma;x)\right\} \frac{\e^{\lambda\left(\frac{R}{R_1} - \frac{R_1}{R}\right)}}
{\e^{\lambda\left(\frac{R}{R_1} - \frac{R_1}{R}\right)}
+ \e^{- \lambda\left(\frac{R}{R_1} - \frac{R_1}{R}\right)}}
+ F_+ R^{n_+} + F_- R^{n_-}\ .
\ee
Here $R_1$ is a constant which is sufficiently small compared with the curvature $R_0$ in the
present universe. On the other hand, we choose a positive constant $\lambda$ to be large enough.
We also choose $F_\pm$ to be small enough so that only the first term dominates when $R\gg R_1$.
Note that the factor $\frac{\e^{\lambda\left(\frac{R}{R_1} -
\frac{R_1}{R}\right)}}
{\e^{\lambda\left(\frac{R}{R_1} - \frac{R_1}{R}\right)}
+ \e^{- \lambda\left(\frac{R}{R_1} - \frac{R_1}{R}\right)}}$ behaves as step function when
$\lambda$ is large:
\be
\label{4.RR26}
\lim_{\lambda\to + \infty}
\frac{\e^{\lambda\left(\frac{R}{R_1} - \frac{R_1}{R}\right)}}
{\e^{\lambda\left(\frac{R}{R_1} - \frac{R_1}{R}\right)}
+ \e^{- \lambda\left(\frac{R}{R_1} - \frac{R_1}{R}\right)}}
= \theta(R - R_1) \equiv \left\{
\begin{array}{ll}
1 & \mbox{when}\ R>R_1 \\
0 & \mbox{when}\ R<R_1 \\
\end{array}\right. \ .
\ee
Then in the early universe and in the present universe, only the first
term dominates and
the $\Lambda$CDM universe could be reproduced. In the future universe where $R\ll R_1$, the factor
$\frac{\e^{\lambda\left(\frac{R}{R_1} - 1\right)}}
{\e^{\lambda\left(\frac{R}{R_1} - 1\right)} + \e^{- \lambda\left(\frac{R}{R_1} - 1\right)}}$
decreases very rapidly and the second terms in
(\ref{4.RR25}) dominate. Then if $w>-1/3$, the universe decelerates again
but if $-1<w<-1/3$, the universe will be accelerating as
 in the quintessence scenario.

Thus, we explicitly demonstrated that the viable $F(R)$ gravity may be
reconstructed so that any requested cosmology may be realized after the
reconstruction. Moreover, one can use the viable $F(R)$ gravity unifying
the early-time inflation with late-time acceleration (and manifesting the
radiation/matter dominance era between accelerations) and passing local
tests in such a scheme. The (small) correction term $F_1(R)$ can be always
constructed so that it slightly corrects (if necessary) the cosmological bounds being
relevant only at the very early/late universe. This scenario opens the way
to extremely realistic description of the universe evolution in $F(R)$
gravity consistent with local tests and cosmological bounds.

\section{Discussion}

In summary, we developed a general scheme for cosmological reconstruction of
modified $F(R)$ gravity in terms of e-folding (or redshift) without use of
auxiliary scalar in intermediate calculations. Using this method, it is
possible to construct the specific modified gravity which contains any
requested FLRW cosmology. The number of $F(R)$ gravity examples is
found where the
following background evolution may be realized:$\Lambda$CDM epoch,
deceleration with subsequent transition to effective phantom
superacceleration leading to Big Rip singularity, deceleration with
transition to transient phantom phase without future singularity,
oscillating universe. It is important that all these cosmologies may be
realized only by modified gravity without use of any dark components
(cosmological constant, phantom, quintessence, etc). In particular we find that the only real valued gravitational action that reproduces an exact $\Lambda$CDM expansion is the Hilbert-Einstein action with a positive cosmological constant. However, it has also shown that approximate solutions that mimic $\Lambda$CDM model at the current time, can well be reproduced in terms of a $F(R)$ function.

It is also shown that our method may be applied to viable $F(R)$ gravities
which
pass local tests and unify the early-time inflation with late-time
acceleration. In this case, the additional reconstruction may be made so
that correction term is not large and it is relevant only in the very
early/very late universe. Hence, the purpose of such additional
reconstruction is only to improve the cosmological predictions if the
original theory does not pass correctly the precise observational
cosmological
bounds. For instance, in this way it is possible to formulate the modified
gravity without finite-time future singularity.

The present reconstruction formulation shows that even if specific
realistic modified gravity
does not pass correctly some cosmological bounds (for instance, does not
lead to correct cosmological perturbations structure) it may be improved
with eventually desirable result. Hence, the successful development of
such method adds very strong argument in
favour of unified gravitational alternative for inflation, dark energy and
dark matter.

\chapter[Viable F(R) cosmology and the presence of phantom fluids]{Viable F(R) cosmology and the presence of phantom fluids}

\footnote{This Chapter is based on: \cite{2009PhRvD..80d4030E}.}Our first aim in this chapter will be to build a reliable cosmological model by using, as starting point, modified gravity theories of the family of the so-called $F(R)$ theories which comprise the class of viable models, e.g., those having an $F(R)$ function such that the theory can pass all known local gravity tests (see \cite{2007PhLB..654....7A,2008PhRvD..77j7501C,2008PhRvD..77d6009C,2009PhRvD..79d4001C,2007PhRvD..76f4004H,2003PhRvD..68l3512N,2007PhLB..657..238N,2008arXiv0801.4843N,2008PhRvD..77b6007N,2007PhRvD..75b3511O,2008PhRvD..77b3503P,2007JETPL..86..157S,2008PhRvD..77b3507T}).
In the next section, we will investigate cosmological evolution as coming from $F(R)$ gravity. We will consider, for $F(R)$, some candidates to produce inflation and cosmic acceleration in a unified fashion. In particular, we investigate in detail the behavior of $F(R)$ gravity as the contribution of a perfect fluid. As a crucial novelty, an additional matter fluid will be included which may play, as we shall see, quite an important cosmological role. In fact, it may decisively contribute to the two accelerated epochs of the Universe, what is to say that we will study a model where dark energy consists of two separate contributions. The possibilities to obtain precision cosmology are enhanced in this way.

Two main cases will be discussed: $F(R)$ cosmology with a constant equation of state (EoS) fluid, and
$F(R)$ cosmology in the presence of a phantom fluid. In the last case, a couple of specific
example will be worked through in detail, namely one with a phantom fluid with constant EoS and, as a second example, a fluid with dynamical EoS of the type proposed in Chapter 3. 

As it is well known, $F(R)$ gravity can be written in terms of a scalar field---quintessence or phantom like---by redefining the function $F(R)$ with the use of a scalar field, and then performing a conformal transformation, what yields to the so-called Einstein frame.  It has been shown that, in general, for any given $F(R)$ the corresponding scalar-tensor theory can, in principle, be obtained, although the solution is going to be very different from one case to another. Also attention will be paid to the reconstruction of $F(R)$ gravity from a given scalar-tensor theory. It is known \cite{2007PhLB..646..105B} that the phantom case in scalar-tensor theory does not exist, in general, when starting from $F(R)$ gravity. In fact, the conformal transformation becomes complex when the phantom barrier is crossed, and therefore the resulting $F(R)$ function becomes complex. We will see that, to avoid this hindrance, a dark fluid can be used in order to produce the phantom behavior in such a way that the $F(R)$ function reconstructed from the scalar-tensor theory continues to be real.
We will prove, in an explicit manner, that an $F(R)$ theory can indeed be constructed from a phantom model in a scalar-tensor theory, but where the scalar field does not behave as a phantom field (in which case the action for $F(R)$ would be complex). Moreover, we will explicitly show that very interesting and quite simple $F(R)$ models crossing the phantom divide can be constructed.

\section{Viable $F(R)$ gravities}

\begin{figure}
 \centering
 \includegraphics[width=3in,height=2in]{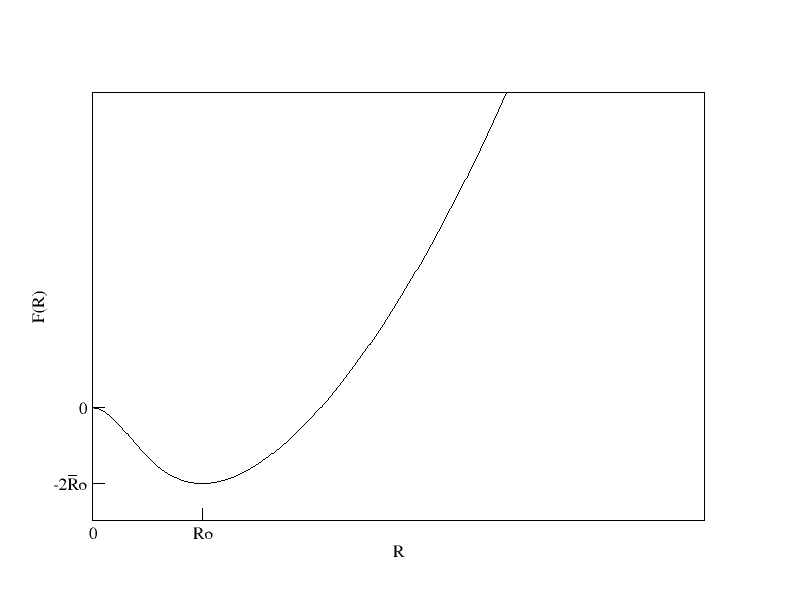}
 \caption{The function $F(R)$ as given by (\ref{5.1.9}) for $n=2$. We see that at the current epoch ($R\sim R_0$), $F(R)$ behaves as a cosmological constant while for $R\rightarrow\infty$ (inflation) its values grow as a power law.}
 \label{5.fig1}
\end{figure}

Our first aim in this chapter will be to construct reliable cosmological models by using, as starting point, modified gravity theories of the family of the so-called $F(R)$ theories which comprise the class of  viable models, e.g., those having an $F(R)$ function such that the theory can pass all known local gravity tests. We now consider the action corresponding to one of these theories which, aside from the gravity part, also contains a matter contribution, namely
\be
S=\int d^4x \sqrt{-g}\left[\frac{1}{2\kappa^2}(R+F(R))+L_m \right]\ .
\label{5.1.1}
\ee
Here $\kappa^2=8\pi G$, and $L_m$ stands for the Lagrangian corresponding to matter of some kind. Note that the first term in (\ref{5.1.1}) is just the usual Hilbert-Einstein action, and the $F(R)$ term can be considered, as it was shown in Chapter 4, as the dynamical part of some kind of perfect fluid, which may constitute an equivalent to the so-called dark fluids. The field equations corresponding to action (\ref{5.1.1}) are obtained by variation of this action with respect to the metric tensor $g_{\mu\nu}$, what yields
\be
R_{\mu\nu} -\frac{1}{2}g_{\mu\nu}R +  R_{\mu\nu} F'(R)- \frac{1}{2} g_{\mu\nu} F(R) + g_{\mu\nu}  \Box F'(R) -  \nabla_{\mu} \nabla_{\nu}F'(R)=\kappa^2T_{\mu\nu}\ .
\label{5.1.2}
\ee
Here the primes denote derivatives with respect to $R$. We  assume a flat FLRW metric, then the Friedmann equations are obtained as the 00 and the ij components. They take the  form
\bea
3H^2=\kappa^2 \rho_m - \frac{1}{2}F(R)+3(H^2+\dot{H})F'(R)-18F''(R)(H^2\dot{H}+H\ddot{H})\ , \nn
-3H^2-2\dot{H}=\kappa^2p_m+\frac{1}{2}F(R)-(3H^2+\dot{H})F'(R)- \Box F'(R)-HF''(R)\dot{R}\ ,
\label{5.1.3}
\eea
where $H(t)=\dot{a}/a$. Note that both equations are written in such a way that the $F(R)$-terms are put on the matter side; thus we may define an energy density and a pressure density for these $F(R)$-terms, as follows
\bea
\rho_{F(R)}=-\frac{1}{2}F(R)+ 3(H^2+\dot{H})F'(R)-18F''(R)(H^2\dot{H}+H\ddot{H})\ , \nn
p_{F(R)}=\frac{1}{2}F(R)-(3H^2+\dot{H})F'(R)- \Box F'(R)-HF''(R)\dot{R}\ .
\label{5.1.4}
\eea
Then, the Friedmann equations (\ref{5.1.3}) take a simple form, with two fluids contributing to the scale factor dynamics. From the energy and pressure densities defined in (\ref{5.1.4}), one can obtain the EoS for the dark fluid, defined in terms of the $F(R)$ components. This is written as
\bea
w_{F(R)}= \frac{\frac{1}{2}F(R)-(3H^2+\dot{H})F'(R)- \Box F'(R)-HF''(R)\dot{R}}{-\frac{1}{2}F(R)+ 3(H^2+\dot{H})F'(R)+\Box F'(R)-\nabla_0\nabla^0F'(R)} \nn
\longrightarrow \qquad p_{F(R)}= - \rho_{F(R)}+2\dot{H}F'(R)-\nabla_0\nabla^0F'(R)-HF''(R)\dot{R}
\ .
\label{5.1.5}
\eea
The EoS, Eq.~(\ref{5.1.5}), defines a fluid that depends on the Hubble parameters and its derivatives, so it may be considered as a fluid with inhomogeneous EoS (see \cite{2006PhLB..634...93C}). In absence of any kind of matter, the dynamics of the Universe are carried out by the $F(R)$-component, which may be chosen so that it reproduces (or at least contributes) to the early inflation and  late-time acceleration epochs. In order to avoid serious problems with known physics, one has to choose the $F(R)$ function in order that the theory contains flat solutions and passes also the local gravity tests. To reproduce the whole history of the Universe, the following conditions on the $F(R)$ function have been proposed (see \cite{2007PhLB..657..238N,2008PhRvD..77b6007N,2008arXiv0801.4843N}):
\begin{enumerate}
\renewcommand{\labelenumi}{\roman{enumi}}
\item) Inflation occurs  under one of the following conditions:
\be
\lim_{R\rightarrow\infty} F(R)=-\Lambda_i
\label{5.1.6}
\ee
or
\be
\lim_{R\rightarrow\infty} F(R)=  \alpha R^n\ .
\label{5.1.7}
\ee
In the first situation (\ref{5.1.6}), the $F(R)$ function behaves as an effective cosmological constant at early times, while the second condition yields accelerated expansion where the scale factor behaves as $a(t)\sim t^{2n/3}$.
\item) In order to reproduce late-time acceleration, we can impose on function $F(R)$ a condition similar to the one above. In this case the Ricci scalar has a finite value, which is assumed to be the current one, so that the condition is expressed as
\be
F(R_0)=-2R_0 \quad F'(R_0)\sim 0\ .
\label{5.1.8}
\ee
\end{enumerate}
Hence, under these circumstances, the $F(R)$ term is able to reproduce the two different accelerated epochs of the universe history. An interesting example that satisfies these conditions has been proposed in Ref. \cite{2007PhRvD..76f4004H}:
\be
F(R)=\frac{\mu^2}{2\kappa^2}\frac{c_1(\frac{R}{c_1})^k+c_3}{c_2(\frac{R}{\mu^2})^k+1}\ .
\label{5.1.8a}
\ee
This model is studied in detail in Ref.~\cite{2009PhRvD..79d4001C}, where it is proven that the corresponding  Universe solution goes, in its evolution, through two different De Sitter points, being one of them stable and the other unstable and which can be identified as corresponding to the current accelerated era and to the inflationary epoch, respectively. Thus, the model (\ref{5.1.8a}) is able to reproduce both accelerated epochs when the free parameters are conveniently chosen---as was done in Ref.~\cite{2009PhRvD..79d4001C}---in such way that the the F(R) theory in (\ref{5.1.8a}) produces two de Sitter epochs and grateful exit from the inflationary stage is achieved. In the same way, from a similar example given in Ref.~\cite{2008PhRvD..77b6007N}, we can consider the function
\be
F(R)=\frac{R^n(\alpha R^n-\beta)}{1+\gamma R^n}.
\label{5.1.9}
\ee
This function, represented in Fig.~\ref{5.fig1}, gives rise to inflation at the early stages of the Universe, assuming the condition (\ref{5.1.7}) holds, while for the current epoch it behaves as an effective cosmological constant. This is explicitly seen in Fig.~\ref{5.fig1}, where function (\ref{5.1.9}) is represented for the specific power $n=2$.

Function (\ref{5.1.9}) leads, at the current epoch,  to a perfect-fluid behavior with an EoS given by $p_{F(R)}\sim -\rho_{F(R)}$. In the next section such kind of $F(R)$ functions will be considered, with the inclusion of a matter Lagrangian in the action (\ref{5.1.1}), and the corresponding cosmological evolution will be studied. We will see that both inflation and the current acceleration can indeed be produced by the $F(R)$ fluid, provided other components are allowed to contribute too.

\section{Cosmological evolution from viable $F(R)$ gravity with a fluid}

In this section we will discuss cosmological evolution as coming from $F(R)$-gravity. We consider the function $F(R)$ as given by  (\ref{5.1.9}). Functions of this kind have been shown to yield viable models which comply with all known local gravity tests (see, e.g., \cite{2008arXiv0801.4843N}), and they are good candidates to produce  inflation and cosmic acceleration in a unified fashion. We will study in detail the behavior of $F(R)$-gravity, by considering it as the contribution of a perfect fluid in the way already explained in the preceding section. As a crucial novelty, an additional matter fluid will be here incorporated, which may play, as we shall see, an important cosmological role. It may decisively contribute to the two accelerated epochs of the Universe, what is to say that we study a model where dark energy consists of two separate contributions. Some examples of phantom evolution will be then discussed, in which the $F(R)$ contribution acts as a cosmological constant and the additional fluid behaves as a phantom field, what opens again interesting new venues.

\subsection{$F(R)$ cosmology with a constant EoS fluid $p_m=w_m\rho_m $}

We consider a Universe governed by action (\ref{5.1.1}), where the $F(R)$ function is given by (\ref{5.1.9}). The matter term is represented by a perfect fluid, whose equation of state is $p_m=w_m\rho_m $ (where $w_m= $cons.) In this case,  by considering this $F(R)$ term as a perfect fluid---with energy and pressure densities given by (\ref{5.1.4})---the Friedmann equations reduce to Eqs.~(\ref{5.1.3}). For simplicity, we will study the case where $n=2$; then, the function $F(R)$ of our specific model reads (the study can be extended to other values of $n$ without much problem)
\be
F(R)= \frac{R^2(\alpha R^2-\beta)}{1+\gamma R^2}\ .
\label{5.2.1}
\ee
First of all, let us explain qualitatively what the aim of this model is. For simplicity, we neglect the contribution of matter, so the first Friedmann equation yields
\[3H^2=-\frac{R^2(\alpha R^2-\beta)}{2(1+\gamma R^2)}+3(H^2+\dot{H})\frac{2R(\alpha\gamma R^4-2\alpha R^2-\beta)}{(1+\gamma R^2)^2}-
 \]
\be
-18F''(R)(H^2\dot{H}+H\ddot{H})\frac{2\alpha\gamma^2 R^6+20\alpha\gamma R^4+6(\beta\gamma-\alpha)R^2-2\beta}{(1+\gamma R^2)^3}\ ,
\label{5.2.1a}
\ee
where $R=6(2H^2+\dot{H})$. It can be rewritten as a dynamical system (see Ref.~\cite{2009PhRvD..79d4001C}):
\be
\dot{H}=C\ , \quad \dot{C}=F_1(H, C)\ ,
\label{5.2.1b}
\ee
and it can be shown that its critical points are those which give a constant Hubble rate ($\dot{H}=0$), i.e., the points that yield a de Sitter solution of the Friedmann equations. We can now investigate the existence of these points for the model (\ref{5.2.1}) by introducing the critical points $H=H_0$ in the Friedmann equation (\ref{5.2.1a})
\be
3H^2_0=-\frac{R^2(\alpha R^2-\beta)}{2(1+\gamma R^2)}+3H^2_0\frac{2R(\alpha\gamma R^4-2\alpha R^2-\beta)}{(1+\gamma R^2)^2}\ .
\label{5.2.1c}
\ee
To simplify, it can be rewritten in terms of the Ricci scalar $R_0=12H^2_0$, what yields
\be
\frac{\gamma}{4}R_0^5-\gamma\beta R_0^4+\frac{\gamma}{2}R_0^3+\frac{1}{4}R_0=0\ .
\label{5.2.1d}
\ee
This can be solved, so that the viable de Sitter points (positive roots) are found explicitly. A very simple study of Eq.~(\ref{5.2.1d}), by using Descartes' rule of signs, leads to the conclusion that Eq.~(\ref{5.2.1d}) can have either two or no positive roots. As shown below, one of these roots is identified as an effective cosmological constant that produces the current accelerated expansion of the Universe, while the second root can produce the inflationary epoch under some initial conditions. Then, the model described by the function (\ref{5.2.1}) is able to unify the expansion history of the Universe. In order to get a grateful exit from the inflationary epoch, stability in the vicinity of the critical points needs to be studied: the corresponding de Sitter point during inflation must be unstable. This can be achieved, as is already known for the function (\ref{5.1.8a}), by choosing specific values of the free parameters. Even in the case of stable dS inflation, exit from it can be achieved by coupling it with matter, by the effect of a small non-local term (or by some other mechanism).

Let us now study the details of this same model at the current epoch, when it is assumed that F(R) produces the cosmic acceleration and where the matter component is taken into account.
Function (\ref{5.2.1}) is depicted in Fig.~\ref{5.fig1}; its minimum is attained at $R=R_0$, which is assumed to be the current value of the Ricci scalar. Further, $F(R)$ as given by (\ref{5.2.1}) behaves as a cosmological constant at  present time. Imposing the condition $\beta\gamma/\alpha>>1$  on the otherwise free parameters, the values of $R_0$ and $F(R_0)$ are then given by  \cite{2008PhRvD..77b6007N}
\be
R_0 \sim \left( \frac{\beta}{\alpha\gamma}\right)^{1/4}\ , \qquad F'(R_0)=0\ , \qquad  F(R_0)= -2\tilde{R_0}\sim -\frac{\beta}{\gamma}\ .
\label{5.2.2}
\ee
For simplicity, we shall study  the cosmological evolution around the present value of the Ricci scalar, $R=R_0$, where (\ref{5.2.1}) can be expressed as $F(R)=-2\tilde{R_0}+f_0(R-R_0)^2+O\left( (R-R_0)^3\right)$, the solution for the Friedmann equations (\ref{5.1.3}) can be written as $H(t)=H_0(t)+\delta H$, where at zero order the solution  is the same as in the case of a cosmological constant, namely
\be
H(t)= \sqrt{\frac{\tilde{R_0}}{3}} \coth\left( \frac{(1+w_m)\sqrt{3\tilde{R_0}}}{2}t\right)\ .
\label{5.2.3}
\ee
As pointed out in Ref.~\cite{2008PhRvD..77b6007N}, the perturbations $\delta H$ around the current point $R=R_0$ may be neglected. Therefore, we can study the evolution of the energy density (\ref{5.1.4}) for the $F(R)$ term as the EoS parameter defined by (\ref{5.1.5}) around the minimum of the $F(R)$ function, which is assumed to be the present value of the Ricci scalar. For such purposes, it is useful to rewrite the Hubble function (\ref{5.2.3}) as a function of the redshift $z$, instead of $t$. The relation between both variables can be expressed as
\be
\frac{1}{1+z}= \frac{a}{a_0}= \left[ A \sinh\left( \frac{(1+w_m)\sqrt{3\tilde{R_0}}}{2}\, t\right)\right]^{\frac{2}{3(1+w_m)}}\ ,
\label{5.2.4}
\ee
where $a_0$ is taken as the current value of the scale parameter, and $A^2=\rho_{0m}a^{-3(1+w_m)}_0$, being $\rho_{0m}$ the current value of the energy density of the matter contribution. Hence, the Hubble parameter (\ref{5.2.3}) is expressed  as a function of the redshift $z$, as
\be
H^2(z)=\frac{\tilde{R_0}}{3}\left[ A^2(1+z)^{3(w_m+1)}+1\right]\ .
\label{5.2.5}
\ee
Thus, the model characterized by the action (\ref{5.1.1}) with the function (\ref{5.2.1}) and a matter fluid with constant EoS, depends on the free parameters ($\alpha,\beta,\gamma$) contained in the expression of $F(R)$ and also on the value of the EoS parameter ($w_m$). When imposing the minimum value for the function $F(R)$ to take place at present time ($z=0$), the free parameters can be adjusted with the observable data. Then we study the behavior of our model close to $z=0$, where the contributions of non-linear terms produced by the function (\ref{5.2.1}) are assumed not to modify the solution (\ref{5.2.5}). In spite of the Hubble parameter being unmodified, for $z$ close to zero, the energy density $\rho_{F(R)}$ will in no way remain constant for small values of the redshift. To study the behavior of the $F(R)$ energy density given by (\ref{5.1.4}) it is convenient to express it as the cosmological parameter, $\Omega_{F(R)}=\frac{\rho_{F(R)}}{3H^2(z)}$, which can be written as
\be
\Omega_{F(R)}(z)=-\frac{F(R)}{6H^2(z)}+\left[ 1+\frac{\dot{H}(z)}{H^2(z)}\right] F'(R)-18F''(R)\left[ \dot{H}(z)+\frac{\ddot{H}(z)}{H(z)}\right]\ ,
\label{5.2.6}
\ee
where the Ricci scalar is given by $R=6[2H^2(z)+\dot{H}(z)]$. This expression (\ref{5.2.6}) can be studied as a function of the redshift. As we are considering the solution (\ref{5.2.5}), which has been calculated near $z=0$, where the $F(R)$ function has a minimum, the second term in the expression (\ref{5.2.6}) is negligible as compared with the other two terms ($F''(R_0)R_0/F'(R_0)\sim\sqrt{\beta\gamma/\alpha}>>1$). This aproximation, as the solution (\ref{5.2.5}), is valid for values of the Ricci scalar close to $R_0$, where the higher derivatives of $F(R)$ are small compared with the function. The aproximation is no longer valid when the F(R) derivatives are comparable with F(R). Then, as discussed above, the free parameters of the model can be fitted by the current observational values of the cosmological parameters, and by fixing the minimum of $F(R)$ to occur for $z\sim 0$.

\begin{figure}[h]
 \centering
 \includegraphics[width=4in,height=3in]{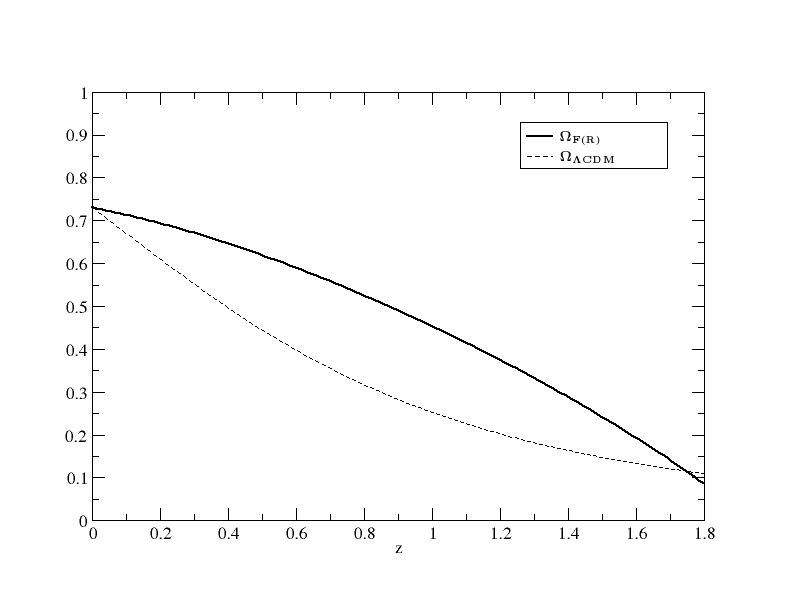}
 \caption{Evolution of the cosmological parameter from dark energy versus redshift, such for $F(R)$ theory as for $\Lambda CDM$ model.}
 \label{5.fig2}
\end{figure}

\begin{figure}[h]
 \centering
 \includegraphics[width=4in,height=3in]{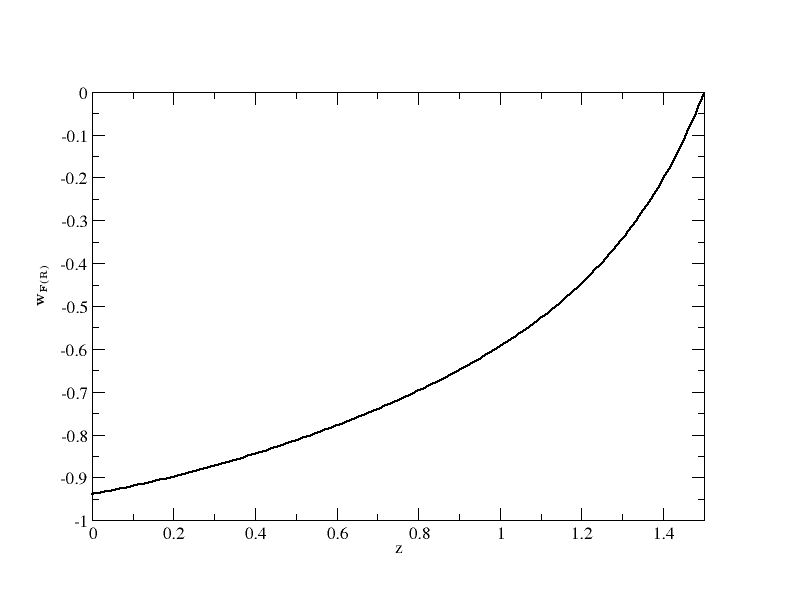}
 \caption{Effective EoS parameter $w_{F(R)}$ versus redshift. It takes values close to $-1$ for $z=0$, and it grows for higher redshifts.}
 \label{5.fig3}
\end{figure}

We shall use the value for the Hubble parameter $H_0=100\, h \ \text{km}\ \text{s}^{-1}\ \text{Mpc}^{-1}$ with $h=0.71\pm 0.03$ and the matter density $\Omega_m^0=0.27 \pm 0.04$ given in Ref.~\cite{2003ApJS..148..175S}. In Fig.~\ref{5.fig2}, the evolution of the $F(R)$ energy density (\ref{5.2.6}) is represented for the model described by (\ref{5.2.1}), where the matter content is taken to be pressureless ($w_m=0$). The cosmological evolution shown corresponds to redshifts from  $z=1.8$ to $z=0$. For redshifts larger than $z=1.8$  perturbations around the solution (\ref{5.2.3}) are non-negligible, and the expression for the Hubble parameter (\ref{5.2.3}) is no more valid. However, in spite of the fact that the evolution shown in Fig.~\ref{5.fig2} is not a complete picture of the $F(R)$ model given by (\ref{5.2.1}), it still provides an illustrative example to compare it to the standard $\Lambda CDM$ model, which is also represented in Fig.~\ref{5.fig2}, around the present time. As shown in Fig.~\ref{5.fig2}, both models have two common points for $z=0$ and  $z=1.74$, while the behavior of each model between such points is completely different from one another. This result shows the possible differences between $F(R)$ models of the type (\ref{5.1.9}) and the $\Lambda CDM$ model, although probably other viable models for $F(R)$ gravity may give a different adjustment to the $\Lambda CDM$ model. Furthermore, the effective EoS parameter for the $F(R)$ fluid, $w_{F(R)}$, given by the expression (\ref{5.1.5}), is plotted in Fig.~\ref{5.fig3}, again as a function of redshift. It is shown there that $w_{F(R)}$ is close to $-1$ for $z=0$, where the $F(R)$ fluid behaves like an effective cosmological constant, while it grows for redshifts up  to $z=1.5$, where it reaches a zero value. According to the analysis of observational dataset from Supernovae (see Ref.\cite{2008ApJ...686..749K}), the results obtained for the evolution of the EoS parameter, represented in Fig.~\ref{5.fig3}, are allowed by the observations.


As a consequence, the $F(R)$ model given by (\ref{5.2.1}), and where  the $F(R)$ fluid behaves as an effective cosmological constant, is able indeed to reproduce the same behavior at present time as the $\Lambda CDM$ model. On the other hand, as was pointed out in the section above, the $F(R)$ model given by (\ref{5.2.1}) also reproduces the accelerated expansion of the inflation epoch, so that the next natural step to undertake with this kind of models should be to study their complete cosmological histories, and the explicit details allowing for a grateful exit from inflation, what should demonstrate their true potential. Also note that, in order to obtain a realistic well behaved model, further analysis should be carried out, as the comparison with the luminosity distance from Supernovae, or the data from CMB surveys, although this is a major task, that will be left for future work. Furthermore, $F(R)$ functions of this kind might even lead to a solution of the cosmological constant problem, by involving a relaxation mechanism of the cosmological constant, as was proposed in Ref.~\cite{2009PhLB..670..246S}. The effective cosmological constant obtained could eventually adjust to the observable value. This will also require a deeper investigation.

\subsection{$F(R)$ cosmology in presence of a phantom fluid}

According to several analysis of observational data (see \cite{2007LNP...720..257P,2005PhRvD..72j3503J}), the effective EoS parameter of the physical theory that governs our Universe should quite probably be less than $-1$, what means that we should be ready to cope with phantom behavior. This possibility has been explored in $F(R)$ theory \cite{2007PhLB..646..105B}, where the possibility to construct an $F(R)$ function that reproduces this kind of behavior has been demonstrated, and where the possible future singularities envisaged, as the Big Rip, so common in phantom models, do take place. In Chapter 4,  an example with phantom behavior that ends in a Big Rip singularity was shown, as well as another phantom model that avoids future singularities. In this last example, the universe has a phantom behavior at large time but there is no Big Rip singularity. However, the main problem of those models is that the corresponding $F(R)$ function is not well constrained at small scales.

We will  now study the behavior of the cosmological evolution when a phantom fluid is introduced that contributes to the accelerated expansion. A phantom fluid can be described, in an effective way, by an EoS $p_{ph}=w_{ph}\rho_{ph}$, where $w_{ph}<-1$. This EoS can be achieved, for example, with a negative kinetic term or in the context of scalar-tensor theory. The microphysical study of this kind of fluids is a problem of fundamental physics that has been studied in several works (see for example \cite{2002PhLB..545...23C,2003PhRvD..68b3509C}), as it is also the study of  stability of the  solutions. But it is not the aim for this chapter to discuss, in this depth, the features of phantom fields, and it will suffice for our purpose to consider the effective EoS that describes a phantom fluid.  The $F(R)$ function we will consider is the same as above, given  by (\ref{5.1.9}), which will contribute as an effective cosmological constant at present time. The motivation for this case study comes, as will be further explored, from the difficulty one encounters in constructing {\it viable} $F(R)$ functions which produce phantom epochs, and this problem has a translation in the scalar-tensor picture in the Einstein frame too. As recent observational data suggest (see \cite{2005PhRvD..72j3503J,2007LNP...720..257P}), the effective EoS parameter for dark energy is around $-1$, so that the phantom case is allowed (and even still favored) by recent, accurate, and extensive observations.\medskip

\noindent{\sf Example 1.} First, we consider a phantom fluid with constant EoS, e.g., $p_{ph}=w_{ph}\rho_{ph}$, where $w_{ph}<-1$ is a constant. As in the case above,  the $F(R)$ function will be given by (\ref{5.2.1}), and it will be assumed that its current value is attained at the minimum $R_0=\left(\beta/\alpha\gamma\right)^{1/4}$. We here neglect other contributions, as the dust term studied previously. Friedmann equations take the form
\be
H^2=\frac{\kappa^2}{2}\rho_{ph}+\frac{\tilde{R}_0}{3}\ , \quad \dot{H}=-\frac{\kappa^2}{3}\rho_{ph}(1+w_{ph})\ .
\label{5.2.7}
\ee
An expanding solution (note that a contracting solution can be obtained from the above equations too) for these equations is given by
\be
H(t)=\frac{3}{2|1+w_{ph}|(t_s-t)}+\sqrt{\frac{\tilde{R_0}}{3}}\ ,
\label{5.2.8}
\ee
where $t_s$ is called the Rip time, that means the instant where a future Big Rip singularity will take place. Although for $t$ much bigger than the present time, the solution (\ref{5.2.8}) is not valid anymore ---because perturbations, due to the derivatives of the function $F(R)$, become large--- and as the Ricci scalar $R=6(2H^2+\dot{H})$ grows with time, this model  will behave as $F(R)\sim R^2$ for large times, which is known to produce accelerated expansion, and whose behavior is  described by 
\be
H(t)\propto \frac{h_0}{t_s-t}\ ,
\label{5.2.8a}
\ee
where $h_0=\frac{4}{3|w_{ph}+1|}$. And the effective EoS parameter for large time is
\be
w_{eff}=-1-\frac{4}{3h_0}\ .
\label{5.2.8b}
\ee

 That is to say, the universe will go through a super-accelerated expansion stage due to the phantom fluid and to the contribution coming from $F(R)$, until it reaches the Big Rip singularity. In this case where no other contribution, as dust matter or radiation, is taken into account, late-time acceleration comes from the phantom behavior when the dark fluid component dominates, while the Universe behaves as dark energy when the $F(R)$ term is the dominant one (as was in absence of this phantom fluid). In other words, the universe would not accelerate as a phantom one alone, but it will as a dark energy fluid with $w_{eff}\geq -1$. This is a fundamental point: the additional dark fluid is essential for a universe described by the $F(R)$ function given in (\ref{5.2.1}) to display a phantom transition.

We can study the evolution of this phantom fluid by using the continuity equation:
\be
\dot{\rho}_{ph}+3H \rho_{ph}(1+w_{ph})=0\ ,
\label{5.2.9}
\ee
which can be solved, at the present time, when the Hubble parameter is expressed by Eq.~(\ref{5.2.8}). Then, the following solution is obtained
\be
\rho_{ph}=\rho_{0ph}\frac{\e^{|1+w_{ph}|\sqrt{3\tilde{R_0}}t}}{(t_s-t)^{9/2}}\ ,
\label{5.2.10}
\ee
where $\rho_{0ph}$ is an integration constant. As we can see, the energy density for the phantom fluid grows with time until the Rip value is reached where the energy density becomes infinite. On the other hand, the evolution of the $F(R)$ term may be studied qualitatively by observing the expression given for the energy density of this fluid in (\ref{5.1.4}), so that this evolution is similar to the one in the above case. That is, as the value of the Ricci scalar increases with time, it is natural to suppose that, in the past, the energy density belonging to $F(R)$  had smaller values than at present, so that the matter dominated epoch could occur when the $F(R)$ fluid and the phantom fluid were much less important than they are now. Again, for this kind of model the $F(R)$ contribution amounts currently to an effective cosmological constant which drives the universe's acceleration. \medskip

\begin{figure}[h]
 \centering
 \includegraphics[width=3in,height=3in]{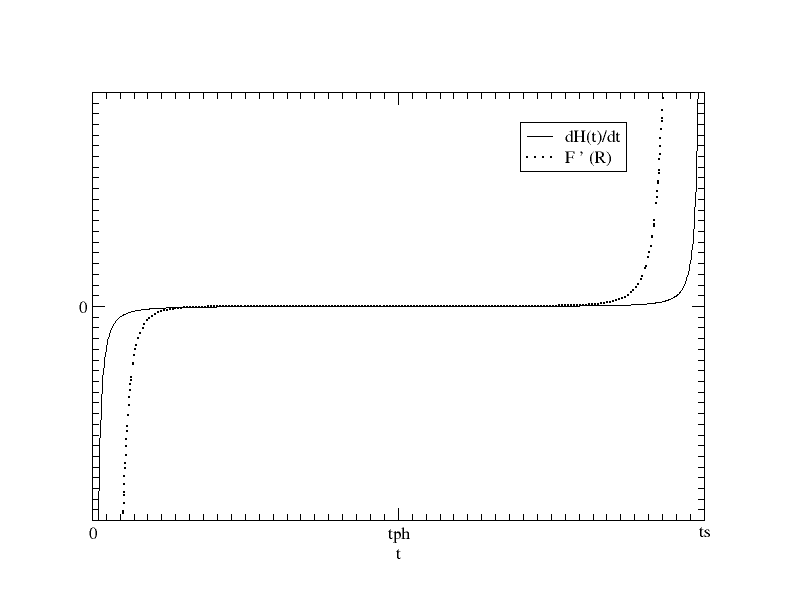}
 \caption{The first derivative of the Hubble parameter is shown here to illustrate the different epochs of the Universe evolution (full curve). Also, the first derivative of the $F(R)$ term with respect to $R$ is represented as a function of time (dotted curve); the constant Hubble parameter region is given by a constant $F(R)$, which plays the role of an effective cosmological constant. Starting at $t=t_{ph}$, the Universe enters into a phantom region dominated by the dark fluid, which ends at $t=t_s$ when the Big Rip singularity takes over.}
 \label{5.fig:4}
\end{figure}

\noindent{\sf Example 2.} As a second example of a phantom fluid, we consider one with a dynamical EoS of the type proposed in Chapter 3. Here, a dark fluid is present which has an inhomogeneous EoS that may depend on the proper evolution of the Universe. This kind of EoS can be derived from the dynamics of an scalar field with some characteristic potential and a variable kinetic term, or either it may be seen as the effective EoS corresponding to the addition of various components that fill up our Universe. An EoS of this kind can be written as
\be
p_{ph}=w\rho_{ph} + g(H, \dot{H}, \ddot{H}...;t)\ ,
\label{5.2.11}
\ee
where $w$ is a constant and $g$ an arbitrary function. The interesting point is that it is possible to specify a function $g$ so that a complete solution of the Friedmann equations is obtained. In this case, our aim is to study a fluid that at present (or in the near future) can behave as phantom; to that purpose  we choose $w=-1$, and the role of the $g$ function will be to determine when exactly the phantom barrier is crossed. Thus, our model will be described by the phantom fluid in (\ref{5.2.11}) and the function $F(R)$ of (\ref{5.2.1}), while the matter component can be neglected. As an example, the EoS for the dark fluid is given by
\be
p_{ph}=-\rho_{ph} - \left( \frac{4}{\kappa^2}h'(t)+p_{F(R)}+\rho_{F(R)}\right)\ ,
\label{5.2.12}
\ee
where the prime denotes  derivative with respect to time, and $p_{F(R)}$ and $\rho_{F(R)}$ are the energy densities defined in (\ref{5.1.4}). As it is shown, through  the EoS defined in (\ref{5.2.12}), the dark fluid will contribute to the acceleration of the Universe as a dark energy, and subsequently as a phantom fluid when it crosses the  barrier $w_{ph}<-1$. We thus see that in this model accelerated expansion comes from two contributions, the $F(R)$ term (\ref{5.2.1}) and the dark fluid one (\ref{5.2.12}), so that the accelerated expansion stage could cross the phantom barrier if the dark fluid dominates and contributes as a phantom one. The difference with respect to the former model is that in the present case the dark fluid changes its behavior in the course of the expansion history (this will be seen again in an example below). As the EoS given in (\ref{5.2.12}) can be rewritten in terms of the Ricci scalar $R$, it can be seen as additional terms to our F(R) function, in order to get the transition to the phantom epoch in the context of F(R) gravity. With the EoS (\ref{5.2.12}), the Friedmann equations can be solved, the following solution being found
\be
H(t)=h(t)\ .
\label{5.2.13}
\ee
Different solutions can be constructed by specifying the function $h(t)$. We are here interested in those solutions that give rise to a phantom epoch; for those cases the following splitting of the Hubble parameter is relevant
\be
H(t)=\frac{H_0}{t}+\frac{H_1}{t_s-t} \ ,
\label{5.2.14}
 \ee

where $H_0$, $H_1$ and $t_s$ (Rip time) are positive constants. This function describes a Universe that starts in a singularity, at $t=0$---which may be identified with the Big Bang one---then evolves to a matter dominated epoch, after which an accelerated epoch starts which is dominated by an effective cosmological constant and, finally, the Universe enters into a phantom epoch that will end in a Big Rip singularity. As in the cases above, the condition (\ref{5.2.2}) remains, that is $F'(R_0)=0$, where $R_0$ is the present value of the Ricci scalar. Fig.~\ref{5.fig:4} is an illustration of how this model works: we see there that the Universe goes through a decelerated epoch until it enters a region where the evolution of its expansion has constant Hubble parameter and the $F(R)$ term behaves as an effective cosmological constant. Finally, at $t_ {ph}$ the Universe enters into a super-accelerated phase which is dominated by the dark fluid of Eq.~(\ref{5.2.12}), until it eventually reaches the Big Rip singularity at $t=t_s$. The aim of this model is that the crossing phantom barrier takes place very  softly, as it is seen in Fig.~\ref{5.fig:4} due to the two contributions to the acceleration of the Universe expansion.
Alternatively, the evolution of the EoS parameter for the dark fluid (\ref{5.2.12}) can be written as
\be
w_{ph}=-1-\frac{\frac{4}{\kappa^2}h'(t)+p_{F(R)}+ \rho_{F(R)}}{\frac{6}{\kappa^2}h^2(t)-\rho_{F(R)}}\ ,
\label{5.2.15}
\ee
and we may look at its asymptotic behavior,
\bea
\text{For} \quad 0<t<<t_{0}\longrightarrow w_{ph}\sim-1+\frac{2(H_0+4)}{3(3-2H^{2}_{0})}\ . \nn
\text{For} \quad t\sim t_0 \longrightarrow w_{ph}\sim-1-\frac{4}{\kappa^2}\dot{H}\ . \nn
\text{For} t_0,t_{ph}<<t<t_s \longrightarrow w_{ph}\sim-1-\frac{2}{9\left(\frac{37H^{2}_1-4H_1+18}{H_1} \right) }\ .
\label{5.2.16}
\eea

We thus see that, at the starting stages of the Universe, the dark fluid contributes to the deceleration of its expansion.  For $t$ close to the present time, $t_0$, it works as a contribution to an effective cosmological constant  and, after $t=t_{ph}$, it gives rise to the transition to a phantom era of the cosmos, which could actually be taking place nowadays at some regions of it. Our hope is that it could even be observable with specific measurements. Finally, for $t$ close to the Rip time, the Universe  becomes completely dominated by the dark fluid, whose EoS is phantom at that time. This model, which is  able to accurately reproduce the dark energy period, may still be modified in such a way that the epoch which is dominated by the effective cosmological constant, produced by the $F(R)$ term and by the dark fluid contribution, becomes significantly shorter. This is supposed to happen when a matter term is included.

To finish this section, the inclusion of a dark fluid that behaves as a phantom one gives rise to a super-accelerated phase, as compared with the case where just the viable $F(R)$ term contributes. In the two examples here studied, we have proven that, while the $F(R)$ term contributes as an effective cosmological constant, the dark fluid contribution produces the crossing of the phantom barrier, and it continues to dominate until the end of the Universe in a Big Rip singularity. In other words, both the contribution of $F(R)$ and of the phantom fluid are needed. This is very
clearly seen before, and the nice thing is that, because
of this interplay, we have shown the appearance of nice properties of
our model that no purely phantom model could have. Specifically,
for our models above, $F(R)$ gravity together with phantom matter,
the effective $w$ value becomes in fact bigger than -1, so that we
are able to show that $F(R)$ gravity can solve the phantom
problem simply by making the phantom field to appear as a normal one.

\section{Scalar-tensor theories and $F(R)$ gravity with a fluid}

We now turn to the study of the solutions given above in the alternative, and more commonly used, scalar-tensor picture. In Refs.~\cite{2005CQGra..22L..35A,2009PhRvD..79h3014B,2007PhLB..646..105B,2006PhLB..634...93C,2007JPhCS..66a2005N}, it was pointed out that  $F(R)$ gravity can be written in terms of a scalar field---quintessence or phantom like---by redefining the function $F(R)$ with the use of a convenient scalar field and then performing a conformal transformation.  The scalar-tensor  theory thus obtained provides a solution which is characterized by this conformal transformation, whose expression depends on the precise form of the $F(R)$ function. It has been shown that, in general, for any given $F(R)$, the corresponding scalar-tensor theory can in principle be obtained, although the solution is going to be very different from case to case. Also attention has been paid to the reconstruction of $F(R)$ gravity from a given scalar-tensor theory. It is also known (see \cite{2007PhLB..646..105B}) that the phantom case in scalar-tensor theory does not allow, in general, the corresponding picture in $F(R)$ gravity. In fact, the conformal transformation becomes complex when the phantom barrier is crossed, and then the resulting $F(R)$ function becomes complex too. To avoid this hindrance, a dark fluid can be used, as in the models of the preceding section, in order to produce the phantom behavior in such a way that the reconstructed $F(R)$ function continues to be {\it real}. This point is important, and will be clearly shown below. Also to be remarked is the fact that scalar-tensor theories, commonly used in cosmology to reproduce dark energy (see Chapter 2 and 3), provide cosmological solutions whose stability should be studied in order to demonstrate the validity of the solution found. This has been investigated, e.g., in Ref. \cite{2002PhLB..545...23C} and requires a very deep and careful analysis. For the purpose of the current chapter, we will here concentrate in the proof the existence of the solution in the scalar-tensor counterpart, and the corresponding stability study will be left for future work.

We start with the construction of the scalar-tensor theory from $F(R)$ gravity. The action (\ref{5.1.1}) can be written as
\be
S=\int d^4x \sqrt{-g}\left[P(\phi)R+Q(\phi) + L_m \right]\ ,
\label{5.3.1}
\ee
which is known as the Jordan frame. Here $F(R)$ has been written in terms of a scalar field. To recover the action (\ref{5.1.1}) in terms of $F(R)$,  the scalar field equation resulting from the variation of the action (\ref{5.3.1}) with respect to $\phi$ is used, which can be expressed as follows
\be
P'(\phi)R+Q'(\phi)=0\ ,
\label{5.3.2}
\ee
where the primes denote derivatives with respect to $\phi$. Then, by solving equation (\ref{5.3.2}) we get the relation between the scalar field $\phi$ and the Ricci scalar, $\phi=\phi(R)$. In this way, the original $F(R)$ function and the action (\ref{5.1.1}) are recovered:
\be
R + F(R)= P(\phi(R))R+Q(\phi(R))\ .
\label{5.3.3}
\ee
Finally, the scalar-tensor picture is obtained by performing a conformal transformation on the action (\ref{5.3.1}). The relation between both frames is given by
\be
g_{E\mu\nu}= \Omega^2g_{\mu\nu}\ , \quad \text{where} \quad \Omega^2= P(\phi)\ ,
\label{5.3.4}
\ee
where the subscript $_E$ stands for Einstein frame. A quintessence-like action results in the Einstein frame
\[
S_E=\int d^4x \sqrt{-g_E} \left[R_E -\frac{1}{2} \omega(\phi) d_{\mu}\phi d^{\mu}\phi -U(\phi) +\alpha(\phi) L_{mE} \right]\ ,
\]
where
\be
\omega(\phi)=\frac{12}{P(\phi)}\left( \frac{d\sqrt{P(\phi)}}{d\phi}\right)^2\ , \quad U(\phi)=\frac{Q(\phi)}{P^2(\phi)}\ \quad \text{and} \quad \alpha(\phi)= P(\phi) \ ,
\label{5.3.5}
\ee
are the kinetic term, the scalar potential and the coupling function, respectively. Hence, by following the steps enumerated above, we can reconstruct the scalar-tensor theory described by the action (\ref{5.3.5}) for a given $F(R)$ gravity. By redefining the scalar field $\phi=R$, and after combining Eqs.~(\ref{5.3.2}) and (\ref{5.3.3}), the form of the two functions $P(\phi)$ and $Q(\phi)$ are found
\be
P(\phi)=1+F'(\phi)\ , \quad Q(\phi)=F'(\phi)\phi -F(\phi) \ .
\label{5.3.6}
\ee
Hence, for a given solution in the Jordan frame (\ref{5.3.1}), the solution in the corresponding quintessence/phantom scalar field scenario---i.e., in the Einstein frame (\ref{5.3.5})---is obtained by the conformal transformation (\ref{5.3.4}), and it is given by
\be
a_E(t_E)= \left[ 1+F'(\phi(t))\right] ^{1/2}a(t)\ \quad \text{where} \quad dt_E= \left[ 1+F'(\phi(t))\right] ^{1/2}dt\ .
\label{5.3.7}
\ee
We will be here interested in the phantom case. With that purpose, we analyze the model described in the above section by the $F(R)$ function (\ref{5.2.1}) and the dark fluid with EoS (\ref{5.2.12}), and whose solution is (\ref{5.2.14}). For simplicity, we restrict the reconstruction to the phantom epoch, when the solution can be written as $H(t)\sim\frac{H_1}{t_s-t}$, and $F(R)\sim\frac{\alpha}{\gamma}R^2$.  Using (\ref{5.3.6}), the function $P(\phi(t))$ takes the form  $P(t)\sim2\frac{\alpha}{\gamma}R^2 =\frac{2\alpha H_1(H_1+1)}{\gamma}\frac{1}{(t_s-t)}$ as a function of time $t$ in the Jordan frame. Then, using (\ref{5.3.7}), the solution in the Einstein frame is found
\be
t_E=-\sqrt{\frac{2\alpha H_1(H_1+1)}{\gamma}}\ln(t_s-t) \longrightarrow a_E(t_E)= \sqrt{\frac{2\alpha H_1(H_1+1)}{\gamma}} \exp\left[2 \sqrt{\frac{\gamma}{2\alpha H_1(H_1+1)}}t_E\right]\ .
\label{5.3.8}
\ee
Through the relation between the time coordinates in both frames, we see that while for the Jordan frame there is a Big Rip singularity, at $t=t_s$, this corresponds in the Einstein frame to $t\rightarrow\infty$, so that the singularity is avoided, and there is no phantom epoch there. By analyzing the scale parameter in the Einstein frame, we realize that it describes a de Sitter Universe, while in the Jordan frame the Universe was described by a phantom expansion. As a consequence, we have shown that a phantom Universe in $F(R)$ gravity may be thoroughly reconstructed as a quintessence-like model, where the phantom behavior is lost completely. This has been achieved in a fairly simple example and constitutes an interesting result.

Let us now explore the opposite way. In this case, a phantom scalar-tensor theory is given and it is $F(R)$ gravity which is reconstructed. As was pointed out in Ref.~\cite{2007PhLB..646..105B}, when a phantom scalar field is introduced, then the corresponding $F(R)$ function---which is reconstructed, close to the Big Rip singularity, by means of a conformal transformation that deletes the kinetic term for the scalar field---is in general complex. As a consequence, there is no correspondence in modified gravity when a phantom scalar produces a Big Rip singularity. However, in our case we will analyze a scalar-tensor theory which includes a phantom fluid that is responsible for the phantom epoch and for the Big Rip singularity. In this situation a real $F(R)$ gravity will be generically reconstructed, as we are going to see.

The action that describes the scalar-tensor theory is
\be
S_E=\int d^4x \sqrt{-g_E} \left[R_E -\frac{1}{2} \omega(\phi) d_{\mu}\phi d^{\mu}\phi -U(\phi) +\alpha(\phi) L_{phE} \right]\ ,
\label{5.3.9}
\ee
where $\alpha(\phi)$ is a coupling function and $L_{phE}$ the Lagrangian for the phantom fluid in the Einstein frame. In our case, we consider a phantom fluid with constant EoS, $p_{phE}=w_{ph}\rho_{phE}$, with $w_{ph}<-1$. The Friedmann equations in this frame are written as
\bea
H_E^2=\frac{\kappa^2}{3}\left(\frac{1}{2}\omega(\phi)\dot{\phi}^2+V(\phi)+\alpha(\phi)\rho_{phE} \right)\ , \nn
\dot{H}_E= -\frac{\kappa^2}{2}\left(\omega(\phi)\dot{\phi}^2+ \alpha(\phi)\rho_{phE}(1+w_{ph})\right)\ .
\label{5.3.10}
\eea
To solve the above equations, it turns out to be very useful to redefine the scalar field as $\phi=t_E$. Then, for a given solution $H_E(t_E)$, the kinetic term for the scalar field can be written as follows
\be
\omega(\phi)=-\frac{\frac{4}{\kappa^2}(\dot{H}_E+3(1+w_{ph})H^2_E)-
(1+w_{ph})V(\phi)}{1-w_{ph}}\ .
\label{5.3.11}
\ee
To reconstruct $F(R)$ gravity, we perform a conformal transformation that deletes the kinetic term, namely
\be
g_{\mu\nu E}=\Omega^2g_{\mu\nu}\ , \quad \text{where} \quad \Omega^2=\exp\left[\pm \sqrt{\frac{2}{3}}\kappa\int d\phi\sqrt{\omega(\phi)}\right]\ .
\label{5.3.12}
\ee
Note that for a phantom scalar field, that is defined by a negative kinetic term, the above conformal transformation would be complex, as remarked in Ref.~\cite{2007PhLB..646..105B} and as will be shown below. Thus, the reconstructed action would be complex, and no $F(R)$ gravity could be recovered. By means of the above conformal transformation, action (\ref{5.3.9}) is given by
\be
S=\int d^4x \sqrt{-g} \left[ \frac{\e^{\left[\pm \sqrt{\frac{2}{3}}\kappa\int d\phi\sqrt{\omega(\phi)}\right]}}{2\kappa^2}R-\e^{\left[\pm 2\sqrt{\frac{2}{3}}\kappa\int d\phi\sqrt{\omega(\phi)}\right]} V(\phi) + L_{ph}\right] \ ,
\label{5.3.13}
\ee
where $L_{ph}$ is the lagrangian for the phantom fluid in the Jordan frame, whose energy-momentum tensor is related with the one in the Einstein frame by $T^{ph}_{\mu\nu}=\Omega^2T^{phE}_{\mu\nu}$,  and where we have chosen a coupling function $\alpha(\phi)=\Omega^{-4}$, for simplicity. By varying now the action (\ref{5.3.13}) with respect to $\phi$, the scalar field equation is obtained
\be
R=\e^{\left[\pm \sqrt{\frac{2}{3}}\kappa\int d\phi\sqrt{\omega(\phi)}\right]}\left( 4\kappa^2 V(\phi) \mp \sqrt{\frac{6}{\omega(\phi)}}V'(\phi) \right)\ ,
\label{5.3.13a}
\ee
which can be solved as $\phi=\phi(R)$, so that by rewriting action (\ref{5.3.13}), the $F(R)$ gravity picture result:
\be
F(R)=\frac{\e^{\left[\pm \sqrt{\frac{2}{3}}\kappa\int d\phi\sqrt{\omega(\phi)}\right]}}{2\kappa^2}R-\e^{\left[\pm 2\sqrt{\frac{2}{3}}\kappa\int d\phi\sqrt{\omega(\phi)}\right]} V(\phi)\ .
\label{5.3.14}
\ee
Hence, for some coupling quintessence theory described by action (\ref{5.3.9}), it is indeed possible to obtain a {\it real} $F(R)$ theory, by studying the system in the Jordan frame through the conformal transformation (\ref{5.3.12}).

To demonstrate this reconstruction explicitly, an example will now be given. As we are interested in the case of a phantom epoch close to the Big Rip singularity, we will start from a solution in the Einstein frame $H_E\sim\frac{1}{t_s-t_E}$, and a scalar potential given by $V(\phi)\sim(t_s-\phi)^n$, with $n>0$. Then, the kinetic term (\ref{5.3.11}) is written as
\be
\omega(\phi)\sim\frac{-\frac{2}{\kappa^2}(2+3(1+w_{ph}))}{1-
w_{ph}}\frac{1}{(t_s-\phi)^2}\ .
\label{5.3.15}
\ee
The solution in the Jordan frame is calculated by performing the conformal transformation (\ref{5.3.12})
\be
(t_s-t_E)=\left[ \left( 1\mp\frac{k}{2}\right)t \right]^{1/(1\mp k/2)}\ \longrightarrow \quad a(t)\sim \left[ \left( 1\mp\frac{k}{2}\right)t \right]^{-\frac{1\pm k}{1\mp k}}\ ,
\label{5.3.16}
\ee
where $k=\sqrt{-\frac{8(1+3(1+w_ph))}{3(1-w_{ph})}}$ and $w_{ph}<-1$. Note that, in this case, we can construct two different solutions, and correspondingly two different $F(R)$ models, depending on the sign selected in Eq.~(\ref{5.3.12}). It is easy to see that the Big Rip singularity is thereby transformed, depending on the case, into an initial singularity (+), or into an infinity singularity(-). By using (\ref{5.3.13a}) and (\ref{5.3.14}), the following function $F(R)$ is recovered
\be
F(R)\sim \frac{(\frac{R}{4\kappa^2})^{\frac{\pm k}{n\pm k}}}{2\kappa^2}R-\left( \frac{R}{4\kappa^2}\right)^{\frac{n\pm 2k}{n\pm k}}\ .
\label{5.3.17}
\ee
To summarize, we have here shown, in an explicit manner, that an $F(R)$ theory can be actually constructed from a phantom model in a scalar-tensor theory  where the scalar field does not behave as a phantom one (in that case the action for $F(R)$  would become complex).  The above reconstruction procedure, where we have taken the $F(R)$ function of (\ref{5.2.1}), can be generalized to other types of modified gravity models.

\section{Discussion}

We have seen in this chapter that the $F(R)$ model given by (\ref{5.2.1}), and where  the $F(R)$ fluid behaves as an effective cosmological constant, is able to reproduce the same behavior, at present time, as the $\Lambda$CDM model. On the other hand, this model gives rise to the accelerated expansion of the inflation epoch too, so that the natural next step to undertake with those models should be to study the complete cosmological history, in particular (what is very important) the explicit details for a grateful exit from inflation, what would demonstrate their actual potential. What is more, $F(R)$ functions of this kind might even lead to a solution of the cosmological constant problem, by involving a relaxation mechanism of the cosmological constant, as was indicated in \cite{2009PhLB..670..246S,2009PhLB..678..427B}). The effective cosmological constant thus obtained could eventually adjust to its precise observable value. This issue is central and deserves further investigation.

We have studied the behavior of the cosmological evolution when a phantom fluid is introduced that contributes to the accelerated expansion of the universe. The $F(R)$ function we have considered is the one given in (\ref{5.1.9}), which contributes as an effective cosmological constant at present time. The motivation for this case study comes, as it will be further explored, from the difficulty one encounters to construct {\it viable} $F(R)$ functions which produce phantom epochs, and this has a representation in the scalar-tensor picture in the Einstein frame. As recent observational data suggest (see \cite{2007LNP...720..257P,2005PhRvD..72j3503J}), the effective EoS parameter for dark energy is around $-1$, so that the phantom case is allowed---and actually favored by recent, extensive observations.

We thus have seen that, at the early stages of the universe history, the dark fluid contributes to the deceleration of its expansion.  For $t$ close to present time, $t_0$, it works as a contribution to an effective cosmological constant and, after $t=t_{ph}$, it gives rise to the transition to a phantom era of the universe, which could actually be taking place right now in some regions of it. Our hope is that it could be actually observable. Finally, for $t$ close to the Rip time, the Universe  becomes completely dominated by the dark fluid, whose EoS is phantom-like at that time. This model, which is able to reproduce the dark energy period quite precisely, may still be modified in such a way that the epoch dominated by the effective cosmological constant, produced by the $F(R)$ term and by the dark fluid contribution, becomes significantly shorter. This is the case when a matter term is included.

The inclusion of a dark fluid with phantom behavior gives rise to a super-accelerated phase, as compared with the case where just the viable $F(R)$ term contributes. In the two examples investigated in the chapter we have proven that, while the $F(R)$ term contributes as an effective cosmological constant, the dark fluid term produces the crossing of the phantom barrier, and it continues to dominate until the end of the universe in a Big Rip singularity. It is for this reason that the contribution of
$F(R)$ and of the phantom fluid are both {\it fundamental}. This has been
clearly explained in the chapter, and the nice thing is that, because
of this interplay, we have shown the appearance of very nice properties of
our model that no purely phantom model could have. This eliminates
in fact some of the problems traditionally associated with phantom
models and makes this study specially interesting. In particular,
for some of our models of $F(R)$ gravity together with phantom matter,
the effective $w$ value becomes in fact bigger than -1, so that we
were able to show that $F(R)$ gravity can solve the phantom
problem simply by making the phantom field to appear as a normal field.

To summarize, we have here shown, in an explicit manner, that an $F(R)$ theory can indeed be constructed from a phantom model in a scalar-tensor theory in which the scalar field does not behave as a phantom one (in the latter case the action for $F(R)$  would be complex). Moreover, very promising $F(R)$ models which cross the phantom divide can be constructed explicitly. The above reconstruction procedure, where we have taken (\ref{5.2.1}) for the $F(R)$ function, can be generalized to other classes of modified gravity models.

\chapter[On $\Lambda$CDM model in modified $F(R,G)$ and Gauss-Bonnet gravitites]{On $\Lambda$CDM model in modified $F(R,G)$ and Gauss-Bonnet gravitites}

\footnote{This Chapter is based on: \cite{2010CQGra..27i5007E,2011GReGr.tmp...14M}}In the preceedings chapters, it was studied and shown how modifications of the gravitational action with dependence on the Ricci scalar, can resolve some of the cosmological problems, specifically the acceleration phases of our Universe. in the present chapter, another kind of mofication of the Hilbert-Einstein action is presented, the so-called Gauss-Bonnet gravity, where the 
gravitational action includes functions of the Gauss-Bonnet 
invariant. This kind of theories have been investigated and may 
reproduce the cosmic history (see Refs. \cite{2009PhRvD..79f3006A,2008JCAP...10..045B,2010EPJC...67..295B,2009PhRvD..79f7504B,2009PhRvD..79l4007C,2006PhRvD..73h4007C,2007PhRvD..75h6002C,2009PhLB..675....1D,2009PhRvD..80f3516D,2009PhRvD..79l1301G,2008GReGr..40.1825G,2007PhRvD..76d4027L,2005PhLB..631....1N,2007PhLB..651..224N,2006JPhA...39.6627N,2006PhRvD..74d6004N,2005PhRvD..71l3509N,2009GReGr..41.2725U,2009JCAP...07..009Z}). 
Here, we show that the $\Lambda$CDM model can be well explained with 
no need of a cosmological constant but with the inclusion of terms 
depending on the Gauss-Bonnet invariant in the action. Even more, it 
is shown that the extra terms in the action coming from the 
modification of gravity could behave relaxing the vacuum energy 
density, represented by a cosmological constant, and may resolve the 
so-called cosmological constant problem.  To study and reconstruct 
the theory that reproduces such a model as well as other kind of 
solutions studied, we shall use the method proposed in Chapter 5, where the FLRW equations are written as functions of 
the so-called number of e-foldings instead of the cosmic time. The 
possible phantom epoch produced by this kind of theories is also 
explored, as well as other interesting cosmological solutions, where 
the inclusion of other contributions as perfect fluids with 
inhomogeneous EoS are studied.

\section{Modified $R+f(G)$ gravity}

We consider the following action, which describes General Relativity 
plus a function of the Gauss-Bonnet term (see Refs.~\cite{2005PhLB..631....1N,2006JPhA...39.6627N}):
\begin {equation}
S=\int 
d^{4}x\sqrt{-g}\left[\frac{1}{2\kappa^{2}}R+f(G)+L_{m}\right]\,, 
\label{6.1.1}
\end{equation}
where $\kappa^2=8\pi G_N$, $G_N$ being the Newton constant, and the 
Gauss-Bonnet invariant is defined as usual:
\begin{equation}
 G=R^2-4R_{\mu\nu}R^{\mu\nu}+R_{\mu\nu\lambda\sigma}R^{\mu\nu\lambda\sigma}\ .
\label{6.I2}
\end{equation}
By varying the action over $g_{\mu\nu}$, the following field equations 
are obtained:
$$
0=\frac{1}{2k^2}(-R^{\mu\nu}+\frac{1}{2}g^{\mu\nu}R)+T^{\mu\nu}+\frac{1}{2}g^{\mu\nu}f(G)-2f_{G}RR^{\mu\nu}+4f_{G}R^{\mu}_{\rho}R^{\nu\rho}
$$
$$
-2f_{G}R^{\mu\rho\sigma\tau}R^{\nu}_{\rho\sigma\tau}-4f_{G}R^{\mu\rho\sigma\nu}R_{\rho\sigma}+2(\nabla^{\mu}\nabla^{\nu}f_{G})R
-2g^{\mu\nu}(\nabla^{2}f_{G})R-4(\nabla_{\rho}\nabla^{\mu}f_{G})R^{\nu\rho}
$$
\begin {equation}
-4(\nabla_{\rho}\nabla^{\nu}f_{G})R^{\mu\rho}+4(\nabla^{2}f_{G})R^{\mu\nu}+4g^{\mu\nu}(\nabla_{\rho}\nabla_{\sigma}f_{G})R^{\rho\sigma}
-4(\nabla_{\rho}\nabla_{\sigma}f_{G})R^{\mu\rho\nu\sigma}\,,
\label{6.1.2}
\end{equation}
where we made the notations $f_G=f'(G)$ and $f_{GG}=f''(G)$. We shall assume throughout a spatially-flat FLRW universe. Then, the field equations give the FLRW equations, 
with the form
\[
0=-\frac{3}{\kappa^{2}}H^{2}+G 
f_{G}-f(G)-24\dot{G}H^{3}f_{GG}+\rho_{m}\,,
\]
\begin {equation}
0=8H^2\ddot{f}_{G}+16H(\dot{H}+H^2)\dot{f}_G+\frac{1}{\kappa^2}(2\dot{H}+3H^2)+f-Gf_G+p_m\,.
\label{6.1.3}
\end{equation}
The Gauss-Bonnet invariant $G$ and  
the Ricci scalar $R$ can be defined as functions of the Hubble parameter as
\begin {equation}
G=24(\dot{H}H^{2}+H^{4}),\quad R=6(\dot{H}+2H^{2}).
\end{equation}
Let us  now rewrite Eq.~(\ref{6.1.3}) by using a new variable,
$N=\ln\frac{a}{a_{0}}=-\ln(1+z)$, i.e. the number of e-foldings, 
instead of the cosmological time $t$, where  $z$ is the redshift. The following expressions are then easily obtained
\begin{equation}
a=a_{0}e^{N},\quad H=\dot{N}=\frac{dN}{dt}\ , \quad
\frac{d}{dt}=H\frac{d}{dN}\ , \quad
\frac{d^2}{dt^2}=H^2\frac{d^2}{dN^2}+HH^\prime\frac{d}{dN}\ ,\quad
H^{\prime}=\frac{dH}{dN}\ .
\label{6.1.6}
\end{equation}
Eq.~(\ref{6.1.3}) can thus be expressed as follows
\begin{equation}
0=-\frac{3}{\kappa^{2}}H^2+24H^{3}(H'+H)f_{G}-f-576H^{6}\left(HH''
+3H'^{2}+4HH'\right)f_{GG}+\rho_{m}\ ,
\label{6.1.7}
\end{equation}
where $G$ and $R$ are now
\begin {equation}
G=24(H^{3}H'+H^{4})\ ,
\quad\dot{G}=24(H^{4}H''+3H^{3}H'^{2}+4H^{4}H')\ , \quad R=
6(HH'+2H^{2})\ .
\label{6.1.8}
\end{equation}
By introducing a new function $x$ as $x=H^2$, we have
\begin {equation}
H'=\frac{1}{2}x^{-1/2}x'\ ,\quad
H''=-\frac{1}{4}x^{-3/2}x'^{2}+\frac{1}{2}x^{-1/2}x''\ .\end{equation}
Hence,  Eq.~(\ref{6.1.7})  takes the form
\begin{equation}
0=-\frac{3}{\kappa^{2}}x+12x(x'+2x)f_{G}-f-24^{2}x\left[\frac{1}{2}x^{2}x''
+\frac{1}{2}xx'^{2}+2x^{2}x'\right]f_{GG}+\rho_{m}\ ,
\label{6.1.9}
\end{equation}
where we have used the expressions
\begin{equation} G=12xx'+24x^{2},\quad
\dot{G}=12x^{-1/2}[x^2x''+xx'^2+4x^{2}x']\  \quad \text{and} \quad 
R=3x'+12x. \label{6.1.10}
\end{equation}
Then, by using the above reconstruction method, any cosmological 
solution can be achieved, by introducing the given Hubble parameter 
in the FLRW equations, which leads to the corresponding Gauss-Bonnet 
action.

\section{Reconstructing $\Lambda$CDM model in $R+f(G)$ gravity}

We are now interested to reconstruct $\Lambda$CDM solution in 
$R+f(G)$ gravity for different kind of matter contributions. We show 
that in GB gravity there is no need of a cc. The cosmological models 
coming from the different versions of modified GB gravity considered 
will be carefully investigated with the help of several particular 
examples where calculations can be carried out explicitly.
 For the 
$\Lambda$CDM model, the Hubble rate is given by
\begin{equation}
H^{2}= \frac{\Lambda}{3}+\frac{\rho_0}{3a^3},
\label{6.LambdaCDM}
\end{equation}
where $\rho_0$ is the matter density (which consists of barionic 
matter and cold dark matter) and $\Lambda$ is the cosmological 
constant. In the rest of this section we put $\kappa^2=1$. 

 For the  
$\Lambda$CDM model, described by the Hubble parameter 
(\ref{6.LambdaCDM}), we can write the derivatives of the scale factor 
as well as the Hubble parameter in the following  useful way:
\[
\dot{a}=\sqrt{\frac{\Lambda a^2}{3}+\frac{\rho_0}{3a}}\ , \quad \ddot{a}=\frac{2\Lambda a^3-\rho_0}{6a^2}\ ,
\]
\begin{equation}
\dot{H}=-\frac{\rho_0}{2a^3}=\frac{3}{2}\left(\frac{\Lambda}{3}-H^{2}\right)\ 
, \quad \ddot{H}=\frac{3\rho_0}{2a^3}\sqrt{\frac{\Lambda 
}{3}+\frac{\rho_0}{3a^3}}=\frac{9}{2}\left(H^{2}- 
\frac{\Lambda}{3}\right)H\ .
\end{equation}
Using these formulas we get
\[
R=4\Lambda+\frac{\rho_0}{a^3}\ , \quad 
G=24\left(\frac{\rho_0}{3a^3}+\frac{\Lambda 
}{3}\right)\left(\frac{\Lambda }{3}-\frac{\rho_0}{6a^3}\right)\ ,
\]
\begin{equation}
\dot{R}=\frac{-3\rho_0}{a^3}\sqrt{\frac{\rho_0}{3a^3}+\frac{\Lambda 
}{3}}\ , \quad 
\dot{G}=\frac{4\rho_0}{a^3}\left(\frac{2\rho_0}{a^3}-\Lambda\right)\sqrt{\frac{\rho_0}{3a^3}+\frac{\Lambda 
}{3}}\ .
\end{equation}
Then, the following relation  between $R$ and $G$ holds:
\begin{equation}
G=-\frac{4}{3}(R^2-9\Lambda R+18 \Lambda^2)\ .
\end{equation}
Let us recall that $x=H^2$. Then, some of the above formulas take the 
form:
\[
\dot{H}=\frac{1}{2}(\Lambda-3x)\ \quad
\ddot{H}=\frac{3}{2}(3x- \Lambda)\sqrt{x}\ , \quad R=3(\Lambda+x)\ ,
\]
\begin{equation}
G=12x(\Lambda -x)\ , \quad \dot{R}=3(\Lambda-3x)\sqrt{x}\ , \quad
\dot{G}=12(3x-\Lambda)(2x-\Lambda)\sqrt{x}\ .
\end{equation}
Note that  the  variable $x$ can be expressed in terms of $R$ or  
$G$ as
\begin{equation}
x=\frac{R}{3}-\Lambda \quad \text{or} \quad 
x=\frac{3\Lambda\pm\sqrt{9\Lambda^2-3G}}{6}\,,
\end{equation}
respectively. The above formulas will be useful to reconstruct the 
$\Lambda$CDM model as well as other cosmological solutions in the 
context of Gauss-Bonnet gravity, as it is shown below.

We write the first Friedmann equation (\ref{6.1.3}) in  the form
\begin {equation}
0=-3H^{2}+12H^2(\Lambda -H^2) 
f_{G}-f-288H^4(3H^2-\Lambda)(2H^2-\Lambda)f_{GG}+\rho_{m}\,,
\end{equation}
or
\begin {equation}
0=-3x+12x(\Lambda -x) 
f_{G}-f-288x^2(3x-\Lambda)(2x-\Lambda)f_{GG}+\rho_{m}\,.
\end{equation}
For further algebra more  convenient is the following form of  this  equation 
\begin {equation}
0=(\rho_{m}-3x-f)(\Lambda-2x)+[48x^2(3x-\Lambda)+x(\Lambda-x)]f_x+24x^2(3x-\Lambda)(\Lambda-2x)f_{xx}\ .
\label{6.ee}
\end{equation}
Now we wish to construct some particular exact solutions of this 
equation.

\subsection*{Case I: Absence of matter, $\rho_m=0$}

First of all, let us  consider the simple case in absence of matter, $\rho_m=0$. Then, the equation (\ref{6.ee}) takes the form
\begin {equation}
0=-(3x+f)(\Lambda-2x)+[48x^2(3x-\Lambda)+x(\Lambda-x)]f_x+24x^2(3x-\Lambda)(\Lambda-2x)f_{xx}\ .
\label{6.C1}
\end{equation}
We can analyze the cases where the cosmological constant term in the 
solution (\ref{6.LambdaCDM}) vanishes and where it is non-zero.
\begin{itemize}
    \item Let $\Lambda=0$. Then Eq. (\ref{6.C1}) reads as
\begin {equation}
0=-2(3x+f)-x(144x-1)f_x+144x^3f_{xx}\ .
\label{6.C2}
\end{equation}
The general  solution of (\ref{6.C2}) is given by
\begin {equation}
f(x)=C_2x^2+C_1x(144x-1)e^{\frac{1}{144x}}+v_1(x)\ ,
\label{6.Sol1}
\end{equation}
where 
\begin{equation}
v_1(x)=-864x\left[\frac{1}{144x}+x\ln{x}+\left(\frac{1}{144x}-x\right)Ei\left(1,\frac{1}{144x}\right)e^{\frac{1}{144x}}\right]\ 
.
\end{equation}
Here  
\begin {equation}
Ei(a,z)=z^{a-1}\Gamma(1-a,z)=\int_{1}^{\infty}e^{-zs}s^{-a}ds\ .
\end{equation}
Then, the solution that reproduces, basically a power-law expansion, is found.

\item Let $\Lambda\neq 0$. Then Eq. (\ref{6.C1}) has a complex solution, 
which has no physical meaning as it gives a complex action.  
\end{itemize}
Hence, it appears that the $\Lambda$CDM model (\ref{6.LambdaCDM}) can 
not be reproduced by $R+f(G)$ gravity in the absence of matter. The 
only solution found, restricted to $\Lambda=0$, does not produce an accelerating expansion. 

\subsection*{Case II: $\rho_m\neq 0$ and $\Lambda = 0$}

We now explore the case when some kind of matter with a particular 
EoS is present in the Universe, but with no cosmological constant 
term in the Hubble parameter described in (\ref{6.LambdaCDM}). We 
explore several examples where different kind of matter 
contributions are considered.
 
\subsubsection*{Example 1} 
Let us now consider the case when $\Lambda=0$ and the evolution of 
the matter density behaves as
\begin {equation}
\rho_m=3H^2=3x.
\end{equation}
In this case the modified Friedmann equation (\ref{6.1.3}) reads as
\begin {equation}
0=2f+x(144x-1)f_x-144x^3f_{xx}.
\label{6.C5}
\end{equation} 
The  general solution of the equation (\ref{6.C5}) is given by
\begin {equation}
f(x)=C_1x^2+C_2x(144x-1)e^{\frac{1}{144x}} .
\end{equation}
This function reproduces the solution (\ref{6.LambdaCDM}) under the conditions imposed above.

\subsubsection*{Example 2} 
Now  we consider a more general case, where the energy density is given by,
\begin {equation}
\rho_m=u(x)\ ,
\end{equation}
where $u(x)$ is some function of $x$. 
In this case the modified Friedmann equation (\ref{6.1.3}) reads as
\begin {equation}
0=2[3x-u(x)+f]+x(144x-1)f_x-144x^3f_{xx}\ .
\end{equation}
Its  general solution is
\begin {equation}
f(x)=C_1x^2+C_2x(144x-1)e^{\frac{1}{144x}}+v_2(x)\ , \label{6.C5b}
\end{equation}
with
\begin {equation}
v_2(x)=288x\left[\left(x-\frac{1}{144}\right)e^{\frac{1}{144x}}J_1-\frac{x}{144}J_2\right]\ ,
\end{equation}
where
\begin {equation}
J_1=\int\frac{3x-u}{x^2}e^{\frac{1}{144x}}dx, \quad J_2=\int\frac{(3x-u)(144x-1)}{x^3}dx\ .
\label{6.C6}
\end{equation}
Then, the solution (\ref{6.C5b}) gives the function of the 
Gauss-Bonnet invariant that reproduces this model for any kind of EoS matter fluid.

\subsection*{Case III: $\rho_m\neq 0$ and $\Lambda\neq 0$}
Let us now explore the most general case for the solution 
(\ref{6.LambdaCDM}) in $R+f(G)$ gravity with a non vanishing matter 
fluid with a given EoS parameter.

We  consider a  Universe filled with a pressureless fluid. By the energy conservation equation, the energy desity can be written as
\begin{equation}
\rho_m=3x+\beta\,. 
\end{equation}
Then Eq. (\ref{6.1.3}) takes the form
\begin {equation}
0=(\beta-f)(\Lambda-2x)+[48x^2(3x-\Lambda)+x(\Lambda-x)]f_x+24x^2(3x-\Lambda)(\Lambda-2x)f_{xx}
\end{equation}
and has the following particular solution:
\begin{equation}
 f(x)=\gamma x^2-\gamma\Lambda x+\beta\,.
\end{equation}
If $\Lambda=0$, then $ f(x)=\gamma x^2+\beta$. Also if $\gamma=0$, 
then the solution takes the form $f=\beta$, which corresponds to the 
cosmological constant. Note that if $\beta=-\Lambda$ then
\begin{equation}
\rho_m=\frac{\rho_{03}}{a^3}=3x-\Lambda, \quad f(x)=\gamma x^2-\gamma\Lambda x-\Lambda\ .
\end{equation}
This gives a solution where the cosmological constant is corrected 
by the contribution from $f(G)$, what may resolve the cosmological 
constant problem.

\section{The $F(R,G)$  model}

Let us now consider a more general model for a class of modified Gauss-Bonnet gravity. 
This can be described by the following action
\begin{equation}
S=\int d^{4}x\sqrt{-g}[\frac{1}{2\kappa^{2}}F(R,G)+L_{m}].
\label{6.F(R,G)action}
\end{equation}
Varying over $g_{\mu\nu}$ the gravity field equations are obtained,
\begin{eqnarray}
0=\kappa^{2}T^{\mu\nu}+\frac{1}{2}g^{\mu\nu}F(G)-2F_{G}RR^{\mu\nu}+4F_{G}R^{\mu}_{\rho}R^{\nu\rho}\nonumber\\
-2F_{G}R^{\mu\rho\sigma\tau}R^{\nu}_{\rho\sigma\tau}-4F_{G}R^{\mu\rho\sigma\nu}R_{\rho\sigma}+2(\nabla^{\mu}\nabla^{\nu}F_{G})R
-2g^{\mu\nu}(\nabla^{2}F_{G})R\nonumber\\
-4(\nabla_{\rho}\nabla^{\mu}F_{G})R^{\nu\rho}-4(\nabla_{\rho}\nabla^{\nu}F_{G})R^{\mu\rho}+4(\nabla^{2}F_{G})R^{\mu\nu}+4g^{\mu\nu}(\nabla_{\rho}\nabla_{\sigma}F_{G})R^{\rho\sigma}\nonumber\\
-4(\nabla_{\rho}\nabla_{\sigma}F_{G})R^{\mu\rho\nu\sigma}-F_{G}R^{\mu\nu}+\nabla^{\mu}\nabla^{\nu}F_{R}-g^{\mu\nu}\nabla^{2}F_{R}.
\label{6.FieldEq}
\end{eqnarray}
In the case of a flat FLRW Universe,  the first FLRW equation yields
\begin {equation}
0=\frac{1}{2}(GF_{G}-F-24H^{3}F_{Gt})+3(\dot{H}+H^{2})F_{R}-3HF_{Rt}+\kappa^{2}\rho_{m}.
\label{6.FriedmEq}
\end{equation}
And from here, using the techniques developed in the previous section, it is plain that explicit $F(R,G)$ functions can be reconstructed for given cosmological solutions.

\subsection*{De Sitter Solutions}

As well known, the de Sitter solution is one of the most important cosmological solutions nowadays, since the current epoch has been observed to have an expansion that behaves approximately as de Sitter. This solution is described by an exponential expansion of the scale factor, which gives a constant Hubble parameter $H(t)=H_0$. By inserting it in the Friedmann equation (\ref{6.FriedmEq}), one finds that any kind of $F(R,G)$ function can possibly admit de Sitter solutions, with the proviso that the following algebraic equation has positive roots for $H_0$
\begin{equation}
0= \frac{1}{2}(G_0F_G(G_0)-F(G_0,R_0))+3H^2_0F_R(R_0)\ ,
\label{6.deSitter}
\end{equation}
being $R_0=12H^2_0$ and $G_0=24H^4_0$; we have here neglected the contribution of matter for simplicity. As it was pointed out  for the case of modified $F(R)$ gravity, the de Sitter points are critical points for the Friedmann equations, what could explain the current acceleration phase as well as the inflationary epoch. This explanation can be extended to the action (\ref{6.F(R,G)action}), so that any kind of function $F(R,G)$ with positive real roots for the equation (\ref{6.deSitter}) could in fact explain the acceleration epochs of the Universe in exactly the same way a cosmological constant does.

\subsection*{Phantom dark energy}

Let us now explore the cosmic evolution described by $H=\e^{mN}$ in the context of the action (\ref{6.F(R,G)action}). This solution  reproduces a phantom behavior, i.e. a superaccelerated expansion that, according to recent observations, our Universe could be in---or either close to cross the phantom barrier. We can now proceed with the reconstruction method, as explicitly shown in the section above, and a $F(R,G)$ function  will be reconstructed. For simplicity, we consider the following subfamily of functions
\begin{equation}
F(R,G)=f_1(G)+f_2(R)\ .
\label{6.3.1a}
\end{equation}
Correspondingly, the Friedmann equation (\ref{6.FriedmEq}) can be split 
into two equations, as
\begin{eqnarray}
 0=-24H^3\dot{G}f_{1GG}+ Gf_{1G}-f_1\ , \nonumber\\
0=-3H\dot{R}f_{2RR}+3(\dot{H}+H^2)f_{2R}-\frac{1}{2}f_2 +\kappa^2\rho_m
\label{6.3.1}
\end{eqnarray}

For this example, the Ricci scalar and the Gauss-Bonnet terms take the following form,
\begin {eqnarray}
G=24(m+1)e^{4mN}=24(m+1)H^4,\nonumber\\ 
R=6(m+2)e^{2mN}=6(m+2)H^2\ .
\label{6.3.2}
\end{eqnarray}
Hence, the first equation in (\ref{6.3.1}) can be written in terms of $G$, as
\begin{equation}
 G^2f_{1GG}-\frac{m+1}{4m}Gf_{1G}+\frac{m+1}{4m}f_1=0\ .
\label{6.3.3}
\end{equation}
This is an Euler equation, easy to solve, and yields
\begin{equation}
f_1(G)=C_1G^{1+\frac{m+1}{4m}}+C_2\ ,
\label{6.3.4}
\end{equation}
where $C_{1,2}$ are integration constants. In the same way, for the case being considered here, the second equation in (\ref{6.3.1}), for $R$, takes the form
\begin{equation}
R^2f_{2RR}-\frac{m+1}{2m}Rf_{2R}+\frac{m+2}{2m}f_2-\frac{\kappa^2(m+2)}{m}\rho_m=0\ .
\label{6.3.5}
\end{equation}
In absence of matter ($\rho_m=0$) this  is also an Euler equation, with solution 
\begin{eqnarray}
f_2(R)=k_1R^{\mu_+}+k_2R^{\mu_-}\, \nonumber\\ 
\mbox{where} \quad \mu_{\pm}=\frac{1+\frac{m+1}{2m}\pm\sqrt{1+\frac{(m+1)^2}{4m^2}-\frac{m+3}{m}}}{2}\ ,
\label{6.3.6}
\end{eqnarray}
and $k_{1,2}$ are integration constants. Then, the complete function F(R,G), given in (\ref{6.3.1a}), is reconstructed (in absence of matter) yielding the solutions (\ref{6.3.4}) and (\ref{6.3.6}). The theory (3.12) belongs to the class of models with positive and negative powers of the curvature introduced in \cite{2003PhRvD..68l3512N}.

Let us now consider the case where matter is included. From the energy conservation equation $\dot{\rho_{m}}+3H(1+w)\rho_{m}=0$ we have that, for a perfect fluid with constant EoS, $p_m=w_m\rho_m$,  the solution is given by $\rho_m=\rho_0 e^{-3(1+w_m)N}$.  By inserting the expression for $R$ (\ref{6.3.2}) into this solution, we get
\begin{equation}
\rho_m=\rho_{m0}=\left(\frac{R}{6(m+2)} \right)^{-\frac{3(1+w_m)}{2m}}\ .
\label{6.3.7}
\end{equation}
In such case the general solution for $f_2$ is given by
\begin{eqnarray}
f_2(R)=k_1R^{\mu_+}+k_2R^{\mu_-}+kR^A\ , \mbox{where,} \nonumber\\   k=\kappa^2\frac{\rho_{m0}(6(m+1))^{-A}}{A(A-1)-1/2m} \quad \mbox{and} \quad A=-\frac{3(1+w_m)}{2m}\ .
\label{6.3.8}
\end{eqnarray}
Hence, we see that the solution for the Hubble parameter $H=\e^{mN}$ can be easily recovered in the context of modified Gauss-Bonnet gravity. Nevertheless, it seems clear that, for more complex examples, one may not be able to solve the corresponding equations analytically and numerical analysis could be required..

\section{Cosmological solutions in pure $f(G)$ gravity}

We have studied so far a theory described by the action (\ref{6.1.1}), 
which is given by the usual Hilbert-Einstein term plus a function of 
the Gauss-Bonnet invariant, that is assumed to become important in 
the dark energy epoch. In this section, we are interested to 
investigated some important cosmic solutions in the frame of a 
theory described only by the Gauss-Bonnet invariant, and whose 
action is given by
\begin{equation}
S=\int d^4x\sqrt{-g}\left[f(G)+L_m\right]\ .
\label{6.D6}
\end{equation}       
In this case, the  FLRW equations are:
\[
0=Gf_G-f-24\dot{G}H^3f_{GG}+\rho_m
\]
\begin{equation}
0=8H^2\ddot{f_G}+16H(\dot{H}+H^2)\dot{f_G}+f-Gf_G+p_m\ .
\label{6.D7}
\end{equation}
We are interested to explore some important solutions from the 
cosmological point of view, as de Sitter and power law expansions.
De Sitter solutions can be easily checked for a given model, as it was shown above. We can explore the de Sitter points 
admitted by a general $f(G)$ by introducing the solution $H(t)=H_0$ 
in the first FLRW equation given in (\ref{6.D7}), which yields
\begin{equation}
0=G_0f_G(G_0)-f(G_0)\ .
\label{6.D9}
\end{equation} 
Here, $G_0=24H_0^2$ and we have ignored the contribution of matter. 
Then, we have reduced the differential equation (\ref{6.D7}) to an 
algebraic equation that can be resolved by specifying a function 
$f(G)$. The de Sitter points are given by the positive roots of this 
equation, which could explain not just the late-time accelerated 
epoch but also the inflationary epoch. The stability of these 
solutions has to be studied in order to achieve a grateful exit in 
the case of inflation, and future predictions for the current cosmic 
acceleration.

\subsection*{Power law solutions}

We are now interested to explore power law solutions for a theory 
described by the action (\ref{6.D6}). This kind of solutions are very 
important during the cosmic history as the matter/radiation epochs 
are described by power law expansions, as well as the possible 
phantom epoch, which can be seen as a special type of these 
solutions. Let us start by studying a Hubble parameter given by
\begin{equation}
H(t)=\frac{\alpha}{t}\rightarrow a(t)\sim t^{\alpha}\ ,
\label{6.D10}
\end{equation}
where we take $\alpha> 1$. Then, by introducing the solution 
(\ref{6.D10}) into the first FLRW equation (\ref{6.D7}), it yields the 
differential equation
\begin{equation}
0=-f(G)+Gf_G+\frac{4G^2}{\alpha-1}f_{GG}\ , \label{6.D11}
\end{equation}
where we have neglected any contribution of matter for simplicity. 
The equation (\ref{6.D11}) is a type of Euler equation, whose solution 
is
\begin{equation}
f(G)=C_1G+C_2G^{\frac{1-\alpha}{4}}\ .
\label{6.D12}
\end{equation}
Thus, we have shown that power-law solutions of the type (\ref{6.D10}) 
correspond to actions with powers on the Gauss-Bonnet invariant, in 
a similar way as in $f(R)$ gravity, where power-law solutions 
correspond to an action with powers on the scalar curvature, $R$, as it was shown above. 

Let us now explore another kind of power-law solutions, where the 
Universe enters a phantom phase and ends in a Big Rip singularity. 
This general class of Hubble parameters may be written as
\begin{equation}
H(t)=\frac{\alpha}{t_s-t}\ ,
\label{6.D13}
\end{equation}
where $t_s$ is the so-called Rip time, i.e. the time when the future 
singularity will take place. By inserting the solution (\ref{6.D13}) 
into the first FLRW equation (\ref{6.D7}), the equation yields
\begin{equation}
0=-f(G)+Gf_G(G)-\frac{4\alpha^2G^2}{1+\alpha}\ , 
\label{6.D14}
\end{equation}
which is also a Euler equation, whose solution is given by,
\begin{equation}
f(G)=C_1G+C_2G^{\frac{1+\alpha}{4\alpha^2}}\ .
\label{6.D15}
 \end{equation}
Thus, we have shown that power law solution of the type 
radiation/matter dominated epochs on one side and phantom epochs on 
the other, are well reproduced in pure $f(G)$ gravity, in a similar 
way as it is in $f(R)$ gravity. 

\section{Conclusions}

We have explored in this chapter several cosmological solutions in the 
frame of Gauss-Bonnet gravity, considering specially the case of an 
action composed of the Hilbert-Einstein action plus a function on 
the Gauss-Bonnet invariant. Also pure $f(G)$ gravity has been 
considered, as well as the possibility of the implication of 
inhomogeneous terms in the EoS of a perfect fluid, which could 
contribute together with modified gravity to the late-time 
acceleration. We have shown that the $\Lambda$CDM model can well be 
explained in this kind of theories, which may give an explanation to 
the cosmological constant problem as the modified gravity terms may 
act relaxing the vacuum energy density. Other kinds of solutions in 
$f(G)$ gravity have been reconstructed. It has been shown that 
$f(G)$ gravity could explain the dark energy epoch whatever the 
nature of its EoS, of type quintessence or phantom, and even the 
inflationary phase.  More complex cosmological solutions would 
require numerical analysis, but our analysis of a few simple cases 
has already shown that $f(G)$ gravity accounts for the accelerated 
epochs and may contribute during the radiation/matter dominated 
eras, and it may explain also the dark matter contributions to the 
cosmological evolution, what will be explored in future works. This kind of modified gravity models which  reproduce dark energy and inflation, can be modeled as an inhomogeneous fluid with a dynamical equation of state, what would be distinguished from other models with a static EoS. Even as perturbations in modified gravity behave different than in General Relativity, it could give a signature of the presence of higher order terms in the gravity action, as the Gauss-Bonnet invariant, when structure formation is studied and simulations are performed, what should be explored in the future. 

\part{On Ho\v{r}ava-Lifshitz gravity and its extension to more general actions in cosmology}
\chapter[Inflation and dark energy in F(R) Ho\v{r}ava-Lifshitz gravity]{Unifying inflation with dark energy in modified F(R) Ho\v{r}ava-Lifshitz gravity}

\footnote{This Chapter is based on: \cite{2010EPJC...70..351E}}The Ho\v{r}ava-Lifshitz quantum gravity \cite{2009PhRvD..79h4008H}
has been conjectured to be renormalizable in four dimensions, at the price of
explicitly breaking Lorentz invariance. Its generalization to an
$F(R)$-formulation, which seems to be also renormalizable in $3+1$ dimensions,
has been considered in Refs.~\cite{2010PhRvD..82f5020C,2010CQGra..27r5021C}, where the Hamiltonian 
structure and FRW cosmology, in a power-law theory, have been investigated for 
such modified $F(R)$ Ho\v{r}ava-Lifshitz gravity. It was also conjectured there 
that it sustains, in principle, the possibility of a unified description of 
early-time inflation and the dark energy epochs.

The purpose of the present chapter is to show a realistic non-linear $F(R)$ gravity
in the Ho\v{r}ava-Lifshitz formulation, with the aim to understand if such theory 
is in fact directly able to predict in a natural way the unification of the two
acceleration eras, similarly as it is done in the convenient version. 
It will be here shown that, for a special choice of parameters, the FRW equations 
do coincide with the ones for the related, convenient $F(R)$
gravity. This means, in particular, that the cosmological history of such 
Ho\v{r}ava-Lifshitz $F(R)$ gravity will be just the same as for its convenient 
version (whereas black hole solutions are generically speaking different).
For the general version of the theory the situation turns out to be more complicated.
Nevertheless, the unification of inflation with dark energy is still possible 
and all local tests can also be passed, as we will prove.

\section{Modified $F(R)$ Ho\v{r}ava-Lifshitz gravity}

In this section, modified Ho\v{r}ava-Lifshitz $F(R)$
gravity is briefly reviewed \cite{2010PhRvD..82f5020C,2010CQGra..27r5021C}. We start by writing a
general metric in the so-called ADM decomposition in a $3+1$ spacetime (for
more details see \cite{2008GReGr..40.1997A,Gravitation1973} and references
therein),
\be
ds^2=-N^2 dt^2+g^{(3)}_{ij}(dx^i+N^idt)(dx^j+N^jdt)\, ,
\label{8.1.1}
\ee
where $i,j=1,2,3$, $N$ is the so-called lapse variable, and $N^i$ is
the shift $3$-vector. In standard general relativity (GR),
the Ricci scalar can be written in terms of this metric, and yields
\be
R=K_{ij}K^{ij}-K^2+R^{(3)}+2\nabla_{\mu}(n^{\mu}\nabla_{\nu}n^{\nu}-n^{\nu}
\nabla_{\nu}n^{\mu})\, ,
\label{8.1.2}
\ee
here $K=g^{ij}K_{ij}$, $K_{ij}$ is the extrinsic curvature, $R^{(3)}$
is the spatial scalar curvature, and $n^{\mu}$ a unit vector
perpendicular to a hypersurface of constant time. The extrinsic
curvature $K_{ij}$ is defined as
\be
K_{ij}=\frac{1}{2N}\left(\dot{g}_{ij}^{(3)}-\nabla_i^{(3)}N_j-\nabla_j^{(3)}
N_i\right)\, .
\label{8.1.3}
\ee

In the original model \cite{2009PhRvD..79h4008H}, the lapse variable $N$ is taken
to be just time-dependent, so that the projectability condition holds
and by using the foliation-preserving diffeomorphisms (\ref{8.1.7}),
it can be fixed to be $N=1$.
As pointed out in \cite{2010PhRvL.104r1302B},  imposing the projectability
condition may cause problems with Newton's law in the Ho\v{r}ava gravity.
On the other hand, Hamiltonian analysis shows that the non-projectable
$F(R)$-model is inconsistent\cite{2010PhLB..693..404C}).
For the non-projectable case, the Newton law
could be restored (while keeping stability) by the ``healthy''
extension of the original Ho\v{r}ava gravity of Ref.~\cite{2010PhRvL.104r1302B}.

The action for standard $F(R)$ gravity can be written as
\be
S=\int d^4x\sqrt{g^{(3)}}N F(R)\, .
\label{8.1.4}
\ee
Gravity of Ref.~\cite{2009PhRvD..79h4008H} is assumed to have different scaling
properties of the space and time coordinates
\be
x^i=b x^i\, , \quad t=b^zt\, ,
\label{8.1.6}
\ee
where $z$ is a dynamical critical exponent that renders the theory
renormalizable for $z=3$ in $3+1$ spacetime dimensions \cite{2009PhRvD..79h4008H}
(For a proposal of covariant renormalizable gravity with dynamical
Lorentz symmetry breaking, see 
GR is recovered when $z=1$. The scaling properties (\ref{8.1.6}) render
the theory  invariant only under the so-called foliation-preserving 
diffeomorphisms:
\be
\delta x^i=\zeta(x^i,t)\, , \quad \delta t=f(t)\, .
\label{8.1.7}
\ee
It has been pointed that, in the IR limit, the full diffeomorphisms
are recovered, although the mechanism for this transition is not
physically clear. The action considered here was introduced
in Ref.~\cite{2010CQGra..27r5021C},
\be
S=\frac{1}{2\kappa^2}\int dtd^3x\sqrt{g^{(3)}}N F(\tilde{R})\, , \quad
\tilde{R}= K_{ij}K^{ij}-\lambda K^2 + R^{(3)}+
2\mu\nabla_{\mu}(n^{\mu}\nabla_{\nu}n^{\nu}-n^{\nu}\nabla_{\nu}n^{\mu})-
L^{(3)}(g_{ij}^{(3)})\, ,
\label{8.1.8}
\ee
where $\kappa$ is the dimensionless gravitational coupling, and where, two
new constants $\lambda$ and $\mu$ appear, which account for the violation
of the full diffeomorphism transformations.
A degenerate version of the above $F(R)$-theory with $\mu=0$ has been proposed and
studied in Ref.~\cite{2010PhRvD..81f4028K,2010PhRvD..82d4004K}.
Note that in the original Ho\v{r}ava gravity theory \cite{2009PhRvD..79h4008H},
the third term in the expression for $\tilde{R}$ can be omitted, as
it becomes a total derivative. The term $L^{(3)}(g_{ij}^{(3)})$ is
chosen to be \cite{2009PhRvD..79h4008H}
\be
L^{(3)}(g_{ij}^{(3)})=E^{ij}G_{ijkl}E^{kl}\, ,
\label{8.1.9}
\ee
where $G_{ijkl}$ is the generalized De Witt metric, namely
\be
G^{ijkl}=\frac{1}{2}\left(g^{ik}g^{jl}+g^{il}g^{jk}\right)-\lambda g^{ij}g^{kl}\, .
\label{8.1.10}
\ee
In Ref.~\cite{2009PhRvD..79h4008H}, the expression for $E_{ij}$ is constructed to
satisfy the ``detailed balance principle'' in order to restrict the
number of free parameters of the theory. This is defined through
variation of an action
\be
\sqrt{g^{(3)}}E^{ij}=\frac{\delta W[g_{kl}]}{\delta g_{ij}}\, ,
\label{8.1.11}
\ee
where the form of $W[g_{kl}]$ is given in Ref.~\cite{2009JHEP...03..020H} for
$z=2$ and  $z=3$. Other forms for
$L^{(3)}(g_{ij}^{(3)})$ have been suggested that abandons the
detailed balance condition but still render the theory power-counting
renormalizable (see Ref.~\cite{2010PhRvD..82f5020C}).

We are interested in the study of (accelerating) cosmological
solutions for the theory described by action (\ref{8.1.8}).
Spatially-flat FRW metric is assumed, which is written in the ADM decomposition as,
\be
ds^2=-N^2dt^2+a^2(t)\sum_{i=1}^3 \left(dx^{i}\right)^2\, .
\label{8.1.14}
\ee
If we also assume the projectability condition,
$N$ can be taken to be just time-dependent and, by using the
foliation-preserving
diffeomorphisms (\ref{8.1.7}), it can be fixed to be unity, $N=1$.
When we do not assume the projectability condition, $N$ depends on both 
the time and spatial coordinates, first. Then, just as an assumption of
the solution, $N$ is taken to be unity.

For the metric (\ref{8.1.14}), the scalar $\tilde{R}$ is
given by
\be
\tilde{R}=\frac{3(1-3\lambda
+6\mu)H^2}{N^2}+\frac{6\mu}{N}\frac{d}{dt}\left(\frac{H}{N}\right)\, .
\label{8.1.15}
\ee
For the action (\ref{8.1.8}), and assuming the FRW metric (\ref{8.1.15}),
the second FRW equation can be obtained by varying
the action with respect to the spatial metric $g_{ij}^{(3)}$, which
yields
\be
0=F(\tilde{R})-2(1-3\lambda+3\mu)\left(\dot{H}+3H^2\right)F'(\tilde{R})-2(1-
3\lambda)
\dot{\tilde{R}}F''(\tilde{R})+2\mu\left(\dot{\tilde{R}}^2F^{(3)}(\tilde{R})
+\ddot{\tilde{R}}F''(\tilde{R})\right)+\kappa^2p_m\, ,
\label{8.1.16}
\ee
here $\kappa^2=16\pi G$, $p_m$ is the pressure of a perfect fluid
that fills the Universe, and $N=1$. Note that this
equation becomes the usual second FRW equation for convenient $F(R)$
gravity (\ref{8.1.4}), by setting the constants $\lambda=\mu=1$.
When we assume the projectability condition,
variation over $N$ of the action (\ref{8.1.8}) yields the following
global constraint
\be
0=\int d^3x\left[F(\tilde{R})-6(1-3\lambda +3\mu)H^2-6\mu\dot{H}+6\mu
H \dot{\tilde{R}}F''(\tilde{R})-\kappa^2\rho_m\right]\, .
\label{8.1.17}
\ee
Now,  using the ordinary conservation equation for the matter fluid
$\dot{\rho}_m+3H(\rho_m+p_m)=0$, and integrating Eq.~(\ref{8.1.16}),
\be
0=F(\tilde{R})-6\left[(1-3\lambda
+3\mu)H^2+\mu\dot{H}\right]F'(\tilde{R})+6\mu H
\dot{\tilde{R}}F''(\tilde{R})-\kappa^2\rho_m-\frac{C}{a^3}\, ,
\label{8.1.18}
\ee
where $C$ is an integration constant, taken to be zero,
according to the constraint equation (\ref{8.1.17}).
If we do not assume the projectability condition, we can directly
obtain (\ref{8.1.18}), which corresponds to the first FRW equation, by 
variation over $N$.
Hence, starting from a given $F(\tilde{R})$ function, and solving 
Eqs.~(\ref{8.1.16}) and (\ref{8.1.17}), a cosmological solution
can be obtained.

\section{Reconstructing FRW cosmology in $F(R)$ Ho\v{r}ava-Lifshitz
gravity}

To start, let us analyze the simple model $F(\tilde{R})=\tilde{R}$,
which cosmology was studied in \cite{2010arXiv1004.2474A,2010CQGra..27d5013B,2010EPJC..tmp..310B,2010CQGra..27g5005B,2009PhRvD..80d3516B,2009JHEP...09..112C,2010PhRvD..81d4006C,2010CQGra..27d5004C,2010JCAP...02..020G,2009PhLB..679..172J,2010PhLB..684..194M,2009PhLB..679....6M,2010PhRvD..82d3506M,2010PhRvD..81d3001N,2009JHEP...09..123P,2010JCAP...01..001P,2010EPJC...67..229S,2010JCAP...06..025S,2009JHEP...10..033S,2009PhRvL.102w1301T,2010arXiv1003.5152W,2009JCAP...07..012W} (for a complete analysis of
cosmological perturbations, see \cite{2010PhRvD..81h4053G}). In such a case, the FRW
equations look similar to GR,
\be
H^2=\frac{\kappa^2}{3(3\lambda-1)}\rho_m\, , \quad
\dot{H}=-\frac{\kappa^2}{2(3\lambda-1)}(\rho_m+p_m)\, ,
\label{8.2.1}
\ee
where, for $\lambda\rightarrow 1$, the standard FRW equations are
recovered. Note that the constant $\mu$ is now irrelevant because, as
pointed out above, the term in front of $\mu$ in (\ref{8.1.8}) becomes a
total derivative. For such
theory, one has to introduce a dark energy source as well as an
inflaton field, in order to reproduce the cosmic and inflationary
accelerated epochs, respectively. It is also important to note that,
for this case, the coupling constant is restricted to be
$\lambda>1/3$, otherwise Eqs.~(\ref{8.2.1}) become
inconsistent. It seems reasonable to think that, for the current epoch,
where $\tilde{R}$ has a small value, the IR limit of the theory is
satisfied $\lambda\sim1$, but for the inflationary epoch, when the
scalar curvature $\tilde{R}$ goes to infinity, $\lambda$ will take a
different value. It has been realized that, for $\lambda=1/3$, the
theory develops an anisotropic Weyl invariance (see \cite{2009PhRvD..79h4008H}),
and thus it takes an special role, although for the present model this
value is not allowed.

We now discuss some cosmological solutions of
$F(\tilde{R})$ Ho\v{r}ava-Lifshitz gravity. The first FRW equation,
given by (\ref{8.1.18}) with $C=0$, can be rewritten as a function of
the number of e-foldings $\eta=\ln\frac{a}{a_0}$, instead of the usual
time $t$. This technique has been developed in Chapter 5 for
convenient $F(R)$ gravity, where it was shown that any $F(R)$
theory can be reconstructed for a given cosmological solution. Here,
we extend such formalism to the Ho\v{r}ava-Lifshitz $F(R)$ gravity. Since
$\frac{d}{dt}=H\frac{d}{d\eta}$ and $\frac{d^2}{d\eta^2}
=H^2\frac{d^2}{d\eta^2}+H\frac{dH}{d\eta}\frac{d}{d\eta}$,
the first FRW equation (\ref{8.1.18}) is rewritten as
\be
0=F(\tilde{R})-6\left[(1-3\lambda+3\mu)H^2+\mu
HH'\right]\frac{dF(\tilde{R})}{d\tilde{R}}
+36\mu H^2\left[(1-3\lambda+6\mu)HH'+\mu H'^2+\mu H''H \right]
\frac{d^2F(\tilde{R})}{d^2\tilde{R}}-\kappa^2\rho_m\, ,
\label{8.D1}
\ee
where the primes denote derivatives with
respect to $\eta$. Thus, in this case there is no restriction on the
values of $\lambda$ or $\mu$.  By using the energy conservation
equation, and assuming a perfect fluid with equation of state (EoS)
$p_m=w_m\rho_m$, the energy density yields
\be
\rho_m=\rho_0 a^{-3(1+w_m)}=\rho_0a_0^{-3(1+w_m)}\e^{-3(1+w_m)\eta}\, .
\label{8.D2}
\ee
As the Hubble parameter can be written as a function of the number of
e-foldings, $H=H(\eta)$, the  scalar curvature in (\ref{8.1.15}) takes the
form
\be
\tilde{R}=3(1-3\lambda+6\mu)H^2+6\mu HH'\, ,
\label{8.D3}
\ee
which can be solved with respect to $\eta$ as $\eta=\eta(\tilde{R})$,
and  one gets an expression (\ref{8.D1}) that gives an equation on
$F(\tilde{R})$ with the variable $\tilde{R}$. This can be 
simplified a bit by writing $G(\eta)=H^2$ instead of the Hubble parameter. In
such case, the differential equation (\ref{8.D1}) yields
\be
0=F(\tilde{R})-6\left[(1-3\lambda+3\mu)G+\frac{\mu}{2}G'\right]\frac{dF(\tilde{R})
}{d\tilde{R}}+18\mu\left[(1-3\lambda+6\mu)GG'+\mu GG''\right]
\frac{d^2F(\tilde{R})}{d^2\tilde{R}}-\kappa^2\rho_0a_0^{-3(1+w)}
\e^{-3(1+w)\eta}\, ,
\label{8.D4}
\ee
and the scalar curvature is now written as
$\tilde{R}=3(1-3\lambda+6\mu)G+3\mu
G'$.  Hence,
for a given cosmological solution $H^2=G(\eta)$, one can resolve
Eq.~(\ref{8.D4}), and the $F(\tilde{R})$ that reproduces such solution
is obtained.

As an example, we consider the Hubble parameter that reproduces the
$\Lambda$CDM epoch. It is expressed as
\be
H^2 =G(\eta)= H_0^2 + \frac{\kappa^2}{3}\rho_0 a^{-3} = H_0^2 +
\frac{\kappa^2}{3}\rho_0 a_0^{-3} \e^{-3\eta} \, .
\label{8.D5}
\ee
where $H_0$ and $\rho_0$ are constant. In General
Relativity, the terms on the rhs of Eq.~(\ref{8.D5}) correspond to an
effective cosmological constant $\Lambda=3H_0^2$ and to cold dark
matter with EoS parameter $w=0$. The corresponding
$F(\tilde{R})$ can be reconstructed by following the same steps as
described above. Using the expression for the scalar curvature
$\tilde{R}=3(1-3\lambda+6\mu)G+3\mu G'$,
the relation between $\tilde{R}$ and $\eta$ is obtained,
\be
\e^{-3\eta}=\frac{\tilde{R}
 -3(1-3\lambda+6\mu)H_0^2}{3k(1+3(\mu-\lambda))}\, ,
\label{8.D7}
\ee
where $k=\frac{\kappa^2}{3}\rho_0 a_0^{-3}$. Then, substituting
(\ref{8.D5}) and (\ref{8.D7}) into Eq.~(\ref{8.D4}), one gets the 
differential expression
\bea
0 &=&
(1-3\lambda+3\mu)F(\tilde{R})-2\left(1-3\lambda+\frac{3}{2}\mu\right)
\tilde{R}
+9\mu(1-3\lambda)H^2_0\frac{dF(\tilde{R})}{d\tilde{R}} \nn
&& -6\mu(\tilde{R}-9\mu H^2_0)(\tilde{R}-3H^2_0(1-3\lambda+6\mu))
\frac{d^2F(\tilde{R})}{d^2\tilde{R}}-R-3(1-3\lambda+6\mu)H^2_0\, ,
\label{8.D8}
\eea
where, for simplicity, we have  considered a pressureless fluid $w=0$
in Eq.~(\ref{8.D4}). Performing the change of
variable $x=\frac{\tilde{R}-9\mu H^2_0}{3H^2_0(1+3(\mu-\lambda))}$,
the homogeneous part of 
Eq.~(\ref{8.D8})  can be easily identified as an hypergeometric
differential equation
\be
0=x(1-x)\frac{d^2 F}{dx^2} + \left(\gamma - \left(\alpha + \beta +
1\right)x\right)\frac{dF}{dx} - \alpha \beta F\, ,
\label{8.D9}
\ee
with the set of parameters $(\alpha,\beta,\gamma)$ being given by
\be
\gamma=-\frac{1}{2}\, , \quad \alpha+\beta=
\frac{1-3\lambda-\frac{3}{2}\mu}{3\mu}\, , \quad
\alpha\beta=-\frac{1+3(\mu-\lambda)}{6\mu}\, .
\label{8.D10}
\ee
The complete solution of Eq.~(\ref{8.D9}) is a Gauss' hypergeometric function
plus a linear term and a cosmological constant coming from the particular
solution of Eq.~(\ref{8.D8}), namely
\be
F(\tilde{R}) = C_1 F(\alpha,\beta,\gamma;x) + C_2 x^{1-\gamma} F(\alpha -
\gamma + 1, \beta - \gamma +
1,2-\gamma;x)+\frac{1}{\kappa_1}\tilde{R}-2\Lambda\, .
\label{8.D11}
\ee
where $C_1$ and $C_2$ are constants, $\kappa_1=3\lambda-1$ and
$\Lambda=-\frac{3H_0^2(1-3\lambda+9\mu)}{2(1-3\lambda+3\mu)}$.
Note that for the exact cosmology (\ref{8.D5}), the classical $F(R)$
gravity was reconstructed and studied in Chapter 5.
In this case, the solution (\ref{8.D11}) behaves similarly
to the classical $F(R)$ theory, except that now the parameters of
the theory depend on $(\lambda,\mu)$, which are
allowed to vary as it was noted above.
One can also explore the solution (\ref{8.D5}) for a particular choice
on the parameters $\mu=\lambda-\frac{1}{3}$, which plays a special role as
it is shown below. In this case, the scalar $\tilde{R}$ turns out to be a
constant,
and Eq.~(\ref{8.D8}), in the presence of a pressureless fluid, has the
solution
\be
F(\tilde{R})=\frac{1}{\kappa_1}\tilde{R}-2\Lambda\, , \quad \text{with}
\quad \Lambda=\frac{3}{2}(3\lambda-1)H_0^2\, .
\label{8.DD11}
\ee
Hence, for this constraint on the parameters,  the only consistent solution 
reduces to the Ho\v{r}ava linear theory with a cosmological constant.

As a further example, we consider the so-called phantom accelerating
expansion.
Currently, observational data do not totally exclude the  possibility that
the Universe could have already crossed the phantom divide, which means
that the effective EoS for dark energy would presently
be slightly less than $-1$. Such kind of system can be easily expressed in GR,
where the FRW equation reads $H^2=\frac{\kappa^2}{3}\rho_\mathrm{ph}$. Here
the subscript ``ph'' denotes the phantom nature of the fluid, which has
an EoS given by $p_\mathrm{ph}=w_\mathrm{ph}\rho_\mathrm{ph}$
with $w_\mathrm{ph}<-1$. By using
the energy conservation equation, the solution for the Hubble
parameter turns out to be
\be
H(t)=\frac{H_0}{t_s-t}\, ,
\label{8.D12}
\ee
where $H_0=-1/3(1+w_\mathrm{ph})$, and $t_s$ is the Rip time which
represents the time still remaining up to the Big Rip singularity.
As in the above example, one can
rewrite the Hubble parameter as a function of the number of
e-foldings; this yields
\be
G(\eta)=H^2(\eta)=H^2_0\e^{2\eta/H_0}\, .
\label{8.D13}
\ee
Then, by using the expression of the scalar curvature, the relation
between $\tilde{R}$ and $\eta$ is given by
\be
\e^{2\eta/H_0}=\frac{R}{H_0(AH_0+6\mu)}\, .
\label{8.D14}
\ee
By inserting (\ref{8.D13}) and (\ref{8.D14}) into the differential
equation (\ref{8.D4}), we get
\be
\tilde{R}^2\frac{d^2F(\tilde{R})}{d\tilde{R}^2}+k_1\tilde{R}
\frac{dF(\tilde{R})}{d\tilde{R}}+k_0F(\tilde{R})=0\, ,
\label{8.D15}
\ee
where
\be
k_1=-\frac{(AH_0+6\mu)((AH_0+3\mu))}{6\mu(AH_0+12\mu)}\, , \quad
k_0=\frac{(AH_0+6\mu)^2}{12\mu(AH_0+12\mu)} \, ,
\label{8.D16}
\ee
here we have neglected any kind of matter contribution for simplicity.
Eq.~(\ref{8.D15}) is an Euler equation, whose solution is well known
\be
F(R)=C_1R^{m_+}+C_2R^{m_-}\, , \quad \text{where} \quad
m_{\pm}=\frac{1-k_1\pm\sqrt{(k_1-1)^2-4k_0}}{2}\, .
\label{8.D17}
\ee
Hence, a $F(\tilde{R})$ HL gravity has been reconstructed that
reproduces the phantom dark epoch with no need of any exotic fluid.
In the same way, any given cosmology may be reconstructed.

\section{Unified inflation and dark energy in modified
Ho\v{r}ava-Lifshitz gravity}

Let us consider here some viable $F(\tilde{R})$ gravities which admit
the unification of inflation with late-time acceleration.
In the convenient $F(R)$ theory, a number of viable models
 which pass all local tests and are able to
unify the inflationary
and the current cosmic accelerated epochs have been proposed, as it was commented in the preceeding chapters. Here we
extend this class of
models to the Ho\v{r}ava-Lifshitz gravity. We consider the action,
\be
F(\tilde{R})=\tilde{R}+f(\tilde{R})\, ,
\label{8.3.1}
\ee
where it is assumed that the term $f(\tilde{R})$ becomes important at
cosmological scales, while for scales compared with the Solar system one
the theory becomes linear on $\tilde{R}$. As an example, we consider
the following function, already analyzed in standard $F(R)$ gravity 
\be
f(\tilde{R})=\frac{\tilde{R}^n(\alpha\tilde{R}^n-\beta)}{1+\gamma
\tilde{R}^n}\, ,
\label{8.3.2}
\ee
where $(\alpha, \beta, \gamma)$ are constants and $n>1$. This theory
reproduces the inflationary and cosmic acceleration epochs in convenient
$F(R)$ gravity, which  is also the case  in the
present theory, as will be shown. During inflation, it is assumed
that the curvature scalar tends to infinity. In this case the model
(\ref{8.3.1}), with (\ref{8.3.2}), behaves as
\be
\lim_{\tilde{R}\rightarrow\infty}F(\tilde{R})=\alpha\tilde{R}^n\, .
\label{8.3.3}
\ee
Then, by solving the FRW equation (\ref{8.1.18}), this kind of function
yields a power-law solution of the type
\be
H(t)=\frac{h_1}{t}\, , \quad \text{where} \quad
h_1=\frac{2\mu(n-1)(2n-1)}{1-3\lambda+6\mu-2n(1-3\lambda+3\mu)}\, .
\label{8.3.4}
\ee
This solution produces acceleration during the inflationary epoch if
the parameters of the theory are properly defined. The acceleration
parameter is given by $\frac{\ddot{a}}{a}=h_1(h_1-1)/t^2$, thus, for
$h_1>1$ the inflationary epoch is well reproduced by the model
(\ref{8.3.2}). On the other hand, the function (\ref{8.3.2}) has a
minimum at $\tilde{R}_0$, given by
\be
\tilde{R}_0 \sim \left( \frac{\beta}{\alpha\gamma}\right)^{1/4}\, ,
\qquad f'(\tilde{R})=0\, , \qquad  f(\tilde{R})= -2\Lambda\sim
-\frac{\beta}{\gamma}\, ,
\label{8.3.5}
\ee
where we have imposed the condition $\beta\gamma/\alpha\gg 1$. Then, at
the current epoch the scalar curvature acquires a small value which
can be fixed to coincide with the minimum (\ref{8.3.5}), such that the
FRW equations (\ref{8.1.16}) and (\ref{8.1.18}) yield
\be
H^2=\frac{\kappa^2}{3(3\lambda-1)}\rho_m+\frac{2\Lambda}{3(3\lambda-1)}\, 
\quad \dot{H}=-\kappa^2\frac{\rho_m+p_m}{3\lambda-1}\, ,
\label{8.3.6}
\ee
which look very similar to the standard FRW equations in GR, except
for the parameter $\lambda$. As has been pointed out, at the current
epoch the scalar $\tilde{R}$ is  small, so the theory is
in the IR limit where the parameter $\lambda\sim1$, and the equations
approach the usual ones for $F(R)$ gravity. Hence, the FRW
equations (\ref{8.3.6}) reproduce the behavior of the well known
$\Lambda$CDM model with no need to introduce a dark energy fluid
to explain the current universe acceleration.

As another example of the models described by (\ref{8.3.1}), we
can considered the function,
\be
f(\tilde{R})=-\frac{(\tilde{R}-\tilde{R}_0)^{2n+1}
+\tilde{R}_0^{2n+1}}{f_0+f_1\left[(\tilde{R}-\tilde{R}_0)^{2n+1}
+\tilde{R}_0^{2n+1}\right]}=-\frac{1}{f_1}+\frac{f_0/f_1}{f_0
+f_1\left[(\tilde{R}-\tilde{R}_0)^{2n+1}+\tilde{R}_0^{2n+
1}\right]}\, .
\label{8.3.7}
\ee
This function could also serve for the unification of inflation and
cosmic acceleration but, in this case, when one takes the limit
$\tilde{R}\rightarrow\infty$, one gets
\be
\lim_{\tilde{R}\rightarrow\infty}F(\tilde{R})=\tilde{R}-2\Lambda_i\, ,
\quad \text{where} \quad \Lambda_i=1/2f_1\, ,
\label{8.3.8}
\ee
where the subscript $i$ denotes that we are in the inflationary
epoch. By inserting this into Eqs.~(\ref{8.1.16}) and (\ref{8.1.18}), the
FRW equations take the same form as in (\ref{8.3.6}). Then, for the
function (\ref{8.3.7}) the inflationary epoch is produced by an
effective cosmological constant, which implies that the parameter
$\lambda>1/3$, or the equations themselves will present
inconsistencies, as it was discussed in the above section. For the
current epoch, it is easy to see that the function (\ref{8.3.7})
exhibits a minimum for $\tilde{R}=\tilde{R}_0$, which implies, as in
the model above, an effective cosmological constant for late time
that can produce the cosmic acceleration. The emergence of matter dominance
before the dark energy epoch can be exhibited, 
in analogy with the case of the convenient theory. Hence, we have 
shown that the model (\ref{8.3.7}) also unifies the cosmic expansion history,
although with different properties during the inflationary epoch as
compared with the model (\ref{8.3.2}). This could be very important
for the precise study of the evolution of the parameters of the theory.

It is also interesting to explore the de Sitter solutions allowed by
the theory (\ref{8.1.8}). By taking $H(t)=H_0$, the FRW equation
(\ref{8.1.18}), in absence of any kind of matter and with $C=0$,
reduces to
\be
0=F(\tilde{R}_0)-6H^2_0(1-3\lambda+3\mu)F'(\tilde{R}_0)\, ,
\label{8.3.9}
\ee
which reduces to an algebraic equation that, for an specific  model,
can be solved yielding the possible de Sitter points allowed by the
theory. As an example, let us consider the model (\ref{8.3.2}), where
Eq.~(\ref{8.3.9}) takes the form
\be
\tilde{R}_0+\frac{\tilde{R}_0^n(\alpha\tilde{R}_0^n-\beta)}{1
+\gamma\tilde{R}_0^n}
+\frac{6H_0^2(-1+3\lambda-3\mu)\left[1+n\alpha\gamma\tilde{R}_0^{3n-1}
+\tilde{R}_0^{n-1}(2\gamma\tilde{R}_0-n\beta)+\tilde{R}_0^{2n-1}(\gamma^2\tilde{R}_0
+2n\alpha)\right]}{(1+\gamma\tilde{R}_0^n)^2}=0\, .
\label{8.3.10}
\ee
Here $\tilde{R}_0=3(1-3\lambda+6\mu)H^2_0$. By specifying the free
parameters of the theory, one can solve Eq.~(\ref{8.3.10}), which
yields several de Sitter points, as the one studied above. They can
be used to explain the coincidence problem, with the argument that
the present will not be the only late-time accelerated epoch
experienced by our Universe. In standard $F(R)$, it was found for
this same model that it contains at least two de Sitter points along
the cosmic history. In the same way, the second
model studied here (\ref{8.3.7}), provides several de Sitter points in
the course of the cosmic history. Note that when $\mu=\lambda-\frac{1}{3}$,
Eq.~(\ref{8.3.9}) turns out to be much more simple, it reduces
to $F(\tilde{R}_0)=0$, where the de Sitter points are the roots.
For example, for (\ref{8.3.10}) we have
$\tilde{R}_0(1+\gamma\tilde{R}_0^n)+\tilde{R}_0^n(\alpha\tilde{R}_0^n-\beta)
=0$,
where the number of positive roots (de Sitter points) depends
on the free parameters of the theory.

Summing up, it has been here shown that, also in $F(\tilde{R})$
Ho\v{r}ava-Lifshitz gravity, the so-called viable models, as
(\ref{8.3.2}) or (\ref{8.3.7}), can in fact reproduce the whole
cosmological history of the universe, with no need to involve any
extra fields or a cosmological constant.

\section{Newton law corrections in $F(\tilde{R})$ gravity}

As is well-known, modified gravity may lead to
violations of  local tests. We explore in this
section how to avoid these violations of Newton's law. It is  known that
$F(\tilde{R})$ theories include scalar particle. This scalar field could
give rise
to a fifth force and to variations of the Newton law, which can be
avoided by a kind of the so-called chameleon mechanism \cite{Khoury:2003rn}.

In the original Ho\v{r}ava gravity, the projectability
condition may cause problems with the Newton law \cite{2010PhRvL.104r1302B} but the
model without
the projectability condition could be inconsistent for the
$F(R)$-model \cite{2010PhLB..693..404C}. The Newton law in the Ho\v{r}ava gravity
may be restored by the ``healthy'' 
extension \cite{2010PhRvL.104r1302B}.
In this section, we do not discuss the gravity sector corresponding to 
the Ho\v{r}ava gravity but we show
that the scalar mode, which also appears in the usual $F(R)$ gravity,
can decouple from gravity and matter, and then the scalar mode does not give
a measurable correction to Newton's law.

To show this, we consider a function of the type (\ref{8.3.1}), and  
rewrite action (\ref{8.1.8}) as 
\be
S=\int
dtd^3x\sqrt{g^{(3)}}N\left[(1+f'(A))(\tilde{R}-A)+A+f(A)\right]\, ,
\label{8.4.1}
\ee
where $A$ is an auxiliary scalar field. It is easy to see that
variation of  action (\ref{8.4.1}) over $A$ gives $A=\tilde{R}$.  
Performing the conformal transformation
$g^{(3)}_{ij}=\e^{-\phi}\tilde{g}^{(3)}_{ij}$, with
$\phi=\frac{2}{3}\ln(1+f'(A))$, action (\ref{8.4.1}) yields
\be
S=\int dtd^3x\sqrt{g^{(3)}}\left[\tilde{K}_{ij}\tilde{K}^{ij}-\lambda\tilde{
K}^2+\left(-\frac{1}{2}+\frac{3}{2}\lambda-\frac{3}{2}\mu\right)
\dot{\tilde{g}}^{ij(3)}\tilde{g}_{ij}^{(3)}\dot{\phi}
+\left(\frac{3}{4}-\frac{9}{4}\lambda+\frac{9}{2}\mu\right)\dot{\phi}^2
 -V(\phi)+\tilde{L}(\tilde{g}^{(3)},\phi)\right]\, ,
\label{8.4.2}
\ee
where $\tilde{L}(\tilde{g}^{(3)},\phi)$ is the conformally transformed
term in (\ref{8.1.9}), and the scalar potential is
\be
V(\phi)=\frac{A(\phi)f'(A(\phi))-f(A(\phi))}{1+f'(A(\phi))}\, .
\label{8.4.3}
\ee
Note that, differently from the convenient $F(R)$, in action
(\ref{8.4.2}) there is a coupling term between the scalar field $\phi$
and the spatial metric $\tilde{g}_{ij}^{(3)}$, which can thus be
dropped, by imposing the following condition on the parameters
\cite{2010PhRvD..82f5020C}:
\be
\mu=\lambda-\frac{1}{3}\, .
\label{8.4.4}
\ee
This condition also renders the theory power-counting renormalizable, 
for the same $z$ as in the original Ho\v{r}ava model.

Let us now investigate the term (\ref{8.1.9}). As already pointed out,
for local scales, where the scalar curvature is assumed to be very
small, the theory enters the IR limit, where such  term could be
written as a spatial curvature,
\be
L^{(3)}(g^{(3)},\phi)\sim R^{(3)}\, .
\label{8.4.5}
\ee
Then, corrections to the Newton law will come from the coupling that
now appears between the scalar field and matter, which makes a test
particle to deviate from its geodesic path, unless the mass of the
scalar field is large enough (since then the effect could be very small).
The precise value can be calculated from
\be
m^2_{\phi}=\frac{1}{2}\frac{d^2V(\phi)}{d\phi^2}
=\frac{1+f'(A)}{f''(A)}-\frac{A+f(A)}{1+f'(A)}\, .
\label{8.4.6}
\ee
In view of that, we can now analyze the models studied in the last section. We are
interested to see the behavior at local scales, as on Earth, where
the scalar curvature is around $A=\tilde{R}\sim 10^{-50} \mathrm{eV}^2$, or
in the solar system, where $A=\tilde{R}\sim 10^{-61} \mathrm{eV}^2$. The
function
(\ref{8.3.2}) and its derivatives can be approximated around these
points as
\be
f(\tilde{R})\sim-\frac{\beta}{\gamma}\, , \quad f'(\tilde{R})\sim
\frac{n\alpha}{\gamma}R^{n-1}\, , \quad
f''(\tilde{R})\sim\frac{n(n-1)\alpha}{\gamma}R^{n-2}\, .
\label{8.4.7}
\ee
Then, the scalar mass for the model (\ref{8.3.2}) is  given
approximately by the expression
\be
m^2_{\phi}\sim\frac{\gamma\tilde{R}^{2-n}}{n(n-1)\alpha}\, ,
\label{8.4.8}
\ee
which  becomes $m^2_{\phi}\sim10^{50n-100}\mathrm{eV}^2$ on Earth and
$m^2_{\phi}\sim10^{61n-122}$ in the Solar System. We thus see that,
for $n>2$, the scalar mass would be sufficiently large in order to
avoid corrections to the Newton law. Even for the limiting case
$n=2$, the parameters $\gamma/\alpha$ can be chosen to be large
enough so that any violation of the local tests is avoided.

For the model (\ref{8.3.7}), the situation is quite similar. For
simplicity, we impose the following condition
\be
f_0\ll f_1\left[(\tilde{R}-\tilde{R}_0)^{2n+1}+
\tilde{R}_0^{2n+1}\right]\sim f_1\tilde{R}^{2n+1}\, .
\label{8.4.9}
\ee
Then, function (\ref{8.3.7}) and its derivatives can be written,
for small values of the curvature, as
\be
f(\tilde{R})\sim -\frac{1}{f_1}+\frac{f_0}{f_1^2R^{2n+1}}, \quad
f'(\tilde{R})\sim-\frac{(2n+1)f_0}{f_1^2\tilde{R}^{2(n+1)}}, \quad
f''(\tilde{R})\sim\frac{2(2n+1)(2n+2)f_0}{f_1^2\tilde{R}^{2n+3}}\, .
\label{8.4.10}
\ee
Using now the expression for the scalar mass (\ref{8.4.6}), this yields
\be
m^2_{\phi}\sim\frac{f_1^2\tilde{R}^{2n+3}}{2(2n+1)(n+1)f_0}
+\frac{1}{f_1}\, .
\label{8.4.11}
\ee
Hence, as $\tilde{R}$ is very small at solar or Earth scales
($\sim10^{-50\sim-61}\mathrm{eV}^2$), and
$\Lambda_i=\frac{1}{2f_1}\sim10^{20\sim38}\mathrm{eV}^2$ is the effective
cosmological constant at inflation, which is much larger than the
other term, it turns out that the second term on the rhs of
(\ref{8.4.11}) will dominate, and the scalar mass will take a value such as
$m^2_{\phi}\sim\frac{1}{f_1}\sim10^{20\sim38}\mathrm{eV}^2$, which is large
enough as compared with the scalar curvature. As a consequence there
is no observable correction to Newton's law.

We have thus shown, in all detail, that the viable models (\ref{8.3.2}) and
(\ref{8.3.7}) do not introduce any observable correction to the Newton
law at small scales. This strongly supports the choice of $F(\tilde{R})$ 
gravity as a realistic candidate for the unified description of the 
cosmological history.

\section{Finite-time future singularities in  $F(\tilde{R})$ gravity}

It is a well-known that a good number of effective 
phantom/quintessence-like dark energy models end their evolution 
at a finite-time future singularity. In the current section, we
study the possible future evolution of the viable $F(R)$ Ho\v{r}ava-Lifshitz
gravity considered above. It has been already proven \cite{2010PhRvD..82f5020C} that power-law
$F(R)$ HL gravities may lead, in its evolution, to a finite-time singularity.
In order to properly define the type of future singularities, let us rewrite the FRW
Eqs.~(\ref{8.1.16}) and (\ref{8.1.18}) in the following way
\be
3H^2=\frac{\kappa^2}{3\lambda-1}\rho_\mathrm{eff}\, , \quad
-3H^2-2\dot{H}=\frac{\kappa^2}{3\lambda-1}p_\mathrm{eff}\, ,
\label{8.5.1}
\ee
where
\bea
&&
\rho_\mathrm{eff}=\frac{1}{\kappa^2F'(\tilde{R})}\left[-F(\tilde{R})
+3(1-3\lambda
+9\mu)H^2 F'(\tilde{R})-6\mu\dot{H}F'(\tilde{R})-6\mu H
\dot{\tilde{R}}F''(\tilde{R})+\kappa^2\rho_m\right]\, , \nn
&&
p_\mathrm{eff}=\frac{1}{\kappa^2F'(\tilde{R})}
\left[F(\tilde{R})-(6\mu\dot{H}
+3(1-3\lambda+9\mu)H^2)F'(\tilde{R})-2(1-3\lambda)
\dot{\tilde{R}}F''(\tilde{R}) \right. \nn
&& \left. \qquad \quad +2\mu\left[\dot{\tilde{R}}^2F^{(3)}(\tilde{R})
+\ddot{\tilde{R}}F''(\tilde{R})\right] +\kappa^2p_m\right]\, .
\label{8.5.2}
\eea
Then, using the expressions for the effective energy and pressure
densities just defined, the list of future singularities, as classified in
Ref.~\cite{2005PhRvD..71f3004N}, can be extended to $F(\tilde{R})$ gravity as follows:
\begin{itemize}
\item Type I (``Big Rip''): For $t\rightarrow t_s$, $a\rightarrow
\infty$ and $\rho_\mathrm{eff}\rightarrow \infty$, $|p|\rightarrow \infty$.
\item Type II (``Sudden''): For $t\rightarrow t_s$, $a\rightarrow
a_s$ and $\rho_\mathrm{eff}\rightarrow \rho_s$, $|p_\mathrm{eff}|\rightarrow
\infty$.
\item Type III: For $t\rightarrow t_s$, $a\rightarrow a_s$
and $\rho_\mathrm{eff}\rightarrow \infty$,
$|p_\mathrm{eff}|\rightarrow \infty$.
\item Type IV: For $t\rightarrow t_s$, $a\rightarrow a_s$ and
$\rho_\mathrm{eff}\rightarrow \rho_s$, $p_\mathrm{eff} \rightarrow p_s$
but higher derivatives of Hubble parameter diverge.
\end{itemize}

To illustrate the possibility of future singularities in viable
$F(\tilde{R})$ gravity, we
explore the model (\ref{8.3.2}), which has been studied in
Ref.~\cite{2010AIPC.1241.1094N} for the convenient $F(R)$ case, where it was shown
that this model is non-singular, in the
particular case $n=2$. Also, it is known that for most models of $F(R)$
gravity, the future singularity  can be cured by adding a term
proportional to $R^2$ (see Ref.~\cite{2005CQGra..22L..35A}) or a non-singular
modified gravity action \cite{2010AIPC.1241.1094N}. However, this is not
possible to do in the Ho\v{r}ava-Lifshitz gravity where, 
even for the simple model
(\ref{8.3.2}) with  $n=2$, a singularity can occur unless some 
restrictive conditions on the parameters are imposed.

In order to study the possible singularities that may occur from the above list,
we consider a Hubble parameter close to one of such singularities,
given by the following expression
\be
H(t)=\frac{h_0}{(t_s-t)^{q}}\, ,
\label{8.5.3}
\ee
where $h_0$ and $q$ are constant. Depending on the value of $q$, the
Hubble parameter (\ref{8.5.3}) gives rise to a particular type of
singularity. Thus,  for $q\geq 1$,
it gives a Big Rip singularity, for $-1<q<0$ a Sudden Singularity,
for $0<q<1$ a Type III singularity and for $q<-1$, it will produce a
Type IV singularity. Using the expression (\ref{8.1.15}) with
(\ref{8.5.3}), the scalar curvature can be approximated, depending on
the value of $q$, as
\be
\label{8.5.4}
R\sim \left\{ \begin{array}{lll}
\frac{3(1-3\lambda+6\mu)h_0^2}{(t_s-t)^{2q}} \quad & \mbox{for}
\quad & q>1 \\
\frac{3(1-3\lambda+6\mu)h_0^2+6\mu q h_0}{(t_s-t)^{2}} \quad &
\mbox{for} \quad & q=1 \\
\frac{6\mu q h_0}{(t_s-t)^{q+1}} \quad & \mbox{for} \quad & q<1
\end{array} \right. \, .
\ee
The model (\ref{8.3.2}) which, as has been shown, unifies
the inflationary and the dark energy epochs, makes the scalar
curvature grow with time so that, close to a possible future
singularity,  the model can be approximated as
$F(\tilde{R})\sim\tilde{R}^n$, where  $n>1$. For
the case $q>1$, it is possible to show, from the FRW 
Eqs.~(\ref{8.1.16}) and (\ref{8.1.18}), that the solution (\ref{8.5.3}) is
allowed for this model just in some special cases: (i) For $n=3/2$
and $\mu=2\lambda-\frac{2}{3}$, which contradicts the decoupling
condition (\ref{8.4.4}) and the Newtonian corrections (where it was found
that $n>2$). \
(ii) When  $\mu=0$ and $\lambda=1/3$, which holds (\ref{8.4.4})
and fixes the values of the parameters, in contradiction with the fact that
fluctuations are allowed for them.

For $q=1$, the Hubble parameter (\ref{8.5.3}) is a natural solution for
this model,  yielding
\be
H(t)=\frac{h_0}{t_s-t} \quad \mbox{with} \quad
h_0=\frac{2\mu(n-1)(2n-1)}{-1+3\lambda-6\mu+2n(1-3\lambda+3\mu)}\, .
\label{8.5.5}
\ee
This model allows for the possibility of occurrence of a Big Rip
singularity, unless the parameters of the theory are fixed. As
 pointed out in Ref.~\cite{2010AIPC.1241.1094N}, for the case $n=2$
the Big Rip singularity can be avoided in standard $F(R)$ gravity,
which can be easily seen by choosing $\lambda=\mu=1$ in the solution
(\ref{8.5.5}). Nevertheless, in $F(\tilde{R})$ gravity, in order to
avoid such singularity, the power $n$ of the model has to be fixed to
the value
\be
n=\frac{1}{2}+\frac{3\mu}{2-6\lambda+6\mu}\, ,
\label{8.5.6}
\ee
which can be interpreted as another constraint on the parameters of
the theory (although, when the decoupling condition (\ref{8.4.4}) is satisfied,
no constraints can be imposed on $n$). In addition, note that $h_0=-h_1$ 
in (\ref{8.3.4}), where
we imposed $h_1>1$ with the aim to reproduce the inflationary epoch,
so that $h_0$ would be negative and the solution (\ref{8.5.5}) would
correspond to the Big Bang singularity. Thus, no future doomsday
will take place.

For the last case, when $q<1$, we  find that the Hubble
parameter (\ref{8.5.3}) can be a consistent solution when Eq.~(\ref{8.4.4})
is satisfied
and $q=(n(1-n)-1)/n(n-2)$, although if we impose $n\geq 2$, as it was found
above, $q<-1$,
which implies a future singularity of the type IV. The exception here is
when $n=2$, that avoids the occurrence of any type of singularity when $q<1$.
Nevertheless, even in the case that, for any reason, some kind of
future singularity would be allowed in the model (\ref{8.3.2}), one must
also take into account possible quantum gravity effects, which may
probably become important when one is close to the singularity. They
have been shown, in phantom models, to prevent quite naturally the
occurrence of the future singularity \cite{2004PhRvD..70d3539E}.

In summary, we have here proven that the theory (\ref{8.3.2}) can
actually be free of future singularities, and thus that it can make a good
candidate for the unification of the cosmic history. Note that, from
the corrections to the Newton law studied in the previous section, as
well as from imposing avoidance of future singularities, we can
fix some of the parameters of the theory. The other constant
parameters that appear in (\ref{8.3.2})---which expresses the algebraic
relation between the powers of the scalar curvature---could be fixed
by comparing the cosmic evolution with the observed data, quite in
the same way as has been successfully done for standard $F(R)$ gravity in
Chapter 6.

\section{Discussion}

In summary, we have here investigated the FRW cosmology of a non-linear modified
Ho\v{r}ava-Lifshitz $F(R)$ gravity theory which has a viable convenient 
counterpart. We have proven that, for a special choice of the parameters, 
the FRW equations are just the same in both theories, and that the cosmic 
history of the first literally coincides with the one for the viable $F(R)$
gravity.
For a more general version of the theory, the unified description of the
early-time inflation and late-time acceleration is proven to be possible
too; however, the details of the cosmological dynamics are here different.
Moreover, corrections to Newton's law are negligible for an extensive region of
the parameter space. We have demonstrated the emergence of possible finite-time future
singularities, and their avoidance, by adding extra higher-derivative terms
which turn out to be qualitatively different, as compare with conventional
$F(R)$ cosmology.

Using the approach shown in Chapter 5, one can construct more
complicated generalizations of modified gravity with anisotropic scaling
properties. For instance, one can obtain non-local or modified Gauss-Bonnet
Ho\v{r}ava-Lifshitz gravities. Technically, this is of course a more involved
task, as compared with the $F(R)$ case. The corresponding cosmology can in 
principle be studied in the same way as in the present chapter, with expected 
qualitatively similar results, owing to the fact that the
convenient cosmological model has been already investigated, in those cases.
It is foreseen that the study of such theories will help us also in the resolution of
the some problems for the theories with broken Lorentz symmetry. For instance,
it is quite possible that a natural scenario of dynamical Lorentz symmetry
breaking, which would reduce gravity to the Ho\v{r}ava limit, may be found in this way, 
what would be certainly interesting.

\chapter[Stability of cosmological solutions in F(R) Ho\v{r}ava-Lifshitz gravity]{Stability of cosmological solutions in F(R) Ho\v{r}ava-Lifshitz gravity}

\footnote{This Chapter is based on: \cite{2010arXiv1011.2090S}}In the previous chapter, we showed how modifications of the original Ho\v{r}ava-Lifshitz action can unify well inflation and late-time acceleration providing a natural explanation on both effects.
In the present chapter, general cosmological solutions of the type of spatially flat FLRW will be studied in the frame of F(R) Ho\v{r}ava-Lifshitz gravity, and its stability is explored. We specially focus on the study of stability of radiation/matter dominated eras, where the Universe expands following a power law, and de Sitter solutions, which can well  shape the accelerated epochs of Universe evolution. We explore  space independent perturbations around these solutions, studying the effects of the extra terms introduced in the field equations by considering $F(\tilde{R})$ gravity. Also the new parameters included in the theory, due to the breaking of the Lorentz invariance, could affect to the Universe evolution. The (in)stability of a cosmological solution gives very important information as the possible exit from one phase of the cosmological history, constraints on the kind of action $F(R)$ and future observational predictions. An explicit example of a $F(\tilde{R})$ action, where an unstable de Sitter solution is found, is studied. This model performs a successful exit from inflation, and produces an instability at the end of matter dominated epoch, such that a phase transition may occur.

\section{Cosmological solutions and its stability in $F(\tilde{R})$ gravity}

In this section, we are interested to study the stability of general cosmological solutions in the frame of $F(\tilde{R})$ Ho\v{r}ava-Lifshitz gravity, with special attention to those cosmological solutions that model the history of the Universe, as de Sitter or power law solutions.  As it was shown in the previous chapter, dark energy and even the unification with the inflationary epoch can be reproduced in this new frame of $F(\tilde{R})$ Ho\v{r}ava-Lifshitz theories. The stability of those solutions plays a crucial role in order to get the transition from one cosmological phase to another.

\subsection{Stability of general flat FLRW cosmological solutions}

Let us start by studying  a general spatially flat FLRW metric.  We will focus below specially on de Sitter and power law solutions of the type $a(t)\propto t^{m}$ as dark energy epoch and the radiation/matter dominated eras are governed by this class of cosmological solutions respectively, so that the implications of the extra geometrical terms coming from $F(\tilde{R})$ could be determinant for the stability and transition during those epochs. We assume a general solution, 
\be
H(t)=h(t) \ .
\label{9.2.7a}
\ee
Then, the scalar curvature $\tilde{R}$ yields,
\be
\tilde{R}_h(t)=  3(1-3\lambda+6\mu)h^2(t)+6\mu \dot{h}(t)\ .
\label{9.2.7}
\ee
Assuming that solution (\ref{9.2.7a}) is satisfied by a particular choice of $F(\tilde{R})$,  the FLRW equation  has to be fulfilled, 
\be
0=F(\tilde{R_h})-6\left[(1-3\lambda
+3\mu)h^2+\mu\dot{h}\right]F'(\tilde{R_h})+6\mu h
\dot{\tilde{R}}_hF''(\tilde{R_h})-\kappa^2\rho_m\, ,
\label{9.2.8}
\ee
where the matter fluid is taken to be a perfect fluid with equation of state $p_m=w_m\rho_m$, with $w_m$ constant. By the energy conservation equation $\dot{\rho}_m+3h(1+w_m)\rho_m=0$, the evolution of the matter energy density can be expressed in terms of the given solution $h(t)$ as,
\be
\rho_{mh}=\rho_0 \e^{-3(1+w_m)\int h(t)dt}\ .
\label{9.2.9}
\ee
being $\rho_0$ an integration constant. We are interested to study the perturbations around the solution $h(t)$. For that purpose, let us expand the function $F(\tilde{R})$ in powers of $\tilde{R}$  around (\ref{9.2.7}), 
\be
 F(\tilde{R})=F_h+F'_h(\tilde{R}-\tilde{R}_h)+\frac{F''_h}{2}(\tilde{R}-\tilde{R}_h)^2+\frac{F^{(3)}_h}{6}(\tilde{R}-\tilde{R}_h)^3+O(\tilde{R}-\tilde{R}_h)^4\ ,
\label{9.2.10}
\ee
where the derivatives of the function $F(\tilde{R})$ are evaluated at $R_h$, given in (\ref{9.2.7}). Note that matter perturbations also contribute to the stability, inducing a mode on the perturbation. Then, we can write the perturbed solution  as,
\be
H(t)=h(t)+\delta(t)\ , \quad \rho_m\simeq\rho_{mh}(1+\delta_m(t))\ .
\label{9.2.11}
\ee
Hence, by introducing the above quantities in the FLRW equation,  the equation for the perturbation $\delta(t)$  becomes (in the linear approximation),
\be
\ddot{\delta}+b\dot{\delta}+\omega^2\delta=\frac{\kappa^2\rho_{mh}}{36\mu^2hF_h''}\delta_m\ ,
\label{9.2.12}
\ee
where,
\[
 b=-\frac{h'}{h}-\frac{1-3\lambda+3\mu}{\mu}h+\frac{1-3\lambda+6\mu}{\mu}+6\left((1-3\lambda+6\mu)h\dot{h}+\mu\ddot{h}\right)\frac{F^{(3)}_h}{F_h''}\ ,
\]
\[
\omega^2=\left[1-3\lambda+6\mu-2(1-3\lambda+3\mu)h\right]\frac{F_h'}{6\mu^2hF_h''}
\]
\be
+(1-3\lambda+6\mu)\left(\frac{-1+3\lambda-3\mu}{\mu^2}h+\frac{-1+h}{\mu h}\dot{h}\right)+\frac{\ddot{h}}{h}+6(1-3\lambda+6\mu)\left(\frac{1-3\lambda+6\mu}{\mu}\dot{h}h+\ddot{h}\right)\frac{F^{(3)}_h}{F_h''}\ .
\label{9.2.12a}
\ee
In this case, the solution for $\delta(t)$ can be split in two branches, one corresponding to the homogeneous part of the equation (\ref{9.2.12}), whose solution will depend on the background theory, i.e. on $F(\tilde{R})$ and its derivatives, and the other one corresponding to the particular solution of the eq.~(\ref{9.2.12}), which represent the solution induced by the matter perturbation, $\delta_m$. Then, the complete solution can be written as,
\be
\delta(t)=\delta_{homg}(t)+\delta_{inh}(t)\ .
\label{9.2.13}
\ee
We are interested on the perturbations induced  by the function $F(\tilde{R})$ and its derivatives, so that we focus on the homogeneous solution, $\delta_{homg}$. By a first qualitative analysis, we can see that the homogeneous part of equation (\ref{9.2.12}) yields exponential or damped oscillating perturbations.  The form of the perturbations will depend completely on the form of the function $F(\tilde{R})$ and its derivatives evaluated at $R_h$. Note that in general, the equation (\ref{9.2.12}) has to be solved by numerical methods.  Nevertheless, we  could assume some restrictions to obtain qualitative information. Let us consider the cases, 
\begin{itemize}

 \item The trivial case,  given by  $F_h'=F_h''=F_h^{(3)}=0$,  makes the perturbation tends to zero, $\delta(t)=0$, and the cosmological solution $h(t)$ is stable. 

\item For $F_h'\neq0$ and $F_h'',F_h^{(3)}\rightarrow0$, the term that dominates in (\ref{9.2.12}) is given by
\be
\omega^2\sim\left[1-3\lambda+6\mu-2(1-3\lambda+3\mu)h\right]\frac{F_h'}{6\mu^2hF_h''}\ .
\label{9.2.13b}
\ee
And the stability of the cosmological solution $h(t)$ depends on the sign of this term, and therefore, on the model $F(\tilde{R})$  and the solution $h(t)$.
\item For $F_h',F_h''\rightarrow0$ but $F_h^{(3)}\neq0$, looking at (\ref{9.2.12a}), the perturbation depends on the value of the last term in the coefficients $b$ and $\omega^2$, which can be approximated to,
\be
b\sim6\left((1-3\lambda+6\mu)h\dot{h}+\mu\ddot{h}\right)\frac{F^{(3)}_h}{F_h''}\ ,\quad \omega^2\sim 6(1-3\lambda+6\mu)\left(\frac{1-3\lambda+6\mu}{\mu}\dot{h}h+\ddot{h}\right)\frac{F^{(3)}_h}{F_h''}\ .
\label{9.2.13a}
\ee
The cosmological solution will be stable in the case that both  coefficients (\ref{9.2.13a}) are greater than zero, what yields a damped oscillating perturbation that decays.

\end{itemize}
However, in general the  equation (\ref{9.2.12}) can not be solved  analytically for arbitrary solutions $h(t)$ and actions $F(\tilde{R})$, and numerical analysis is required.  Nevertheless, by imposing certain conditions on $F(\tilde{R})$ as above,  qualitative information can be obtained. In order to perform a deeper analysis, some specific solutions $h(t)$ are studied below, as well as an explicit example of $F(\tilde{R})$.

\subsection{de Sitter solutions in $F(\tilde{R})$  gravity}

Let us consider one of the simplest but most important solution in cosmology, de Sitter (dS) solution. As dark energy and  inflation can be shaped (in its simplest form) by a dS solution, its stability becomes very important, specially in the case of inflation, where a successful exit is needed to end  the accelerated phase occurred during the  early Universe. In general,  standard $F(R)$ gravity contains several de Sitter points, which represent critical points (see \cite{2009PhRvD..79d4001C}). The analysis can be extended to  $F(\tilde{R})$ Ho\v{r}ava-Lifshitz gravity, where the de Sitter solution $H(t)=H_0$, with $H_0$ being a constant, has to satisfy the first equation FLRW equation,
\be
0=F(\tilde{R}_0)-6H^2_0(1-3\lambda+3\mu)F'(\tilde{R}_0)\, ,
\label{9.2.1}
\ee
where we have taken $C=0$ and  assumed absence of any kind of matter. The scalar $\tilde{R}$ is given in this case by,
\be
\tilde{R}_0=3(1-3\lambda+6\mu)H_0^2\ .
\label{9.2.2}
\ee
Then, the positive roots of equation (\ref{9.2.1}) are the de Sitter points allowed by a particular choice of $F(\tilde{R})$. By assuming a de Sitter solution, we expand $F(\tilde{R})$  as a series of powers of the scalar $\tilde{R}$ around $\tilde{R}_0$, 
\be
 F(\tilde{R})=F_0+F'_0(\tilde{R}-\tilde{R}_0)+\frac{F''_0}{2}(\tilde{R}-\tilde{R}_0)^2+\frac{F^{(3)}_0}{6}(\tilde{R}-\tilde{R}_0)^3+O(\tilde{R}^4)\ .
\label{9.2.3}
\ee
Here, the primes denote derivative respect $\tilde{R}$ while the subscript $0$ means that the function $F(\tilde{R})$ and its derivatives are evaluated at $\tilde{R}_0$. Then, by perturbing the solution,  the Hubble parameter can be writing as,
\be
H(t)=H_0+\delta(t)\ .
\label{9.2.4}
\ee
Using  (\ref{9.2.3}), and the perturbed solution (\ref{9.2.4}) in the first FLRW equation, the equation for the perturbation yields,
\[
0=\frac{1}{2}F_0-3H_0^2(1-3\lambda +3\mu)-3H_0\left[\left((1-3\lambda)F_0'+6F_0''H_0^2(-1+3\lambda-6\mu)(-1+3\lambda-3\mu)\right)\delta(t) \right.
\]
\be
\left.+6F_0''\mu H_0(-1+3\lambda-3\mu)\dot{\delta}(t)-12F_0''\mu^2\ddot{\delta}(t)\right]\ .
\label{9.2.5}
\ee
Here we have restricted the analysis to the linear approximation on $\delta$ and its derivatives. Note that the first two terms in the equation (\ref{9.2.5}) can be removed because of equation (\ref{9.2.1}), which is assumed to be satisfied, and equation (\ref{9.2.5}) can be rewritten in a more convenient form as,
\be
\ddot{\delta}(t)+\frac{H_0(1-3\lambda+9\mu)}{2\mu}\dot{\delta}(t)+\frac{1}{12\mu^2}\left[(3\lambda-1)\frac{F_0'}{F_0''}-6H_0^2(1-3\lambda+6\mu)(1-3\lambda+3\mu)\right]\delta(t)=0\ .
\label{9.2.6}
\ee
Then, the perturbations on a  dS solution will depend completely on the model, specifically on the derivatives of  $F(\tilde{R})$, as well as on the parameters  $(\lambda,\mu)$. The instability becomes large if the term in front of $\delta(t)$ (the frequency) in the equation (\ref{9.2.6}) becomes negative and the perturbation grows exponentially, while if we have a positive frequency, the perturbation  behaves as a damped harmonic oscillator. During  dark energy epoch, when the scalar curvature is very small,  the IR limit of the theory can be assumed, where  $\lambda=\mu\sim1$, and the frequency  depends completely on the value of $\frac{F_0'}{F_0''}$. In order to avoid large instabilities during the dark energy phase, the condition  $\frac{F_0'}{F_0''}>12H_0^2$ has to be fulfilled. Nevertheless, when the scalar curvature is large, the IR limit is not a convenient approach, and the perturbation depends also on the values of $(\lambda,\mu)$. If we assume a very small $F_0''$,  the frequency in the equation (\ref{9.2.6})  dominates compared to the other terms, and by assuming $\lambda>1/3$, the stability of the solution will depend on the sign of $\frac{F_0'}{F_0''}$, being stable when it is positive.

\subsection{Stability of radiation/matter eras: Power law solutions}

In this section, an important class of cosmological solutions is considered, the power law solutions, which are described by the Hubble parameter,
\be
H(t)=\frac{m}{t} \quad \rightarrow \quad a(t)\propto t^m\ .
\label{9.2.14}
\ee
In the context of General Relativity, this class of solutions are generated by a perfect fluid with equation of state parameter $w=-1+\frac{2}{3m}$, and the matter/radiation dominated epochs are approximately described by  (\ref{9.2.14}). Also phantom epochs can be described by this class of solutions when $m<0$ . Let us study the stability for the Hubble parameter (\ref{9.2.14}), and how the inclusion of extra terms in the action and the new parameters ($\lambda, \mu$) may affect the stability of the solution (\ref{9.2.14}).  As in the above section,  the perturbation equation (\ref{9.2.12}) can not  be solved analytically in general, although under some restrictions  we can obtain important qualitative information about the stability of the solution. Then, by assuming an $F(\tilde{R})$ that approximately does not deviate from Hilbert-Einstein action during radiation/matter dominated epochs, the  second and third derivatives can  be neglected $F''_h,F^{(3)}_h\sim0$  (as they must become important only during  dark energy epoch and/or inflation). In such a case, the coefficient in front of $\delta(t)$ in the eq.~(\ref{9.2.12}) is approximated as,
\be
\omega^2\sim\left[1-3\lambda+6\mu-2(1-3\lambda+3\mu)h(t)\right]\frac{F_h'}{6\mu^2h(t)F_h''}\ .
\label{9.2.15}
\ee
Then, the value of the frequency $\omega^2$ depends on the time, such that the stability may change along the phase. For small values of $t$, the frequency takes the form $\omega^2\sim-2(1-3\lambda+3\mu)\frac{6\mu^2F_h'}{F_h''}$,  and  assuming $\lambda\sim\mu$, the perturbations will grow exponentially when   $\frac{F_h'}{F_h''}>0$, and  the solution becomes unstable.  While for large $t$, the frequency can be approximated as $\omega^2\sim(1-3\lambda+6\mu)\frac{F_h'}{6\mu^2h(t)F_h''}$ and the instability will be large if $\frac{F_h'}{F_h''}<0$, and  a phase transition may occur.

\section{Example of a viable $F(\tilde{R})$ model}

Let us consider an explicit model of  $F(\tilde{R})$ gravity in order to apply the analysis about the stability performed above. We are interested to study  the stability of radiation/matter dominated eras as well as de Sitter solutions. The model considered here, a class of the so-called viable models, was proposed in Ref.~\cite{2008PhRvD..77b6007N}, and a variant was studied in Chapter 6 in the context of standard gravity and generalized to Ho\v{r}ava-Lifshitz gravity in Chapter 9. Let us write the action
\be
F(\tilde{R})=\chi\tilde{R}+\frac{\tilde{R}^n(\alpha\tilde{R}^n-\beta)}{1+\gamma\tilde{R}^n}\, ,
\label{9.3.1}
\ee 
where ($\chi, \alpha, \beta, \gamma$) is a set of constant parameters of the theory. In standard gravity, this model can reproduce well late-time acceleration with no need of a cosmological constant or any kind of exotic field, as well as also  inflation, such that the unification of both epochs of the Universe history under the same mechanism can be performed (see Ref.~\cite{2008PhRvD..77b6007N}). For simplicity,  we assume $n=2$ in (\ref{9.3.1}) for our analysis. The radiation/matter dominated epochs, which can be described by the class of solutions given in (\ref{9.2.14}),  could suffer a phase transition to the era of dark energy due to the instabilities caused by the second term of the action (\ref {9.3.1}). Then, we are interested to study the possible effects produced by the presence of these extra geometric terms during the cosmological evolution. By assuming the solution (\ref{9.2.14}), and following the steps described in the above section, the stability is affected by the derivatives of the function (\ref{9.3.1}) evaluated at $h(t)=m/t$. We are interesting in large times, when the end of matter dominated epoch has to occur. At that moment the derivatives of $F(\tilde{R})$ can be approximated as,
\be
F_h'\rightarrow \chi\ , \quad F_h''\rightarrow -2\beta\ , \quad F_h^{(3)}\rightarrow0\ .
\label{9.3.2}
\ee 
Here for simplicity, we have assumed  $0<\beta<<1$. By means of the analysis performed in the previous section, we can conclude that the linear perturbation $\delta(t)$  grows exponentially, and the radiation/matter dominated phase becomes unstable for large times, what may produce the transition to another different phase. Then, the  $F(\tilde{R})$ function (\ref{9.3.1}) can explain perfectly the end of matter dominated epoch with no need of the presence of a cosmological constant. \\

Let us now study the stability of de Sitter solutions. It is known that the above model (\ref{9.3.1}) may contain several de Sitter solutions, which are the solutions of the algebraic equation (\ref{9.2.1}),
\be
\tilde{R}_0+\frac{\tilde{R}_0^n(\alpha\tilde{R}_0^n-\beta)}{1
+\gamma\tilde{R}_0^n}
+\frac{6H_0^2(-1+3\lambda-3\mu)\left[1+n\alpha\gamma\tilde{R}_0^{3n-1}
+\tilde{R}_0^{n-1}(2\gamma\tilde{R}_0-n\beta)+\tilde{R}_0^{2n-1}(\gamma^2\tilde{R}_0
+2n\alpha)\right]}{(1+\gamma\tilde{R}_0^n)^2}=0\, .
\label{9.3.3}
\ee 
This  equation has to be solved numerically, even for the simple case studied here $n=2$. Nevertheless, one of the de Sitter points of this model is defined by  a minimum of the second term in the action (\ref{9.3.1}). By assuming the constraint on the parameters $\beta\gamma/\alpha\gg 1$,  the minimum that represents a de Sitter point is given by,
\be
\tilde{R}_0 \sim \left( \frac{\beta}{\alpha\gamma}\right)^{1/4}\, ,
\qquad F'(\tilde{R}_0)=\chi\, , \qquad  F(\tilde{R}_0)= \tilde{R}_0-2\Lambda\ , \quad \text{where} \quad \Lambda\sim
\frac{\beta}{2\gamma}\, .
\label{9.3.5}
\ee
Then, by evaluating the derivatives of (\ref{9.3.1}) around $\tilde{R}_0$ and by the equation (\ref{9.2.6}) the perturbation $\delta$ can be calculated. Note that the stability condition for  de Sitter solution,  given by $\frac{F_0'}{F_0''}>12H_0^2$, is not satisfied for this case as $F_0''>>F_0'$, such that the de Sitter point (\ref{9.3.5}) is  unstable. By solving eq.~(\ref{9.2.6}), the perturbation is given by exponential functions,
\be
\delta(t)=C_1\e^{a_+t}+C_2\e^{a_-t}\ , \quad \text{with} \quad a_{\pm}=\frac{H_0(1-3\lambda+3\mu)}{2\mu}\ .
\label{9.3.6}
\ee
Hence, the model (\ref{9.3.1}) is unstable around the de Sitter point (\ref{9.3.5}), what  predicts the exit from an accelerated phase in the near future, providing a natural explanation about the end of the inflationary epoch, or a future prediction about the end of dark energy era. However,  the theory described by (\ref{9.3.1}) may contain more de Sitter points, given by the roots of equation (\ref{9.3.3}), which may be stable. Then, a deeper analysis has to be performed to study the entire Universe evolution for this model of $F(\tilde{R})$ gravity.

\section{Discussions}
At the present chapter, we have analyzed spatially flat FLRW cosmology for  nonlinear Ho\v{r}ava-Lifshitz gravity. Basically we have  extended  standard $F(R)$ gravity to Ho\v{r}ava-Lifshitz theory, which reduces to the first one in the IR limit (where we assume that the parameters $(\lambda,\mu)$  are reduced to unity). The stability of this general class of solutions has been studied and it is shown that it depends mainly on the choice of the function $F(\tilde{R})$ and in part on the values of the parameters $(\lambda,\mu)$. For large times, when the scalar curvature  is very small,  the main effect of the perturbation on a cosmological solution is caused by the value of the derivatives of $F(\tilde{R})$. It is shown that in general, the perturbation equation can not be solved analytically, even in the linear approach. Nevertheless, under some restrictions, important information is obtained, and the (in)stability of the different phases of the Universe history can be studied. For specific values of the derivatives of $F(\tilde{R})$, a given solution can becomes (un)stable, which means a major constraint on models. By analyzing an explicit example in Sect. IV, where an $F(\tilde{R})$ function of the class of viable models is considered, we have found that this kind of theories can well explain the end of matter dominated epoch, and reproduces late-time acceleration. We have shown that for this specific example, there is a de Sitter point that becomes unstable, what  predicts the end of this de Sitter epoch, providing a  natural explanation of the end of inflationary era. However, as this model (and in general every $F(\tilde{R})$ model) may contain several de Sitter solutions, where some of them can be stable, a further analysis of the phase space  has to be performed to connect the different regions of the Universe history. \\
Hence, the analysis made here provides a general approach for the study of spatially flat FLRW solutions in the frame of higher order Ho\v{r}ava-Lifshitz gravities, which can restrict the class of functions $F(\tilde{R})$ allowed by the observations, and it gives a natural explanation of the end of inflation and matter dominated epoch, shaping the Universe history in a natural way.

\chapter[U(1) Invariant F(R) Ho\v{r}ava-Lifshitz Gravity]{U(1) Invariant F(R) Ho\v{r}ava-Lifshitz Gravity}

\footnote{This Chapter is based on: \cite{2010arXiv1012.0473K}}The HL gravity is based on an idea that the Lorentz symmetry is restored in IR
limit of given theory and can be absent at high energy regime of given theory.
Explicitly,  Ho\v{r}ava considered systems whose scaling at short
distances exhibits a strong anisotropy between space and time,
In $(D+1)$ dimensional space-time in order to have power counting
renormalizable theory  requires that $z\geq D$. It turns out however that
the symmetry group of given theory is reduced from the full diffeomorphism
invariance of General Relativity  to the foliation preserving diffeomorphism
The common  property of all modified theories of gravity is that whenever
the group of symmetries is restricted (as in Ho\v{r}ava-Lifshitz gravity)
one more degree of freedom appears that is a spin-0 graviton.
An existence of this mode could be dangerous for all these
theories (for review, see \cite{2008PhyU...51..759R}).
For example, in order to have the theory compatible
with observations one has to demand
that this scalar mode decouples in the IR regime.
Unfortunately, it seems that this might not be the case.
It was shown that the spin-0 mode is not stable in
the original version of the HL theory \cite{2009PhRvD..79h4008H}
as well as in  SVW generalization \cite{2009JHEP...10..033S}.
Note that in both of these two versions, it was all
assumed the projectability condition that means that the lapse function $N$
depends on $t$ only. This assumption has a fundamental consequence for the
formulation of the theory since there is no local form of the Hamiltonian
constraint but the only global one.
Even if these instabilities indicate to problems with the projectable version
of the HL theory  it turns out that this is not the end of the whole story.
Explicitly, these instabilities are all found around the Minkowski background.
Recently, it was indicated that the de Sitter space-time is stable in the SVW
setup \cite{2010MPLA...25.2267H} and hence it seems to be reasonable to consider de
Sitter background as the natural vacuum of projectable version of the HL gravity.
This may be especially important for the theories with unstable flat space
solution.

On the other hand there is the second version of the HL gravity where the
projectability condition is not imposed
so that $N=N(\textbf{x},t)$. Properties of
such theory were extensively studied in
\cite{2010arXiv1010.5531B,2009JHEP...10..029B,2010PhRvL.104r1302B,2010PhRvD..81f4002H,2010PhRvD..82f4011K,2009JHEP...08..015L,2010PhRvD..82d4011P}.
It was shown recently in \cite{2010arXiv1007.3503B} that  so called
healthy extended version of such theory could really be an interesting
candidate for the quantum theory of reality without ghosts and without
strong coupling problem despite its unusual Hamiltonian structure
\cite{2010PhRvD..82d4004K,2010JHEP...07..038K}.
Nevertheless, such theory is not free from its own internal problems.

Recently Ho\v{r}ava and Melby-Thompson \cite{2010PhRvD..82f4027H}
proposed very interesting way to eliminate the spin-0 graviton.
They considered the projectable version of the HL gravity together
with extension of the foliation preserving diffeomorphism
to include a local $U(1)$ symmetry.
The resulting theory is then called as non-relativistic  covariant theory of
gravity.
It was argued there \cite{2010PhRvD..82f4027H} that the presence
of this new symmetry forces the coupling constant $\lambda$ to be equal
to one.
However, this result was questioned in \cite{2010arXiv1009.4885D}
where an alternative formulation of non-relativistic general covariant
theory of gravity was presented.
Furthermore, it was shown 
that the presence of this new symmetry implies that the spin-0 graviton
becomes non-propagating and the spectrum of the linear fluctuations
around the background solution coincides with the fluctuation spectrum
of General Relativity.
This  construction was also extended to the case of RFDiff invariant
HL gravities \cite{2010arXiv1007.3503B,2010PhRvD..82l4011K} in \cite{2010arXiv1011.1857K}
where it was shown that the number of physical degrees of
freedom coincides with the number of physical degrees of freedom in General
Relativity.

The goal of this chapter is to extend above construction to the case of
$F(\tR)$ HL gravities. 
$F(\tR)$ HL gravity can be considered as natural
generalization of covariant $F(R)$ gravity as it was pointed out in the above chapters, where current interest to $F(R)$
gravity is caused by several important reasons.
Additionally, it is expected that such modification may
be helpful for resolution of internal inconsistency problems of the HL theory.
Indeed, we will present the example of $F(\tR)$ HL gravity which has
stable de Sitter solution but unstable flat space
solution. In such a case, the original scalar graviton problem formally
disappears because flat space is not vacuum state. Hence, there is no sense
to study propagators structure around flat space.
The complete propagators structure should be investigated around
de Sitter solution which seems to be the candidate for vacuum space.


\section{On flat space solutions in $F(\tilde{R})$ gravity}

Let us now study flat space solutions in $F(\tilde{R})$ HL  gravity. In this
section we restrict to the case of $3+1$ dimensional space-time.  A general
metric in the  ADM decomposition in a $3+1$ space-time  is given by,
\be
ds^2=-N^2 dt^2+g^{(3)}_{ij}(dx^i+N^idt)(dx^j+N^jdt)\, ,
\label{10.2.7}
\ee
where $i,j=1,2,3$, $N$ is the so-called lapse variable,
and $N^i$ is the shift $3$-vector.
For flat space  the variables from the metric (\ref{10.2.7}) take the values,
\be
N=1\, , \quad N_i=0 \quad \mbox{and} \quad g_{ij}=g_0\delta_{ij}\, ,
\label{10.2.8}
\ee
where $g_0$ is a constant. Then, the scalar curvature
$\tilde{R}=0$, and so that our theory has flat space solution, the function
$F(\tilde{R})$  has to satisfy,
\be
F(0)=0\, .
\label{10.2.9}
\ee
Hence, we assume  the condition (\ref{10.2.9}) is satisfied, otherwise the theory
has no flat space solution.
We are interested to study the stability of such solutions for a general
$F(\tilde{R})$, by perturbing the metric in vacuum (\ref{10.2.8}), this yields,
\be
N=1+\delta_N(t) \quad \mathrm{and} \quad
g_{ij}=g_0(1+\delta_g(t))\delta_{ij}\, .
\label{10.2.10}
\ee
For simplicity, we restrict the study on the time-dependent perturbations (no
spatial ones) and on the diagonal terms of the metric. Note that, as we are
assuming the projectability condition, by performing a transformation of
the time coordinate, we can always rewrite $N=1$.
Then, the study of the perturbations is focused on the spatial
components of the metric $ g_{ij}$, which can be written in a more
convenient way as,
\be
g_{ij}=g_0 \left(1+\int \delta(t)\right)\delta_{ij}
\sim g_0\e^{\int \delta(t)dt}\delta_{ij}\, .
\label{10.2.11}
\ee
By inserting (\ref{10.2.11}) in the first field equation, obtained by
the variation of the action on $N$, it yields at lowest order on $\delta(t)$,
\be
12\mu^2F_0''\ddot{\delta}(t)\delta(t)-6\mu^2F_0''\dot{\delta}^2(t)
+(-1+3 \lambda)F_0'\delta^2(t) =0\, .
\label{10.2.12}
\ee
Here the derivatives $F_0'$, $F_0''$ are evaluated on $\tilde{R}=0$.
Then, the perturbation $\delta$ will depend completely on the kind of theory
assumed. We can study some general cases by imposing conditions on the
derivatives of $F(\tilde{R})$.
\begin{itemize}
\item For the  case $F_0''=0$, it gives $\delta(t)=0$,
such that at lowest order the flat space is completely stable for this case.
\item For $F_0'=0$ and $F_0''\neq0$, the differential equation (\ref{10.2.12})
has the solution,
\be
\delta(t)=C_1t\left(1+\frac{C_1t}{C_2}\right)+C_2
\label{10.2.13}
\ee
where $C_{1,2}$ are integration constants.
Then, for this case, the perturbations grow as the power of the time coordinate, and flat
space becomes unstable.
\end{itemize}

Hence, depending on the theory, the flat space solution will be stable or
unstable, which becomes very important as it could be used to distinguish the
theories or analyze their consistency.

\subsection{A simple example}

Let us now discuss a simple example.
We consider the function,
\be
F(\tilde{R})=\kappa_0\tilde{R}+\kappa_1\tilde{R}^n\, ,
\label{10.2.14}
\ee
where $\kappa_{0,1}$ are coupling constants and $n>1$.
Note that this family of theories satisfies the condition (\ref{10.2.9}).
The values of first and second derivatives evaluated in the solution depend on
the value of $n$ in (\ref{10.2.14}),
\be
F'(0)=\kappa_0\, , \quad
F''(0)=\kappa_1n(n-1)\tilde{R}^{n-2}\, .
\label{10.2.15}
\ee
Then, we can distinguish between the cases,
\begin{itemize}
\item For $n\neq2$, we have $F''(0)=0$,
and by the analysis performed above, it follows that flat space is stable
\item For $n=2$, we have $F''(0)=\kappa_1$, and flat space is unstable.
\end{itemize}
Hence, we have shown  that the stability of solution
depends completely on the details of the theory.

We can now analyze the de Sitter solution.
Using the the analysis performed in the above chapter, we can find the de Sitter
points allowed by the class of theories given in Eq.~(\ref{10.2.14}),
\be
\frac{3}{2}H_0^2(3\lambda-1)\kappa_0
+\frac{\kappa_1\left(3H_0^2(1-3\lambda+6\mu)\right)^n(1-3\lambda
+6\mu-2n(1-3\lambda+3\mu))}{2(1-3\lambda+6\mu)}=0\, .
\label{10.2.16}
\ee
Resolving the Eq.~(\ref{10.2.16}),  de Sitter solutions are obtained.
For simplicity, let us consider  $n=2$, in such a case the
equation $(\ref{10.2.16})$ has two roots for $H_0$ given by,
\be
H_0= \pm
\frac{\sqrt{\kappa_0(3\lambda-1)}}{3\sqrt{\kappa_1(1-3\lambda+6\mu)(-1+3\lambda-2\mu)}}
\, .
\label{10.2.17}
\ee
As we are interested in de Sitter points, we just consider the positive root in
(\ref{10.2.17}). Then, the stability of such de Sitter point can be analyzed by
studying the derivatives of the function $F(\tilde{R})$ evaluated in $H_0$.
The stability will depend on the value of $\frac{F_0'}{F_0''}$, which
for this case yields,
\be
\frac{F_0'}{F_0''}=\frac{\kappa_0}{2\kappa_1} +12H_0^2\, .
\label{10.2.18}
\ee
In the IR limit of the theory ($\lambda\rightarrow1$,
$\mu\rightarrow1$), the condition for the stability of de Sitter points
$\frac{F_0'}{F_0''}>12H_0^2$  is clearly satisfied by (\ref{10.2.18}).
Even in the non IR limit, $F_0''=2\kappa_1$
and assuming $\kappa_1\ll 1$, we have that,
\be
\frac{F_0'}{F_0''}=3H_0^2(1-3\lambda+6\mu)
+\frac{\kappa_0}{2\kappa_1}\, .
\label{10.2.19}
\ee
As this term is positive, we have that the instabilities will oscillate and be
damped, such that the de Sitter point becomes stable.

Thus, we presented the example of the $F(\tilde{R})$ theory where flat space
is unstable solution and de Sitter space is stable solution. The problem of
scalar graviton does not appear in this theory because one has to analyze
the spectrum of theory around de Sitter space which is real vacuum.
Indeed, flat space is not stable and cannot be considered as the vacuum
solution.
Of course, deeper analysis of de Sitter spectrum structure of the theory is
necessary.
Nevertheless, as we see already standard $F(\tilde{R})$ gravity suggests
the way to resolve the pathologies which are well-known in
Ho\v{r}ava-Lifshitz gravity.


\section{$U(1)$ invariant $F(\tilde{R})$ Ho\v{r}ava-Lifshitz gravity \label{10.fourth}}

Our goal is to see whether it is possible to extend the gauge symmetries
for above action as in \cite{2010PhRvD..82f4027H}.
As the first step we introduce two non-dynamical fields $A,B$ and
rewrite the action for $F(\tilde{R})$ into the form
\begin{equation}
S_{F(\tilde{R})}= \frac{1}{\kappa^2}\int dt d^D\bx
\sqrt{g}N (B(\tilde{R}-A)+F(A)) \, .
\end{equation}
It is easy to see that solving the equation of motion with respect to
$A,B$ this action reduces into the original one. On the other hand
when we perform integration by parts we obtain the action in the form
\begin{eqnarray}
\label{10.SFtR}
S_{F(\tilde{R})} =\frac{1}{\kappa^2}\int dt d^D\bx \left( \sqrt{g}N B(
K_{ij}\mG^{ijkl}K_{kl} -\mV(g)-A) 
\nonumber \right. \\
\left. +\sqrt{g}N F(A) -2\mu \sqrt{g}N\nabla_n B
K  + 2\mu \partial_i B \sqrt{g}g^{ij}
\partial_j N \right) \, , 
\end{eqnarray}
where we ignored the boundary terms and where
\begin{equation}
\nabla_n B=\frac{1}{N} (\partial_t B-N^i\partial_i B) \, .
\end{equation}
Let us now introduce $U(1)$ symmetry where the shift function transforms as
\begin{equation}
\label{10.deltaNi}
\delta_\alpha N_i(\bx,t)= N(\bx,t)\nabla_i\alpha(\bx,t) \, .
\end{equation}
It is important to stress that as opposite to the case of pure
Ho\v{r}ava-Lifshitz gravity the kinetic term is multiplied with $B$ that is
space-time dependent and hence it is not possible to perform similar
analysis as in \cite{2010PhRvD..82f4027H}.
This procedure frequently uses the integration by parts and the fact that
covariant derivative annihilates metric tensor together with the crucial
assumption that $N$ depends on time only. Now due to the presence of $B$
field we have to  proceed step by step with the construction of the action
invariant under (\ref{10.deltaNi}).
As the first step note that under (\ref{10.deltaNi}) the kinetic term
$S^\mathrm{kin}=\frac{1}{\kappa^2}\int dt d^D\bx \sqrt{g}K_{ij}\mG^{ijkl}K_{kl}$
transforms as
\[
\delta_\alpha S^\mathrm{kin} =-\frac{2}{\kappa^2}\int dt
d^D\bx\sqrt{g}NB K_{ij}\mG^{ijkl}\nabla_i \nabla_j
\alpha \, .
\]
In order to compensate this variation of the action we introduce new
scalar field $\nu$ that under (\ref{10.deltaNi}) transforms as
\begin{equation}
\label{10.deltaNinu}
\delta_\alpha \nu(t,\bx)=\alpha(t,\bx)
\end{equation}
and add to the action following term
\begin{equation}
S_\nu^{(1)}= \frac{2}{\kappa^2}\int dt d^D\bx\sqrt{g}NB
K_{ij}\mG^{ijkl}\nabla_i \nabla_j \nu \, .
\nonumber 
\end{equation}
Note that under (\ref{10.deltaNinu}) this term transforms as
\[
\delta_\alpha S_\nu^{(1)}= \frac{2}{\kappa^2}\int dt
d^D\bx\sqrt{g}NB K_{ij}\mG^{ijkl}\nabla_k \nabla_l \alpha 
 - \frac{2}{\kappa^2}\int dt d^D\bx\sqrt{g}NB \nabla_i\nabla_j
\alpha \mG^{ijkl}\nabla_k \nabla_l \nu \, .
\]
so that we add the second term into the action
\begin{equation}
S_{\nu}^{(2)}= \frac{1}{\kappa^2}\int dt d^D\bx \sqrt{g}N B
\nabla_i\nabla_j\nu \mG^{ijkl} \nabla_k\nabla_j \nu \, .
\end{equation}
As a result, we find that $S^\mathrm{kin}+S_\nu^{(1)}+ S_\nu^{(2)}$ is
invariant under (\ref{10.deltaNi}) and (\ref{10.deltaNinu}).

As the next step we analyze the variation of the $B$-kinetic part of
the action $S^{B\mathrm{kin}}=-\frac{2\mu}{\kappa^2}
\int dt d^D\bx  \sqrt{g} N \nabla_n B K$ under the variation (\ref{10.deltaNi})
\be
\label{10.alphaBn}
\delta_\alpha S^{B\mathrm{kin}}
=\frac{2\mu}{\kappa^2} \int dt d^D\bx
\sqrt{g}\alpha \nabla^i(\nabla_i B
K) +\frac{2\mu}{\kappa^2} \int dt d^D\bx
\sqrt{g}N \alpha\nabla_i\nabla_j(
g^{ij}\nabla_n B) \ .
\ee
We see that in order to cancel this variation it is appropriate to add
following expression into the action
\begin{eqnarray}
S^{\nu-B}&=& -\frac{2}{\kappa^2}\mu \int dt d^D\bx
\sqrt{g}N\nu\nabla^i(\nabla_i B K) -\frac{2}{\kappa^2}\mu\int dt d^D\bx
\sqrt{g}N \nu\nabla_i\nabla_j[ g^{ij}\nabla_n B]
\nonumber \\
&& +\frac{2}{\kappa^2}\mu \int dt d^D\bx \sqrt{g}N \nabla^k \nu \nabla_k
B \nabla_i\nabla^i\nu \, . \nonumber 
\end{eqnarray}
Then it is easy to see that $S^{B\mathrm{kin}}+S^{\nu-B}$ is invariant under
(\ref{10.deltaNi}) and (\ref{10.deltaNinu}).
Collecting all these results we find following $F(\hat{R})$ HL action that
is invariant under (\ref{10.deltaNi}) and (\ref{10.deltaNinu})
\begin{eqnarray}
\label{10.SFtR2}
S_{F(\tilde{R})} &=&\frac{1}{\kappa^2}\int dt d^D\bx \left( \sqrt{g}N B(
K_{ij}\mG^{ijkl}K_{kl} -\mV(g)-A)  \nonumber \right. \\
&& \left. +\sqrt{g}N F(A) -2\mu \sqrt{g}N\nabla_n B K + 2\mu
\partial_i B \sqrt{g}g^{ij} \partial_j N \right) \nonumber \\
&& - 2\mu \int d^D\bx dt \sqrt{g}\nu\nabla^i(\nabla_i B K)
-2\mu\int d^D\bx dt \sqrt{g}N \nu\nabla_i\nabla_j(
g^{ij}\nabla_n B)  \nonumber \\
&& + 2\mu \int dt d^D\bx \sqrt{g} N \nabla^i \nu \nabla_i B \nabla^
j\nabla_j\nu \nonumber \\
&& + 2\int d^D\bx\sqrt{g}NB K_{ij}\mG^{ijkl}\nabla_k \nabla_l \nu
+\int d^D\bx \sqrt{g}B \nabla_i\nabla_j\nu \mG^{ijkl}
\nabla_k\nabla_l \nu \, .
\end{eqnarray}
Note that we can write this action in suggestive form
\begin{eqnarray}
\label{10.SFtR3}
S_{F(\tilde{R})} &=& \frac{1}{\kappa^2}\int dt d^D\bx \left( \sqrt{g}N B(
(K_{ij}+\nabla_i\nabla_j\nu)\mG^{ijkl}(K_{kl}+\nabla_k\nabla_l \nu)
-\mV(g)-A) \nonumber \right. \\
&& \left. +\sqrt{g}N F(A) -2\mu
\sqrt{g}N(\nabla_n B+\nabla^i \nu\nabla_i B)
g^{ij}(K_{ji}+\nabla_j\nabla_i\nu) + 2\mu
\partial_i B \sqrt{g}g^{ij} \partial_j N \right)
\nonumber \\
\end{eqnarray}
or in even more suggestive form by introducing
\begin{equation}
\hN_i=N_i-N\nabla_i \nu  \, ,
\quad \hK_{ij}=\frac{1}{2N}(\partial_t g_{ij}
 -\nabla_i \hN_j-\nabla_j \hN_i)
\end{equation}
so that
\begin{eqnarray}
\label{10.SFtRfinal}
S_{F(\tilde{R})} &=& \frac{1}{\kappa^2}\int dt d^D\bx \left( \sqrt{g}N B(
\hK_{ij}\mG^{ijkl}\hK_{kl} -\mV(g)-A) \nonumber \right. \\
&& \left. +\sqrt{g}N F(A) -2\mu \sqrt{g}N\hnabla_n B
g^{ij}\hK_{ji}  + 2\mu \partial_i B \sqrt{g}g^{ij}
\partial_j N \right)
\end{eqnarray}
is formally the same as the $F(\tR)$ Ho\v{r}ava-Lifshitz gravity action.
Clearly this action is invariant under arbitrary $\alpha=\alpha(t,\bx)$.
Moreover, such an introduction of $U(1)$ symmetry is trivial and does not
modify the physical properties of the theory.
This is nicely  seen from the fact that $\nu$ appears in the action in the
combination with $N_i$ through $\bar{N}_i$ where $\nu$ plays the role of
the St\"{u}ckelberg field. In order to get physical content of given symmetry
we follow \cite{2010PhRvD..82f4027H} and \cite{2010arXiv1009.4885D}
and introduce following  term into action
\begin{equation}
\label{10.Snuk}
S^{\nu,k} = \frac{1}{\kappa^2}\int dt d^D\bx  \sqrt{g}
B\mG(g_{ij})(\mA-a) \, ,
\end{equation}
where
\begin{equation}
a=\dot{\nu}-N^i\nabla_i\nu + \frac{N}{2}
\nabla^i \nabla_i \nu \, ,
\end{equation}
where $\dot{X}\equiv \frac{dX}{dt}$.
Note $a$ transforms under $\alpha$ variation as
\[
a'(t,\bx) = a(t,\bx)+\dot{\alpha}(t,\bx)-N^i(t,\bx)
\nabla_i\alpha(t,\bx) \, .  \nonumber 
\]
Then it is natural to suppose that
$\mA$ transforms under $\alpha$-variation as
\begin{equation}
\mA'(t,\bx)=\mA(t,\bx)+\dot{\alpha}(t,\bx)
 - N^i(t,\bx)\nabla_i \alpha(t,\bx)
\end{equation}
so that we immediately see that $\mA-a$ is invariant under $\alpha$-variation.
The function $\mG$ can generally depend on arbitrary combinations of metric $g$
and matter field and we only demand that it should
be invariant under foliation preserving diffeomorphism
 and under (\ref{10.deltaNi}) and (\ref{10.deltaNinu}).
For our purposes it is, however, sufficient to restrict ourselves to the models
where $\mG$ depends on the spatial curvature $R$ only.
Now one observes that the equation of motion for $\mA$ implies the constraint
\begin{equation}
B\mG(R)=0
\end{equation}
that for non-zero $B$ implies the condition $\mG(R)=0$.
Note that this condition is crucial for elimination of the
scalar graviton when we study fluctuations around flat background. We demonstrate
this important result in the next section.

Finally note that it is possible to integrate out $B$ and $A$ fields from the
actions (\ref{10.SFtRfinal}) and (\ref{10.Snuk}) that leads to
\begin{eqnarray}
\label{10.SFbarRf}
S_{F(\bar{R})}&=& \frac{1}{\kappa^2} \int dt d^D\bx \sqrt{g}N
F(\bar{R}) \, , \nonumber \\
\bar{R}&=&\hK_{ij}\mG^{ijkl}\hK_{kl}-\mV(g)+
\frac{2\mu}{\sqrt{g} N} \left\{ \partial_t \left( \sqrt{g} \bar K \right)
 - \partial_i \left( \sqrt{g} N^i \bar{K} \right) \right\}+
\frac{1}{N}\mG(\mA-a) \,  .
\nonumber \\
\end{eqnarray}
This finishes the construction of $U(1)$ invariant  $F(\bar{R})$ HL
theory action.

\subsection{Lagrangian for the scalar field}

Now we extend above analysis to the action for the matter field with the
following general form of the scalar field action
\begin{equation}
\label{10.Smatt}
S_\mathrm{matt}=-\int dt d^D\bx \sqrt{g}N X \, ,
\end{equation}
where
\begin{equation}
X=-(\nabla_n\phi)^2 + F(g^{ij}\partial_i\phi\partial_j\phi) \, .
\end{equation}
where $F(x)=X+\sum_{n=2}^z X^n$ and where we defined
\begin{equation}
\nabla_n\phi=\frac{1}{N}(\partial_t\phi-N^i \partial_i \phi) \, .
\end{equation}
Note that this general  form of the scalar field action is consistent with
the anisotropy of target space-time as was shown in \cite{2010JHEP...02..068C,2009NuPhB.821..467K,2010JHEP...01..122K,2010arXiv1001.0490M,2009arXiv0910.0411K,2010PhRvD..81f5013R}.

Now we try to extend above action in order to make it invariant under (\ref{10.deltaNi}).
Note that under this variation the scalar field  action (\ref{10.Smatt}) transforms as
\be
\label{10.deltaSmatt}
\delta_\alpha S_\mathrm{matt}= -2\int dt d^D\bx \sqrt{g}N
\nabla^i\alpha \nabla_i\phi \nabla_n\phi
\ee
using
\begin{equation}
\label{10.deltaX} \delta_\alpha X=
2\nabla^i\alpha\nabla_i\phi \nabla_n\phi \, .
\end{equation}
We compensate the variation (\ref{10.deltaSmatt}) by introducing
additional term into action
\be
\label{10.Smattnu}
S_{\mathrm{matt-}\nu} = -2\int dt d^D\bx \sqrt{g}\nu N \nabla^i
(\nabla_i\phi\nabla_n\phi)
+ \int dt d^D\bx \sqrt{g}
\nabla^i \nu \nabla^j \nu \nabla_i\phi \nabla_j\phi 
\ee
Then the action (\ref{10.Smattnu}) transforms under
(\ref{10.deltaNi}) and (\ref{10.deltaNinu}) as
\begin{equation}
\delta_\alpha S_{\mathrm{matt-}\nu}=
2\int dt d^D\bx  \sqrt{g}N
\nabla^i \alpha \nabla_i\phi\nabla_n\phi
\end{equation}
that compensates the variation (\ref{10.deltaSmatt}).

In the same way one can analyze more general form of
the scalar action
\begin{equation}
\label{10.Smatgen}
S_\mathrm{matt}=-\int dt d^D\bx \sqrt{g}N K(X) \, .
\end{equation}
In order to find the generalization of the action (\ref{10.Smatgen})
which is invariant under (\ref{10.deltaNi}) we introduce two
auxiliary fields $C$, $D$ and write the action (\ref{10.Smatgen}) as
\begin{equation}
\label{10.Smattex}
S_\mathrm{matt}=-\int dt d^D\bx \sqrt{g}N [K(C)+D(X-C)] \, .
\end{equation}
Clearly this action transforms under (\ref{10.deltaNi}) as
\be
\label{10.deltaSmattG}
\delta_\alpha S_\mathrm{matt}= -\int dt d^D\bx
\sqrt{g}N D \delta_\alpha X= -2\int  dt d^D\bx
\sqrt{g}N D \nabla^i\alpha \nabla_i\phi \nabla_n\phi \, ,
\ee
where relation (\ref{10.deltaX}) is used. It is easy to see that the
variation of the  following term
\be
\label{10.SmattnuG}
S_{\mathrm{matt}-\nu} = 2\int dt d^D\bx \sqrt{g} N D
\nabla^i\nu \nabla_i\phi\nabla_n\phi
+ 
\int dt d^D\bx  \sqrt{g} D \nabla^i \nu \nabla^j \nu \nabla_i\phi
\nabla_j\phi 
\ee
compensates the variation (\ref{10.deltaSmattG}).
Finally note that (\ref{10.Smattex}) together
with (\ref{10.SmattnuG}) can be written in more elegant form
\be
\label{10.Smattin}
S_\mathrm{matt} = -\int  dt d^D\bx  \sqrt{g}N
[K(C)+D(\bar{X}-C)]= -\int dt d^D\bx
\sqrt{g}N K(\bar{X}) \, , 
\ee
where
\begin{eqnarray}
\bar{X}&=&-(\bar{\nabla}_n\phi)^2+
F(g^{ij}\partial_i\phi\partial_j\phi) \, , \nonumber \\
\bar{\nabla_n}\phi &=& \frac{1}{N}
(\partial_t \phi-\bar{N}^ i\nabla_i \phi)=\frac{1}{N}
(\partial_t \phi-N^i\nabla_i + N\nabla^i\nu\nabla_i \phi) \, .
\nonumber 
\end{eqnarray}
In this section we constructed $U(1)$-invariant scalar field action in
the form which closely follows the original construction presented in
\cite{2010PhRvD..82f4027H}.
In the next section  more elegant approach to the construction
of $U(1)$ invariant $F(\bar{R})$ HL gravity and the scalar
field action is given.


\section[Fluctuations around flat background in $U(1)$ invariant $F(\tR)$]{Study of fluctuations around flat background in $U(1)$ invariant $F(\tR)$ HL
Gravity \label{10.sixth}}

Let us analyze the spectrum of fluctuations in case of $U(1)$
invariant $F(\tR)$ HL gravity for the special case $\mu=0$.
For simplicity we assume that $F(\bar{R})$ Ho\v{r}ava-Lifshitz gravity has flat
space-time as its solution with the background values of the fields
\begin{equation}
g^{(0)}_{ij}=\delta_{ij} \, , \quad N^{(0)}=1 \, , \quad N_i^{(0)}=0 \, ,
\quad  \mA^{(0)}=0 \, , \quad \nu^{(0)}=0 \, .
\end{equation}
Note that for the flat background the equation of motion for $B$ and $A$ takes
the form
\begin{equation}
\label{10.eqABflat}
\mV(g^{(0)})-A^{(0)}=0 \, , \quad B^{(0)}-F'(A^{(0)})=0
\end{equation}
that implies that $A^{(0)}$, $B^{(0)}$ are constants. In order to find the
spectrum of fluctuations we expand all fields  up to linear order around this
background
\begin{eqnarray}
&& g_{ij}=\delta_{ij}+\kappa h_{ij} \, , \quad
N_i=\kappa n_i \, , \quad  N=1+\kappa n \,  , \nonumber \\
&& A=A^{(0)}+\kappa a \, , \quad B=B^{(0)}+\kappa b \, , \quad
\mA=\mA^{(0)}+\kappa\tilde{\mA} \, ,
\quad \nu=\nu^{(0)}+ \kappa\tilde{\nu} \, . \nonumber 
\end{eqnarray}
Since $n$ depends on $t$ only, its equation of motion gives one integral
constraint. This constraint does not affect the number of local degrees of
freedom.
For that reason it is natural  to consider the equation of motion for
$h_{ij}$, $n_i$ and $\nu$ only.
We further decompose the field $h_{ij}$ and $n_i$
into their irreducible components
\begin{equation}
h_{ij}=s_{ij}+\partial_i w_j+\partial_j
w_i +\left(\partial_i\partial_j -\frac{1}{D}\delta_{ij}
\partial^2 \right)M+\frac{1}{D}\delta_{ij}h \, ,
\end{equation}
where the scalar $h=h_{ii}$ is the trace part of $h_{ij}$ while $s_{ij}$
is symmetric, traceless and transverse
\begin{equation}
\partial^i s_{ij}=0  \, , \partial^i= \delta^{ij}\partial_j \
\end{equation}
and $w_i$ is transverse
\begin{equation}
\partial^i w_i=0 \, .
\end{equation}
In the same way we decompose $n_i$
\begin{equation}
n_i=u_i+\partial_i C
\end{equation}
with $u_i$ transverse $\partial^i u_i=0$.
In what follows we fix the spatial diffeomorphism symmetry by fixing the gauge
\begin{equation}
w_i=0 \, , \quad  M=0 \, .
\end{equation}
We begin with the equation of motion for $N_i$
\begin{equation}\label{10.eqNi}
2\nabla_j (B\mG^{ijkl}\hK_{kl}) +B\mG(R)\nabla_i\nu=0 \, .
\end{equation}
and for $\nu$
\[
2\nabla_i\nabla_j (BN\mG^{ijkl}\hK_{kl}) + \frac{1}{\sqrt{g}}
\left(\frac{d}{dt}(\sqrt{g}B \mG)  -\nabla_i (\sqrt{g} BN^i \mG)
 -\frac{1}{2}\nabla_i\nabla^i (\sqrt{g}B N\mG)\right)=0 \, .
\]
Combining these equations and using
the equation of motion for $\mA$ one obtains
\begin{equation}
\label{10.eqmA}
\mG(R)=0
\end{equation}
Then
\begin{equation}
\label{10.nu2} B\frac{d}{dt}\mG - BN^i\nabla_i \mG
-\frac{N}{2}(B\nabla_i\nabla^i \mG + 2 \nabla_i B\nabla^i \mG) =0 \ .
\end{equation}
Further, in the linearized approximation the equation of motion
(\ref{10.eqNi}) takes the form
\be
\label{10.eqNilin}
 -2B_0(1-\lambda)\partial^i (\partial_k\partial^k C)
+B_0\partial_k\partial^k u^i+\frac{1}{D} (1-D\lambda)\partial^i
\dot{h}+ 2(1-\lambda)B_0\partial^i (\partial_j\partial^j \tilde{\nu}) =0
\ee
using the fact that $\mG(R_0)=0$ and also
$\partial^i s_{ij}=\partial^j s_{ij}=0$ and $\delta^{ij}s_{ji}=0$.
Let us now focus on the solution of the
constraint $\mG(R)=0$ in the linearized approximation.
Let $R^{(D)}_0$ is the solution of the
equation of motion and let us consider
the perturbation around this equation.
These perturbations have to obey the equation
\[
\mG(R_0+\delta R) = \mG(R_0)+\frac{d\mG}{dR}\delta R
= \frac{d\mG}{dR}(R_0)\delta R=0 \, .
\]
We see that in order to eliminate the scalar graviton we have to demand that
$\frac{d\mG}{dR}(R_0)\neq 0$.
To proceed further  note that
\begin{equation}
\delta R_{ij}= \frac{1}{2}[\nabla^{(0)}_i\nabla^{(0)k}
h_{jk}+ \nabla_j^{(0)}\nabla^{k(0)}h_{ik} -\nabla_k^{(0)}\nabla^{k(0)}h_{ij}
 -\nabla^{(0)}_i\nabla^{(0)}_j h]
\end{equation}
where $\nabla^{(0)}$ is the covariant derivative calculated using the
background metric $g_{ij}^{(0)}$.
Then in the flat background it follows
\[
\delta R = \frac{\kappa}{D}(1-D)\partial^k\partial_k h 
\]
so that the condition $\delta R=0$ implies $\partial_k\partial^k h=0$.
Then $h=h(t)$. However, in this case the fluctuation mode does not obey the
boundary conditions that we implicitly assumed. Explicitly we demand that all
fluctuations vanish at spatial infinity.
For that reason one should demand that $h=0$.
Then it is easy to see that the equation (\ref{10.nu2}) is trivially solved.
In the linearized approximation the equation of motion for $g_{ij}$ takes the form
\begin{eqnarray}
\label{10.eqgij}
&& \frac{1}{2}(\ddot{s}_{ij}+\frac{1}{D}\delta_{ij}(1-\lambda D) \ddot{h}-\partial_i
\dot{u}_j-2\partial_i\partial_j \dot{C}+ 2\partial^i\partial^j \tilde{\nu}-2\lambda
\delta^{ij}\partial_k\partial^k \tilde{\nu} +2\lambda \delta_{ij}\partial_k\partial^k
\dot{C}) \nonumber \\
&& - \frac{\delta \mathcal{V}_2}{ g_{ij}} +\frac{d\mG}{dR}[
\partial^i\partial^j  (\tilde{\mA}- \tilde{a}) +\delta^{ij}\partial_k\partial^k(\tilde{\mA}
 - \tilde{a})] =0
\end{eqnarray}
where
\begin{equation}
\tilde{a}=\frac{d\tilde{\nu}}{dt} + \frac{1}{2}\partial_k\partial^k \tilde{\nu} \, .
\end{equation}
Note that we have not fixed the $U(1)$ gauge symmetry yet. It turns out that
it is natural to fix it as
\begin{equation}
\nu=0 \, .
\end{equation}
Then the trace of the equation (\ref{10.eqgij}) is equal to
\be
\label{10.traceeqg}
(1-\lambda D)\partial_k\partial^k \dot{C} + \delta^{ij}\frac{\delta \mathcal{V}_2}
{\delta g_{ij}} -\frac{d\mG}{dR}(1-D)\partial_i\partial^i
\tilde{\mA} =0 \, .  
\ee
Let us again consider the equation of motion (\ref{10.eqNilin}) and take its
$\partial^i$.
Then using the fact that  $\partial^i u_i=0$ one gets the
condition $C=f(t)$ that again with suitable boundary conditions implies
$C=0$. However then inserting this result in (\ref{10.eqNilin}) we find
\begin{equation}
\partial_k\partial^k u_i=0
\end{equation}
that also implies $u_i=0$.

To proceed further we now assume that $\mV_2=-R$ so that
$\frac{\delta \mV_2}{\delta g_{ij}}=-\frac{1}{2}\partial_k
\partial^k s_{ij}-\frac{D-2}{2D} (\partial_i\partial_j -\delta_{ij}\partial^k)h$.
Clearly the trace of this equation is proportional
to $h$ and hence it implies following equation for $\tilde{a}$
\begin{equation}
\partial_k\partial^k \tilde{\mA}=0 \, .
\end{equation}
Imposing again the requirement that $\tilde{\mA}$ vanishes at spatial
infinity we find that the only solution of given equation is $\tilde{\mA}=0$.
Finally the equation of motion for $g_{ij}$ gives following result
\begin{equation}
\ddot{s}_{ij}+\partial_k\partial^k s_{ij}=0 \, .
\end{equation}
In other words, it is demonstrated that under assumption that $U(1)$ invariant
$F(\bar{R})$ HL gravity has flat space-time as its
solution it follows that the perturbative spectrum contains the
transverse polarization of the graviton only.
Clearly, this result may be generalized for general version of
theory with arbitrary parameter $\mu$.

Finally we consider the linearized equations of motion for $A$ and $B$. In
case of $A$ one gets
\begin{equation}
\label{10.eqa}
 -b+F''(A_0)a=0
\end{equation}
while in case of $B$ we obtain
\begin{equation}
\label{10.eqb}
 -\frac{d\mathcal{V}}{dR}(R_0) \delta R-a=0
\end{equation}
Using the fact that $\delta R\sim h=0$ we get from (\ref{10.eqb}) and from (\ref{10.eqa})
\begin{equation}
a=b=0 \, .
\end{equation}
In other words there are no fluctuations corresponding to the scalar fields $A$ and $B$.
One can compare this situation with the conventional $F(R)$ gravity where the
mathematical equivalence of the theory with the Brans-Dicke theory implies the
existence of propagating scalar degrees of freedom.
In our case, however, the fact that $U(1)$ invariant $F(\bar{R})$
HL gravity is invariant under the foliation preserving diffeomorphism
allows us to consider theory without kinetic term for $B$ ($\mu=0$).

Thus, it seems U(1) extension of $F(R)$ HL gravity may lead to solution of
the problem of scalar graviton.


\section{Cosmological Solutions of
$U(1)$ Invariant $F(\tR)$ HL gravity \label{10.seventh}}

Let us investigate the flat FLRW cosmological
solutions for the theory described by
action for $F(\tilde{R})$. Spatially-flat
FLRW metric is now assumed,  $ds^2=-N^2dt^2+a^2(t)\sum_{i=1}^3
\left(dx^{i}\right)^2$, and we
choose the gauge fixing condition for
the local $U_\Sigma (1)$ symmetry as
follows, \be \label{10.U(1)14} \nu = 0\, .
\ee In the flat FLRW metric,
one gets $N_i=0$ and $\bar{R}$ and
$\bar{K}_{ij}$ only depend on the
cosmological time $t$. Therefore, the
constraint equation  can
be satisfied trivially. The FLRW equations 
are identical with the corresponding equations in the $F(\tilde{R})$-gravity  in absence of $U(1)$ symmetry given by (\ref{8.1.16}) and (\ref{8.1.18}) . Hence, $U(1)$ extension does not influence
the FLRW cosmological dynamics.
Let us consider the theory which admits a de Sitter universe solution. We now
neglect the matter contribution by putting $p_m=\rho_m=0$.
Then by assuming $H=H_0$, first FLRW equation (\ref{8.1.18})  gives
\be
\label{10.HLF18}
0 = F\left( 3 \left(1 - 3 \lambda + 6 \mu \right) H_0^2 \right)
 - 6 \left(1 - 3\lambda + 3\mu\right) H_0^2 F'\left( 3 \left(1
 - 3 \lambda + 6 \mu \right) H_0^2 \right) \, ,
\ee
as long as the integration constant vanishes ($C=0$) in (\ref{8.1.18}).
We now consider the following model:
\be
\label{10.HLF22}
F\left(\bar R\right) \propto \bar R + \beta \bar R^2
+ \gamma \bar R^3\, .
\ee
Then Eq. (\ref{10.HLF18}) becomes
\bea
\label{10.HLF23}
0 &=& H_0^2 \left\{ 1 - 3\lambda + 9\beta \left(1 - 3\lambda
+ 6 \mu \right) \left( 1 - 3\lambda + 2\mu \right) H_0^2 \right. \nn
&& \left. + 9\gamma \left(1 - 3\lambda + 6 \mu \right)^2
\left( 5 - 15\lambda + 12 \mu \right) H_0^4 \right\}\, ,
\eea
which has the following two non-trivial solutions,
\be
\label{10.HLF24}
H_0^2 = - \frac{ \left( 1 - 3\lambda + 2\mu
\right) \beta }{2 \left(1 - 3\lambda + 6 \mu \right) \left( 5 - 15\lambda
+ 12 \mu \right) \gamma} \left( 1 \pm \sqrt{
1 - \frac{4 \left(1 - 3\lambda \right)\left( 5 - 15\lambda + 12 \mu
\right) \gamma} { 9 \left( 1 - 3\lambda
+ 2\mu \right)^2 \beta^2} } \right)\, ,
\ee
as long as the r.h.s. is real and positive. If
\be
\label{10.HLF25}
\left| \frac{4 \left(1 - 3\lambda \right)\left( 5 - 15\lambda + 12 \mu
\right) \gamma} { 9 \left( 1 - 3\lambda + 2\mu \right)2 \beta^2} \right|
\ll 1\, ,
\ee
one of the two solutions is much smaller than the other solution.
Then one may regard that the larger solution corresponds to the inflation
in the early universe and the smaller one to the late-time acceleration.

More examples of $F\left(\bar R\right)$ theory which can contain more
than one dS solution, such that inflation and dark energy epochs can be
explained under the same mechanism,
as it was pointed in the above chapters, may be considered. First of all, as
generalization of the model (\ref{10.HLF22}), a general polynomial
function may be discussed
\be
F\left(\bar R\right)=\sum_{n=1}^m \alpha_n\bar R^n\, ,
\label{10.D1}
\ee
Here ${\alpha_n}$ are coupling constants. Using the equation
(\ref{10.HLF18}), it yields the algebraic equation,
\be
0=\sum_{n=1}^m \alpha_n\bar R^n_0-2\frac{1-3\lambda+3\mu}{1-3\lambda+6\mu}\bar
R_0\sum_{n=1}^m n\alpha_n\bar R^{n-1}_0\, .
\label{10.D2}
\ee
By a qualitative analysis, one can see that the number of positive real roots,
i.e., of the de Sitter points, depends completely on the sign of the coupling
constants ${\alpha_n}$.
Then, by a proper choice, $F\left(\bar R\right)$
gravity can well explain dark energy and inflationary epochs in a unified
natural way. Even it could predict the existence of more than two accelerated
epochs, which could resolve the coincidence problem.

Let us now consider an explicit example
\be
F\left(\bar R\right)=\frac{\bar R}{\bar R(\alpha \bar
R^{n-1}+\beta)+\gamma}\, ,
\label{10.D3}
\ee
where ${\alpha, \beta, \gamma, n}$ are  constants. By introducing this
function in (\ref{10.HLF18}), it is straightforward to show that for the
function (\ref{10.D3}), there are several de Sitter solutions. In order to
simplify this example, let us consider the case  $n=2$, where the equation
(\ref{10.HLF18}) yields,
\be
\gamma-3\gamma\lambda-3\beta
H_0^2(1-3\lambda+6\mu)^2+27\alpha
H_0^4(-1+3\lambda-4\mu)(1-3\lambda+6\mu)^2=0
\, .
\label{10.D4}
\ee
The solutions are given by
\bea
&& H_0^2 = \frac{1}{18\alpha(1-3\lambda+6\mu)^2(-1+3\lambda-4\mu)}
\left\{ \beta(1-3\lambda+6\mu)^2 \right. \nn
&& \left. \quad \pm\sqrt{(1-3\lambda+6\mu)^2
\left[12\alpha\gamma(-1+3\lambda)(-1+3\lambda-4\mu)
+\beta^2(1-3\lambda+6\mu)^2\right]} \right\} \, .
\label{10.D5}
\eea
Then, by a proper choice of the free
parameters of the model, two positive roots of the equation
(\ref{10.D4}) are solutions. Hence, such a model can explain
inflationary and dark energy epochs in unified manner.

\section{Discussion \label{10.eighth}}

In summary, in this chapter we aimed to resolve (at least, partially) the
inconsistency problems of the projectable HL gravity.
First of all, it is demonstrated that some versions of $F(R)$ HL gravity may
have stable de Sitter solution and unstable flat space solution.
As a result, the spectrum analysis showing the presence of scalar graviton
is not applied. The whole spectrum analysis should be redone for de Sitter
background.

Second, $U(1)$ extension of $F(R)$ HL gravity is formulated.
The analysis of fluctuations of $U(1)$ invariant $F(\bar{R})$ HL
gravity is performed.
It is shown that like in case of $U(1)$ HL gravity the
scalar graviton ghost does not emerge.
This opens good perspectives for consistency of such class of models.
It is also interesting that spatially-flat FLRW equations for $U(1)$ invariant
$F(\bar{R})$ gravity turn out to be just the same as for the one without
$U(1)$ symmetry. This indicates that all (spatially-flat FLRW) cosmological
predictions of viable conventional $F(R)$ gravity are just the same as for
its HL counterpart (with special parameters choice).

\part{Other aspects of FLRW Universes}
\chapter[Cardy-Verlinde formula and entropy bounds near future singularities]{Generalizing Cardy-Verlinde formula and entropy bounds near future singularities}

\footnote{This Chapter is based on: \cite{2010EPJC...69..563B}}In the above chapters,  several alternatives to explain the dark energy epoch have been analised, and even the inflationary phase by means mainly of modifications of gravity, which can be modelled as effective fluids. The study of phantom-like or quintessence-like effective dark fluids
opens for a number of new phenomena which are typical for such a DE
universe.
For instance, it is known that phantom DE may drive the future
universe to a so-called finite-time Big Rip singularity (for earlier
works on this, see Refs.\cite{2004astro.ph.12619A,2005PhRvD..72f4017A,2005CQGra..22..143B,2006GReGr..38.1609B,2005hep.th....5186C,2002PhLB..545...23C,2003PhRvL..91g1301C,2003PhRvL..91u1301C,2004MPLA...19.2479C,2006AnPhy.321..771D,2006PhRvD..74d4022D,2004PhRvD..70d3539E,2005PhRvD..71j3504E,2002IJMPD..11..471F,2005PhLB..623...10G,2004PhLB..586....1G,2004hep.th....8225G,2005PhRvD..72l5011G,2005PhLB..606....7H,2005PhRvD..71h4011L,2002JHEP...08..029M,2003PhLB..562..147N,2003PhLB..571....1N,2004MPLA...19.1509S,2005PhLB..624..147S,2004PhLB..586....5S,2006MPLA...21..231Z}). From another
side, quintessence-like DE may  bring the future universe to a
milder future singularity (like the sudden singularity
\cite{2006PhRvD..73j1301B,2008JCAP...10..045B,2010EPJC...67..295B,2004CQGra..21L..79B,2005CQGra..22.1563B,2008PhLB..659....1B,2008IJMPD..17.2269B,2010JCAP...04..016B,2009PhRvD..79l4007C,2005CQGra..22.4913C,2005JGP....55..306C,2004PhRvD..70l1503F,2006PhRvD..74f4030F,2008PhRvD..77l3513K,2004PhLB..595....1N,2004PhRvD..70j3522N,2008PhRvD..78d6006N,2006PhRvD..74d3514S,2005PhRvD..71h4024S,2006CQGra..23.3259T} where the effective energy-density
is finite).
Actually, the study of Ref.\cite{2005PhRvD..71f3004N} shows that there
are four different types of future finite-time singularities where the
Type I singularity corresponds to the Big Rip, the sudden singularity is of
Type II, etc. The universe looks quite strange near to the
singularity where curvature may grow up so that
quantum gravity effects may be dominant \cite{2004PhRvD..70d3539E}.
In any case, the study of the universe under critical conditions (for
instance, near a future singularity) may clarify the number of
fundamental issues relating seemingly different physical
theories.

Some time ago \cite{2000hep.th....8140V} it was shown that the first FLRW equation for a
closed FLRW universe may have a more fundamental origin than what is
expected from standard General Relativity. It was demonstrated that
this equation may be rewritten so as to describe the universe entropy in terms
of total energy and Casimir energy (the so-called Cardy-Verlinde (CV)
formula). Moreover, it turns out that the corresponding formula has a
striking correspondence with the Cardy formula for the entropy
of a two-dimensional conformal field theory (2d CFT).
Finally, the formula may be rewritten as a dynamical entropy bound
from which a number of  entropy bounds, proposed earlier,
follow. The connection between the  standard gravitational equation and the 2d CFT
 dynamical entropy bound indicates a very deep relation
between gravity and thermodynamics. It raises  the question about
to which extent  the CV formula is universal. Problems of this
sort are natural to study when the universe is under critical
conditions, such as near a singularity.

The present chapter is devoted to a study of the universality of the
CV formula and the corresponding dynamical entropy bound in a DE
universe filled with a generalized fluid, especially near the
singularity regime. Generalization of the CV formula for a
multicomponent fluid with interactions, assuming the  EoS to be
inhomogeneous, is presented.  It
is shown that the standard CV formula with correct power (square
root) is restored only for some very special cases. The dynamical
entropy bound for such fluids near  all the four types of the
future singularity is considered. It is demonstrated that this
dynamical entropy bound is most likely violated near the
singularity (except from some cases of Type II and Type IV
singularity). This situation is not qualitatively changed even if
account is taken  of quantum effects in conformally invariant
theory. Using the formalism of modified gravity to describe an
effective dark fluid, the corresponding CV formula is constructed
also for $F(R)$-gravity. The corresponding dynamical entropy bound
is derived. It is shown that the bound is satisfied for a  de
Sitter universe solution. Further discussion and outlook  is given
in the discussion section.

\section[Generalization of CV formula in FLRW Universe]{Generalization of Cardy-Verlinde formula in FLRW Universe for various types of fluids}

This section is devoted to consideration of the Cardy-Verlinde
formula for more general scenarios than those considered in
previous works (see \cite{2000hep.th....8140V,2002PhLB..531..276Y}). We consider here a
$(n+1)$-dimensional spacetime described by the FLRW metric, written
in comoving coordinates as 
\be ds^2=-dt^2+\frac{a(t)^2
dr^2}{1-kr^2}+r^2 d\Omega^2_{n-1}\ , \label{11.1} \ee 
where $k=-1, 0, +1$  for an open, flat, or closed spatial
Universe, $a(t)$ is taken to have units of lenght, and $d\Omega^2_{n-1}$ is the metric of an $n-1$ sphere. By inserting the
metric (\ref{11.1}) in the Einstein field equations the FLRW equations
are derived, 
\be H^2 = \frac{16\pi
G}{n(n-1)}\sum^m_{i=1}\rho_i-\frac{k}{a^2}\, ,\quad \dot{H} =
-\frac{8\pi G}{n-1}\sum^m_{i=1}(\rho_i + p_i) + \frac{k}{a^2}\ .
\label{11.3} \ee 
Here $\rho_i={E_i}/{V}$ and $p_i$ are the
energy-density and pressure of the matter component $i$ that fills
the Universe. In this chapter we consider only the $k=1$ closed
Universe. Moreover, we assume an equation of state (EoS) of the
form $p_i=w_i\rho_i$ with $w_i$ constant for each fluid, and
assume at first no interaction between the different components.
Then, the conservation law for energy has the form \be
\dot{\rho_i}+nH \left(\rho_i+p_i\right)=0\ , \label{11.4} \ee and by
solving (\ref{11.4}), we find that the $i$ fluid depends on the scale
factor as \be \rho_i\propto a^{-n(1+w_i)}\ . \label{11.5} \ee Let us
now review the case of Ref.\cite{2002PhLB..531..276Y}, where just one fluid with
EoS $p=w\rho$ and $w= \text{constant}$ is considered. The total
energy inside the comoving volume $V$, $E=\rho V$, can be written
as the sum of an extensive part $E_\mathrm{E}$ and a subextensive
part $E_\mathrm{C}$, called the Casimir energy, and  takes the
form: \be E(S,V)=E_\mathrm{E}(S,V)+\frac{1}{2}E_\mathrm{C}(S,V)\ .
\label{11.6} \ee Under a rescaling of the entropy
($S\rightarrow\lambda S$) and the volume ($V\rightarrow\lambda
V$), the extensive and subextensive parts of the total energy
transform as \be E_\mathrm{E}(\lambda S,\lambda V) = \lambda
E_\mathrm{E}(S,V)\ , \quad E_\mathrm{C}(\lambda S,\lambda V) =
\lambda^{1-2/n} E_\mathrm{C}(S,V)\ . \label{11.7} \ee Hence, by
assuming that the Universe satisfies the first law of
thermodynamics, the term corresponding to the Casimir energy
$E_\mathrm{C}$ can be seen as a violation of the Euler identity
according to the definition in Ref.\cite{2000hep.th....8140V}: \be
E_\mathrm{C}=n(E+pV-TS) \ . \label{11.8} \ee Since the total energy
behaves as $E\sim a^{-nw}$ and by the definition (\ref{11.6}), the
Casimir energy also goes as $E_\mathrm{C}\sim a^{-nw}$. The FLRW
Universe expands adiabatically, $dS=0$, so the products
$E_\mathrm{C}a^{nw}$ and $E_\mathrm{E}a^{nw}$ should be
independent of the volume $V$, and be just a function of the
entropy. Then, by the rescaling properties (\ref{11.7}), the
extensive and subextensive part of the total energy can be written
as functions of the entropy only \cite{2002PhLB..531..276Y}, \be E_\mathrm{E}=
\frac{\alpha}{4\pi a^{nw}}S^{w+1}\ , \quad
E_\mathrm{C}=\frac{\beta}{2\pi a^{nw}}S^{w+1-2/n}\ . \label{11.9} \ee
Here $\alpha$ and $\beta$ are undetermined constants. By combining
these expressions with (\ref{11.6}), the entropy of the Universe is
written as a function of the total energy $E$ and the Casimir
energy $E_\mathrm{C}$, \cite{2002PhLB..531..276Y}, \be S=\left(\frac{2\pi
a^{nw}}{\sqrt{\alpha\beta}}\sqrt{E_\mathrm{C}(2E-E_\mathrm{C})}
\right)^{\frac{n}{n(w+1)-1}}\ , \label{11.10} \ee which for $w=1/n$
(radiation-like fluid) reduces to \cite{2000hep.th....8140V}, \be
S=\frac{2\pi
a}{\sqrt{\alpha\beta}}\sqrt{E_\mathrm{C}(2E-E_\mathrm{C})}\ ,
\label{11.19} \ee which has the same form as the Cardy formula given
in Ref.\cite{1986NuPhysB..Cardy}. The first FLRW equation (\ref{11.3}) can be
rewritten as a relation between thermodynamics variables, and
yields \be
S_\mathrm{H}=\frac{2\pi}{n}a\sqrt{E_\mathrm{BH}(2E-E_\mathrm{BH})}\
, \quad \mbox{where} \quad S_\mathrm{H}=(n-1)\frac{HV}{4G}\, ,
\quad E_\mathrm{BH}=n(n-1)\frac{V}{8\pi Ga^2}\ . \label{11.20} \ee It
is easy to check that for the bound proposed in
Ref.\cite{2000hep.th....8140V}, $E_\mathrm{C}\leq E_\mathrm{BH}$, the
equation for the entropy (\ref{11.19}) coincides with the first FLRW
equation (\ref{11.20}) when the bound is reached. We will see below
that when there are several fluid components, the same kind of
expression as in Ref.\cite{2000hep.th....8140V} cannot be found. Nor is there
the same correspondence with the FLRW equation when the bound is
saturated.

\subsection*{Multicomponent Universe}

If $m$ fluids are considered with arbitrary EoS, $p_i=w_i\rho_i$,
the expression for the total entropy is simple to derive just by
following the same method as above. The total entropy is given by
the sum of the entropies for each fluid, \be S=\sum_{i=1}^m S_i =
\sum_{i=1}^m \left(\frac{2\pi
a^{nw_i}}{\sqrt{\alpha\beta}}\sqrt{E_{iC}(2E_{i}-E_{iC})}
\right)^{\frac{n}{n(w_i+1)-1}}\ . \label{11.11} \ee This expression
cannot be reduced to one depending only on the total energy unless
very special conditions on the nature of the fluids are assumed.
Let us   for simplicity  assume that  there are
only two fluids with EoS given by $p_1=w_1\rho_1$ and
$p_2=w_2\rho_2$, $w_1$ and $w_2$ being constants. We can
substitute the fluids by an effective fluid described by the EoS
\be p_\mathrm{eff}=w_\mathrm{eff}\rho_\mathrm{eff}\ , \quad
\text{where} \quad w_\mathrm{eff}=
\frac{p_1+p_2}{\rho_1+\rho_2}=w_1+\frac{w_2-w_1}{1+\rho_1/\rho_2}\
, \label{11.12} \ee and $p_\mathrm{eff}=\frac{1}{2}(p_1+p_2)$,
$\rho_\mathrm{eff}=\frac{1}{2}(\rho_1+\rho_2)$. Then, by using the
energy conservation equation (\ref{11.4}), we find $\rho_1\sim
(a/a_0)^{-n(1+w_1)}$ and $\rho_2\sim (a/a_0)^{-n(1+w_2)}$, where $a_0$ is assumed to be the value of the scale factor at the time $t_0$. The effective
EoS parameter $w_\mathrm{eff}$ can be expressed as a function of
the scale factor $a(t)$ \be
w_\mathrm{eff}=w_1+\frac{w_2-w_1}{1+(a/a_0)^{n(w_2-w_1)}}\ . \label{11.13}
\ee The total energy inside a volume $V$ becomes \be E_T=E_1+E_2
\propto (a/a_0)^{-nw_1}+(a/a_0)^{-nw_2}\ . \label{11.16} \ee As the
energy is proportional to two different powers of the scale factor
$a$, it is not possible to write it as a function of
the total entropy only. As a special case, if the EoS parameters are $w_1=w_2=w_\mathrm{eff}$, the formula for the entropy
reduces to (\ref{11.10}), and coincides with the CV formula  when $w_\mathrm{eff}=1/n$. \\

As another  case, one might consider that for some epoch of the cosmic history,  $w_1 \gg w_2$. Taking also   $a>>a_0$, we could then approximate  the total energy  by the
function $E_T \propto a^{-nw_2}$. From  (\ref{11.6}) the
Casimir energy would also depend on the same power of $a$,
$E_\mathrm{C}\propto a^{-nw_2}$. The expression (\ref{11.10}) is again recovered with $w=w_2$. \\

Thus in general, when a multicomponent FLRW Universe is assumed, the formula for the total entropy does
not resemble the Cardy formula, nor does it correspond to the FLRW
equation when the Casimir bound is reached.  It becomes possible to
reconstruct the formula (\ref{11.19}), and establish the correspondence with the
Cardy formula, only if we make specific choices for the EoS of the fluids.

\subsection*{Interacting fluids}

As a second case we now consider a Universe, described by the metric (\ref{11.1}),
 filled with two interacting fluids. One can write the
energy conservation equation for each fluid as
\be
\dot{\rho_1}+n H(\rho_1+p_1)= Q\ , \quad
\dot{\rho_2}+n H(\rho_2+p_2)= -Q\ ,
\label{11.22}
\ee
where $Q$ is a function that accounts for the
energy exchange between the fluids. This kind of interaction has
been discussed previously in studies of dark energy and dark
matter. The effective EoS parameter is given by the same expression
(\ref{11.12}) as before. With a
specific choice for the coupling function $Q$, the equations
(\ref{11.22}) may be solved. One can in principle find the
dependence of the energy densities $\rho_{1,2}$ on the scale
factor $a$,
\be
\rho_1= a(t)^{-n(1+w_1)}\left(C_1+\int
a^{n(1+w_1)}Q(t)dt \right) \ , \quad
\rho_2= a(t)^{-n(1+w_2)}\left(C_2-\int a^{n(1+w_2)}Q(t)dt \right)\ ,
\label{11.24}
\ee
where $C_1$ and $C_2$ are integration constants.
In general it is not possible to reproduce the CV
formula, and the result will be a sum of  different
contributions, similar to the entropy expression given in (\ref{11.11}).
However, for the case where the effective EoS parameter (\ref{11.13})
is a constant, the expression for the entropy will be given by
 equation (\ref{11.10}) as before. This condition only holds when $w_\mathrm{eff}=w_1=w_2$, where the situation is thus  equivalent to the one-fluid case, and the entropy reduces to the CV formula when $w_\mathrm{eff}=1/n$.   \\

Let us  consider a simple choice for the function $Q$ that leads to the CV formula for a certain limit. Let
 $Q=Q_0 a^mH$, where $m$ is a positive number,
$Q_0$ is a constant, and $H(t)$  the Hubble parameter.
Then the integral in (\ref{11.24}) is easily calculated,
and the energy densities depend on the scale factor according to
\be
\rho_1=C_1a^{-n(1+w_1)}+k_1a^m\ , \quad
\rho_2=C_2a^{-n(1+w_2)}+k_2a^m\ ,
\label{11.24c}
\ee
where
$k_{1,2}=Q_0/(n(1+w_{1,2})+m)$. If we restrict ourselves to the
regime where $a\gg C_{1,2}$ such that the first terms in the expressions
for $\rho_{1,2}$ are negligible, the effective EoS parameter becomes
\be
w_\mathrm{eff}=w_1+\frac{w_2-w_1}{1 + \frac{k_1}{k_2}}\ .
\label{11.24d}
\ee
Then, the entropy of the universe is given by (\ref{11.10}) with
$w=w_\mathrm{eff}$. The CV formula can be reproduced only with very  specific
choice of the free parameters, just as above.

We have thus shown that in general  a formula for the entropy of the type (\ref{11.10}) cannot be reconstructed for interacting fluids. Coincidence with the Cardy formula is obtained  if the effective EoS parameter is  radiation-like,
$w_\mathrm{eff}=1/n$. Then the expression for the entropy turns out to be
in agreement with the formula (\ref{11.19}), corresponding to
the first FLRW equation (\ref{11.20})
when the Casimir energy reaches the bound
$E_\mathrm{C}=E_\mathrm{BH}$.

\subsection*{Inhomogeneous EoS fluid and bulk viscosity}

Let us now explore the case of an $n+1$-dimensional Universe filled with
a fluid satisfying an inhomogeneous EoS. This kind of EoS,  generalizing  the perfect fluid model, has been
considered in  previous chapters as a way to describe effectively the
dark energy.
We assume an EoS expressed as a function of the scale factor,
\be
p=w(a)\rho+g(a)\ .
\label{11.28}
\ee
This EoS fluid could be taken to correspond to  modified gravity,
or to bulk viscosity. By introducing (\ref{11.28})
in the energy conservation equation (\ref{11.4}) we obtain
\be
\rho'(a)+\frac{n(1+w(a))}{a}\rho(a)=-n\frac{g(a)}{a}\ .
\label{11.29}
\ee
Here we have performed a variable change $t=t(a)$ such that
the prime over $\rho$ denotes derivative with respect to the
scale factor $a$. The general solution of this equation is
\be
\rho(a)=\e^{-F(a)}\left(K -n\int\e^{F(a)}\frac{g(a)}{a}da\right)
\quad \text{where} \quad F(a)=\int^a \frac{1+w(a')}{a'}da'\ ,
\label{11.30}
\ee
and $K$ is an integration constant. As shown above,
only for some special choices of the functions $w(a)$ and $g(a)$,
the formula (\ref{11.19}) can be recovered. Let us assume, as  an example,  that
$w(a)=-1$ and $g(a)=-a^m$, with $m= $ constant.
Then, the energy density behaves as $\rho\propto a^m$. Hence, by following the same steps as described above, the
extensive and subextensive energy go as $a^{m+n}$, and by imposing
conformal invariance and the rescaling properties (\ref{11.7}), we
calculate the dependence on the entropy to be
\be
E_\mathrm{E}=
\frac{\alpha}{4\pi na^{-(m+n)}}S^{-m/n}\, , \quad
E_\mathrm{C}=\frac{\beta}{4\pi na^{-(m+n)}}S^{-(2n+m/n)}\ .
\label{11.31}
\ee
The expression for the entropy is easily
constructed by combining these two expressions and substituting
the extensive part by the total energy. This gives us the same expression as in (\ref{11.10}) with $w=-(n+m)/n$. Note that for
$m=-(1+n)$, the formula (\ref{11.19}) is recovered and also its
correspondence with the CFT formula. However for a generic power
$m$, the CV formula cannot be reconstructed, like
the cases studied above. Only for some special choices does the correspondence
work, leading to the identification between the FLRW equation and
the Cardy formula. \\

\section{On the cosmological bounds near future singularities}

In Ref.\cite{2000hep.th....8140V}, Verlinde proposed a new universal bound on
cosmology based on a restriction of the Casimir energy $E_C$; cf.
 his entropy formula (\ref{11.19}). This new bound postulated was
\be
E_\mathrm{C}\leq E_\mathrm{BH}\ ,
\label{11.B1}
\ee
where $E_\mathrm{BH}=n(n-1)\frac{V}{8\pi Ga^2}$. It was
deduced by the fact that in the limit when the Universe passes
between strongly and weakly self-gravitating regimes, the
Bekenstein entropy $S_\mathrm{B}=\frac{2\pi a}{n}E$ and the
Bekenstein-Hawking entropy $S_\mathrm{BH}=(n-1)\frac{V}{4Ga}$,
which define each regime, are equal. This bound could be
interpreted to mean that the Casimir energy  never  becomes  able
to reach sufficient energy, $E_\mathrm{BH}$, to form a black hole
of the size of the Universe. It is easy to verify that the strong
($Ha\geq 1$) and weak ($Ha\leq 1$) self-gravity regimes have the
following restrictions on the total energy,
\bea
E &\leq &
E_\mathrm{BH}\ \quad \text{for} \quad Ha\leq 1 \ , \nn
E &\geq &
E_\mathrm{BH}\ \quad \text{for} \quad Ha\geq 1\ .
\label{11.B2}
\eea
From here it is easy to calculate the bounds on the entropy of
the Universe in the case when the Verlinde formula (\ref{11.19}) is
valid; this is (as  shown in the above sections) for an
effective radiation dominated Universe $w_\mathrm{eff}\sim 1/n$.
The bounds for the entropy deduced in Ref.\cite{2000hep.th....8140V} for
$k=1$ are
\bea
S &\leq& S_\mathrm{B}\ \quad \text{for} \quad Ha\leq 1 \ , \nn
S &\geq& S_\mathrm{H}\ \quad \text{for} \quad Ha\geq 1\ ,
\label{11.B3}
\eea
where $S_\mathrm{B}$ is the Bekenstein entropy defined
above, and $S_\mathrm{H}$ is the Hubble entropy given by
(\ref{11.20}).
Note that for the strong self-gravity regime, $Ha\geq 1$,
the energy range is $E_\mathrm{C}\leq E_\mathrm{BH}\leq E$.
According to the formula (\ref{11.19}) the maximum entropy is
reached when the bound is saturated, $E_\mathrm{C}=E_\mathrm{BH}$.
Then $S=S_\mathrm{H}$, such that the FLRW equation coincides
with
the CV formula, thus indicating  a connection with
CFT.  For the weak regime, $Ha\leq 1$, the range of energies
goes as $E_\mathrm{C}\leq E\leq E_\mathrm{BH}$ and the maximum
entropy is reached earlier, when $E_\mathrm{C}=E$, yielding the
result $S=S_\mathrm{B}$. The entropy bounds can be extended to more
general cases, corresponding to an arbitrary EoS parameter $w$. By taking
the bound (\ref{11.B1}) to be universally valid one can easily
deduce the new entropy bounds for each regime, from the expression
of the entropy (\ref{11.10}). These new bounds, discussed in
Ref.\cite{2002PhLB..531..276Y}, differ from the ones given in (\ref{11.B3}), but
still establish a bound on the entropy as long as the bound on
$E_\mathrm{C}$ expressed in (\ref{11.B1}) is taken to be valid. The
entropy bounds can be related through  the first FLRW equation,
yielding the following quadratic expression (for $k=1$),
\be
S^2_\mathrm{H}+(S_\mathrm{B}-S_\mathrm{BH})^2=S^2_\mathrm{B}\ .
\label{11.B3a}
\ee
We would like to study
what happens to the bounds,  particularly to the fundamental
bound (\ref{11.B1}), when the cosmic evolution is close to a future
singularity; then the effective fluid  dominating the
cosmic evolution could have an unusual EoS. As  shown below,
for some class of future singularities such a bound could soften
the singularities in order to avoid  violation of the
\textit{universal} bound (\ref{11.B1}). It  could be interpreted
to mean that  quantum effects become important when the bound is reached.
 However, as the
violation of the bound could happen long before the singularity
even in the presence of  quantum effects, it could be a signal
of  breaking of the universality of the bound (\ref{11.B1}). Let
us first of all give a list of the possible future cosmic
singularities, which can be classified according to
Ref.\cite{2005PhRvD..71f3004N} as
\begin{itemize}
\item Type I (``Big Rip''): For $t\rightarrow t_s$, $a\rightarrow
\infty$ and $\rho\rightarrow \infty$, $|p|\rightarrow \infty$.
\item Type II (``Sudden''): For $t\rightarrow t_s$, $a\rightarrow
a_s$ and $\rho\rightarrow \rho_s$, $|p|\rightarrow \infty$.
\item Type III: For $t\rightarrow t_s$, $a\rightarrow a_s$
and $\rho\rightarrow \infty$, $|p|\rightarrow \infty$.
\item Type IV: For $t\rightarrow t_s$, $a\rightarrow a_s$ and
$\rho\rightarrow \rho_s$, $p \rightarrow p_s$
but higher derivatives of Hubble parameter diverge.
\end{itemize}
Note that the above list was suggested in the case of a flat FLRW
Universe.
As we consider in this chapter a closed Universe ($k=1$), we should
make an analysis to see
if the list of singularities given above is also valid in this case.
It is straightforward to see that all the singularities listed above
can be reproduced for a particular choice of the effective EoS.
To show how the cosmic bounds behave for each type of singularity, we
could write an explicit
solution of the FLRW equations, expressed as a function of time
depending on free parameters
that will be fixed for each kind of singularity. Then, the Hubble
parameter may be written as follows
\be
H(t)=\sqrt{\frac{16\pi G}{n(n-1)}\rho -\frac{1}{a^2}}=H_1(t_s-t)^m
+H_0\ ,
\label{11.B4}
\ee
where $m$ is a constant properly chosen for each type of singularity.
Note that this is just a solution that ends in the singularities
mentioned above, but there are other solutions which also reproduce
such singularities.  We will study how the cosmic bounds behave near
each singularity listed above. As pointed out in
Ref.\cite{2004PhRvD..70d3539E,2005PhRvD..71f3004N}, around a singularity
 quantum effects could become important as the curvature of the
Universe grows and diverges in some of the cases. In other words, approaching the finite-time future singularity the curvature grows and universe reminds the early universe where quantum gravity effects are dominant ones because of extreme conditions.
Then one has to take into account the role of such quantum gravity effects which should define the behaviour of the universe just before the singularity. Moreover, they may act so that to prevent the singularity occurrence. In a sense, one sees the return of quantum gravity era.
However, the consistent quantum gravity theory does not exists so far.
Then, in order to estimate the influence of quantum effects to universe near to singularity one can use the effective action formulation.
We will apply the effective action produced by conformal anomaly (equivalently, the effective fluid with pressure/energy-density corresponding to conformal anomaly ones) because of several reasons.
It is known that at high energy region (large curvature) the conformal invariance is restored so one can neglect the masses. Moreover, one can use large N approximation to justify why large number of quantum fields may be considered as effective quantum gravity. Finally, in the account of quantum effects via conformal anomaly we keep explicitly the graviton (spin 2) contribution. The conformal anomaly $T_A$ has the
following well-known form
\be
T_A=b\left(F+\frac{2}{3}\square R\right) +b'G+b''\square R\ .
\label{11.B4a}
\ee
Here we assume for simplicity a 3+1 dimensional spacetime. Then, $F$
is the square of a 4D Weyl tensor and $G$ is the Gauss-Bonnet
invariant,
\be
F=\frac{1}{3}R^2-2R_{ij}R^{ij}+R_{ijkl}R^{ijkl}\ ,\quad
G=R^2-4R_{ij}R^{ij}+R_{ijkl}R^{ijkl}\, .
\label{11.B4b}
\ee
The coefficients $b$ and $b'$ in (\ref{11.B4a}) are described by the
number of $N$ scalars, $N_{1/2}$ spinors, $N_{1}$ vector fields,
$N_2$ gravitons and $N_\mathrm{HD}$ higher derivative conformal
scalars.
They can be written as
\be
b=\frac{N+6N_{1/2}+12N_{1}+611N_{2}-8N_\mathrm{HD}}{120(4\pi)^2}\
,\quad
b'=-\frac{N+11N_{1/2}+62N_{1}+1411N_{2}-28N_\mathrm{HD}}{360(4\pi)^2}\
.
\label{11.B4c}
\ee
As $b''$ is arbitrary  it can be shifted by a finite
renormalization of the local counterterm. The conformal anomaly $T_A$
can be written as $T_A=-\rho_A+3p_A$, where $\rho_A$ and $p_A$ are
the energy and pressure densities respectively. By using (\ref{11.B4a})
and the energy conservation equation $\rho_A+3H(\rho_A+p_A)=0$, one
obtains the following expression for $\rho_A$
\cite{2001IJMPA..16.3273N,2005PhRvD..71f3004N},
\bea
\rho_A &=& -\frac{1}{a^4}\int dt a^4 HT_A \nn
&=& -\frac{1}{a^4}\int dt
a^4H\left[-12b\dot{H}^2+24b'(-\dot{H}^2+H^2\dot{H}+H^4)
�-(4b+6b'')(\dddot{H}+7H\ddot{H}+4\dot{H}^2+12H^2\dot{H}) \right]\ .
\label{11.B4d}
\eea
The quantum corrected FLRW equation is given by
\be
H^2=\frac{8\pi G}{3}(\rho+\rho_A)-\frac{1}{a^2}\ .
\label{11.B4e}
\ee
We study now how the bounds behave around the singularity in
the classical case when no quantum effects are added, and then
include the conformal anomaly (\ref{11.B4a}) quantum effects in the
FLRW equations. We will see that for some cases the violation of the
cosmic bound can be avoided.

\subsection*{Big Rip Singularity}

This type of singularity has been very well studied and has become
very popular
as it is a direct consequence in the majority of the cases when the
effective EoS parameter is less than $-1$, the so-called phantom
case.
Observations currently indicate that the phantom barrier could have
already been crossed
or it will be crossed in the near future, so a lot of attention has
been paid to this case.
It can be characterized by the solution (\ref{11.B4}) with $m\leq -1$,
and this yields
 the following dependence of the total energy density on the scale
factor near the singularity,
when $a\gg 1$, for a closed Universe ($k=1$),
\be
\rho=\frac{n(n-1)}{16\pi G}H^2+\frac{1}{a^2}\sim a^{-n(1+w)} \quad
\text{for} \quad t\rightarrow t_s\ ,
\label{11.B5}
\ee
where we have chosen $H_1=2/n|1+w|$ with $w<-1$, $m=-1$ and $H_0=0$
for clarity.
This solution drives the Universe to a Big Rip singularity for
$t\rightarrow t_s$,
where the scale factor diverges. If the singularity takes place, the
bound (\ref{11.B1})
has to be violated before this happens. This can be seen from
equation (\ref{11.B5}),
as the Casimir energy behaves as $E_\mathrm{C}\propto a^{n|w|}$ while
the Bekenstein-Hawking energy
goes as $E_\mathrm{BH}\propto a^{n-2}$.
Then, as $w<-1$, the Casimir energy grows faster than the BH energy,
so close to the singularity
where the scale factor becomes very big, the value of $E_\mathrm{C}$ will be
much higher than $E_\mathrm{B}$, thus violating
the bound (\ref{11.B1}).
Following the postulate from Ref.\cite{2000hep.th....8140V} one could interpret
the bound (\ref{11.B1})
as the limit where General Relativity and Quantum Field Theory
converge,
such that when the bound is saturated quantum gravity effects should
become important.
QG corrections could help to avoid the violation of the bound and may
be the Big Rip singularity occurrence.
As this is just a postulate based on the CV formula, which is only
valid for special cases
as  shown in the sections above, the bound on $E_\mathrm{C}$
could not be valid for any kind of fluid.

Let us now include the
conformal anomaly (\ref{11.B4a}) as a quantum effect that becomes
important around the Big Rip. In such a case there is a phase
transition and the Hubble evolution will be given by the solution of
the FLRW equation (\ref{11.B4e}). Let us  approximate to get some
qualitative results,  assuming $3+1$ dimensions.
Around $t_s$ the curvature is large, and
$|\rho_A|>>(3/\kappa^2)H^2+k/a^2$. Then  $\rho\sim-\rho_A$, and from
(\ref{11.B4d})we get
\be
\dot{\rho}+4H\rho=H\left[-12b\dot{H}^2+24b'(-\dot{H}^2+H^2\dot{H}+H^4)
�-(4b+6b'')(\dddot{H}+7H\ddot{H}+4\dot{H}^2+12H^2\dot{H}) \right]\ .
\label{11.B5a}
\ee
We assume that the energy density, which diverges in the
classical case, behaves now as
\be
\rho\sim (t_s-t)^{\lambda}\ ,
\label{11.B5b}
\ee
where $\lambda$ is some negative number. By using the energy
conservation equation $\dot{\rho}+3H(1+w)\rho=0$, the Hubble
parameter goes as $H\sim 1/(t_s-t)$. We can check if this
assumption is correct in the presence of  quantum effects by
inserting both results in Eq. (\ref{11.B5a}). We get
\be
\rho\sim 3H^4(-13b+24b')\ .
\label{11.B5c}
\ee
Hence as $b>0$ and $b'<0$,  $\rho$ becomes negative,
which is an unphysical result. Thus  $\rho$ should not
go to infinity in the presence of the quantum correction. This
is the same result as obtained in Ref.\cite{2005PhRvD..71f3004N} where
numerical analysis showed that the singularity is moderated by the
conformal anomaly, so that the violation of the bound that naturally
occurs in the classical case can be avoided/postponed when quantum
effects are included.

\subsection*{Sudden singularity}

This kind of singularity is also problematic with respect to the bounds, but as
the energy density $\rho$ does not diverge,
the violation of the bound may be avoided for some special choices.
The sudden singularity can be described
by the solution (\ref{11.B4}) with $0<m<1$, and constants $H_{0,1}>0$.
Then the scale factor goes as
\be
a(t)\propto \exp\left[ -\frac{H_1}{m+1}(t_s-t)^{m+1}+H_0t\right] \ ,
\label{11.B6}
\ee
which gives $a(t)\sim \e^{h_0t}$ (de Sitter) close to $t_s$. From the
first FLRW equation the total energy density becomes
\be
\rho=H^2(t)+\frac{1}{a^2}=\left[H_1(t_s-t)^m+H_0 \right]^2
+ \exp\left[ 2\frac{H_1}{m+1}(t_s-t)^{m+1}-2H_0t\right]\ ,
\label{11.B6a}
\ee
which tends to a constant $\rho\sim H_0^2+\e^{-2H_0t_s}$ for
$t\rightarrow t_s$.
Then the Casimir energy grows as $E_\mathrm{C}\propto H^2_0
a^{n}+a^{n-2}$, while $E_\mathrm{BH}\propto a^{n-2}$ close to $t_s$.
The BH energy grows slower than the Casimir energy, and the bound is
violated for a finite $t$.
However, by an specific choice of the coefficients, the violation of
the bound (\ref{11.B1}) could be avoided.
For $H_0=0$, and by some specific coefficients, the bound could be
obeyed.
In general, it is very possible that $E_\mathrm{C}$ exceeds
its bound. In the presence of quantum corrections, the singularity
can be avoided but the bound can still be violated, depending on the
free parameters for each model. We may assume that in the presence of
the conformal anomaly for $n=3$, the energy density grows as
\cite{2005PhRvD..71f3004N}
\be
\rho=\rho_0+\rho_1(t_s-t)^{\lambda}\ ,
\label{11.B6b}
\ee
where $\rho_0$ and $\rho_1$ are constants, and $\lambda$ is now a
positive number. Then the divergences on the higher derivatives of
the Hubble parameter can be avoided, as  is shown in
Ref.\cite{2005PhRvD..71f3004N}. Nevertheless, $E_\mathrm{C}$ still
grows faster than $E_\mathrm{BH}$, such that the Universe has to be
smaller than a critical size in order to hold the bound (\ref{11.B1}) as
 is pointed in Ref.\cite{2002PhLB..531..276Y} for the case of a vacuum dominated
universe.

\subsection*{Type III singularity}

This type of singularity is very similar to the Big Rip, in spite of
the scale factor $a(t)$ being finite at the singularity.
The solution (\ref{11.B4}) reproduces this singularity by taking
$-1<m<0$. The scale factor goes as
\be
a(t)= a_s\exp \left[-\frac{H_1}{m+1}(t_s-t)^{m+1} \right]\ ,
\label{11.B7}
\ee
where for simplicity we take $H_0=0$. Then, for $t\rightarrow t_s$,
the scale factor $a(t)\rightarrow a_s$.
To see how $E_\mathrm{C}$ behaves near the singularity, let us write
it in terms of the time instead of the scale factor,
\be
E_\mathrm{C}\propto a_s^nH_1^2 (t_s-t)^{2m}+ a_s^{n-2}\ ,
\label{11.B8}
\ee
where $m<0$.
Hence, the Casimir energy diverges at the singularity, while
$E_\mathrm{BH}\propto a_s^{n-2}$ takes a finite value
for the singularity time $t_s$, so the bound is clearly violated long
before the singularity.
Then, in order to maintain the validity of the bound (\ref{11.B1}), one
might assume, as in the Big Rip case,
that GR is not valid near or at the bound. Even if  quantum
effects are included, as was pointed in Ref.\cite{2005PhRvD..71f3004N},
 for this type of singularity the energy density diverges more
rapidly than in the classical case, so that the bound is also
violated in the presence of  quantum effects.

\subsection*{Type IV singularity}

For this singularity, the Hubble rate behaves as
\be
\label{11.III_1}
H = H_1(t) + \left(t_s - t\right)^\alpha H_2(t)\ .
\ee
Here $H_1(t)$ and $H_2(t)$ are regular function and do not vanish at
$t=t_s$.
The constant $\alpha$ is not integer and larger than $1$.
Then the scale factor behaves as
\be
\label{11.III_2}
\ln a(t) \sim \int dt H_1(t) + \int dt \left(t_s - t\right)^\alpha
H_2(t)\ .
\ee
Near $t=t_s$, the first term dominates and every quantities like
$\rho$, $p$, and $a$ etc. are finite and
therefore the bound (\ref{11.B1}) would not be violated near the
singularity.

\subsection*{Big Bang singularity}

When the matter with $w\geq 0$ coupled with gravity and dominates,
the scale factor behaves as
\be
\label{11.BB1}
a \sim t^{\frac{2}{n(1+w)}}\ .
\ee
Then there appears a singularity at $t=0$, which may be a Big Bang
singularity.
Although the Big Bang singularity is not a future singularity, we
may consider the bound (\ref{11.B1}) when $t\sim 0$.
Since $n(1+w)>2$, the energy density behaves as
$\rho \sim a^{-n(1+w)}$ and therefore the Casimir energy behaves as
$E_\mathrm{C} \sim a^{-nw}$. On the other hand, we find
$E_\mathrm{BH} \sim a^{n-2}$.
Then when $n> 2$ or when $n \geq 2$ and $w>0$, $E_\mathrm{C}$
dominates when $a\to 0$, that is, when $t\to 0$,
and the bound (\ref{11.B1}) is violated. This tells us, as expected, that
 quantum effects become important
in the early universe.

Above, we  have thus explored what happens near the future cosmic
singularities.
We have seen that in general, and with some very special exceptions
on the case of Type II and Type IV,
the bound will be violated if one assumes the validity of GR close to
the singularity. Even if  quantum corrections are assumed, it seems that the bound will  be violated, although in the Big Rip case the singularity may be avoided when quantum effects are incorporated.
It is natural to suggest, in accordance with  Verlinde, that
the bound on the Casimir energy means a finite range for the validity of the classical theory. When this kind of theory becomes saturated,
 some other new quantum gravity effects
have to be taken into account.
We  conclude that the universality of the bound (\ref{11.B1}) is
not clear and may hold just
for some specific cases, like the radiation dominated Universe.

\section{$F(R)$-gravity and the Cardy-Verlinde formula}

We specify here a modified $F(R)$-gravity modeled  as an effective
fluid and construct the corresponding CV formula for it.
The action that describes $F(R)$-gravity is given by
\be
S=\frac{1}{2\kappa^2}\int d^{n+1}x\sqrt{-g} (F(R) +L_m)\ ,
\label{11.F1}
\ee
where $L_m$ represents the matter Lagrangian and $\kappa^2=8\pi G$.
The field equations are obtained by varying the action (\ref{11.F1}) with
respect to the metric $g_{\mu\nu}$,
\be
R_{\mu\nu}F'(R)-\frac{1}{2}g_{\mu\nu}F(R)+g_{\mu\nu}\Box
F'(R)-\nabla_{\mu}\nabla_{\nu}F'(R)
=\kappa^2T^{(m)}_{\mu\nu}\ .
\label{11.F2}
\ee
Here $T^{(m)}_{\mu\nu}$ is the energy-momentum tensor for the matter
 filling the Universe,
and we have assumed a $1+3$ spacetime for simplicity. For closed
$3+1$ FLRW Universe, the modified FLRW equations are expressed as
\bea
\frac{1}{2}F(R)-3(H^2+\dot{H})F'(R)+3HF''(R)\dot{R}=\kappa^2
\rho_{m}\ , \nn 
-\frac{1}{2}F(R)+\left[3H^2+\dot{H}+\frac{2}{a^2}\right] F'(R)-[(\partial_{tt}F'(R))+2H(\partial_t F'(R))]=\kappa^2 p_m\ , 
\label{11.F3} \eea
where  primes denote derivatives respect to $R$ and  dots with
respect to $t$. These equations can be rewritten in order to be
comparable with those of standard GR. For such a propose the
geometric terms can be presented as an effective energy-density
$\rho_{F(R)}$ and a pressure $p_{F(R)}$, \bea
H^2+\frac{1}{a^2}=\frac{\kappa^2}{3F'(R)}\rho_m
+\frac{1}{3F'(R)}\left[\frac{RF'(R)-F(R)}{2}-3H\dot{R}F''(R)\right]
\ , \nn 2\dot{H}+3H^2+\frac{1}{a^2}=-\frac{\kappa^2}{F'(R)}p_m
�-\frac{1}{F'(R)}\left[\dot{R}^2F'''(R)+2H\dot{R}F''(R)+\ddot{R}F''(R)
+\frac{1}{2}(F(R)-RF'(R)) \right]\ . \label{11.F4} \eea Then, an EoS
for the geometric terms can be defined as
$p_{F(R)}=w_{F(R)}\rho_{F(R)}$. We can define an effective
energy-density $\rho=\rho_m/F'(R)+\rho_{F(R)}$ and pressure
$p=p_m/F'(R)+p_{F(R)}$. Hence, for some special cases the formula
for the entropy developed in the second section can be obtained in
$F(R)$-gravity (for an early attempt deriving a  CV formula in a
specific version of $F(R)$-gravity, see \cite{2004PhRvD..70d3520B}). For
example, for an $F(R)$ whose solution gives $\rho\propto
a^{-3(1+w_\mathrm{eff})}$, the formula for the entropy (\ref{11.10})
is recovered although in general, as in the cases studied above,
no such expression can be given. On the other hand, one could
assume that the geometric terms do not contribute to the matter
sector. Supposing a constant EoS matter fluid, the expression for
the entropy is given by (\ref{11.10}), although the cosmic Cardy
formula (\ref{11.20}) has not the same form and in virtue of the
modified first FLRW equation (\ref{11.F3}) the form of the Hubble
entropy $S_\mathrm{H}$, the total energy $E$, and the Bekenstein
energy $E_\mathrm{BH}$, will be very different. It is not easy to
establish correspondence between two such approaches. Note that
using the effective fluid representation the generalized CV
formula may be constructed for any modified gravity.

Now we consider the case where $F(R)$ behaves as
\be
\label{11.FBB1}
F(R) \sim R^\alpha \ ,
\ee
when the curvature is small or large.
Then if the matter has the EoS parameter $w>-1$, by solving
(\ref{11.F3}) we find
\be
\label{11.FBB2}
a \sim \left\{ \begin{array}{cl}
t^{\frac{2\alpha}{n(1+w)}} & \mbox{when}\ \frac{2\alpha}{n(1+w)}> 0
\\
(t_s - t)^{\frac{2\alpha}{n(1+w)}} & \mbox{when}\
\frac{2\alpha}{n(1+w)}< 0
\end{array} \right. .
\ee
Then there may appear a singularity at $t=0$, which corresponds to
the Big Bang singularity,
or at $t=t_s$, which corresponds to the Big Rip singularity.
Since the Casimir energy behaves as $E_\mathrm{C} \sim a^{-nw}$ but
$E_\mathrm{BH} \sim a^{n-2}$,
only when $t\to 0$, $E_\mathrm{C}$ dominates in case that $n> 2$ and
$w\geq 0$ or in case that $n \geq 2$ and $w>0$.
Even in the phantom phase where $\frac{2\alpha}{n(1+w)}< 0$, the
bound (\ref{11.B1}) is not violated. \\
Let us now consider de Sitter space solution in $F(R)$-gravity
(for review of CV formula in dS or AdS spaces, see \cite{2002NuPhB.628..375C,2002PhLB..525..331C}). As
was pointed in the above part of the present thesis, almost every function
$F(R)$ admits a de Sitter solution.  In this case the formula for the entropy
(\ref{11.10}) can be reproduced for $w=-1$, and even the universal
bound (\ref{11.B1}) can hold by taking a critical size of the
Universe. The formula that relates the cosmic bounds in
(\ref{11.B3a}) is easily obtained also in $F(R)$-gravity for a de
Sitter solution. In such a case one can identify \be
S_\mathrm{H}=\frac{H_0V}{2G}\ , \quad S_\mathrm{B}=\frac{a
V}{24G}\frac{F(R_0)}{F'(R_0)}\ , \quad
S_\mathrm{BH}=\frac{V}{2Ga}\ , \label{11.F6} \ee which corresponds to
the first FLRW equation written as
$S^2_\mathrm{H}+(S_\mathrm{B}-S_\mathrm{BH})^2=S^2_\mathrm{B}$.
Thus, one can conclude that dynamical entropy bounds are not
violated for modified gravity with de Sitter solutions. Note that
quantum gravity effects may be presented also as an effective
fluid contribution. In case when de Sitter space turns out to be
the solution, even with the account of quantum gravity the above
results indicate that dynamical cosmological/entropy bounds are
valid. In other words, the argument indicates the universality of
dynamical bounds. It seems that their violation is caused only by
future singularities if they are not cured by quantum gravity
effects. Note that a large number of modified gravity theories do
not contain  future singularities; they are  cured by higher
derivatives terms.

\section{Discussions}

In summary, we have derived a generalized CV formula for
multicomponent, interacting fluids, generalized in the sense that
an inhomogeneous EoS  was assumed. We
also considered  modified $F(R)$-gravity, using its fluid
representation. We showed  that for some special cases the formula
is reduced to the standard CV formula expressing the
correspondence with 2d CFT theory. The dynamical entropy bound for
all above cases was found. The universality of dynamical entropy
bound near  all four types of the future singularity, as well as
the initial Big Bang singularity, was investigated. It was proved
that except from some special cases of Type II and Type IV
singularity the dynamical entropy bound is violated near the
singularity. Taking into  account  quantum effects of conformally
invariant matter does not improve the situation.

One might think that the dynamical entropy bound is universal and that its
violation simply indicates that the situation will be changed with the
introduction of  quantum gravity effects. However,  arguments given below indicate
that it is not the case and that the future singularity is the domain
where all known physical laws and equations are not valid.
Indeed, taking account of quantum effects such as done in section VI does not
improve the situation with respect to non-universality of the dynamical entropy bound.
From another side, it was shown that the dynamical entropy bound is valid
for the de Sitter  solution. Having in mind that quantum gravity
corrections may  always be presented as a generalized effective fluid,
one sees that the dynamical cosmological bound is not valid near the
singularity (even when account is taken of Quantum Gravity). It is only when
modified gravity (with or without
quantum corrections) is regular in the future,  where the future universe is
asymptotically de Sitter, that the dynamical bound
remains valid. Hence, the problem of non-universality of
dynamical entropy bound is related to the more fundamental question about
the real occurrence of a future singularity. It remains a challenge to find
any observational indications for the structure of  the future universe.

\chapter[Casimir effects near Big Rip singularity]{Casimir effects near Big Rip singularity}

\footnote{This Chapter is based on: \cite{2010GReGr..42.1513B,2010OAJ.....3...73G}}Here, it is investigated the analytical properties of the cosmic expansion, especially near Big Rip singularity, for a comic fluid with and without bulk viscosity, and with an oscillating equation of state. We study the effects of the inclusion of a Casimir-induced term in the energy-momentum tensor, and its behavior near the future singularity. 
Cosmic fluids - whether considered in the early or in the
late epochs - are usually taken to be nonviscous, but there
are two viscosity  (shear and bulk) coefficients naturally
occurring in general  linear hydrodynamics; the linear
approximation meaning physically that one is considering  only
first order deviations from thermal equilibrium. The shear
viscosity coefficient is evidently of importance when dealing with
flow near solid surfaces, but it can  be crucial also under
boundary-free conditions such as in isotropic turbulence. In later
years it has  become more common to take into account viscosity
properties of the cosmic fluid, however.  Because of assumed
spatial isotropy in the fluid the shear viscosity is usually left
out; any anisotropic deviations like those encountered in the
Kasner universe become rather quickly smoothed out. Thus only the
bulk viscosity coefficient, called $\zeta$, remains in the
energy-momentum tensor of the fluid. One should here note,
however,  that at least in the plasma region in the early universe
the value of the shear viscosity as derived from kinetic theory is
greater than the bulk viscosity by many orders of magnitude. Cf.,
for instance, Refs.~\cite{1994AstroSpaceSci..Brevik..Heen,1977PhysLettB..CaderniEtal}.

Early treatises on viscous cosmology are given in Refs.~\cite{1990AstroSpaceSci..Gron,1987PhysLettA..PadmanabhanEtal}.  Also in \cite{2005PhLB..619....5C}, it is
considered viscous dark energy and phantom evolution using
Eckhart's theory of irreversible thermodynamics. As discussed in Refs.~\cite{2006PhRvD..73d3512C,2005PhRvD..72b3003N}, a dark fluid with a time dependent bulk
viscosity can be considered as a fluid with an inhomogeneous
equation of state. Some other recent papers on viscous cosmology
are Refs.~\cite{2009PhLB..680..355F,2009PhRvD..79j3521L}.

A special branch of viscous cosmology is to investigate how the
bulk viscosity can influence the future singularity, commonly
called the Big Rip, when the fluid is in the phantom state
corresponding to the thermodynamic parameter $w$ being less than
$-1$.  Some recent papers  in this direction are
Refs.~\cite{2006GReGr..38.1317B,2008EPJC..56.579Brevik,2008GravCosm..14.32Brevik,2008arXiv0811.1129B,2005IJMPD..14.1899B,2008arXiv0812.2549G}.
In particular, as first pointed out in Ref.~\cite{2005GReGr..37.2039B}, the
presence of a bulk viscosity proportional to the  expansion can cause the fluid to pass from the quintessence region
into the phantom region and thereby inevitably lead to a future
singularity.

The purpose of the present chapter is to generalize these viscous
cosmology theories in the sense that we take into account the {\it
Casimir effect}. We shall model the Casimir influence in a different way than in the preceding chapter, by writing
the total Casimir energy in the same form as for a perfectly
conducting shell, identifying the cosmological "shell" radius
essentially with the cosmic scale factor. This is a very simple,
though natural, approach to the problem. The approach is of the
same kind as that followed in an earlier quantum cosmology paper
dealing with the  expanding FLRW universe in the nonviscous case
\cite{2002hep.th...10286B}. Other ways  of
treating the influence from the Casimir effect in cosmology; the
reader may consult, for instance,
Refs.~\cite{1992ComMatPhys...148.139,2006JPhA...39.6299E,2003PhRvD..67f3515E,2008GrCo...14...17G,2006PhRvA..73d2102S}.

\section{Formalism}

We start with the standard  spatially flat FLRW metric. We let subscript 'in' refer to
 present time  quantities, and choose $t_{in}=0$. The scale factor
$a(t)$ is normalized such that $a(0) \equiv a_{in}=1$. The
equation of state is taken as
\begin{equation}
p=w\rho, \label{12.2}
\end{equation}
with $w$ constant. As  mentioned, $w<-1$ in the phantom region
(ordinary matter is not included in the model). The bulk viscosity
$\zeta$ is taken to be  constant. The energy-momentum tensor of
the fluid can be expressed as
\begin{equation}
 T_{\mu\nu}=\rho U_{\mu}U_{\nu}+ \tilde{p}(g_{\mu\nu}+U_{\mu}U_{\nu})\ ,
\label{12.3}
\end{equation}
where  $U_{\mu}$ is the comoving four-velocity  and
$\tilde{p}=p-\zeta\theta$ is the effective pressure,
$\theta=3H=3\dot{a}/a$ being the scalar expansion. The Friedmann
equations take the form
\begin{equation}
\theta^2=24\pi G\rho, \label{12.4}
\end{equation}
\begin{equation}
\dot{\theta}+\frac{1}{6}\theta^2=-4\pi G\tilde{p}. \label{12.5}
\end{equation}
The energy conservation equation leads to
\begin{equation}
\dot{\rho}+(\rho+p)\theta=\zeta \theta^2. \label{12.6}
\end{equation}
From the above equations the differential equation for the scalar
expansion, with $w$ and $\zeta$ as free parameters, can be derived
as
\begin{equation}
\dot{\theta}+\frac{1}{2}(1+w)\theta^2-12\pi G\zeta \theta =0.
\label{12.7}
\end{equation}
The solution is, when subscript zero signifies that the Casimir
effect is so far left out,
\begin{equation}
 \theta_0(t)=\frac{\theta_{in} e^{t/t_c}}{1+\frac{1}{2}(1+w)\theta_{in}t_c(e^{t/t_c}-1)}\
 .
\label{12.8}
\end{equation}
Here  $\theta_{in}$ is the initial (present-time) scalar expansion
and $t_c$ the 'viscosity time'
\begin{equation}
t_c=\frac{1}{12\pi G\zeta}. \label{12.9}
\end{equation}
Since $(1+w)<0$ by assumption, it follows from Eq.~(\ref{12.8}) that
the future singularity occurs at a rip time $t_{s0}$ given by
\begin{equation}
t_{s0}=t_c\ln \left[
1-\frac{2}{1+w}\frac{1}{\theta_{in}t_c}\right]. \label{12.10}
\end{equation}
This means that in the nonviscous limit $\zeta \rightarrow 0$,
\begin{equation}
t_{s0} =-\frac{2}{1+w}\frac{1}{\theta_{in}}, \label{12.11}
\end{equation}
which is independent of $\zeta$. At the other extreme, in the high
viscosity limit $t_c \rightarrow 0$,
\begin{equation}
t_{s0} = t_c\ln \left[
\frac{-2}{1+w}\frac{1}{\theta_{in}t_c}\right], \label{12.12}
\end{equation}
showing that $t_{s0}$ becomes small. The fluid is quickly driven
into the Big Rip singularity.

\section{The Casimir effect included}

As mentioned above, a simple and  natural way of dealing with the
Casimir effect in cosmology is to relate it to the single length
parameter in the ($k=0$) theory, namely the scale factor $a$. It
means effectively that we should put the Casimir energy $E_c$
inversely proportional to $a$. This is in accordance with the
basic property of the Casimir energy, viz. that it is a measure of
the stress in the region interior to the "shell" as compared with
the unstressed region on the outside. The effect is evidently
largest in the beginning of the universe's evolution, when $a$ is
small. At late times, when $a\rightarrow \infty$, the Casimir
influence should be expected to fade away. As we have chosen $a$
nondimensional, we shall introduce an auxiliary length $L$ in the
formalism. Thus we adopt in model in which
\begin{equation} E_c=\frac{C}{2La}, \label{12.13}
\end{equation}
where $C$ is a nondimensional constant. This is the same form as
 encountered for the case of a perfectly conducting shell
 \cite{1978Ann.Phys..115..388}. In the last-mentioned case, $C$ was found to
 have the value
\begin{equation}
C=0.09235. \label{12.14}
\end{equation}
The expression (\ref{12.13}) is  of the same form as adopted in
Ref.~\cite{2001PhRvD..64h8701B,2000PhRvD..62f4005B}). It is strongly
related to the assumptions made by Verlinde when dealing with the
holographic principle in the universe \cite{2000hep.th....8140V}, and explored in the previous chapter. 

In the following we shall for definiteness assume $C$ to be
positive, corresponding to a repulsive Casimir force, though $C$
will not necessarily be required to have the value (\ref{12.14}). The
repulsiveness is a characteristic feature of conducting shell
Casimir theory, following from electromagnetic field theory under
the assumption that dispersive short-range effects are left out
(\cite{2001PhRvD..64h8701B,1978Ann.Phys..115..388}). Another assumption
that we shall make, is that $C$ is small compared with unity. This
is physically reasonable, in view of the conventional feebleness
of the Casimir force.

The expression (\ref{12.13}) corresponds to a Casimir pressure
\begin{equation}
p_c=\frac{-1}{4\pi (La)^2}\frac{\partial E_c}{\partial
(La)}=\frac{C}{4\pi L^4a^4}, \label{12.15}
\end{equation}
and leads consequently to a Casimir energy density $\rho_c \propto
1/a^4$.

The Casimir energy-momentum tensor
\begin{equation}
T_{\mu\nu}^c=\rho_c U_\mu U_\nu+p_c(g_{\mu\nu}+U_\mu U_\nu),
\label{12.16}
\end{equation}
together with the Casimir equation of state $p_c=w_c\rho_c$, yield
the energy balance
\begin{equation}
\frac{\dot{\rho}_c}{\rho_c}+(1+w_c)\theta=0, \label{12.17}
\end{equation}
having the solution $\rho_ca^{3(1+w_c)}=$ constant. To get $\rho_c
\propto 1/a^4$ we must have $w_c=1/3$. The Casimir contributions
to the pressure and energy density become accordingly
\begin{equation}
p_c=\frac{C}{8\pi L^4a^4}, \quad \rho_c=\frac{3C}{8\pi L^4a^4}.
\label{12.18}
\end{equation}
The Friedmann equations now become
\begin{equation}
\theta^2=24\pi G\left( \rho+ \frac{3C}{8\pi L^4a^4}\right),
\label{12.19}
\end{equation}
\begin{equation}
\dot{\theta}+\frac{1}{6}\theta^2=-4\pi G\left( w\rho -\zeta \theta
+\frac{C}{8\pi L^4a^4}\right), \label{12.20}
\end{equation}
while the energy conservation equation preserves its form,
\begin{equation}
\dot{\rho}+(1+w)\rho \theta =\zeta \theta^2. \label{12.21}
\end{equation}
Note again that  we are considering the {\it dark energy fluid}
only, with density $\rho$ and thermodynamical parameter $w$. The
ordinary matter fluid (dust) is left out.

Solving $\rho$ from Eq.~(\ref{12.19}) and inserting into
Eq.~(\ref{12.21}) we obtain as governing equation for the scalar
expansion
\begin{equation}
\dot{\theta}+\frac{1}{2}(1+w)\theta^2-12\pi G\zeta \theta = -
(1-3w)\frac{3GC}{2L^4a^4}\ . \label{12.22}
\end{equation}
It is convenient to introduce the constant $\alpha$, defined as
\begin{equation}
\alpha=-(1+w)>0, \label{12.23}
\end{equation}
and also to define the quantity $X(t)$,
\begin{equation}
X(t)=1-\frac{1}{2}\alpha \theta_{in}t_c(e^{t/t_c}-1), \label{12.24}
\end{equation}
which satisfies
\begin{equation}
X(0)=1, \quad X(t_{s0})=0. \label{12.25}
\end{equation}
In view of the assumed smallness of $C$ we now make a Stokes
expansion for $\theta$ to the first order,
\begin{equation}
\theta(t)=\theta_0(t)+C\theta_1(t)+O(C^2), \label{12.26}
\end{equation}
using henceforth the convention that subscript zero refers to the
$C=0$ case. The zeroth order solution is
\begin{equation}
\theta_0(t)=\theta_{in}X^{-1}e^{t/t_c}, \label{12.27}
\end{equation}
in accordance with Eq.~(\ref{12.8}). It corresponds to the zeroth
order scale factor
\begin{equation}
a_0(t)=X^{-\frac{2}{3\alpha}}, \label{12.28}
\end{equation}
satisfying $a_0(0)=1$ as before.

As the right hand side of Eq.~(\ref{12.22}) is already of order $C$,
we can replace $a(t)$ with $a_0(t)$ in the denominator. Thus we
get the following equation for the first order correction
coefficient $\theta_1$:
\begin{equation}
\dot{\theta}_1-\left(\alpha \theta_{in}X^{-1}e^{t/t_c}+12\pi
G\zeta
\right)\theta_1=-(1-3w)\frac{3G}{2L^4}X^{\frac{8}{3\alpha}}.
\label{12.29}
\end{equation}
We impose the same initial condition for the scalar expansion as
in the $C=0$ case: $\theta(0)=\theta_0(0) \equiv \theta_{in}$. It
means according to Eq.~(\ref{12.26}) that $\theta_1(0)=0$.

The homogeneous solution of Eq.~(\ref{12.29}), called $\theta_{1h}$,
may be written
\begin{equation}
\theta_{1h}(t)=\exp \left[ \int_0^t(\alpha
\theta_{in}X^{-1}e^{t/t_c}+12\pi G\zeta)dt \right], \label{12.30}
\end{equation}
satisfying $\theta_{1h}(0)=1$. The the full solution becomes
\begin{equation}
\theta_1(t)=-(1-3w)\frac{3G}{2L^4}\theta_{1h}(t)\cdot \int_0^t
\frac{X^{\frac{8}{3\alpha}}}{\theta_{1h}} dt, \label{12.31}
\end{equation}
satisfying $\theta_1(0)=0$. The two terms on the right hand side
of Eq.~(\ref{12.26}), $\theta_0(t)$ and $C\theta_1(t)$, are
accordingly determined.

\subsection{ Analytic approximation for low viscosity fluid}

 Although in general the expression for $\theta(t)$ has to be calculated
numerically, the main features of the solution can be shown
already analytically. Consistent with the assumed smallness of $C$
we need not distinguish between the rip time $t_s$ corresponding
to $C\neq 0$ and the rip time $t_{s0}$ corresponding to $C=0$. Let
us assume for mathematical simplicity the low-viscosity limit
\begin{equation}
\theta_{in}t_c \gg 1, \label{12.32}
\end{equation}
being physically  the most important case also. It corresponds to
$t_{s0}/t_c =2/(\alpha \theta_{in}t_c) \ll 1$. Then,
\begin{equation}
X(t)\sim \frac{1}{2}\alpha \theta_{in}(t_{s0}-t)=\frac
{t_{s0}-t}{t_{s0}}, \label{12.33}
\end{equation}
\begin{equation}
\theta_0(t) \sim \frac{2}{\alpha}\,\frac{1}{t_{s0}-t}, \label{12.34}
\end{equation}
\begin{equation}
a_0(t) \sim \left(
\frac{t_{s0}}{t_{s0}-t}\right)^{\frac{2}{3\alpha}}. \label{12.35}
\end{equation}
Both $\theta_0(t)$ and $a_0(t)$ diverge ($\alpha >0$ by
assumption). Using Eq.~(\ref{12.33})
 we can calculate $\theta_1(t)$ from Eq.~(\ref{12.31}),
\begin{equation}
\theta_1(t) =-(1-3w)\frac{9G}{2L^4}\,\frac{\alpha
t_{s0}}{8+9\alpha}\,\frac{1-(1-t/t_{s0})^{\frac{8}{3\alpha}+3}}{(1-t/t_{s0})^2}.
\label{12.36}
\end{equation}
From the expansion (\ref{12.26}) we thus obtain for the scalar
expansion to the first order in $C$,
\begin{equation}
\theta(t)=\frac{2}{\alpha
t_{s0}}\frac{1}{1-t/t_{s0}}\left\{1-(1-3w)\frac{9GC}{4L^4}\,\frac{\alpha^2t_{s0}^2
}{8+9\alpha}\,\frac{1-(1-t/t_{s0})^{\frac{8}{3\alpha}+3}}{1-t/t_{s0}}\right\}.
\label{12.37}
\end{equation}
The viscosity is absent in this expression. This is as we would
expect in view of the low-viscosity approximation.

The expression (\ref{12.37}) cannot, however, be valid near the
singularity. The reason is that the Taylor expansion in $C$ in
Eq.~(\ref{12.26}) is not applicable at $t=t_{s0}$. The solution
(\ref{12.37}) can  be applied safely as long as $t$ stays
considerably smaller than $t_{s0}$.  By making a first order
expansion in $t/t_{s0}$ of the expression between the curly
parentheses  we can write the solution in simplified form as
\begin{equation}
\theta(t)=\frac{2}{\alpha
t_{s0}}\frac{1}{1-t/t_{s0}}\left\{1-(1-3w)\frac{3GC\alpha
t_{s0}}{4L^4}t\right\}, \quad t \ll t_{s0}. \label{12.38}
\end{equation}
As $(1-3w)>0$ this means that $\theta(t)$ becomes slightly reduced
by the Casimir term. The repulsive Casimir force causes the energy
density $\rho$ in Eq.~(\ref{12.19}) to be slightly smaller than in
the $C=0$ case.

To deal with the conditions close to the singularity, we have to
go back to the governing equations themselves.

Let now $t_{\zeta s}$ denote the singularity time in the presence
of viscosity. We thus get
\begin{equation}
t_{s\zeta}=t_c\ln
\left(1+\frac{2}{\alpha}\frac{1}{\theta_{in}t_c}\right).
\label{12.39}
\end{equation}
It corresponds to $\theta(t_{s\zeta})=\infty$. We see that
$t_{s\zeta}$ is always less than the singularity time for the
nonviscous case,
\begin{equation}
t_{s\zeta}<t_{s,\zeta=0}=\frac{2}{\alpha \,\theta_{in}}.
\label{12.40}
\end{equation}
For the scalar expansion we find close to the singularity, again
assuming for simplicity low viscosity so that $\theta_{in}t_c \gg
1$ \cite{2008EPJC..tmp..169B},
\begin{equation}
\theta(t)=\frac{2/\alpha}{t_{s\zeta}-t}, \quad t\rightarrow
t_{s\zeta}. \label{12.41}
\end{equation}
In turn, this corresponds to
\begin{equation}
a(t) \sim \frac{1}{(t_{s\zeta}-t)^{2/3\alpha}}, \quad t\rightarrow
t_{s\zeta}, \label{12.42}
\end{equation}
\begin{equation}
\rho(t)\sim \frac{1}{(t_{s\zeta}-t)^2}, \quad t\rightarrow
t_{s\zeta}. \label{12.43}
\end{equation}

\subsection{On the nonviscous case}

It may finally be worthwhile to consider the entirely nonviscous
case, while keeping $C>0$. Let us analyze first of all, a simple example when $w(t)=w_0$ is a constant. 
Setting $\zeta=0$ we get from
Eq.~(\ref{12.22}) the governing  equation for the scale factor $a$:
\begin{equation}
a^3\ddot{a}+\frac{1}{2}(1+3w)a^2{\dot{a}}^2=-(1-3w)\frac{GC}{2L^4}.
\label{12.44}
\end{equation}
And the general solution for this equation results:
\be
t-t_s=\int \frac{ada}{\sqrt{ka^{1-3w_0}+\frac{GC}{L^4}}}\ ,
\label{12.D.3}
\ee
where $k$ is an integration constant. When we have a phantom field with $w_0<<-1$, the solution takes the form:
\be
a(t)=B\left[\left(1+\frac{2}{\e^{\sqrt{kB}(t_s-t)}-1} \right)^2 -1  \right]^{1/1-3w_0}\ ,
\label{12.D.4}
\ee 
here $B=\frac{GC}{kL^4}$. Note that for $t\rightarrow t_s$, the scale factor given by ($\ref{12.D.4}$) diverges, such that the so-called Big Rip singularity takes place. On the other hand, if we neglect the Casimir contribution $0<C<<1$, the solution for the scale factor yields
\be
a(t)\sim (t_s-t)^{\frac{2}{3}(1+w_0)}=(t_s-t)^{-\frac{2}{3}|1+w_0|}\ ,
\label{12.D.5}
\ee
which is the solution in absence of the Casimir term. This solution grows slower than in the above case when the Casimir contribution is taken into account (\ref{12.D.4}). Also note that  the Rip time $t_s$ for each case is different if the Casimir contribution is neglected in the second case, due to $t_s$ depends on the content from the Universe as it can be seen by the first Friedmann equation $t_s-t_0=\int^{\infty}_{a_0} \frac{da}{H_0 a(\sum_i \Omega^i(a))}$, where $t_0$ denotes the current time and $\Omega^i(a)$ is the cosmological parameter for the component $i$ of the Universe  \\
Let us now to analyze the equation (\ref{12.22}) with null viscosity when $w=w(t)$  is a periodic function given by,
\be
\label{12.T8}
w= -1 + w_0 \cos \omega t\ .
\ee
By inserting  (\ref{12.T8}) into (\ref{12.22}), the equation turns much more complicated than the constant case studied above. As the Casimir contribution is assumed to be very small, we can deal the equation (\ref{12.22}) by perturbation methods. Hence, the solution might be written as 
\be
\theta(t)= \theta_0 +C\theta_1 + O(C^2)\ .
\label{12.D.6}
\ee
And by inserting (\ref{12.D.6}) into (\ref{12.22}), the zero and first order in perturbations can be separated, and it yields the following two equations:
\[
\dot{\theta_0}+\frac{1}{2}w_0\cos\omega t \theta_0=0\ , 
\]
\be
\dot{\theta_1}+w_0\cos\omega t \theta_0 \theta_1=-\frac{3G}{2L^4}(4w_0\cos\omega t)\frac{1}{a^4_0(t)}\ , 
\label{12.D.7}
\ee
where $a_0(t)=\exp\left(\int dt\frac{2\omega}{3(w_1+w_0\sin\omega t)} \right)$ is the solution for the zero order. We suppose for simplicity $w_1 \sim w_0$, such that a Big Rip singularity appears when $1+\sin\omega t_s=0$. As we are interested in the possible effects close to the singularity, we can expand the trigonometric functions  as series around $t_s$. Then, the second equation in (\ref{12.D.7}) takes the form: 
\[
\dot{\theta_1}+\frac{4}{3\omega^2}\frac{1}{t_s-t}\theta_1=
\]
\be
=-\frac{3G}{2L^4}(4+\frac{3w_0}{\omega}(t_s-t)) \e^{-\frac{16}{3\omega (t_s-t)} }\ ,
\label{12.D.8}
\ee
whose solution is given by
\[
\theta_1(t)= k(t_s-t)^{\frac{4}{3\omega^2}}+(t_s-t)^{\frac{4}{3\omega^2}}g(1/t_s-t)\ ,
\]
where
\be
g(x)=\frac{3GC}{2L^4B}\left[ 4x^A\e^{-Bx}- \frac{4(4A+B)}{B}\int x^{A-1}\e^{-Bx} dx \right] \ .
\label{12.D.9}
\ee
Here $A=\frac{4-6\omega^2}{3\omega^2}$ and $B=\frac{16}{3\omega}$. Then, we can analyze the behavior from the contribution to the solution at first order (\ref{12.D.9}) when the Universe is close to the singularity. When $t\rightarrow t_s$ ($x\rightarrow \infty$), the function (\ref{12.D.9}) goes to zero, what means that the Casimir contribution has no effect at the Rip time.

Thus, we have  demonstrated, that dynamical Casimir effect gives no essential contribution to (phantom oscillating) dark energy dynamics near to Rip singularity.

\section{Concluding remarks}

Considerable attention has recently been devoted to the behavior
of the dark energy fluid near the future singularity. Various
possibilities have been contemplated. In addition to the
references given above, we
may refer also to the papers \cite{2003PhRvL..91g1301C,2003PhLB..562..147N}. It has even been
suggested that the future singularity can be avoided via quantum
gravity effects. Thus in Ref.~\cite{2004PhRvD..70d3539E} it is shown how
the universe may turn into a de Sitter phase.

Let us finally summarize our results above:

-  \quad If $\zeta>0$ and $C=0$, the rip singularity time $t_{s0}$ is
given by Eq.~(\ref{12.10}). In particular, if $\zeta \rightarrow 0$
then Eq.~(\ref{12.11}) holds.

-  \quad If $\zeta>0$ and $C>0$, the scalar expansion $\theta(t)$ is
given by the first-order series (\ref{12.26}), with $\theta_0(t)$ and
$\theta_1(t)$ given by Eqs.~(\ref{12.27}) and (\ref{12.31}). In the
low-viscosity case $\theta_{in}t_c \gg 1$, $\theta(t)$ is given by
the series (\ref{12.37}) which, however, is not applicable near the
singularity as $\theta(t)$ is not analytic in $C$ at the
singularity.

-  \quad Near the singularity, the Casimir effect fades away and the
viscous rip time $t_{s\zeta}$ is given by Eq.~(\ref{12.39}).
Corresponding values for $\theta(t),a(t)$ and $\rho(t)$ near the
singularity follow from Eqs.~(\ref{12.41}) - (\ref{12.43}).

-  \quad If $\zeta=0$ and $C>0$, the Rip time is usually longer than in the viscous case. The singularity occurs even with the presence of the Casimir term, which does not affect to the evolution, as it can be seen in the examples studied.

\chapter[Conclusions and perspectives]{Conclusions and perspectives}

Let us summarize and analyze the models and results discussed along the present thesis. Possible perspectives for the future theoretical physics in cosmology, in the frame of the theories studied here, are discussed here. As commented in the introduction, the main aim of this work is to show how different approaches could resolve the problem of dark energy and shape the entire evolution of the  history of the Universe, from inflation to current epoch. The different answers and intrinsic questions related to each model have been analyzed, as none of them is free of its proper unsolved questions. The possibility to distinguish between different theories is discussed, where the observations as well as possible predictions have to play a fundamental role.

First of all, we have discussed cosmological models where scalar fields (non-)minimally coupling are included. It has been shown that a single scalar-field, with a canonical or phantom kinetic term, is enough to shape the entire Universe evolution, where the appearance of different accelerated epochs, inflation and current phase, looks to be a natural consequence of the behavior of the scalar field. In this sense, several examples have been provided, where different solutions for the Hubble parameter is explored, from de Sitter-like solutions to solutions with a rapidly dynamical EoS for the scalar field. In general, models with phantom fields  contain future singularities, specially the so-called Big Rip singularity, where the scale factor diverges. It has been shown that a phantom scalar field could own, among other microscopical problems that are not discussed here, divergent instabilities during the transition from canonical to phantom phase due to the null kinetic term at the phantom barrier, where a discontinuity is produced. This problem  can be cured by introducing more than one scalar field, where each field behaves as canonical or phantom but they do not have a transition phase. Then, with several scalar fields the unification of late-time acceleration and inflation is achieved, and even when the effective EoS parameter for the Universe crosses the phantom barrier. The reconstruction of the inflationary epoch when there is more than one scalar field has been  performed in detail, where a type of slow-roll inflation is reconstructed, what provides constraints on the free parameters of the model. \\
In theories of the type of Brans-Dicke theory, where an scalar field is coupling directly to the Ricci scalar in the action, we have shown that dark energy can also be easily reproduced in this way. This kind of theories can lead to severe corrections of the Newtonian law at local scales. Nevertheless, this kind of problems can be avoided by the chameleon mechanism. In this sense, we have considered here a coupling of the type $1+f(\phi)$, where the function $f(\phi)$ can be constrained to be null at local scales (and GR is recovered) by means of the chameleon mechanism. Another important purpose for studying these theories was to compare the solutions in the original (Jordan) frame and in the Einstein frame, both related by a conformal transformation. Several examples are provided, and it is shown a very interesting example, where pure de Sitter cosmology in the Einstein frame, transforms into a completely different solution (as expected) in the Jordan frame, where the solution contains a singularity. Hence, we can see that a regular solution in one frame does not correspond to a regular one in the other, but singular points of the spacetime in the other frame can be obtained.\\
On the other hand, oscillating cosmologies have been explored by effective dark fluids (those with non-static and inhomogeneous EoS that can be effective descriptions of scalar fields or modified gravities). It is shown that the current accelerated era can be seen as a phase of the Universe evolution that is periodically repeated, and which started with the birth of the Universe and the period of inflation. This could provide an answer to the so-called ``coincidence problem'' as it predicts a periodicity of the behavior of the expansion, whose period can be fitted with the estimated age of the Universe. The end could be an eternal oscillating Universe, or depending on the parameters of the model, a final future singularity. It has been shown that such behavior can be well implemented by scalar fields or modified gravities. Also interacting dark energy is considered, at least in an effective way with no microscopic details, where a proper interaction could drive the accelerated expansion, such that a signal of a possible interaction between dust matter and an unknown dark energy fluid could be distinguished from other models.\\
Of course, many questions could be discussed in greater
detail, a more realistic matter content should be taken into
account, and the universe expansion
history described in a  more precise and  detailed way.
After all, we live in an era of more and more precise
cosmological tests. Nevertheless, the
 effective description of the cosmic expansion history
presented here in terms of scalar fields or effective dark fluids seems quite promising. Using it in
more  realistic contexts in which, of
course, technical details become much more complicated, appears
to be quite possible. Nevertheless, scalar fields seem more as a toy or effective descriptions models than realistic models.

Non-minimally scalar-tensor theories with a null kinetic term are mathematically equivalent to $f(R)$ gravities (though this is surely to be restricted to the classical level). This modification of GR, whose starting point is a generalization of the Hilbert-Einstein action, can easily reproduced any cosmic solution. Here, we have shown that from a scalar-tensor theory, there is a corresponded $f(R)$ action, such that the reconstruction of modified gravity theories of this type can be easily achieved given a cosmic solution and using an auxiliary scalar field. The main success of $f(R)$ gravities is that it can naturally explain the dark energy epoch with no need of an extra exotic field, and even could act relaxing the  vacuum  energy expected value, what could give an answer to the cosmological constant problem. Several examples of this kind of reconstruction have been given, where unification of inflation and late-time acceleration seems to be quite possible in $f(R)$ gravities. One has to point out that in higher order gravities, the presence of several de Sitter solutions is quite natural, such that accelerated phases along the Universe evolution  can be seen as normal epochs that depending on each de Sitter point could be a (un)stable phase. \\
By using of a new reconstruction technique, based on rewriting the FLRW equations as functions of the number of e-foldings instead of the cosmic time, several realistic cosmologies have been reconstructed. It has been shown that an exact solution of Hubble parameter, corresponded to an evolution of the kind  of $\Lambda$CDM model, yields an action composed by a divergent hypergeometric function of the Ricci scalar and a cosmological constant, what actually is reduced for the physical case to the Einstein-Hilbert action plus a cc, what means that exact $\Lambda$CDM model can be reproduced physically only by GR plus a cosmological constant, as it is natural. However, actions can be constructed for $f(R)$ gravity, which in spite of not providing an exact $\Lambda$CDM behavior for the Hubble parameter, can mimic a cosmological constant at the current epoch (low redshift). One of the examples provided here,  shows  that after a matter dominated epoch, the Universe enters in an accelerated expansion phase where the effective EoS parameter $w\sim-1$, and then it crosses the phantom barrier ending in a future Big Rip singularity. \\
It is well known that higher order theories of gravity may lead to important corrections to gravitational law at local scales (Ref.~\cite{1978GReGr..9.353}). It is remarkable that even particular modified gravity does not fulfill some cosmological bounds. This may be always achieved using new reconstruction techniques via a corresponding reconstruction at very early universe. Nevertheless, it has been shown that the so-called viable models can avoid problems at small scales, where GR is recovered, such that the effects of higher order terms in the field equations become important at cosmological scales (some examples of viable $f(R)$ gravities can be found in Refs.\cite{2008PhRvD..77j7501C,2007PhRvD..76f4004H,2007PhLB..657..238N,2007JETPL..86..157S}). In this sense, implementations of this type of models can be reconstructed as it has been shown using the same technique, where some corrections to these viable models can be introduced to incorporate observational requirements, and not only unified cosmological history is reproduced but may also achieve to explain dark matter effects (see Ref.~\cite{2009A&A...505...21C}). \\
One type of these viable models has been studied, where it is showed that can reproduce quite well inflation and late-time acceleration with a power-law phase between them (radiation/matter dominated epochs). Even, it was shown that de Sitter solution for inflation can be unstable providing a successful exit from this early phase. At low redshifts the cosmological parameter associated to the extra terms in the action behaves very similarly to the one by a cosmological constant, but where the effective EoS parameter associated to these geometrical terms has a dynamical behavior, as it is natural, moving from zero at redshifts $z\sim1.4$ to -1 for redshifts $z\sim0$, and  it mimics a cosmological constant at the current epoch. In the same model,  the effects of the presence of a phantom fluid in the model have been explored. \\
On the other hand, it is well known that phantom scalar models in the Einstein frame do not have a physical correspondence in the Jordan frame, where $F(R)$ gravity is defined. Nevertheless, the presence of a phantom fluid can solve this problem, where the scalar field in the Einstein frame owns a canonical kinetic term, and the phantom behavior corresponds solely to the fluid. \\
Also modified gravities where Gauss-Bonnet invariants are evolved have been studied.  As equivalent to $f(R)$ theories, Gauss-Bonnet gravity can easily reproduce unification of inflation and late-time acceleration. A model has been considered, where the action is given by $R+f(G)$, i.e. GR action plus a correction coming from a function of the Gauss-Bonnet invariant. This kind of models can easily reproduce $\Lambda$CDM model with no need of dark energy. Also the pure Gauss-Bonnet gravity has been explored, where several examples of important cosmological solutions have been given. Finally, general models of the type $f(R,G)$ have been considered, where similar results are obtained.\\
Hence, we have showed that modified gravities can be well reconstructed to reproduce the entire cosmological evolution and keep unchanged the behavior of gravity at local scales, where GR is well tested. Probably the next step in this study would be to know how they differ from other kinds of models. The study of the evolution of density perturbations  may provide some differences that could be measured in the future (see Ref.~\cite{2008PhRvD..77l3515D}). Even in the regions of the spacetime where the curvature is very large, as in the neighborhood of a black hole, some corrections on the solution from GR could  be found (see Refs.~\cite{2008CQGra..25h5004C,2009PhRvD..80l4011D}).

The recently proposal Ho\v{r}ava-Lifshitz gravity has been also studied, which seems to be power-counting renormalizable by losing the invariance under full diffeomorphisms. In an equivalent way as the extension from GR to $f(R)$ gravity,  HL gravity has been extended to more general actions.  A generalization of HL action shows that unification of accelerated epochs of Universe evolution can be performed. In this sense, reconstruction of cosmological solutions have been explored showing that any cosmic solution can be reconstructed from a given modification of HL gravity $f(\tilde{R})$. Even it seems that the so-called viable models in standard $f(R)$ gravity can be extended to HL gravity, where its properties remained unchanged: inflation and dark energy epochs are well reproduced while the scalar mode related to $f(\tilde{R})$ is neglible in the Newtonian corrections. The construction of an action free from future singularities has been performed. Also the study on the stability of solutions of the FLRW equations have been performed, where important information about the restrictions of the gravitational theories is obtained. It was found that $f(\tilde{R})$ gravity can well explain the end of the matter dominated epoch and reproduce the entire cosmic history.  
Then, the theory can well reproduce the Universe evolution and it seems a good promise as a candidate for a quantum field theory of gravity, in spite of the fact that serious problems are  unresolved yet. The breakdown of Lorentz invariance supposes that different observers will measure different speeds of light as well as the field equations will be different for all of them. However, it is assumed that in the IR limit the full diffeomorphism is recovered $z=1$. The main problem of the theory is  the existence of a spin-0 mode that introduces severe instabilities around the Mikowski background. However, it has been shown here that for some specific actions  $f(\tilde{R})$, one can find unstable flat solutions but stable de Sitter spacetime, which can be considered as the natural vacuum solution of the theory instead of Mikowski spacetime. On the other hand, a different approach  has been performed in order to recover general covariance in the IR limit  (see Ref.~\cite{2010PhRvD..82f4027H}) by means of the incorporation of a $U(1)$ symmetry into the theory. Here we have extended such theory to more general actions, where the incorporation of higher terms can account for the effects of cosmic expansion.  \\
Hence, we have shown that in the frame of HL gravity, accelerated expansion can be well reproduced as well as other features coming from standard $f(R)$ gravity. However, as it was pointed out, HL gravities contain several problems, and in spite of the new proposals, the problem to recover general covariance in a natural way does not have a consistent solution yet, but interesting attempts have been explored here. Also, as any new science,  predictive power has to be provided, so that the breakdown of covariance should be measured in some limit. Nevertheless, it seems that the energy scales dealt with in the laboratories are far from being able to probe this.

Finally, in the last part of the thesis, future singularities have been studied, where semiclassical effects are taken into account to show its possible effects near the singularities. Accounting for a cosmological Casimir effect seems not to  cure the occurrence of the Big Rip singularity, but it could affect the cosmological evolution as an effective radiation fluid, although its density probably has decayed nowadays. Even by taking into account the conformal anomaly, it does not seem to affect future singularities. Moreover, it has been proven that for a cosmology that contains future singularities, universal bounds on the entropy are violated, even much before reaching the singularity, what predicts the non-validity of such bounds. \\
On the other hand, generalizations of the CV formula have been performed, where it seems that its correspondence with a 2d CFT is only valid for some special cases, and can not be generalized. Nevertheless, the CV formula gives a direct way to the derivation of dynamical bounds for universe and black hole entropy bounds.

\bibliographystyle{hplain}
\bibliography{RefSaez1}

\newpage
\
\newpage

\end{document}